\documentclass{aa}
\usepackage{graphicx}
\usepackage[varg]{txfonts}
\usepackage{natbib,twoopt}
\usepackage{nccmath}
\usepackage{nicefrac}
\usepackage{float}
\usepackage{multirow}
\usepackage{placeins}
\usepackage{textcomp}
\usepackage{chngcntr}
\usepackage{afterpage}
\counterwithout{figure}{section}
\counterwithout{table}{subsection}
\usepackage[breaklinks=True, colorlinks, citecolor=blue]{hyperref} 

\bibpunct{(}{)}{;}{a}{}{,}             
\makeatletter
  \newcommandtwoopt{\citeads}[3][][]{\href{http://adsabs.harvard.edu/abs/#3}%
    {\def\hyper@linkstart##1##2{}%
     \let\hyper@linkend\@empty\citealp[#1][#2]{#3}}}
  \newcommandtwoopt{\citepads}[3][][]{\href{http://adsabs.harvard.edu/abs/#3}%
    {\def\hyper@linkstart##1##2{}%
     \let\hyper@linkend\@empty\citep[#1][#2]{#3}}}
  \newcommandtwoopt{\citetads}[3][][]{\href{http://adsabs.harvard.edu/abs/#3}%
    {\def\hyper@linkstart##1##2{}%
     \let\hyper@linkend\@empty\citet[#1][#2]{#3}}}
  \newcommandtwoopt{\citeyearads}[3][][]%
    {\href{http://adsabs.harvard.edu/abs/#3}
    {\def\hyper@linkstart##1##2{}%
     \let\hyper@linkend\@empty\citeyear[#1][#2]{#3}}}
\makeatother

\bibpunct{(}{)}{;}{a}{}{,} 

\newfont{\gwpfont}{cmssq8 scaled 1000}
\newcommand{\xifu}{{\gwpfont X-IFU}}

\begin{document}

\title{\textit{Athena} X-IFU synthetic observations of galaxy clusters to probe the chemical enrichment of the Universe}

\author{E.~Cucchetti\inst{\ref{inst1}} \and E.~Pointecouteau\inst{\ref{inst1}}  \and P.~Peille\inst{\ref{inst2}}  \and N.~Clerc\inst{\ref{inst1}}  \and E.~Rasia\inst{\ref{inst3}}  \and V.~Biffi\inst{\ref{inst3},\ref{inst5}} \and S.~Borgani\inst{\ref{inst3},\ref{inst5},\ref{inst6}}  \and L.~Tornatore\inst{\ref{inst3}} \and K.~Dolag\inst{\ref{inst7},\ref{inst8}} \and M.~Roncarelli\inst{\ref{inst9},\ref{inst10}}  \and M. Gaspari\inst{\ref{inst12},}\thanks{\textit{Einstein and Spitzer} Fellow} \and S.~Ettori\inst{\ref{inst10},\ref{inst11}} \and E.~Bulbul\inst{\ref{inst14}} \and T.~Dauser\inst{\ref{inst15}}  \and J.~Wilms\inst{\ref{inst15}} \and F.~Pajot\inst{\ref{inst1}}  \and D.~Barret\inst{\ref{inst1}}}

\institute{
IRAP, Université de Toulouse, CNRS, CNES, UPS, (Toulouse), France
\label{inst1} 
\and CNES, 18 Avenue Edouard Belin 31400 Toulouse France \label{inst2}
\and INAF, Osservatorio Astronomico di Trieste, via Tiepolo 11, I-34131, Trieste, Italy \label{inst3}
\and Dipartimento di Fisica dell'Università di Trieste, Sezione di Astronomia, via Tiepolo 11, I-34131, Trieste, Italy \label{inst5}
\and INFN - National Institute for Nuclear Physics, Via Valerio 2, I-34127, Trieste Italy \label{inst6}
\and University Observatory Munich, Scheinerstr. 1, D-81679, Munich, Germany \label{inst7}
\and Max Plank Institut für Astrophysik, Karl-Schwarzschield Strasse 1, 85748 Garching bei Munchen, Germany \label{inst8}
\and Dipartimento di Fisica e Astronomia, Università di Bologna, via Gobetti 93 I-40127 Bologna, Italy \label{inst9}
\and INAF, Osservatorio di Astrofisica e Scienza dello Spazio, via Pietro Gobetti 93/3, 40129 Bologna, Italy \label{inst10}
\and INFN, Sezione di Bologna, viale Berti Pichat 6/2, I-40127 Bologna, Italy \label{inst11}
\and Department of Astrophysical Sciences, Princeton University, Princeton, NJ 08544, US \label{inst12}
\and Harvard-Smithsonian Center for Astrophysics, 60 Garden Street, Cambridge, MA, 02138, USA  \label{inst14}
\and Dr. Karl Remeis-Observatory and Erlangen Centre for Astroparticle Physics, Sternwartstr. 7, 96049 Bamberg, Germany  \label{inst15}
}

\date{Received 2018}

\abstract{
Answers to the metal production of the Universe can be found in galaxy clusters, notably within their Intra-Cluster Medium (ICM). The X-ray Integral Field Unit (\xifu) on board the next-generation European X-ray observatory \textit{Athena} (2030s) will provide the necessary  leap forward in spatially-resolved spectroscopy required to disentangle the intricate mechanisms responsible for this chemical enrichment. In this paper, we investigate the future capabilities of the \xifu\ in probing the hot gas within galaxy clusters. From a test sample of four clusters extracted from cosmological hydrodynamical simulations, we present comprehensive synthetic observations of these clusters at different redshifts (up to $z \leq 2$) and within the scaled radius $R_{500}$ performed using the instrument simulator SIXTE. Through 100\,ks exposures, we demonstrate that the \xifu\ will provide spatially-resolved mapping of the ICM physical properties with little to no biases ($\lessapprox 5$\%) and well within statistical uncertainties. The detailed study of abundance profiles and abundance ratios within $R_{500}$ also highlights the power of the \xifu\ in providing constraints on the various enrichment models. From synthetic observations out to  $z =2$,  we also quantify its  ability to track the chemical elements across cosmic time with excellent accuracy, and thereby to investigate  the evolution of metal production mechanisms as well as the link to the stellar initial mass-function. Our study demonstrates the unprecedented capabilities of the \xifu\ in unveiling the properties of the ICM but also stresses the data analysis challenges faced by future high-resolution X-ray missions such as \textit{Athena}.}
\keywords{Galaxies: abundances - Galaxies: intra-cluster medium - Galaxies: fundamental parameters - Instrumentation: \textit{Athena}/X-IFU - Methods: numerical - Techniques: imaging spectroscopy - X-rays: galaxies: clusters}

\maketitle

\section{Introduction}

Metals and other heavy elements in the intra-cluster medium (ICM) represent a fossil record of the chemical evolution of the Universe. Trapped in the dark matter (DM) potential of galaxy clusters \citep{White1993Cluster}, they remain unaltered within the optically-thin collisionless thermal plasma. Elements originate within stars or through supernov\ae\ (SN), before being spread by stellar winds or by the SN explosions. Hence, the chemical enrichment of a given cluster relates to the integrated star formation history of the cluster, as well as to the overall stellar initial mass function (IMF). The abundances and spatial distribution of metals in the ICM can also be connected to its dynamical history and to the mechanical action of AGN (Active Galactic Nuclei) outflows or jets \citep[e.g.][]{Gaspari2011jets}. 

Most of the low-mass elements (C, O, Mg, Si, S) are produced by end-of-life massive stars ($\geq 10 M_{\odot}$) undergoing core-collapse supernov\ae\ (SN$_{\text{cc}}$) \citep[see][for a review]{Nomoto2013Nucleo}. The evolution of SN$_{\text{cc}}$-related enrichment through time is dictated by the initial mass and metallicity of the progenitor star. High-mass elements, from Si-like elements (Al, Si, S, Ca, Ar) to Fe and Ni, are on the other hand the result of thermonuclear reactions occurring during the explosion of white dwarfs (type Ia supernov\ae\ -- SN$_{\text{Ia}}$) \citep{Hillebrandt2013SNIa}. Although the mechanisms of these explosions -- either via accretion of a companion star onto the white dwarf \citep{Whelan1973Binaries} or via mergers of binary systems \citep{Webbink1984Mergers} -- is still poorly understood \citep[see][]{Maoz2014SNIa}, the time scale of these events, related to longer-living low-mass stars, suggests a later enrichment across cosmic time. Traces of other elements (C, N, Ne, Na) can also be produced when low- and intermediate-mass stars (typically $\leq 6 M_{\odot}$) enter their Asymptotic Giant Branch (AGB) phase \citep{Iben1983AGB}. The individual study of these phenomena based on detailed observations of nearby SN is difficult due to their rarity. Rather than a direct study on stellar populations, the detailed spectroscopic study of the ICM is an interesting alternative probe to test metal production models up to the early periods of the Universe.

Beyond the first steps in high-resolution X-ray spectroscopy \citep{Canizares79,Canizares82} and despite the lack of spatial resolution \citep{Peterson06Cooling}, the advent of high-resolution grating instruments such as XMM-\textit{Newton}/RGS \citep{denHerder2001RGS} and \textit{Chandra}/HETG \citep{Canizares2005HETG} drastically changed our view of the ICM enrichment, by giving access for the very first time to a large number of atomic lines \citep{dePlaa2007Abundances, dePlaa2011Abundances, Molendi2016Ab, Werner2007abundances}. Clusters therefore became excellent laboratories to test plasma physics and the chemical enrichment models up to the present epoch \citep[see][for a review]{Werner2008Review}. Despite limited spectral resolutions, CCD-type instruments have also been pushed to the maximum of their abilities to benefit of their spatial resolution in investigating the spatial distribution of chemical elements in the ICM  \citep{deGrandi09Abundances,Mernier2017Radial,Mernier2016I, Mernier2016II}. 

The perspective of micro-calorimeter-based imaging spectrometers, such as the soft X-ray spectrometer (SXS) on board \textit{Hitomi} \citep{Takahashi2016Hitomi}, opened new possibilities in studying the ICM: from the spatial scales of the enrichment (sources of production, processes of mixing and dispersion) to the kinematics of the hot gas (turbulence, shocks -- \citealt{Hitomi2016Quiescent, Hitomi2017Temperature, Hitomi2017Resonant, Hitomi2017Gas}), which complement the indirect estimates via surface brightness and warm gas tracers \citep[e.g.][]{Churazov2012Bright, Gaspari2013Turb, Hofmann2016kine, Gaspari2018kine}. Unfortunately, the short lifetime of the SXS only gave a glimpse of its potential. These renewed capabilities in galaxy cluster observation now rely on future missions, such as the X-ray Recovery Imaging and Spectroscopy Mission (\textit{{\color{red} XRISM}}) \citep{Ishisaki2018XARM} or the Advanced Telescope for High-ENergy Astrophysics (\textit{Athena}) \citep{Nandra2013Athena}. Namely, the X-ray Integral Field Unit (\xifu) on board the future European X-ray observatory \citep{Barret2016XIFU, Pajot2018XIFU}, will provide narrow-field observations (5$^{\prime}$ in equivalent field-of-view diameter) over the 0.2 -- 12\,keV bandpass, with a required 5$^{\prime\prime}$ spatial resolution and an unprecedented spectral resolution of 2.5\,eV (required up to 7\,keV).   

Investigating the chemical enrichment of the Universe is one of \textit{Athena}'s prime science objectives \citep{Ettori2013Athena, Pointecoutea13Evolution} {\color{black} which drives top-level performances of the telescope . In addition to the spectral resolution of the \xifu , which will allow to resolve faint atomic lines of less abundant elements, this science objective drives the need for a high effective area of the telescope along with a well-calibrated low energy band, required to accurately resolve lines of light elements such as C ($\geq 0.2$\,keV).
Number of breakthroughs on the study of chemical species and their evolution should in fact come from measurement in the low-energy band, where the effective area is the highest.} The fine spectroscopic capabilities {\color{black} of the \xifu\ in this energy band} will probe the production and circulation of metals within galaxy clusters across cosmic time, up to a redshift of $z \leq 2$ and a distance of $R_{500}$\footnote{$R_{500}$ is the radius including a density contrast of 500 times the critical density of the Universe, $\rho_c={3 H(z)^2}/{8 \pi G}$, at the given redshift $z$} from the cluster's center. By accurately measuring the abundances of the most common elements (e.g. O, Si, S, Fe) , the \xifu\ will be capable of constraining the number of time-integrated SN$_{\text{Ia}}$ and SN$_{\text{cc}}$ products. For the first time, the spatially-resolved measurements of less abundant elements (e.g., C, Al, S, Ca) as well as rare elements (e.g. Mn, Cr, and Ti) will provide insights on the initial metallicity of the SN$_{\text{Ia}}$ progenitors, and therefore on their formation mechanisms. {\color{black} The science of the chemical enrichment is a driver of the performance of the instrument, which needs to be assessed before launch.}

In this paper, we investigate the feasibility of recovering the physical parameters of the ICM through \xifu\ observations. A careful attention is given to the different enrichment mechanisms and their evolution over time. We use a sample of four simulated galaxy clusters with different masses studied at different redshifts, obtained via hydrodynamical cosmological simulations \citep{Rasia2015Simu, Biffi2017Clusters}. These objects are passed as input to a dedicated end-to-end (E2E) simulation pipeline of the \xifu\ instrument, based on the simulator SIXTE \citep{Wilms2014SIXTE}. In the next section (Sec. \ref{sec:sim}), we present the properties of the sample of simulated clusters. This is followed by a detailed description of our simulation pipeline (Sec.~\ref{sec:e2e}). The data analysis, post-processing procedures and results validation are in turn described (Sec.~\ref{sec:dat}). The outputs of our synthetic observations obtained through the pipeline for the four local clusters are then used (Sec.~\ref{sec:icm}) to infer the main properties of the sample and study its enrichment. This investigation is also extended to higher redshift values (Sec.~\ref{sec:evo}) to look into the \xifu\ abilities to capture the evolution of abundances through cosmic time. Finally, results and outcomes of our study are discussed (Sec.~\ref{sec:dis}). 

\section{Generation of the cluster sample}
\label{sec:sim}

The sample of four clusters of galaxies analysed in this study is taken from \cite{biffi2018} and includes two massive and two smaller systems, to bracket a broad mass range across the considered redshift values (Table~\ref{table:1}). In both mass bins, we choose a cool-core (CC) and a non-cool-core cluster (NCC), defined based on their pseudo-entropy profiles as described in \cite{Leccardi2010CC}. This small sample gives a view of part of the expected cluster population planned to be investigated by the \xifu. The objects are part of a larger set of 29 Lagrangian regions extracted from a parent cosmological DM-only simulation and re-simulated at higher resolution including baryons~\cite[see][]{bonafede2011}. The parent cosmological volume is $1~h^{-1}\rm{\,Gpc}$ per side and adopts a $\Lambda$-CDM cosmological model with $\Omega_{\rm M }= 0.24$, $\Omega_{\rm b} = 0.04$, $H_0 = 72$ km s$^{-1}$ Mpc$^{-1}$ (i.e., $H_0 = h \times H_{100}$, where $h = 0.72$ and $H_{100} = 100$ km s$^{-1}$ Mpc$^{-1}$), $\sigma_8 = 0.8$ and $n_s = 0.96$, consistently with WMAP-7 constraints given in \cite{komatsu2011}. 

\begin{table*} [!t]
\centering 
\caption{Properties of the simulated clusters at different redshift values in their evolution.} 
\begin{tabular}{c c c c c c} 
\hline\hline \\[-0.8em]
Name & & C1 & C2& C3 & C4 \\[0.2em] 
Type & &  CC & NCC & CC & NCC \\[0.2em]
\hline \\[-0.8em] 
\multirow{3}{*}{$z = 0.105$} & \multicolumn{1}{c}{$R_{500}$ (kpc $h^{-1}$)} &  \multicolumn{1}{c}{723} &  \multicolumn{1}{c}{799} & \multicolumn{1}{c}{1027} &  \multicolumn{1}{c}{1009} \\[0.2em]
							  & \multicolumn{1}{c}{$M_{500}$ ($M_{\odot}$ $h^{-1}$)} &  \multicolumn{1}{c}{$2.39 \times 10^{14}$} &  \multicolumn{1}{c}{$3.22 \times 10^{14}$} &  \multicolumn{1}{c}{$6.86 \times 10^{14}$} &  \multicolumn{1}{c}{$6.51 \times 10^{14}$} \\[0.2em]
							 & \multicolumn{1}{c}{$T_{500}$ (keV)} &  \multicolumn{1}{c}{3.22} &  \multicolumn{1}{c}{4.20} &  \multicolumn{1}{c}{6.47} &  \multicolumn{1}{c}{6.36} \\[0.2em]
\hline \\[-0.8em]
\multirow{3}{*}{$z = 0.5$} & \multicolumn{1}{c}{$R_{500}$ (kpc $h^{-1}$)} & \multicolumn{1}{c}{552} & \multicolumn{1}{c}{676} & \multicolumn{1}{c}{694} & \multicolumn{1}{c}{715} \\[0.2em]
							  & \multicolumn{1}{c}{$M_{500}$ ($M_{\odot}$ $h^{-1}$)} & \multicolumn{1}{c}{$1.55 \times 10^{14}$} & \multicolumn{1}{c}{$2.84 \times 10^{14}$} & \multicolumn{1}{c}{$3.08 \times 10^{14}$} &  \multicolumn{1}{c}{$3.38 \times 10^{14}$} \\[0.2em]
							  &\multicolumn{1}{c}{$T_{500}$ (keV)} &  \multicolumn{1}{c}{2.88} &  \multicolumn{1}{c}{4.06} &  \multicolumn{1}{c}{4.58} &  \multicolumn{1}{c}{4.84} \\[0.2em]
\hline \\[-0.8em]
\multirow{3}{*}{$z = 1.0$} & \multicolumn{1}{c}{$R_{500}$ (kpc $h^{-1}$)} & \multicolumn{1}{c}{389} & \multicolumn{1}{c}{396} & \multicolumn{1}{c}{446} & \multicolumn{1}{c}{570} \\[0.2em]
									 & \multicolumn{1}{c}{$M_{500}$ ($M_{\odot}$ $h^{-1}$)} & \multicolumn{1}{c}{$0.92 \times 10^{14}$} & \multicolumn{1}{c}{$0.97 \times 10^{14}$} & \multicolumn{1}{c}{$1.38 \times 10^{14}$} & \multicolumn{1}{c}{$2.89 \times 10^{14}$} \\[0.2em]
									 & \multicolumn{1}{c}{$T_{500}$ (keV)} &  \multicolumn{1}{c}{2.41} &  \multicolumn{1}{c}{2.47} &  \multicolumn{1}{c}{3.12} &  \multicolumn{1}{c}{4.41} \\[0.2em]
\hline \\[-0.8em]
\multirow{3}{*}{$z = 1.48$}  & \multicolumn{1}{c}{$R_{500}$ (kpc $h^{-1}$)} & \multicolumn{1}{c}{269} & \multicolumn{1}{c}{289} & \multicolumn{1}{c}{345} & \multicolumn{1}{c}{351}\\[0.2em]
									  & \multicolumn{1}{c}{$M_{500}$ ($M_{\odot}$ $h^{-1}$)} & \multicolumn{1}{c}{$0.50 \times 10^{14}$} & \multicolumn{1}{c}{$0.62 \times 10^{14}$} & \multicolumn{1}{c}{$1.07 \times 10^{14}$} & \multicolumn{1}{c}{$1.12 \times 10^{14}$}\\[0.2em]
							  		  & \multicolumn{1}{c}{$T_{500}$ (keV)} &  \multicolumn{1}{c}{1.76} &  \multicolumn{1}{c}{2.10} &  \multicolumn{1}{c}{3.07} &  \multicolumn{1}{c}{3.08} \\[0.2em]									  
\hline \\[-0.8em]
\multirow{3}{*}{$z = 2.0$} & \multicolumn{1}{c}{ $R_{500}$ (kpc $h^{-1}$)} & \multicolumn{1}{c}{174} & \multicolumn{1}{c}{181} & \multicolumn{1}{c}{220} & \multicolumn{1}{c}{215} \\[0.2em]
									 & \multicolumn{1}{c}{$M_{500}$ ($M_{\odot}$ $h^{-1}$)} & \multicolumn{1}{c}{$0.23 \times 10^{14}$} & \multicolumn{1}{c}{$0.25 \times 10^{14}$} & \multicolumn{1}{c}{$0.45 \times 10^{14}$} & \multicolumn{1}{c}{$0.42 \times 10^{14}$} \\[0.2em]
									& \multicolumn{1}{c}{$T_{500}$ (keV)} &  \multicolumn{1}{c}{1.19} &  \multicolumn{1}{c}{1.55} &  \multicolumn{1}{c}{2.24} &  \multicolumn{1}{c}{1.86} \\[0.2em]
\hline \hline
\end{tabular}
\label{table:1} 
\end{table*} 

The zoom-in simulations were performed with a version of the Tree-PM Smoothed-Particle-Hydrodynamics (SPH) code \texttt{GADGET-3}~\citep{springel2005}, including an improved hydrodynamical scheme \citep{beck2016} and a variety of physical processes describing the evolution of the baryonic component \citep[see][for more details]{Rasia2015Simu}.  Briefly, these comprise metallicity-dependent radiative cooling \citep{wiersma2009}, star formation and stellar feedback (thermal supernova feedback and galactic winds, see \citealt{springel2003}), cold and hot gas accretion onto super-massive black holes powering AGN thermal feedback~(\citealt{steinborn2015}; modeling the action of cold accretion \citealt{Gaspari2017Feed}), and metal enrichment \citep{tornatore2004,Tornatore2007Enrichment} from SN$_{\text{Ia}}$, SN$_{\text{cc}}$, and AGB stars. Specifically, we assume the IMF by \cite{chabrier2003}, the mass-dependent life times by~\cite{padovani1993} and stellar yields by \cite{Thielemann2003} for SN$_{\text{Ia}}$, \cite{WoosleyWeaver1995} and~\cite{romano2010} for SN$_{\text{cc}}$, and \cite{karakas2010} for AGB stars.

In our model of chemical enrichment, we follow the production and evolution of 15 chemical species: H, He, C, Ca, O, N, Ne, Mg, S, Si, Fe, Na, Al, Ar, and Ni. {\color{black} These elements are the individual species traced in the simulations. Although these do not cover the full spectrum of interest (lacking e.g., Mn or Cr, which are important tracers of the enrichment as recently shown in \citealt{Hitomi2017Enrich, Simionescu2018Perseus}), the variety of abundances provides a good starting point for a meaningful study on the ICM and the demonstration of the \xifu\ capabilities in this view.} For every gas particle in the simulation, we trace the chemical composition and the fraction of each metal that is  produced by the three enrichment sources~\cite[i.e.\ SN$_{\text{Ia}}$, SN$_{\text{cc}}$, AGB; see][for more details]{Biffi2017Clusters,biffi2018}. Each object is analysed at different redshifts, $z = 0.105$, $0.5$, $1$, $1.48$, and $2$, to assess the enrichment through time. Table~\ref{table:1} provides the characteristic radius, $R_{500}$, along with the mass, $M_{500}$, and the mass-weighted temperature, $T_{500}$, of the associated sphere of radius $R_{500}$ for the entire cluster sample. 

For each SPH particle, the output quantities provided by \texttt{GADGET-3} are used as input for our simulation.  These include the position of the particle, $\underline{x}$, its 3D velocity in the observer's frame, $\underline{v}$, its mass density, $\rho$, its mass, $m$, its temperature, $T$, and the individual masses of the 15 individual chemical species X, $\mu_{\rm X}$, tracked in the simulations. The gas density $n$ of each SPH particle is obtained by dividing $\rho$ by $m$. The mass of each element of atomic mass number $A_{\rm X}$ is converted into abundances $Z_{\rm X}$, expressed in solar metallicity units assuming the solar fractions $Z_{\odot, \rm X}$ from \citet{Anders1989Solar}. Abundances can be therefore written as

\begin{equation}
Z_{\text{X}}= \frac{1}{Z_{\odot,\rm X}} \times  \frac{\mu_{\rm X}}{\mu_{\rm H} \times A_{\rm X}}
\end{equation}
with $\mu_{\rm H}$ the hydrogen mass of the particle.

\section{End-to-end simulations}
\label{sec:e2e}

In this section, we detail the set-up of the pipeline used for the synthetic X-IFU observations, as well as the physical assumptions made in the simulations.

\subsection{Synthetic \xifu\ observations}

Simulations of the cluster data set are carried out using the \xifu\ end-to-end (E2E) simulator SIXTE\footnote{\url{www.sternwarte.uni-erlangen.de/research/sixte/}} \citep{Wilms2014SIXTE}, which creates realistic \xifu\ observations. SIXTE uses as an input a specific SIMPUT file \citep{Schmid2013Simput} containing either all the emission spectra of the particles or directly a photon list, with the time, coordinates on the sky and energy of the emitted photons. This second approach is preferred for our simulations, as it exerts a lower computational demand, induced in the former by the unparallelised random generation of photons currently implemented in SIXTE \citep[an example of the first approach is given in][]{Roncarelli2018XIFU}.  SIXTE outputs are generated not only considering the instrumental spatial and spectral responses, but also incorporate other features from the detectors such as their geometry, vignetting and internal particle background.

\subsubsection{Photon list generation}

Each simulated cluster comes as a list of SPH particles, which may emit X-ray photons. To generate the photon list used in the E2E simulation, the particle emission is modelled by a collisional diffuse thermal plasma using the \texttt{APEC} code \citep{Smith2001APEC}.  More specifically, the \texttt{vvapec} model on \texttt{XSPEC} \citep{Arnaud1996XSPEC} is adopted, as it can be parametrised according to the particles physical properties listed above, notably the individual abundances of each element. The corresponding atomic database used for the emission model is derived from \texttt{ATOMDB v3.0.9}. For the galactic absorption, the \texttt{wabs} model \citep{MorrisonWabs} is preferred for computational speed, although more accurate absorption models do exist (e.g. \texttt{TBabs}, \citealt{Wilms2000tbabs}). For all four clusters we fixed the column density to  $n_{\rm H}=0.03\times 10^{22}$\,cm$^{-3}$, which is a representative value for the latitudes at which most clusters shall be observed with \xifu\ \citep{Kalberla05Abs}. Abundances are set to solar as per \cite{Anders1989Solar} and atomic cross-sections are taken as per \cite{Verner1996XSec}. The overall flux, ${F}$ (in counts/s/cm$^2$), of each particle is computed using the \texttt{vvapec} normalisation $\mathcal{N}$ (emission-measure weighted by the distance in units of cm$^{-5}$):

\begin{equation}
\mathcal{N}=\frac{10^{-14}}{4\pi[D_{\text{A}}(1+z)]^2} \int n_{\text{e}} n_{\text{p}} dV
\label{eq:norm}
\end{equation}
where $D_\text{A}$ is the angular distance of the particle computed from its redshift $z$ (derived from the speed of the particle and the cluster mean redshift), and $V$ the particle volume. We consider a full ionisation of the intra-cluster gas with  $n_{\text{e}} = 1.2~n_{\text{p}}$ ($n_{\text{e}}$ and $n_{\text{p}}$ being the densities of electrons and protons, respectively). These emission spectra are considered as probability density distribution function and normalised accordingly over the instrumental energy bandpass (i.e., 0.2 to 12\,keV). For a fixed exposure time $\Delta t$, photons are drawn from the afferent probability distribution following a Poisson statistic of parameter $F\, \Delta t\, A$, where $A$ is the total mirror area ({\color{black} taken at 1.4\,m$^{2}$ at 1\,keV,} energy dependence of the effective area is included later on in SIXTE via the ancillary response function -- ARF {\color{black} as explained below}). Each newly created photon is added to the photon list with the sky coordinates of its parent particle (right ascension/declination).

The output product of this stage is a ``complete'' photon list (with their true energy and position) at the entrance of the telescope. This list is computed once for each cluster, and contains a large  number of simulated photons ($\geq 1$Ms). It is then sampled randomly by SIXTE to achieve smaller lists for more typical exposure times (e.g., 100\,ks).

\subsubsection{Observational setup}
\label{subsub:obs}
For each simulation, we consider an exposure time $\Delta t = 100$\,ks over the entire \xifu\ field-of-view. The ``complete'' photon list is used as input for the \texttt{xifupipeline} function of SIXTE, which samples the photon list accounting for the energy-dependence of the effective area to create the event list seen by the \xifu\ detector over $\Delta t$. The pipeline accounts for the most up-to-date responses of the current baseline of the telescope (i.e., 15-row mirror modules corresponding to a mirror effective area of 1.4\,m$^2$ at 1\,keV\footnote{RMF: \path{athena_xifu_rmf_v20171107.rmf}  | ARF: \path{athena_xifu_15row_onaxis_pitch249um_v20171107.arf}}) and for a hexagonal detector array of 3832 micro-calorimeter pixels, more specifically Large Pixel Array 2 (LPA2) pixel configuration, developed for the \xifu\ and described in \citet{Smith2016Pix}. Pile-up, telescope point spread function, vignetting and detector geometry effects are also included as function of the pixels corresponding off-axis angles. Finally, we verified that given the low count rates of our clusters ($\leq\,1$\,cts/s/pix), pile-up and cross-talk over the observation can be neglected \citep[see][]{denHartogCtk2018, PeilleCtk2018}.

For each cluster, we simulate enough pointings to map the cluster spatially up to at least $R_{500}$ (as required in the current science objectives for the \xifu). This translates, for local clusters, into at least seven pointings. The corresponding event lists are then merged during post-processing to obtain a single event file. 

\subsubsection{Foreground/Background components}
\label{subsub:back}

In addition, we account for the contribution of different foreground and background sources to ensure more representative observations.
\begin{table} [t]
\begin{center} 
\caption{Parametrisation of the galactic foreground model used in the simulation with a \texttt{apec + phabs*apec} model}
\label{table:2} 
\small
\begin{tabular}{c c c c} 
\hline\hline\\[-0.8em] 
Model & Parameter & Unit & Value \\[0.2em]  
\hline \\[-0.8em]  
\texttt{apec} & $T$ & keV & 0.099 \\
\texttt{apec} & $Z$ & & 1 \\
\texttt{apec} & $z$ & & 0 \\
\texttt{apec} & Norm & cts/s/amin$^{2}$ & $1.7 \times 10^{-6}$\\
\texttt{phabs} & $n_{\text{H}}$ & $10^{22}$ cm$^{-3}$ & 0.018 \\
\texttt{apec} & $T$ & keV & 0.225 \\
\texttt{apec} & $Z$ & &1 \\
\texttt{apec} & $z$ & & 0 \\
\texttt{apec} & Norm & cts/s/amin$^{2}$ & $7.3 \times 10^{-7}$ \\
\hline \hline
\end{tabular}
\end{center}
\normalsize
\end{table}

\begin{description}
\item{\textbf{Astrophysical foreground:}} The foreground emission is caused by the X-ray emission of the local bubble in which the solar system is embedded and by the Milky Way hot gaseous content. This component can be modelled by the sum of a non-absorbed and absorbed thermal plasma emission as specified in \cite{McCammon2002Bkg} and parametrised as per \citet[][see Table~\ref{table:2}]{Lotti2014Bkg}. An additional normalisation constant over the entire model is used for versatility purposes, resulting in a total foreground model reading as  \texttt{constant*(apec + phabs*apec)} in XSPEC. This component is folded into SIXTE using a SIMPUT file. \\
\begin{figure*}[t]
\centering
\includegraphics[width=0.495\textwidth, trim={50cm 0 22cm 8cm}, clip]{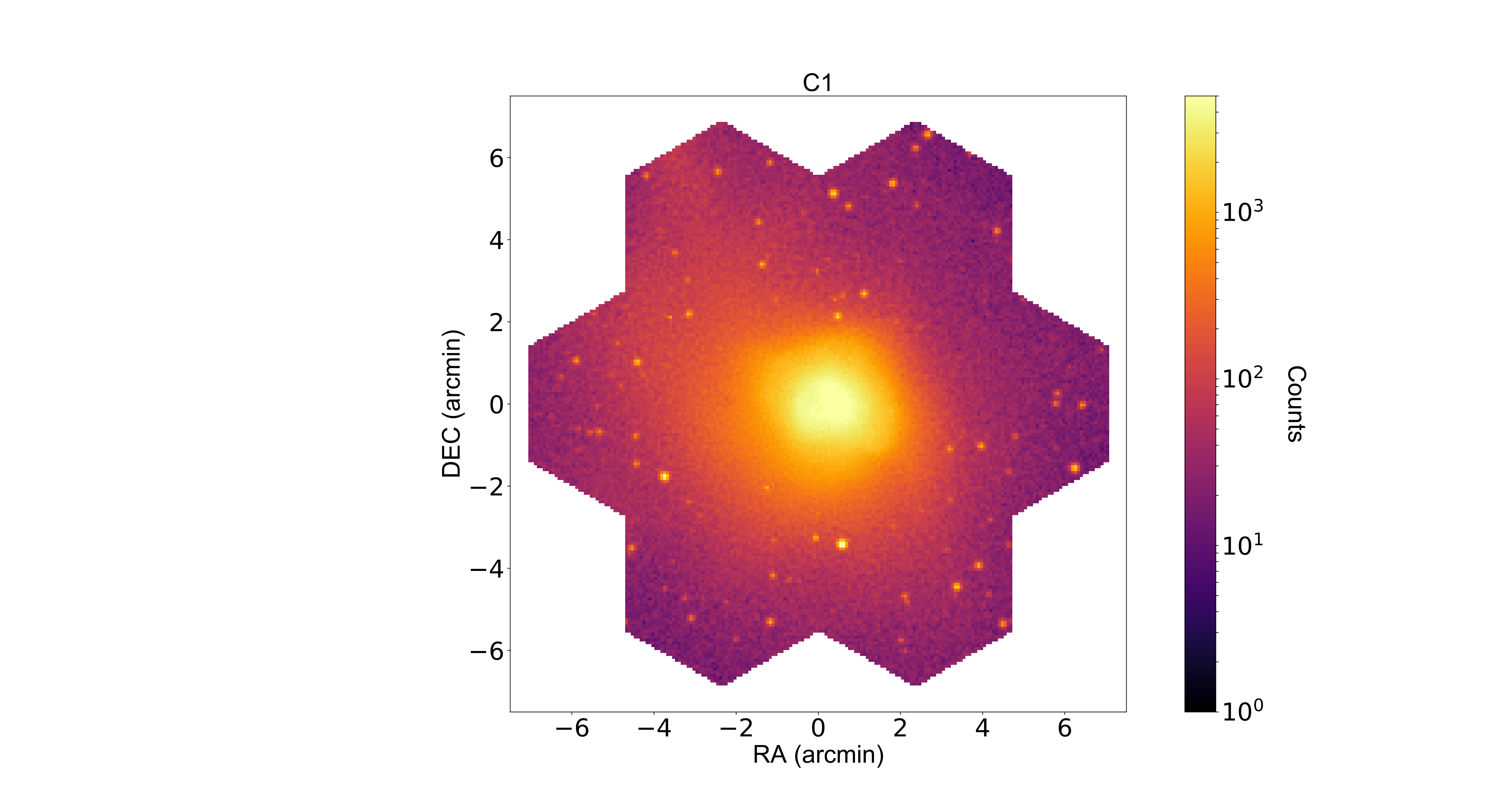}
\includegraphics[width=0.495\textwidth, trim={50cm 0 22cm 8cm}, clip]{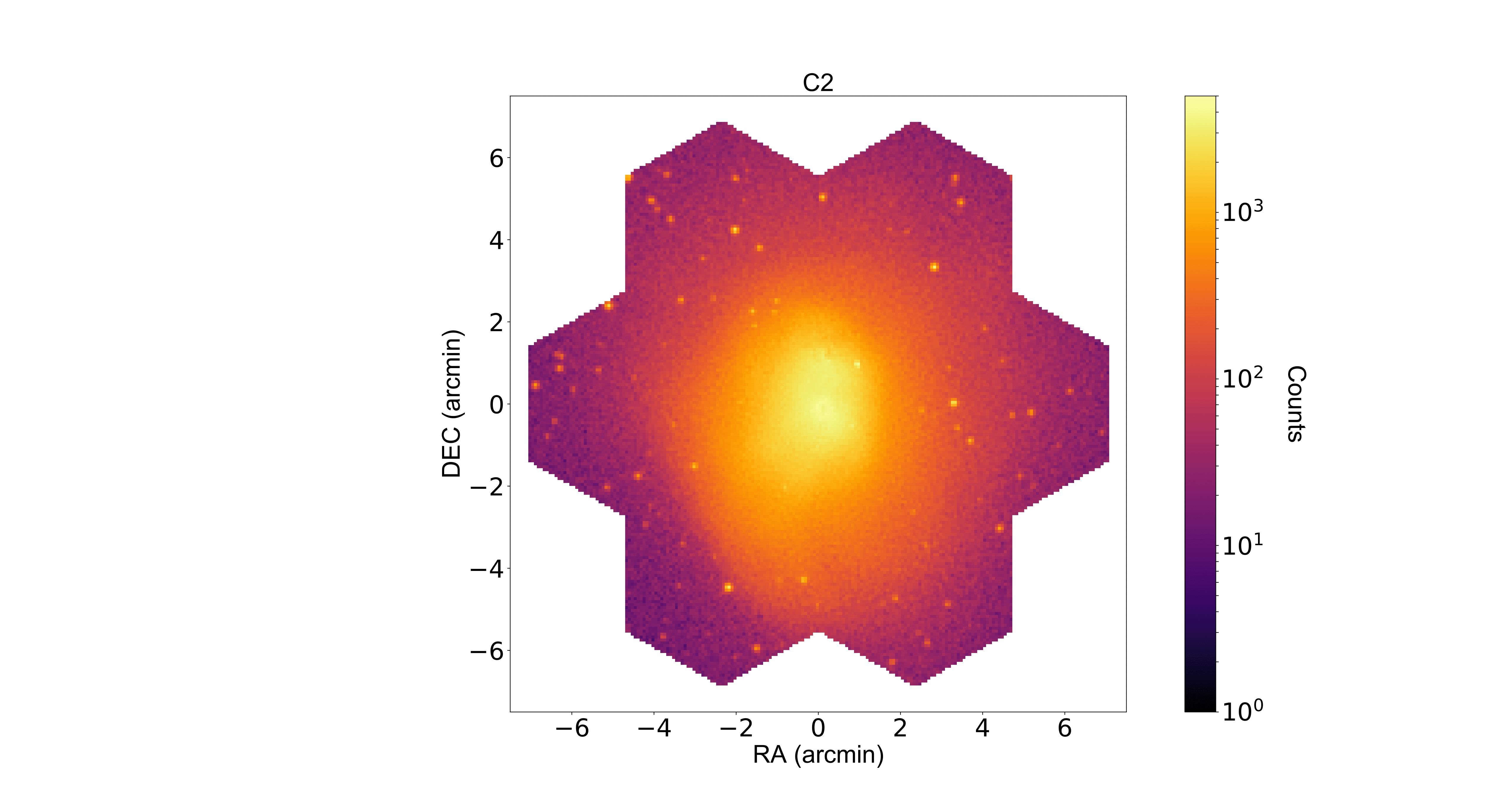}
\includegraphics[width=0.495\textwidth, trim={50cm 0 22cm 8cm}, clip]{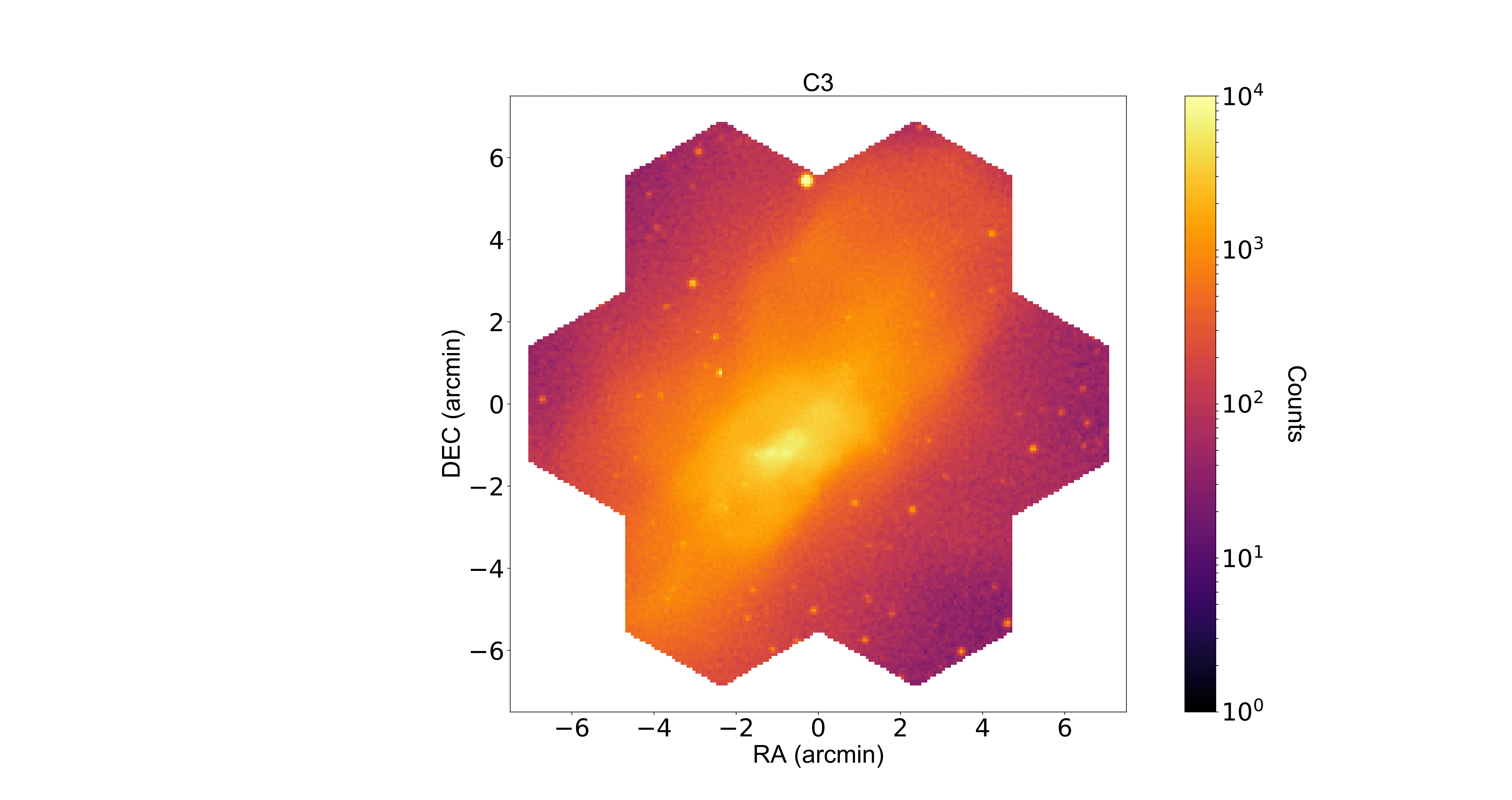}
\includegraphics[width=0.495\textwidth, trim={50cm 0 22cm 8cm}, clip]{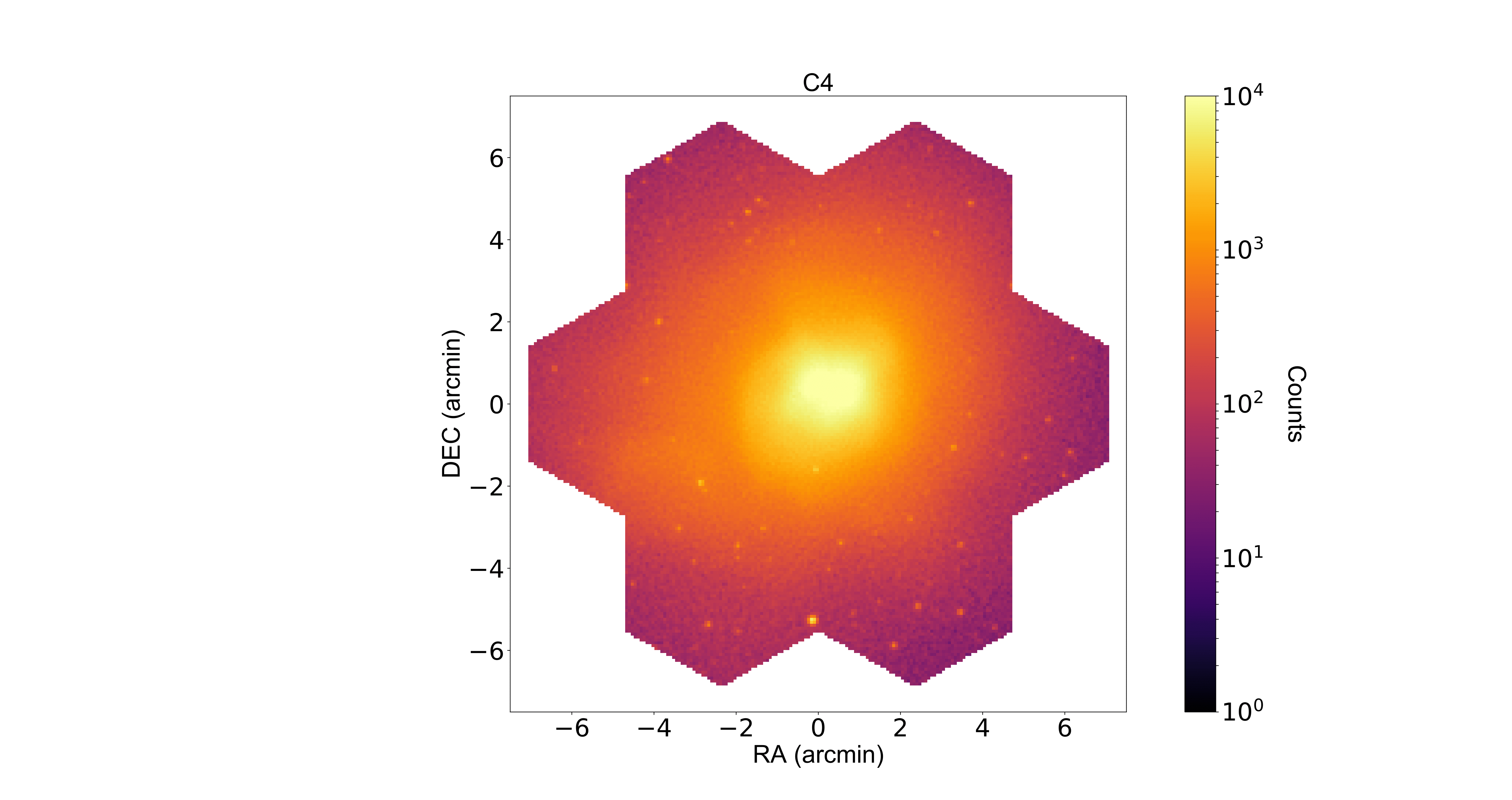}
\caption{Maps in number of counts per \xifu\ pixel (249\,\textmu m pitch) for our clusters 1 to 4 (see Table~\ref{table:1}) at redshift $z = 0.1$. Each  mosaic is made of 7 \xifu\ pointings of 100\,ks each.}
\label{fig:clusters}
\end{figure*}

\item{\textbf{Cosmic X-ray background:}} The cosmic X-ray background (CXB) component is due to the contributions of AGNs, star forming galaxies and active stars along the line-of-sight \citep{Lehmer2012AGN}. A fraction of these sources will be resolved by the instrument as a function of its spatial resolution, and will be excised from the observations. Given the requirement on the spatial resolution for \textit{Athena}/\xifu\ (5$^{\prime\prime}$), 80\% of the total flux of these point sources in terms of the integral of their log(N)/log(S) distribution should be resolved by the instrument \citep{Moretti2003CXB}. For 100\,ks exposure times, this translates into limiting fluxes of $\sim\,3\,\times\,10^{-16}$\,ergs/s/cm$^{2}$ for the \xifu. As the number of star forming sources is at least an order of magnitude lower at this flux, we only considered the AGN contribution in this study.  The unresolved fraction of these point sources results in a diffuse background component, which is classically fitted using an absorbed power-law model during post-processing \citep{McCammon2002Bkg}. 

\hspace{0.5cm} AGN point sources are not included in the inputs derived from the hydrodynamical simulations. Instead, to generate realistic CXB data, we draw a list of AGN sources with associated X-ray spectra by sampling the luminosity function of \cite{Hasinger2005AGN} in the luminosity-redshift space, given the boundary conditions $L_X\,\geq\,10^{42}$\,erg/s unabsorbed 0.5~--~2\,keV rest-frame luminosity, $0<z<5$ and the size of the cosmological volume encompassed within a field-of-view. Each source is associated with a spectral energy distribution following templates described in \cite{Gilli2007nH}, according to a distribution of power-law indexes and intrinsic absorption column densities related to various levels of obscuration, as described in \cite{Gilli1999AGN, Gilli2007nH}. Spatial distributions are fully random in the sky plane (no clustering). More details on the procedure can be found in 
\cite{Clerc2018Bkg}. This component is also included in the simulations using a SIMPUT file. \\

\item{\textbf{Instrumental background}:} The instrumental particle background is caused by interactions of high-energy cosmic rays/protons with the instrument structure, which create secondary particles in the soft X-ray band. Both primary and secondary particles can hit the detectors and be recorded as regular events. The \xifu\ design includes an onboard  Cryogenic Anti-Coincidence detector \citep{Macculi2016Cryo}, which will ensure the required level of $5\times 10^{-3}$\,counts/s/keV/cm$^{2}$  over the 2 -- 10\,keV energy band \citep{Lotti2017NXB}. The generation of this component is directly implemented within SIXTE.
\end{description}

For each of these three components, we associate a  flag on the photons in the event file. Thus, these specific events can be respectively masked to study background effects on the observations (notably of the internal particle background) or selected exclusively to generate background maps. Throughout this study, we assume that these background components have no systematic uncertainty. Systematic effects of the background knowledge on the observations are discussed Section \ref{sec:dis} and are considered in more ample detail elsewhere \citep{Cucchetti2018NXB}.



\section{Data processing}
\label{sec:dat}
In this section we describe the post-processing approach used in our simulations and its validation. 

\subsection{Source contamination}

From the event list output of SIXTE, a first selection is made on the grading of the events, which is conditioned by the frequency of detection in a given pixel and defines the spectral resolution of the event. In practice, the grading procedure will occur in-flight using the onboard event processor depending on the time separation between events in the same pixel \citep{PeilleCtk2018}, similarly to the strategy implemented on the SXS \citep{Seta2012grades}. In this case, grading occurs automatically within the simulator and is available in the event list. Only high-resolution events corresponding to $\Delta E=2.5$\,eV, the required spectral resolution for \xifu, were used. As the count rates of the clusters are low over the entire field-of-view ($\leq 1$\,count/s/pix), almost all events (throughput $\geq99\%$) are high-resolution photons. Events with lower grading values are discarded in the rest of the study. 

Using the selected events, we reconstructed raw brightness maps in counts as presented on Fig.~\ref{fig:clusters}. Beside the ICM, emission from other sources either present in the hydrodynamical simulations (i.e., strongly-emitting particles or clumps) or in the CXB can be observed. For the CXB, following the assumption by \cite{Moretti2003CXB}, we start by selecting the brightest simulated sources (with fluxes above $\sim 3 \times 10^{-16}$\,ergs/s/cm$^{2}$, see Sec.~\ref{subsub:back}). After simulating CXB-only pointings to find their coordinates on the detector, these sources are excised automatically from the brightness maps by finding all the corresponding pixels above a $2\sigma$ threshold in counts with respect to the average count in neighbouring pixels in the event list. This way, the diffuse emission of the cluster is accounted for when masking a pixel. The average cut-off flux of the sources will be higher than $\sim 3 \times 10^{-16}$ over the full map. The lower the emission (e.g., in the outskirts), the closer the limiting flux of the excised point-sources will be to the threshold flux. Although this process would not be possible for real event files, we adopted this strategy to avoid potential biases related to specific point-source detection algorithms. Once these sources are removed, a final visual inspection is performed to remove any residual unexcised AGNs as well as any remaining visible point-like source which may be related to the hydrodynamical simulations.

\subsection{Spatial binning}
\label{s:bin}
For the considered exposure time of 100\,ks, single pixels do not always capture enough photons to allow the measurement of chemical abundances. We therefore group them into regions to increase the signal-to-noise ratio ($\textrm{S/N}$) adequately.

Two methods were considered to spatially bin the pixels: the \texttt{contbin} tool developed by \cite{Sanders2006Contour} and an adapted 2D Voronoï tessellation method by \cite{Cappellari2003Voronoi}. Both methods were tested and a comparison showed no visible difference in accuracy of the pipeline (see also Appendix \ref{app:1} or \citealt{Sanders2006Contour} for a more detailed comparison). Unlike Voronoï tessellation, contour binning offers better results in describing the radial-like shape of the cluster emission, and was therefore retained for this study. The \texttt{contbin} binning scheme can be directly applied to our cluster count maps to compute regions of equal $\textrm{S/N}$. The real exposure map (constant here) and the spatial mask of excised sources are also used as inputs of the algorithm. In addition, we chose to fix the aspect-ratio of the binning regions (so called ``constrain filling-factor'') to 2, to avoid long radial filamentary regions, which could artificially mix spatially-uncorrelated structures (especially in low count rates areas such as the outskirts of the cluster).

The binning procedure operates on the count maps, which are dominated by the ICM bremsstrahlung emission, i.e., the continuum of the spectral energy distribution. We further optimised our pixel binning in view of our scientific objective of measuring chemical abundances and to account for smaller local surface brightness variations atop the bulk of the cluster emission due to e.g., clumps or bubbles. To do so, we first divided the surface brightness map into annuli centred on the brightest part of the cluster and containing roughly the same number of counts ($\sim$ 300\,000). For each of these annular regions, the total spectrum of the annulus is fitted over the entire bandpass (0.2~--~12\,keV) with a continuum-only \texttt{vvapec} model (metal abundances set to zero) to estimate the number of counts in the continuum for a given annulus $C_{\rm continuum~only}$. Although slightly overestimating counts (due to the presence of lines), this simple approach converges quickly. When abundances are left free then set to zero after the fit, strong lines are sometimes spaced by values of the order of the energy resolution of the instrument (due to strong bulk motion or clumping within the cluster), causing fits to converge on local minima or not at all. Even after convergence, differences with the previous method do not exceed $\leq 3\%$. The total number of counts on a pixel~$i$ ($C_i$) in the annulus is then rescaled in the following manner:
\begin{equation}
\bar{C_i} = \left(1 - \frac{C_{\rm continuum~only}}{C_{\rm total~annulus}} \right) C_{i}
\end{equation}

\begin{figure}[tb]
\centering
\includegraphics[width=0.495\textwidth, trim={50cm 0 22cm 8cm}, clip]{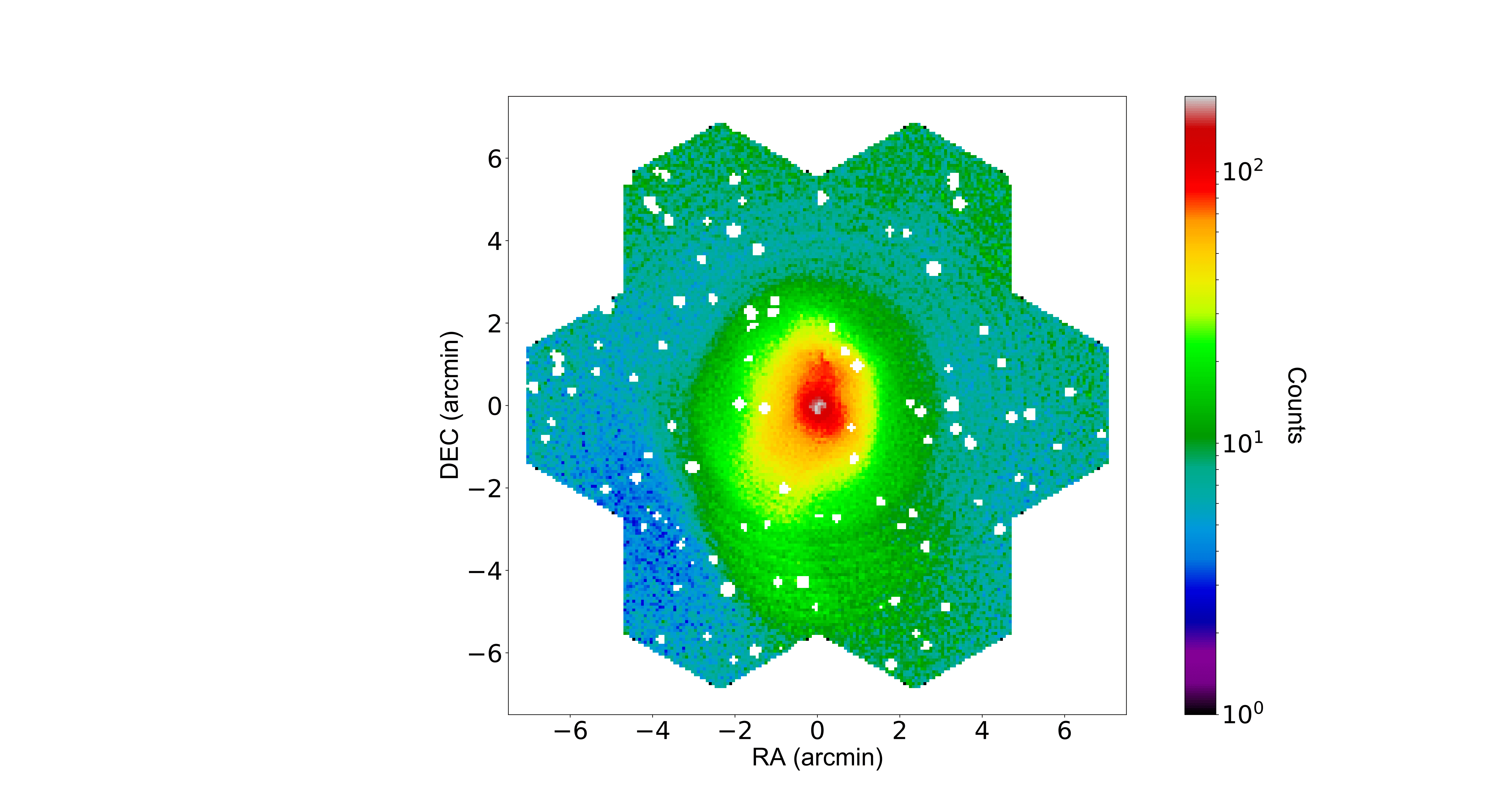}
\caption{Example of a continuum subtracted count map for cluster C2 used for spatial binning. Each of the $\sim 27\,000$ pixels was rescaled as explained in Sec.~\ref{s:bin} to enhance density contrasts in the cluster.} 
\label{fig:data_processing}
\end{figure}

The resulting template image for cluster C2 can be seen Fig.~\ref{fig:data_processing}.  This approach allows to define regions over a continuum subtracted image, enhancing the brightness fluctuations with respect to the azimuthal continuum model due to chemical elements emission and other local density contrasts. This template image is used exclusively for spatial binning. Accounting for its statistics, we request a S/N level of 30 (900 counts in regions) to \texttt{contbin}. The resulting spatial regions are used over the full count maps to compute the corresponding spectra. Given a count ratio of $\sim 100$ between the template map and the full surface brightness map, we ensure that each spectrum has a high statistical significance i.e., $\textrm{S/N} \sim 300$. This represents a total of at least 90\,000 counts per region, i.e.~3~counts in each instrumental channel, allowing high significance regions for the fits. 

This binning approach is used throughout this paper when statistics allows it, i.e. in the case of local clusters ($z \leq 0.5$). For high-redshift clusters, count rates are often insufficient to ensure statistically independent regions with a high enough S/N ratio. In this case, we followed the formulation of the \textit{Athena} science objectives on chemical evolution of the Universe (specified in the introduction) and considered two radial annular regions, over 0~--~0.3\,$R_{500}$ and over 0.3 -- 1\,$R_{500}$.

\subsection{Accounting for vignetting effects}

Despite the narrow field-of-view of the \xifu, the \textit{Athena} telescope will introduce a slight vignetting effect in the observations (between 4 and 8\% of the counts over the energy bandpass). Vignetting is simulated within SIXTE using tabulated values derived from ray-tracing simulations of the mirror assemblies \citep{Willingale2014SPIE} and needs to be accounted for before attempting a spectral fit. To do so, for each binned region, the baseline \xifu\ effective area of every single pixel is multiplied over the energy bandpass by the same vignetting function implemented in SIXTE accounting for the off-axis position of the pixel. The specific response function of the entire region is then obtained by averaging, in each energy channel, the response of each of pixels weighted by their respective number of counts.  

\subsection{Temperature and metal maps}

The resulting unbinned spectra are fitted within XSPEC, using a log-normal likelihood minimisation \citep[i.e., C-statistics adapted from][] {Cash1979}, which is well adapted for Poissonian data sets (i.e., channels with low number of counts). An absorbed  single-temperature thermal plasma model (i.e., \texttt{wabs*vvapec}) was fitted to the spectrum accumulated in each region. The column density $n_{\rm H}$ is fixed to its input value, while temperature, abundances of metals traced in the input numerical simulations, (see Sec.~\ref{sec:sim}), redshift and normalisation are left free. As a first approach, all parameters (metallicity for each of the 12 chemical elements  $Z_{\rm X}$, temperature, redshift and normalisation) are fit simultaneously over the 0.2 -- 12\,keV energy band (broadband fit). The background components are accounted as an additional model, as described in Sec~\ref{subsub:back}. Their spectral shapes is assumed to be perfectly known whilst the normalisation of each three component is set free (either the model norm for the CXB and instrumental background, or the multiplicative constant factor for the foreground, see Sec.~\ref{subsub:back}). Using the best fit results provided by XSPEC, we can construct spatial maps for each physical parameter over each full synthetic observation. 

\subsection{Input parameter maps}
\label{sub:inp}
To estimate the goodness of the fit, the output maps are compared to weighted input maps, reconstructed from the input numerical simulations using the same spatial binning regions, $j$, than the outputs. The adopted weighting scheme depends on the considered parameters. For instance, emission-measure-weighted quantities are expected to be more representative than mass-weighted schemes \citep{Biffi2013mass}, especially for abundances. The value of the input emission-measure-weighted parameter $P$ in the region $j$ therefore reads as:

\begin{equation}
P_j= \dfrac{\Sigma_{i}\, P_i\, \mathcal{N}_i}{\Sigma_{i}\, \mathcal{N}_i} 
\label{eq:wei}
\end{equation}
where $\mathcal{N}_i$ is given by Eq.~\ref{eq:norm} for each SPH particle $i$ contributing to the spatial region $j$. 

For the temperature, it has been long known that emission-measure or mass-weighted schemes do not match fitted quantities \citep{Gardini2004Spectral, Rasia2006Spectral, Rasia2008Spectral}. A known method to account for this bias is to use the so-called {\color{black} spectroscopic} temperature weighting, introduced by \citet{Mazzotta2004Spectral}, which translates here into:

\begin{equation}
T_j= \dfrac{\Sigma_{i}\, \mathcal{N}_i\, T_i^{+0.25}}{\Sigma_{i}\, \mathcal{N}_i\, T_i^{-0.75}} 
\label{eq:spec}
\end{equation}
Notably, we verified that the use of {\color{black} spectroscopic} temperature input maps indeed reduced the biases of the fitted temperature with respect to the emission-measure-weighted input maps (see also Appendix \ref{app:2}). We note that this method is particularly suited to high-temperature regions ($\geq 3$ keV), well represented in the central parts of our clusters (see also Fig.~\ref{fig:maps} -- \textit{Upper central panel}), but may be limited towards the cluster outskirts or in the case of cooler systems. Although the low-temperature regions could be processed more accurately with the extended method presented in \cite{Vikhlinin2006Temp}, since our outskirts regions are very large (low statistics) and relative differences remain within statistical error bars of the XSPEC fit, we use the implementation of the {\color{black} spectroscopic} temperature presented in \citet{Mazzotta2004Spectral} for all our regions.

\subsection{Assessment of systematics}
\label{sub:sys}
For a physical parameter $P$, the goodness of the reconstruction is evaluated using two methods. First, the deviation of the fitted value is evaluated in terms of its relative error distribution with respect to the weighted input value (i.e., $\Delta P_j=(P_{j}-P_{\text{in}, j})/P_{\text{in}, j}$) over the various spatial regions $j$. A priori, if no systematic effect is present in the pipeline, the relative error distribution for $P$ should be Gaussian (if the number of regions is sufficiently high) and centred, with a standard deviation $\sigma_{\Delta P}$ of the same order as the averaged normalised value of the statistical error, $\mu_{\rm fit}$, derived from the XSPEC fits (see Appendix~\ref{app:2} for a more generic estimation for non-Gaussian cases). This approach is mainly used to determine the presence of biases in the reconstruction by fitting the relative error distribution using a Gaussian, accounting for the corresponding errors $\sigma_{j}$ derived from XSPEC. For this fit, clear outliers with relative errors above 100\% (1 or 2 regions overall) are removed for consistency. As a second test, we compute in each region $j$ the value $\chi_j=(P_{j}-P_{\text{in}, j})/\sigma_{j}$, which shows the goodness of the fits in terms of the fitting errors and should follow a $\chi$ distribution. By computing the reduced value of the distribution, $\chi_{\rm red}^2$, we can estimate the overall goodness of the fit with respect to the statistical errors derived from XSPEC. 
\begin{figure*}[p]
\centering
\includegraphics[width=0.34\textwidth, trim={50cm 2cm 23cm 8cm}, clip]{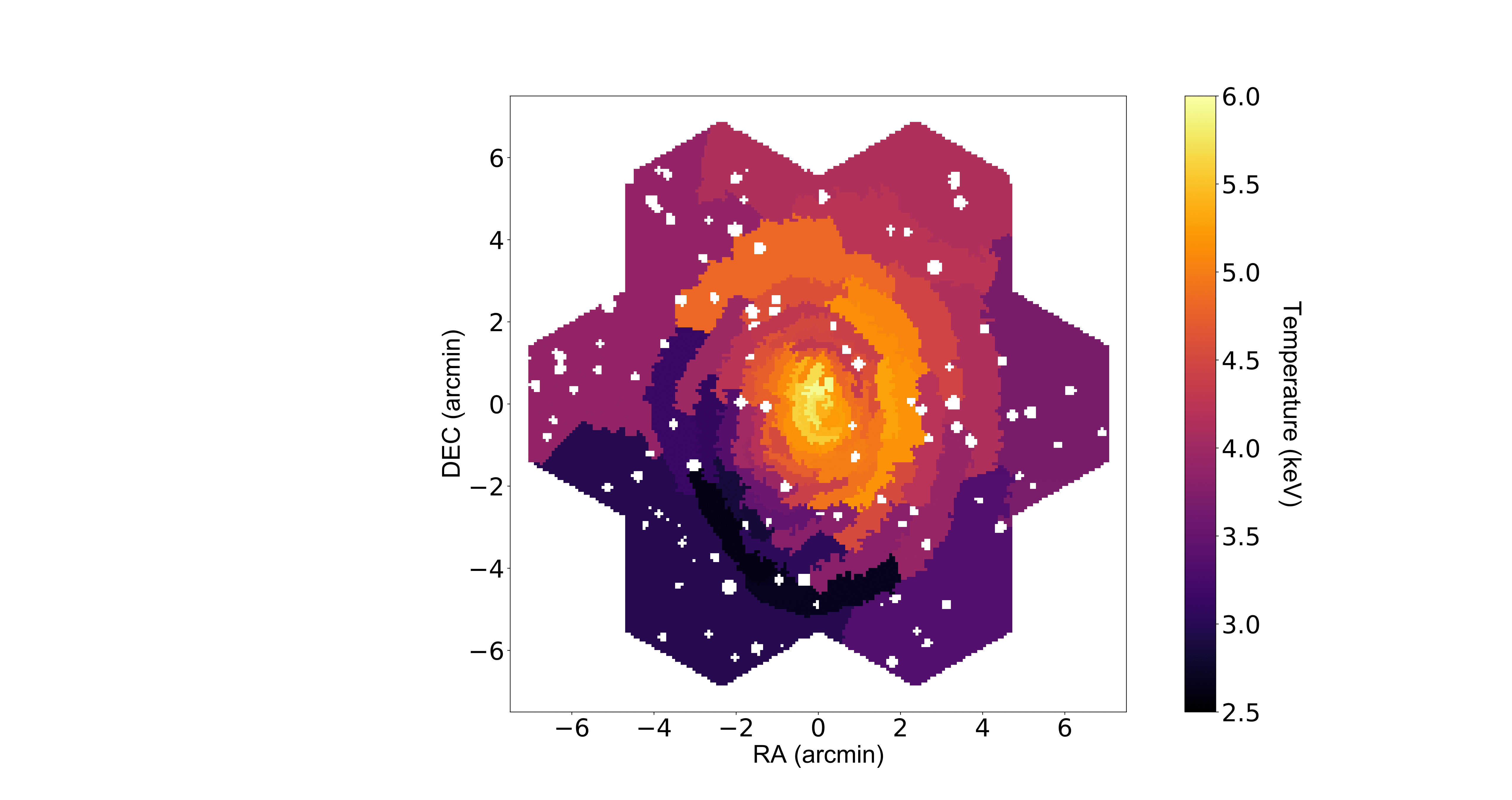}
\vspace{0.6cm}
\includegraphics[width=0.34\textwidth, trim={50cm 2cm 23cm 8cm}, clip]{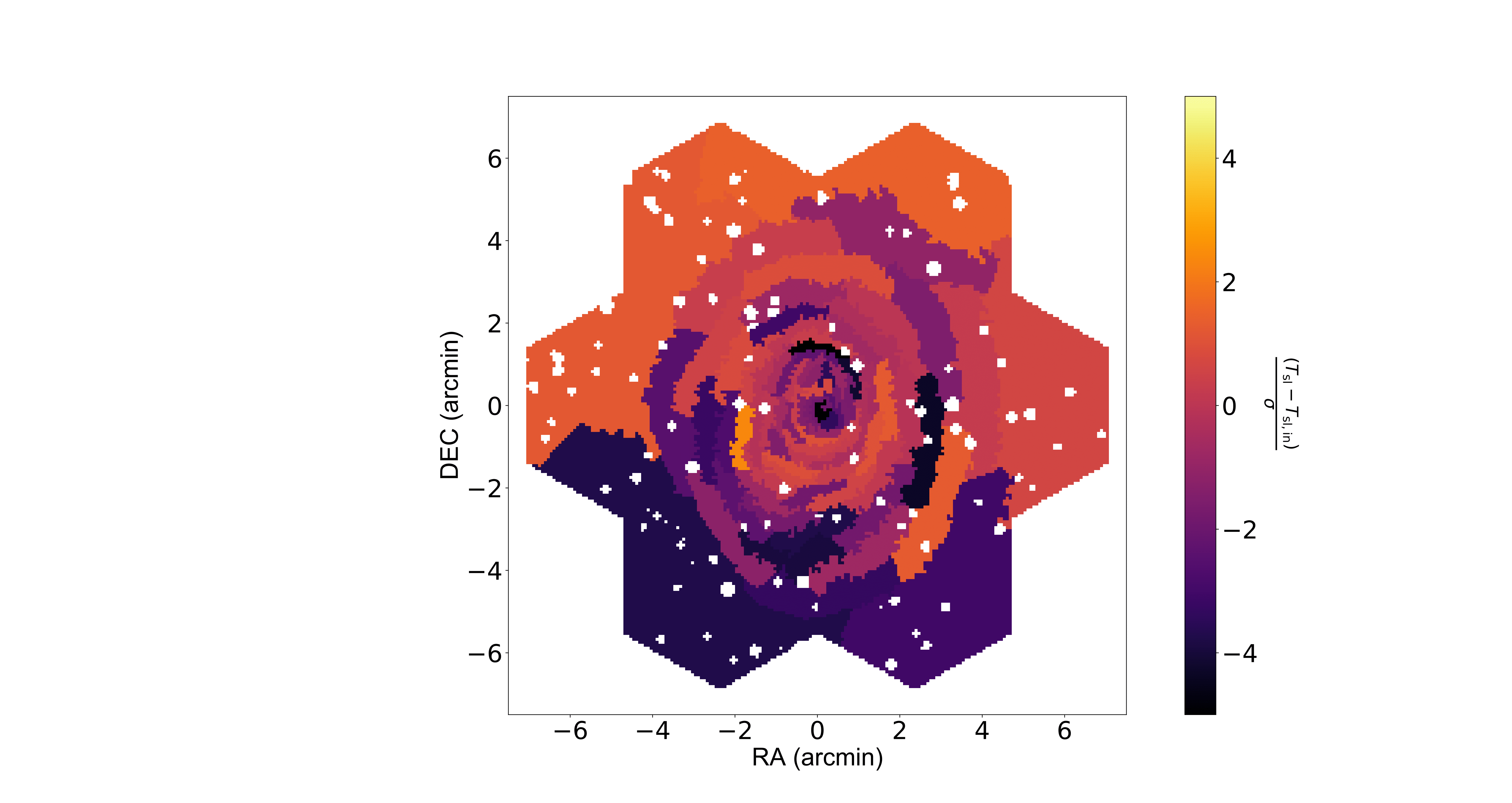}
\includegraphics[width=0.31\textwidth, trim={0 0 0 0.5}, clip]{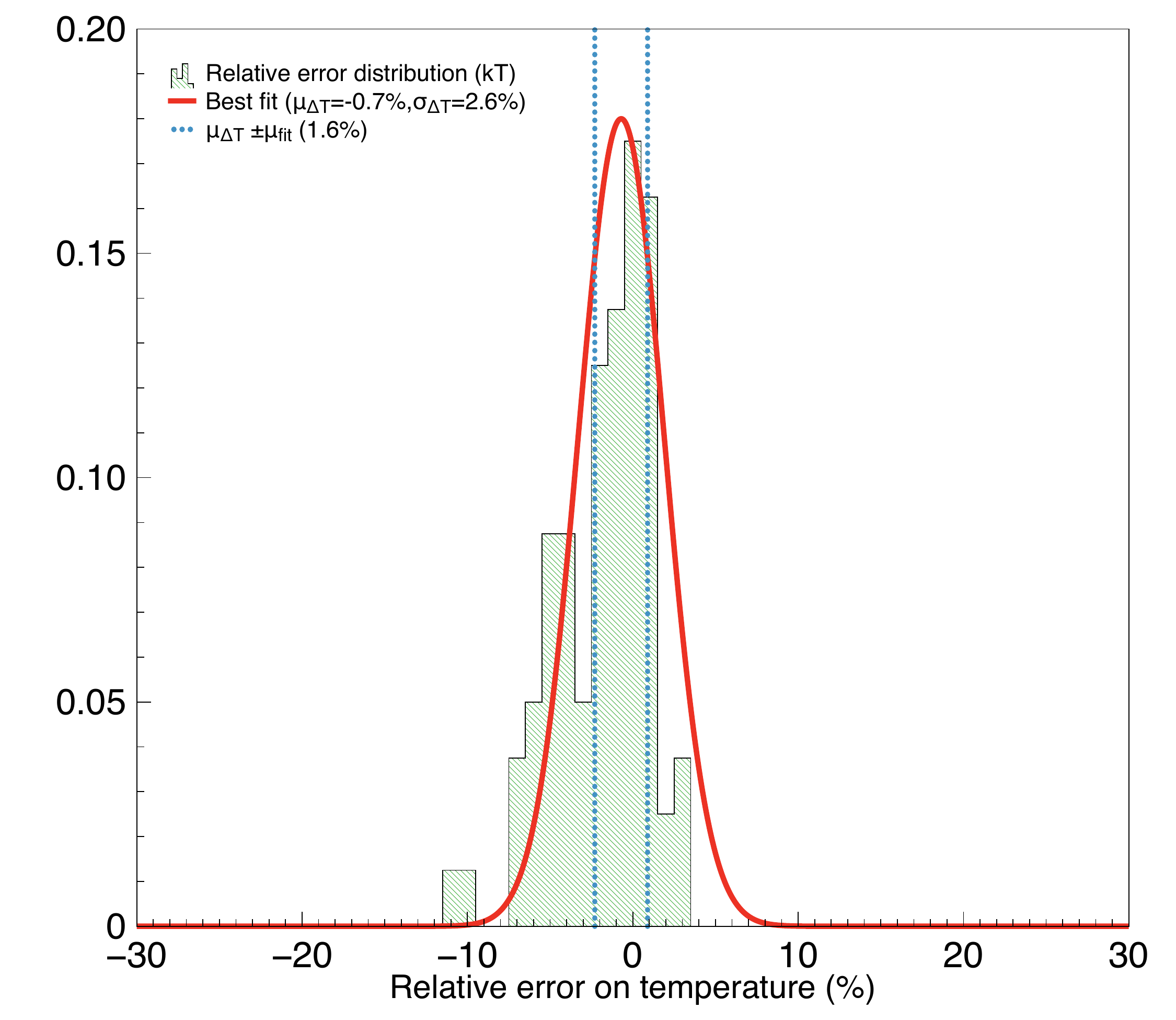}

\includegraphics[width=0.34\textwidth, trim={50cm 2cm 23cm 8cm}, clip]{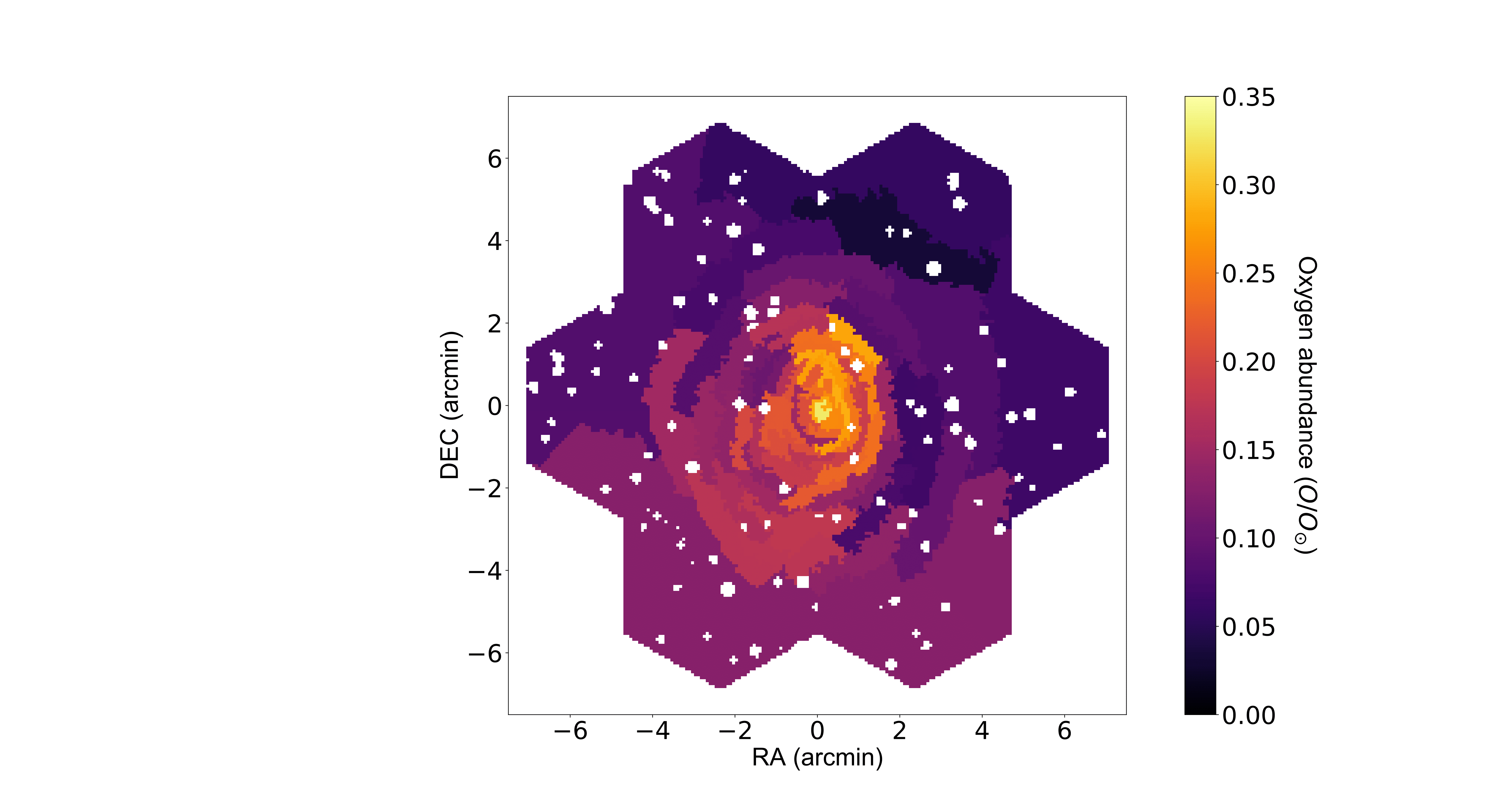}
\vspace{0.6cm}
\includegraphics[width=0.34\textwidth, trim={50cm 2cm 23cm 8cm}, clip]{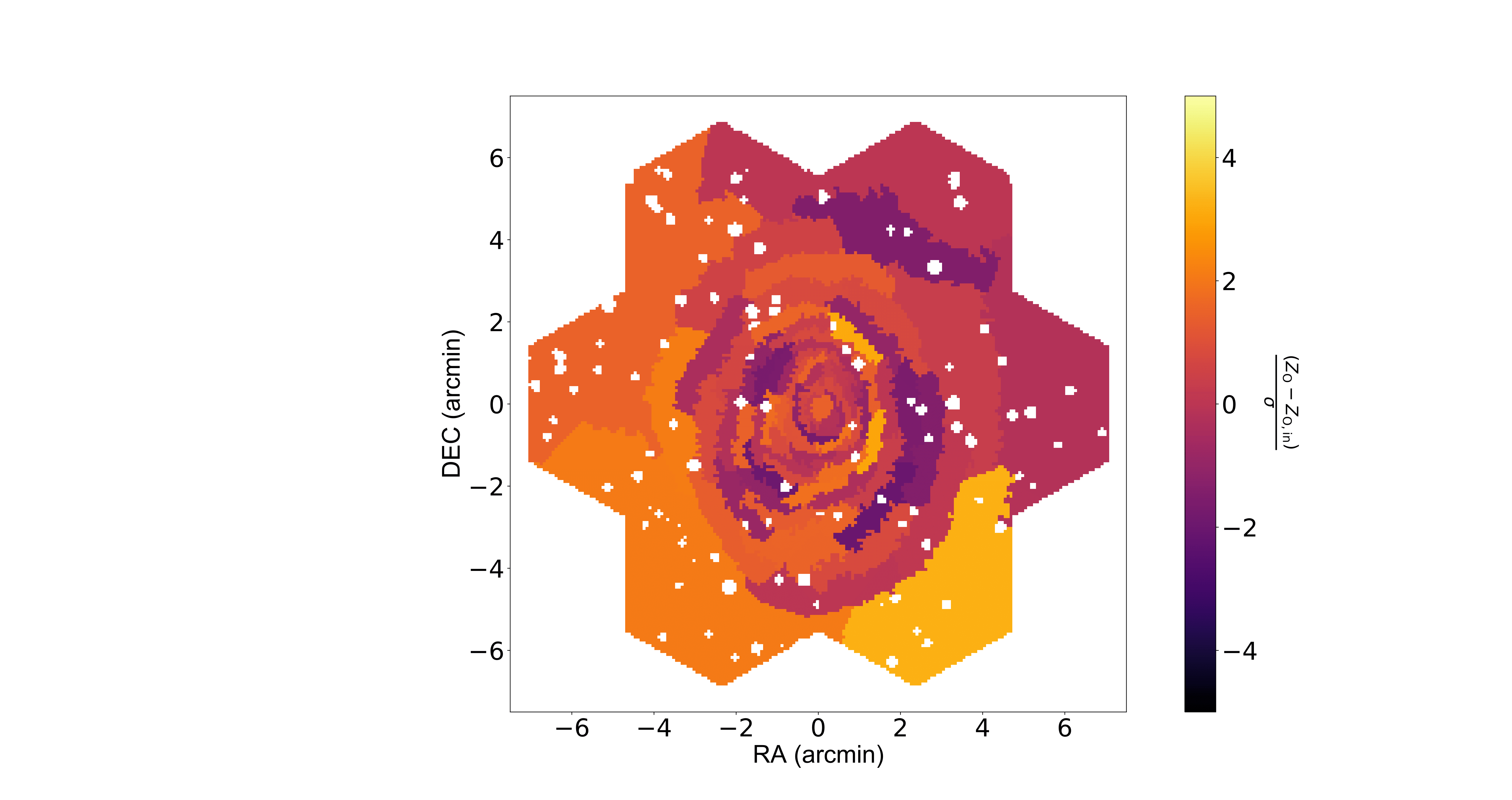}
\includegraphics[width=0.31\textwidth, trim={0 0 0 0.5}, clip]{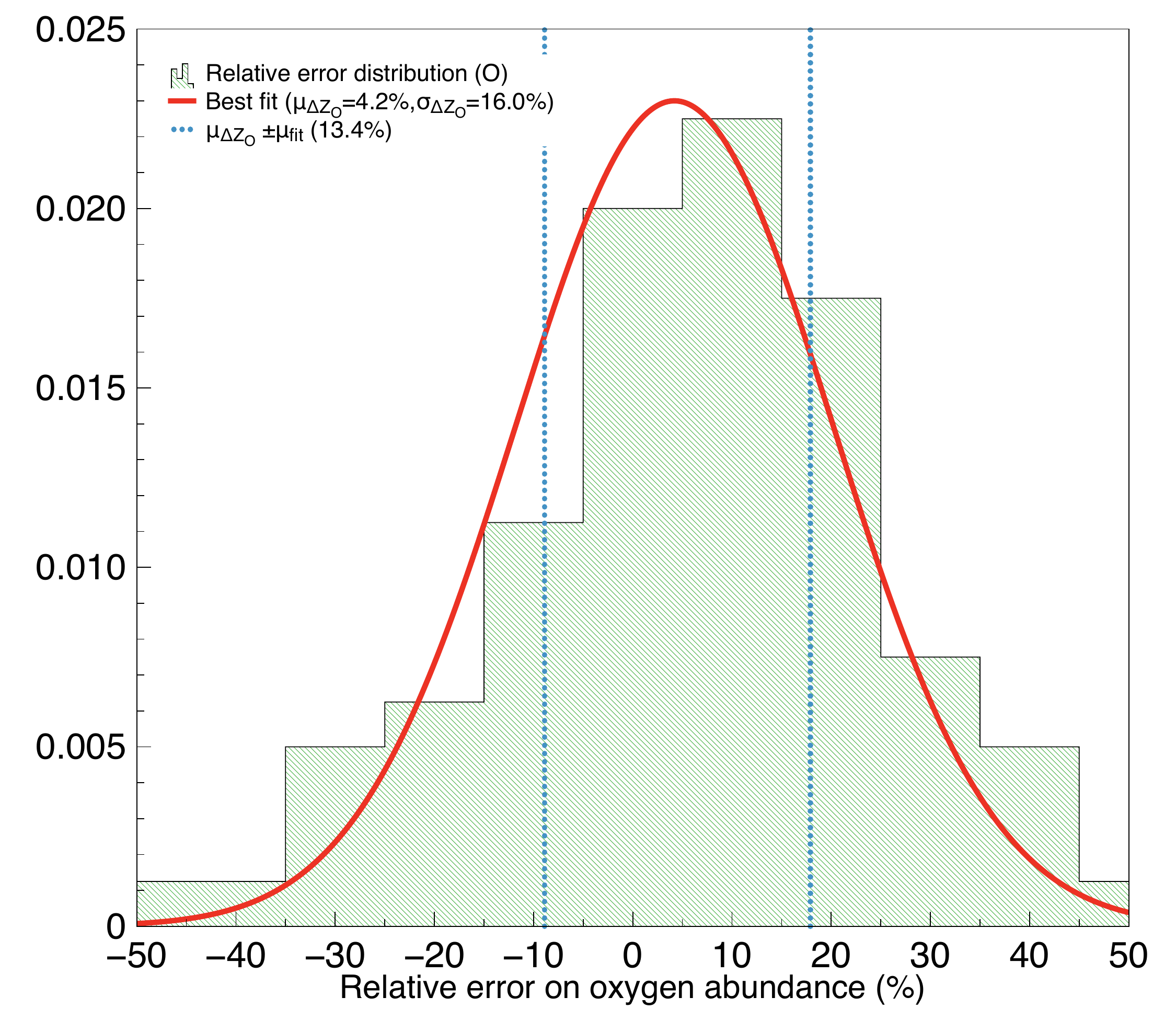}

\includegraphics[width=0.34\textwidth, trim={50cm 2cm 23cm 8cm}, clip]{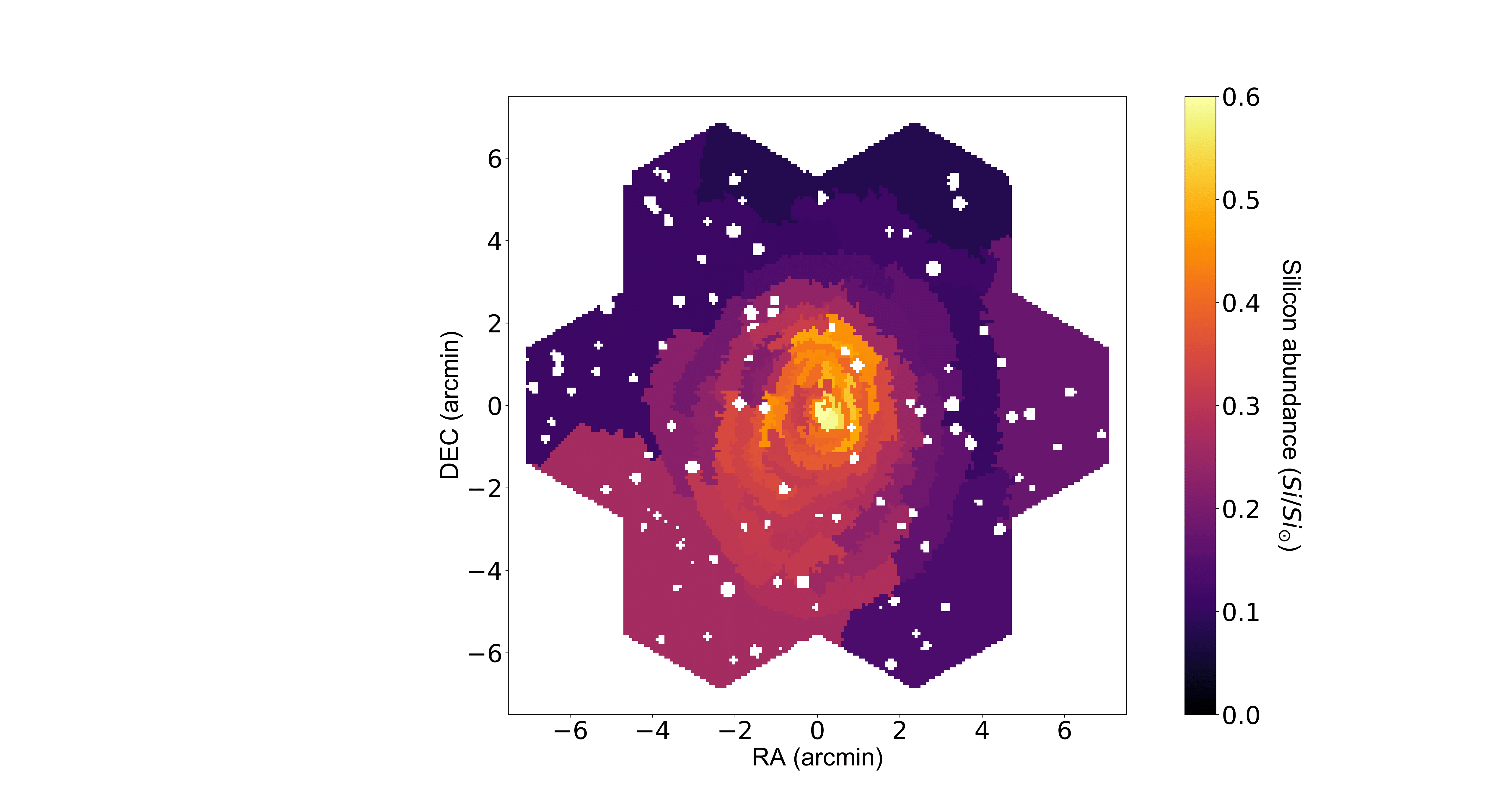}
\vspace{0.6cm}
\includegraphics[width=0.34\textwidth, trim={50cm 2cm 23cm 8cm}, clip]{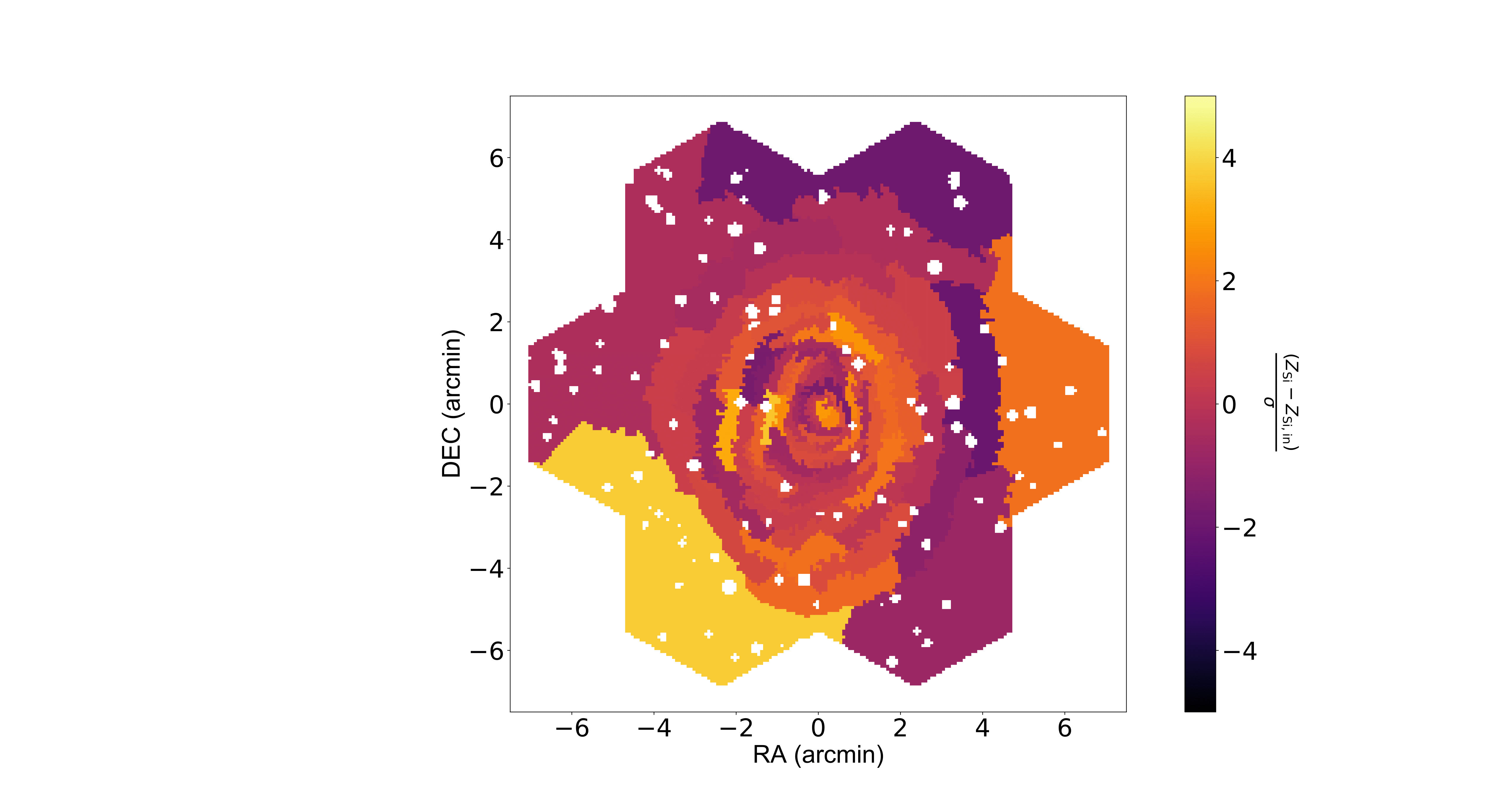}
\includegraphics[width=0.31\textwidth, trim={0 0 0 0.5}, clip]{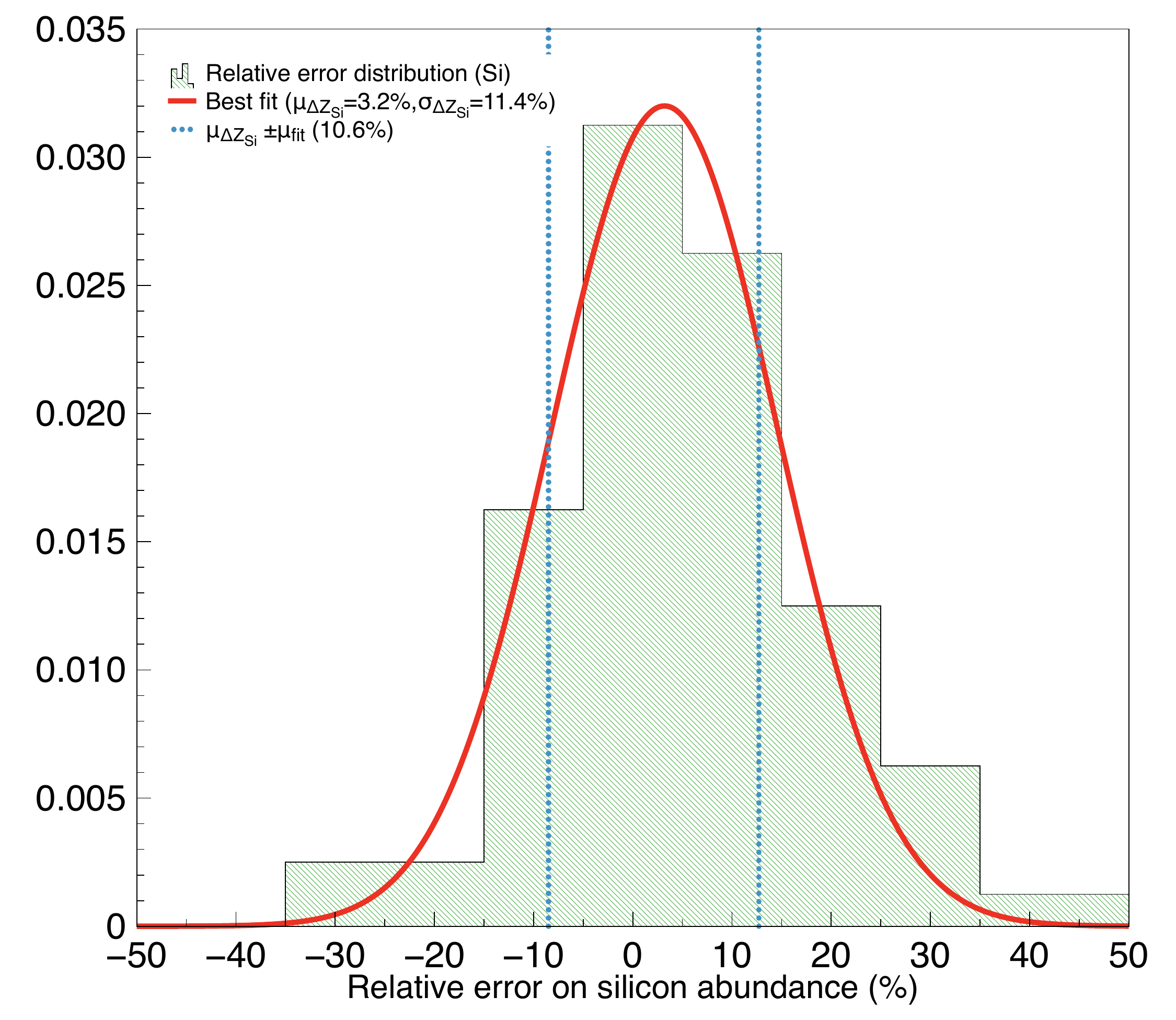}

\includegraphics[width=0.34\textwidth, trim={50cm 2cm 23cm 8cm}, clip]{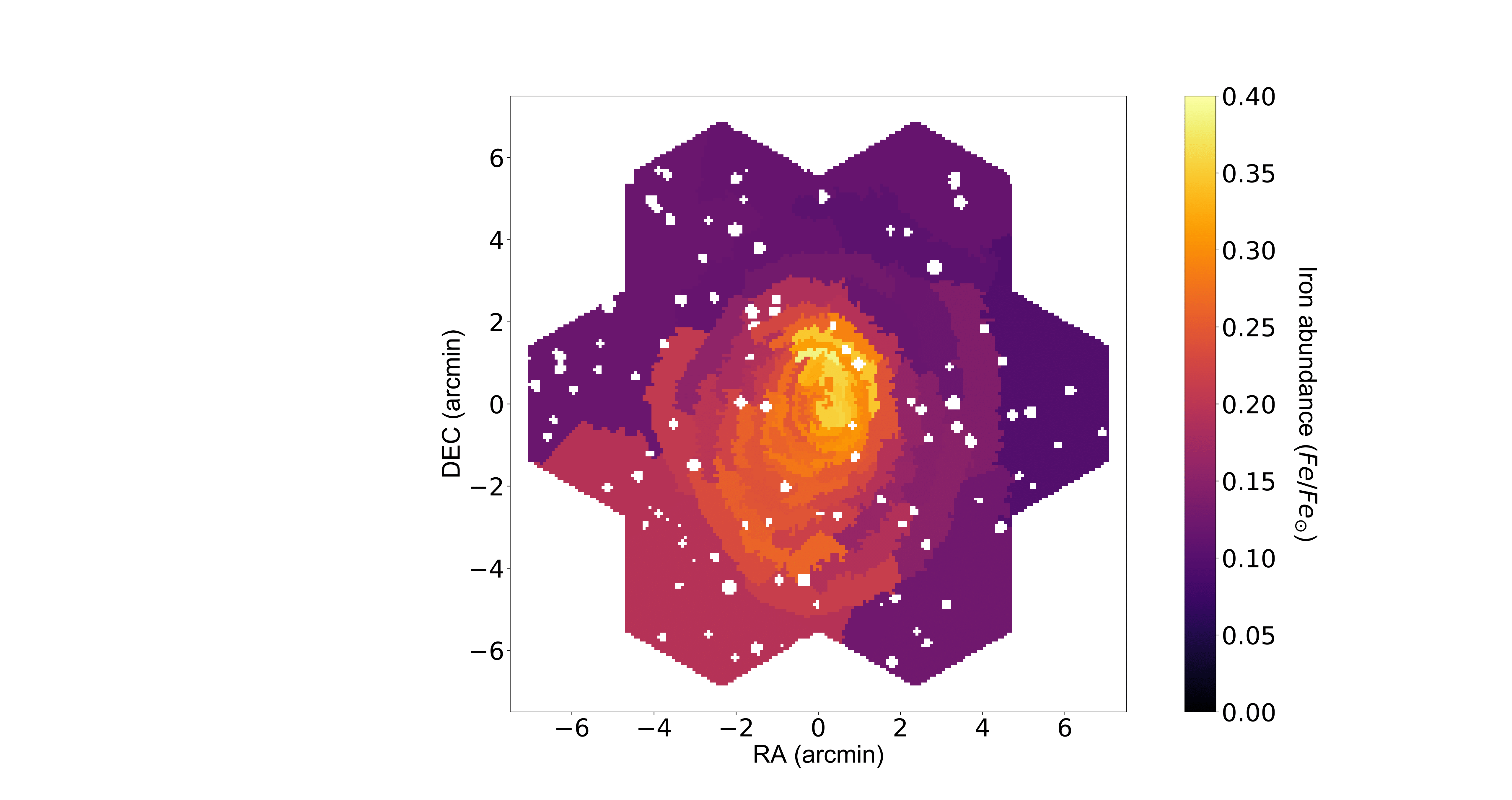}
\vspace{0.6cm}
\includegraphics[width=0.34\textwidth, trim={50cm 2cm 23cm 8cm}, clip]{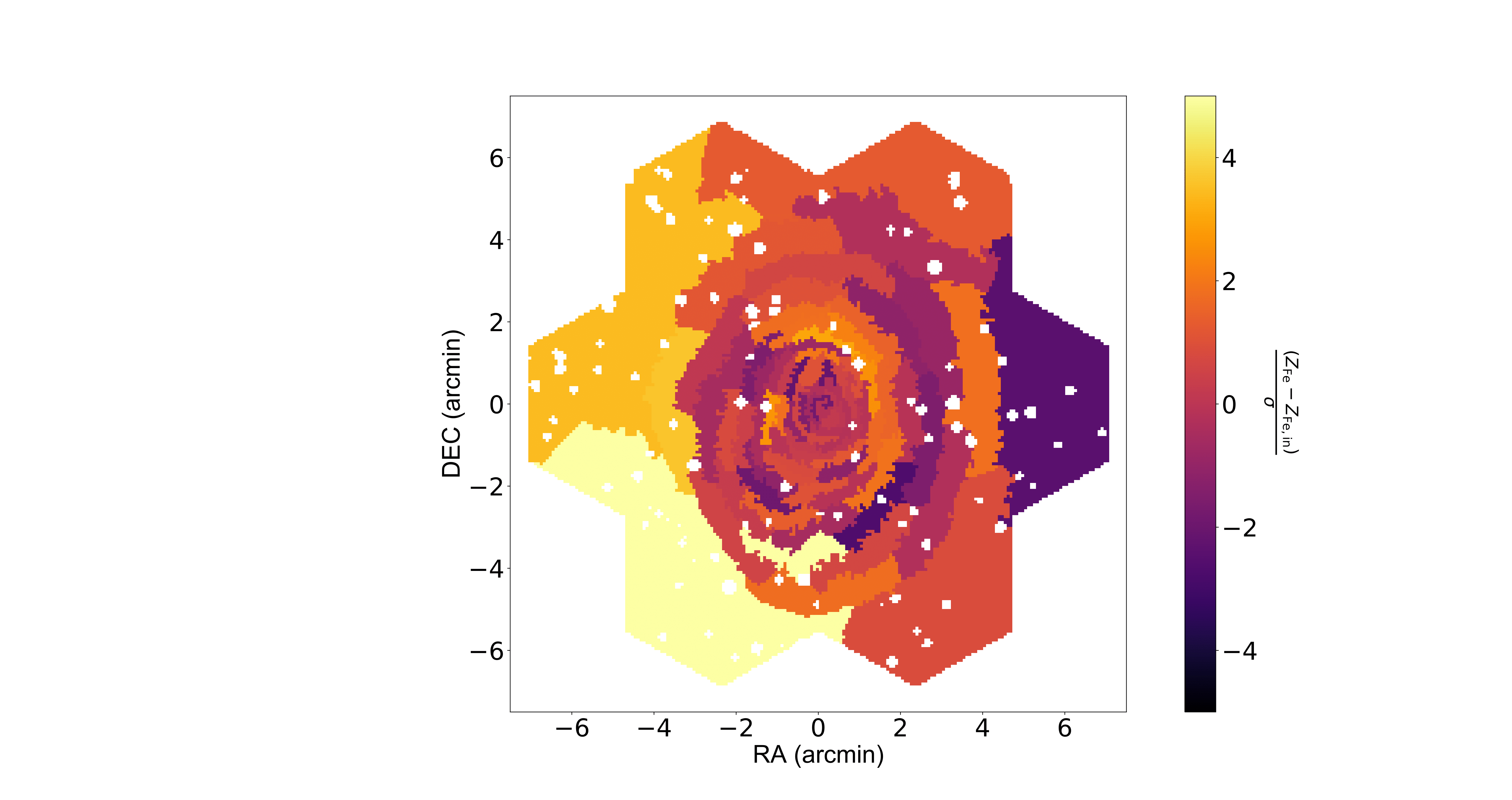}
\includegraphics[width=0.31\textwidth, trim={0 0 0 0.5}, clip]{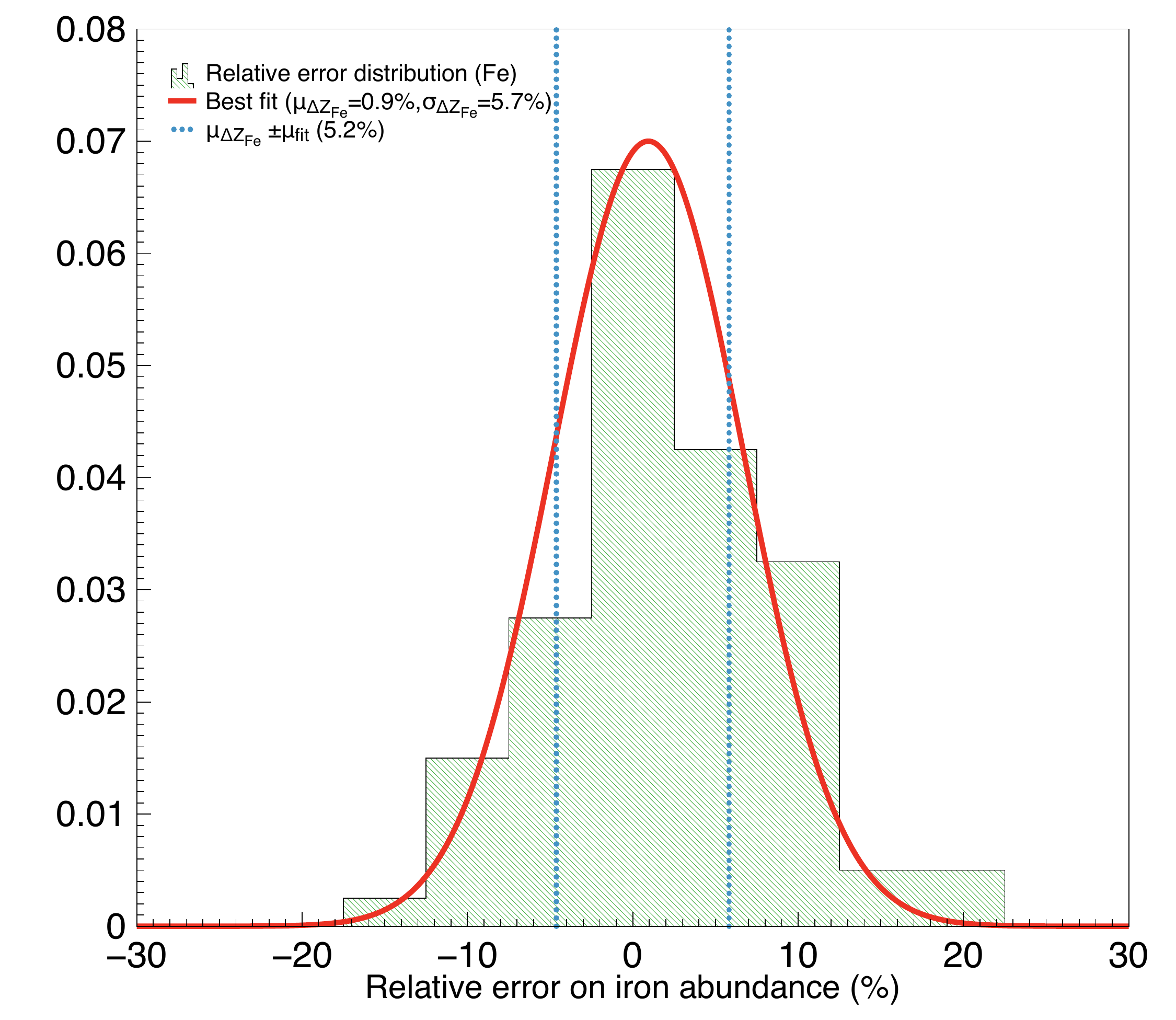}
\caption{\textit{(Left)} Reconstructed ICM parameters maps for C2 at $z \sim 0.1$ using the multi-band fit presented in Sec.~\ref{sub:sys} {\color{black} with a $\textrm{S/N} \sim 300$ ($\sim$ 90\,000~counts per spatial bin)}. \textit{(Middle)} Distribution within the regions $j$ of $(P_j - P_{\rm in, j})/\sigma_j$ indicating the goodness of the fit for each region in terms of $\sigma_{j}$ (see Sec.~\ref{sub:sys}). \textit{(Right)} Relative error distribution across all spectral regions (green histogram). The red solid line pictures the Gaussian best fit. The vertical blue dashed lines are set at the mean value of the fit errors (see Sec.~\ref{sub:sys}). \textit{(From top to bottom)} {\color{black} spectroscopic} temperature $T_{\rm sl}$ (in keV), emission-measure-weighted abundances of oxygen (O), silicon (Si) and iron (Fe) (with respect to solar).}
\label{fig:maps}
\end{figure*}

\afterpage{\clearpage}
\begin{figure*}[p]
\centering
\includegraphics[width=0.335\textwidth, trim={50cm 2cm 23cm 8cm}, clip]{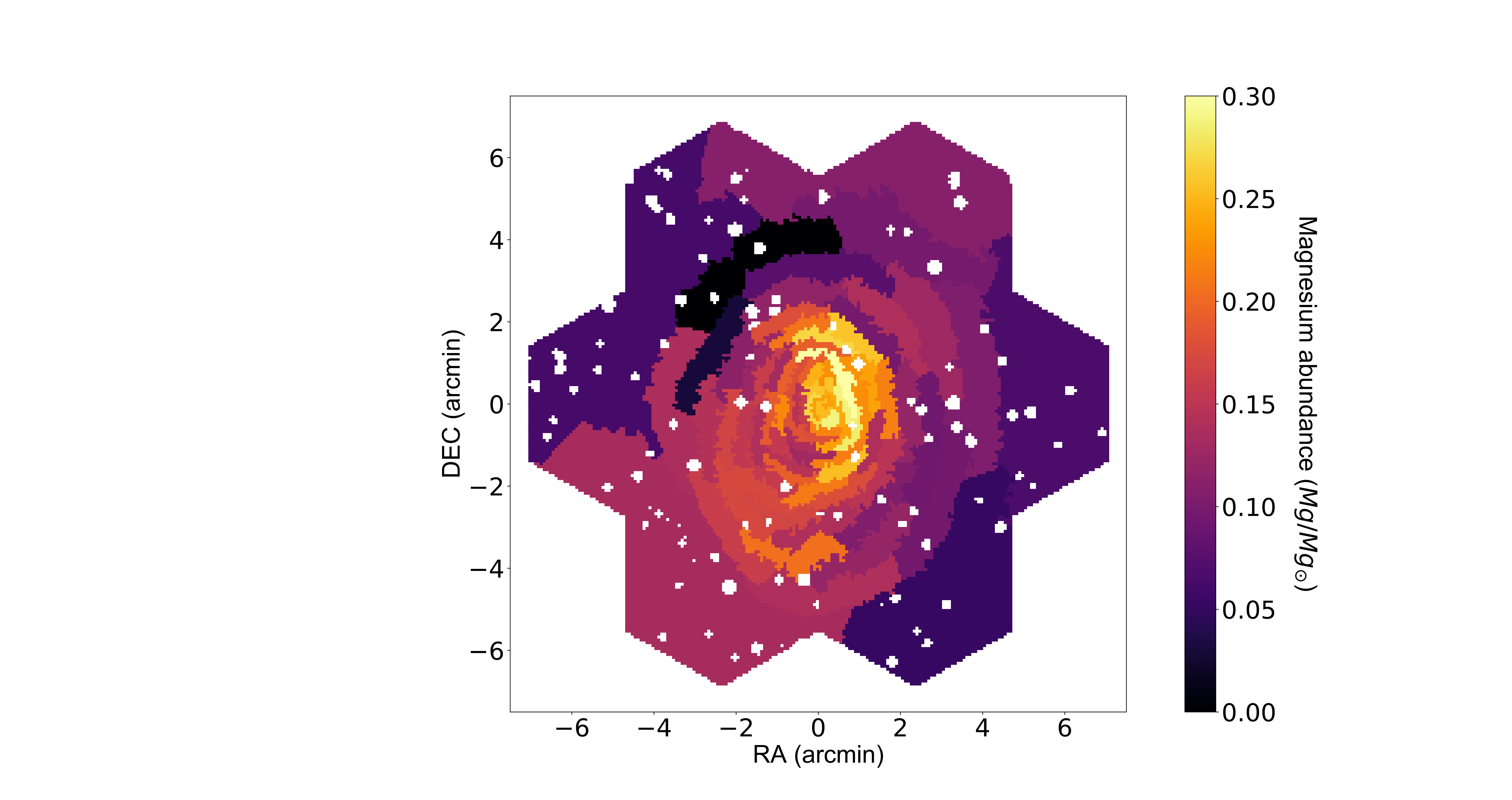}
\vspace{0.6cm}
\includegraphics[width=0.335\textwidth, trim={50cm 2cm 23cm 8cm}, clip]{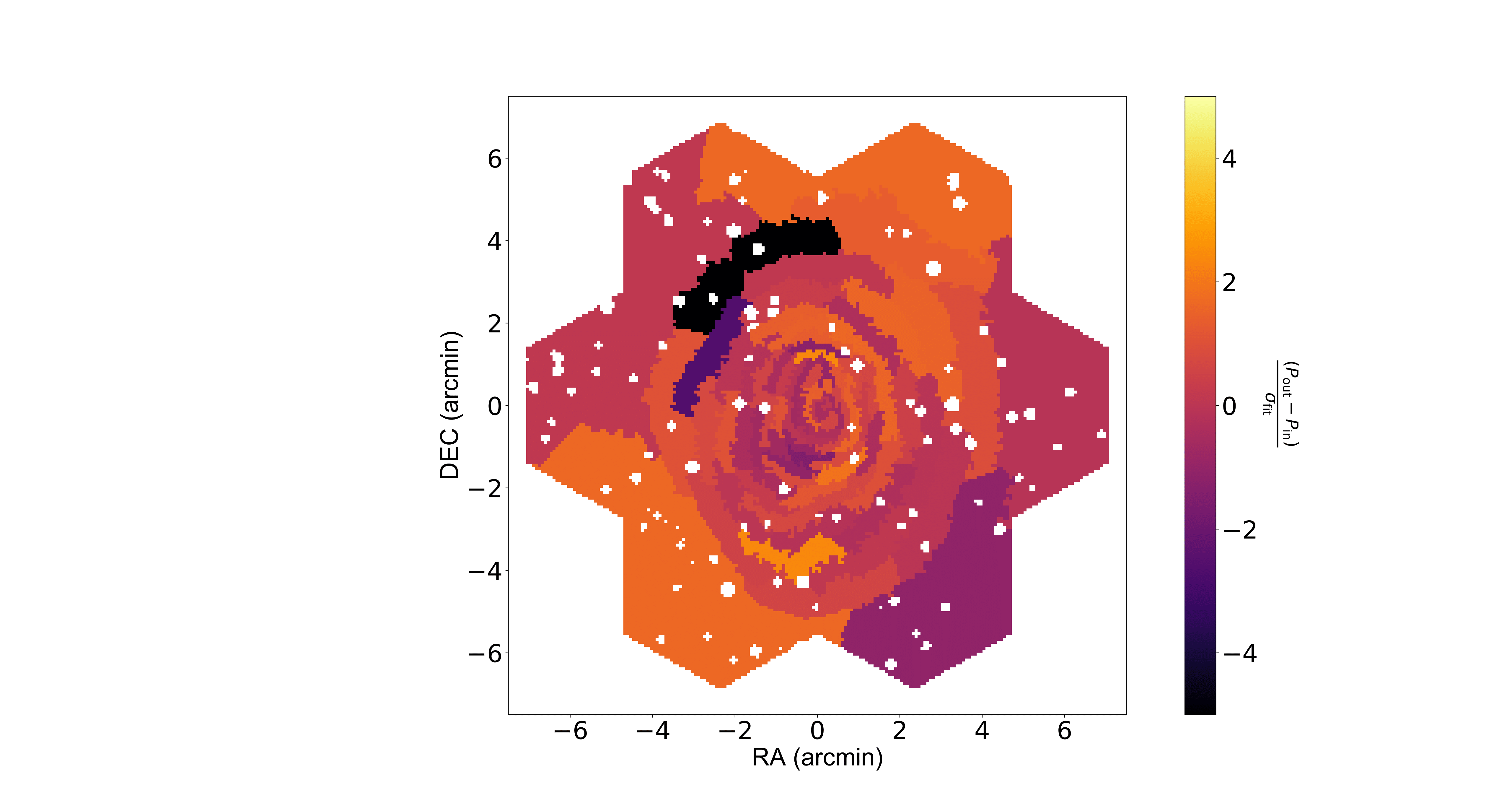}
\includegraphics[width=0.31\textwidth, trim={0 0 0 0.5}, clip]{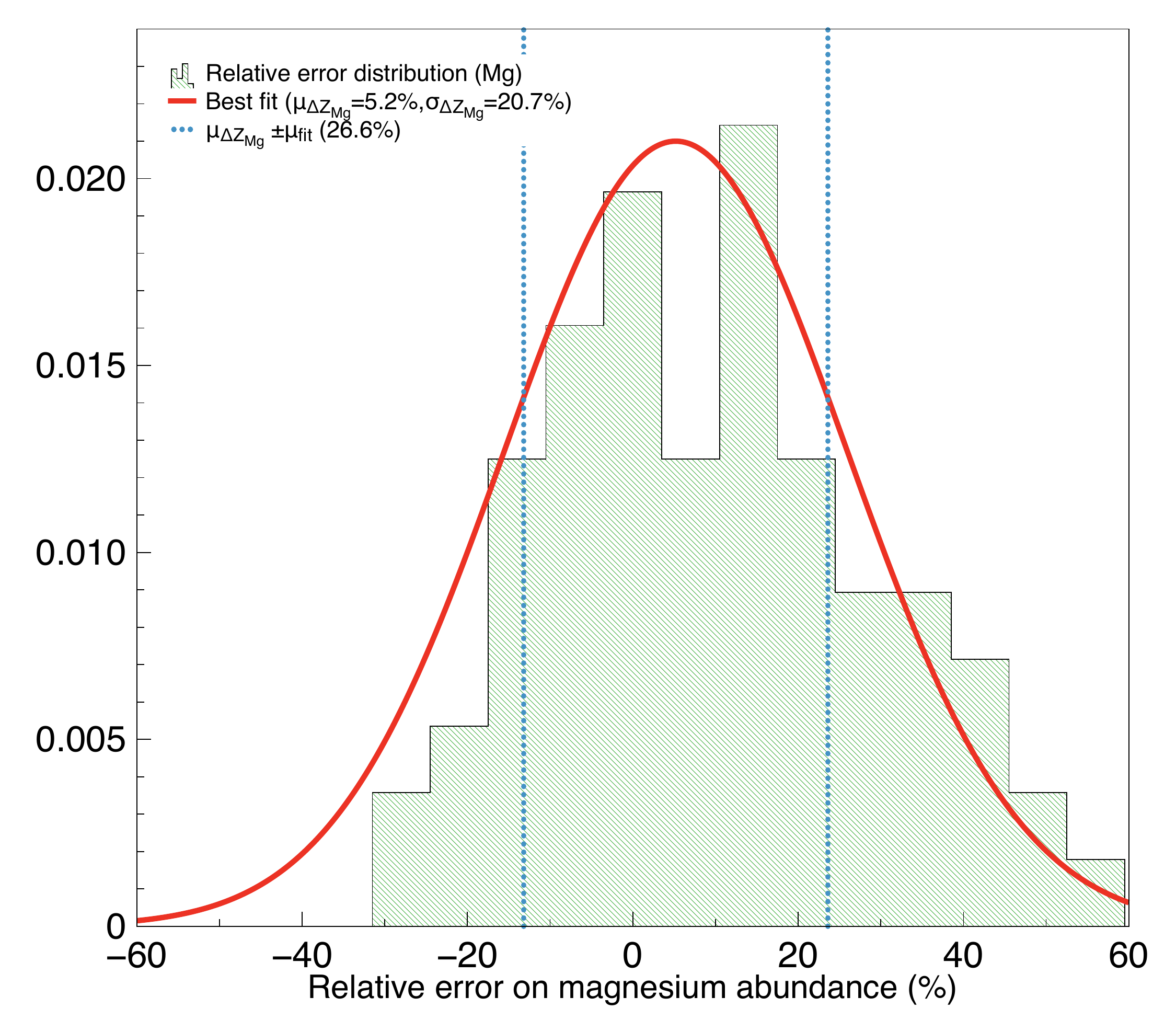}

\includegraphics[width=0.34\textwidth, trim={50cm 2cm 23cm 8cm}, clip]{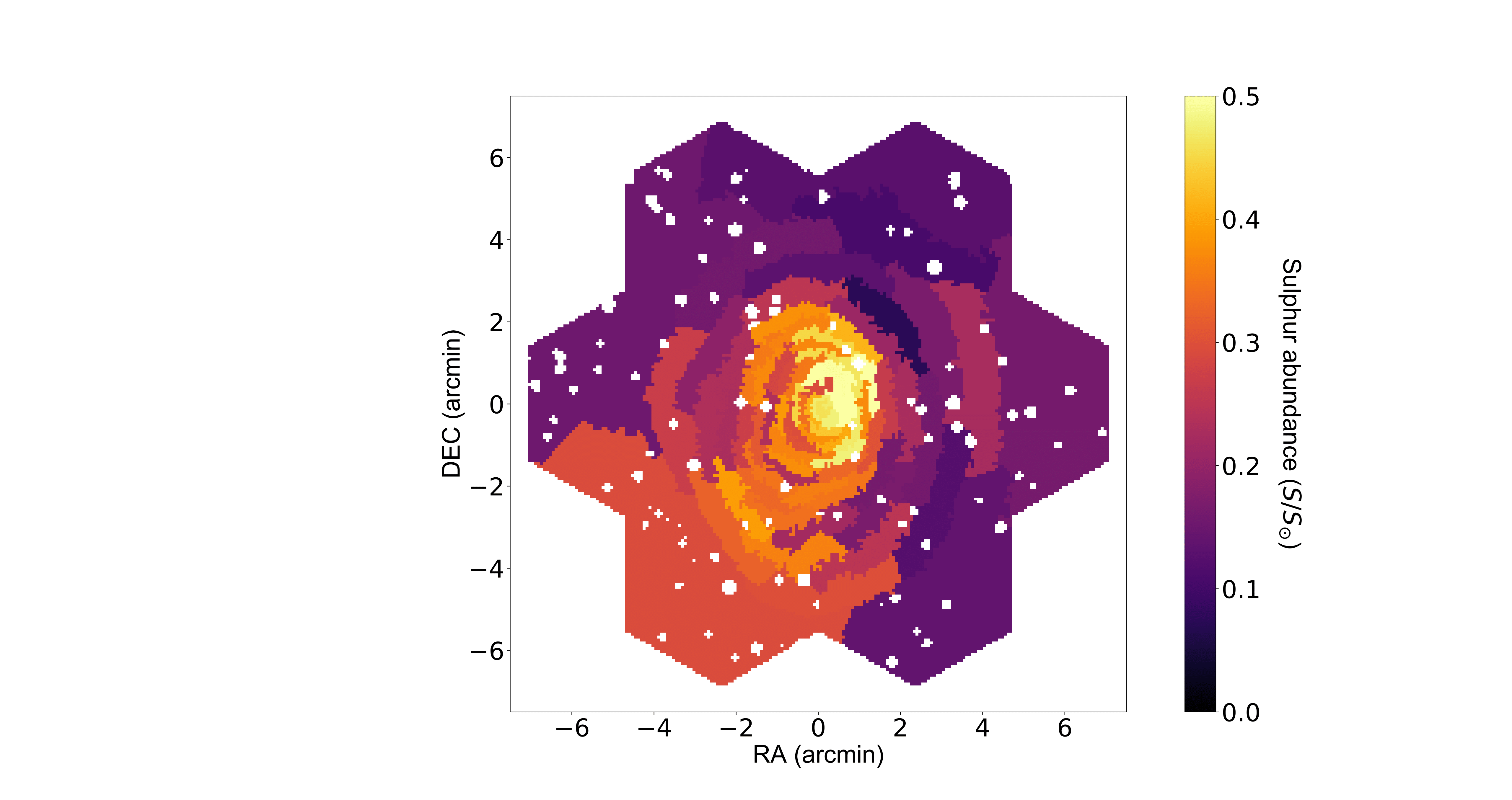}
\vspace{0.6cm}
\includegraphics[width=0.34\textwidth, trim={50cm 2cm 23cm 8cm}, clip]{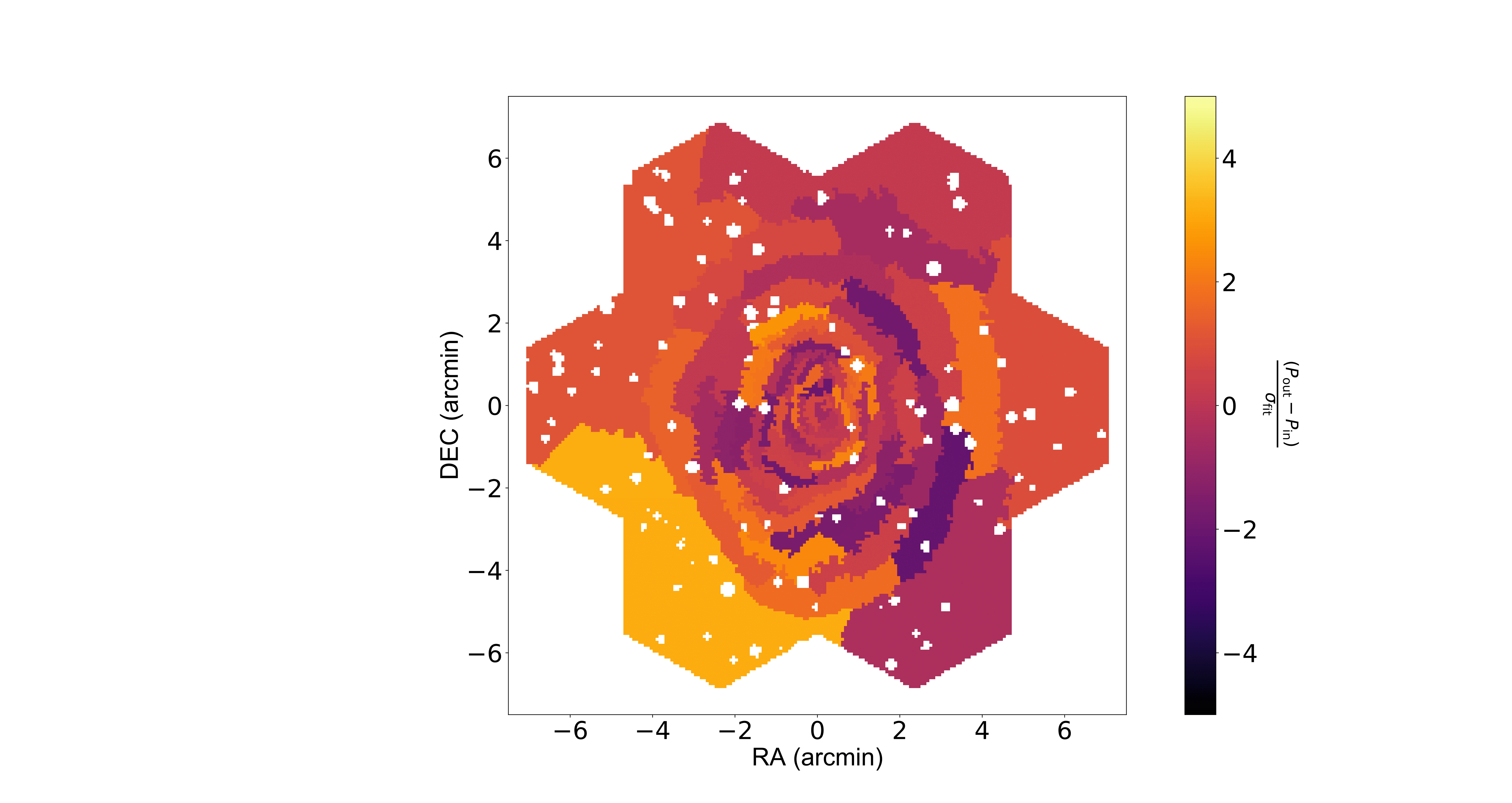}
\includegraphics[width=0.31\textwidth, trim={0 0 0 0.5}, clip]{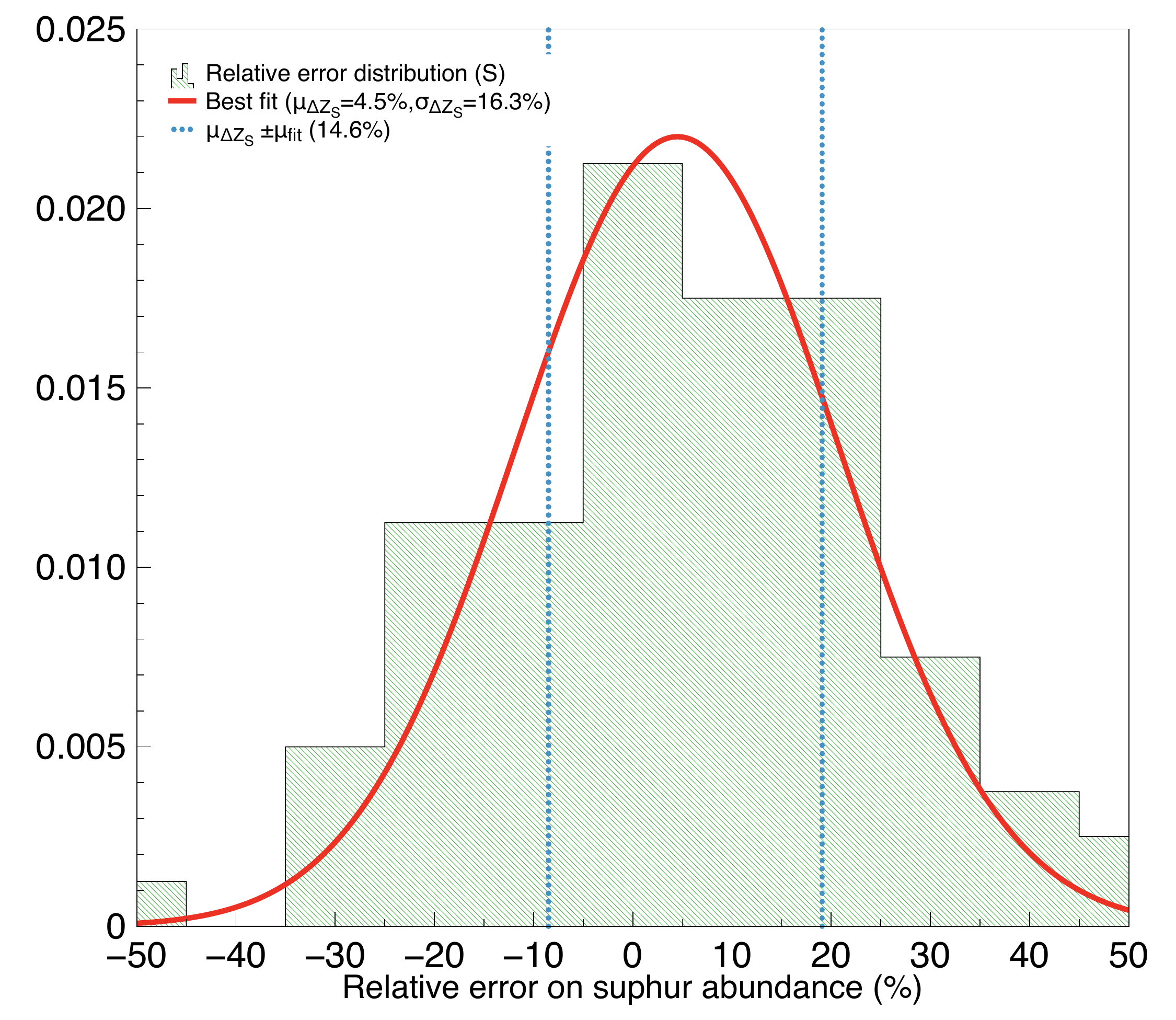}

\includegraphics[width=0.34\textwidth, trim={50cm 2cm 23cm 8cm}, clip]{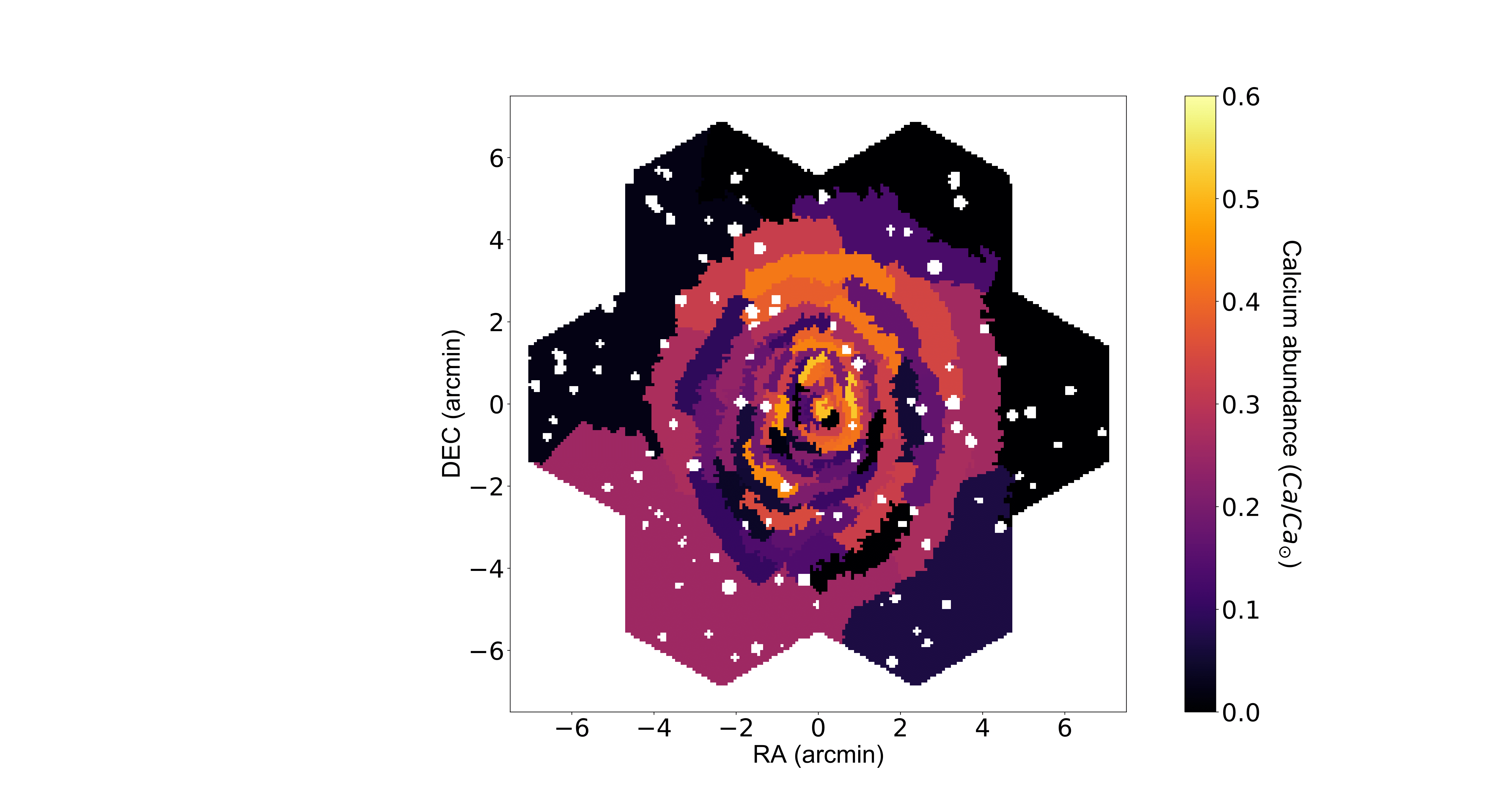}
\vspace{0.6cm}
\includegraphics[width=0.34\textwidth, trim={50cm 2cm 23cm 8cm}, clip]{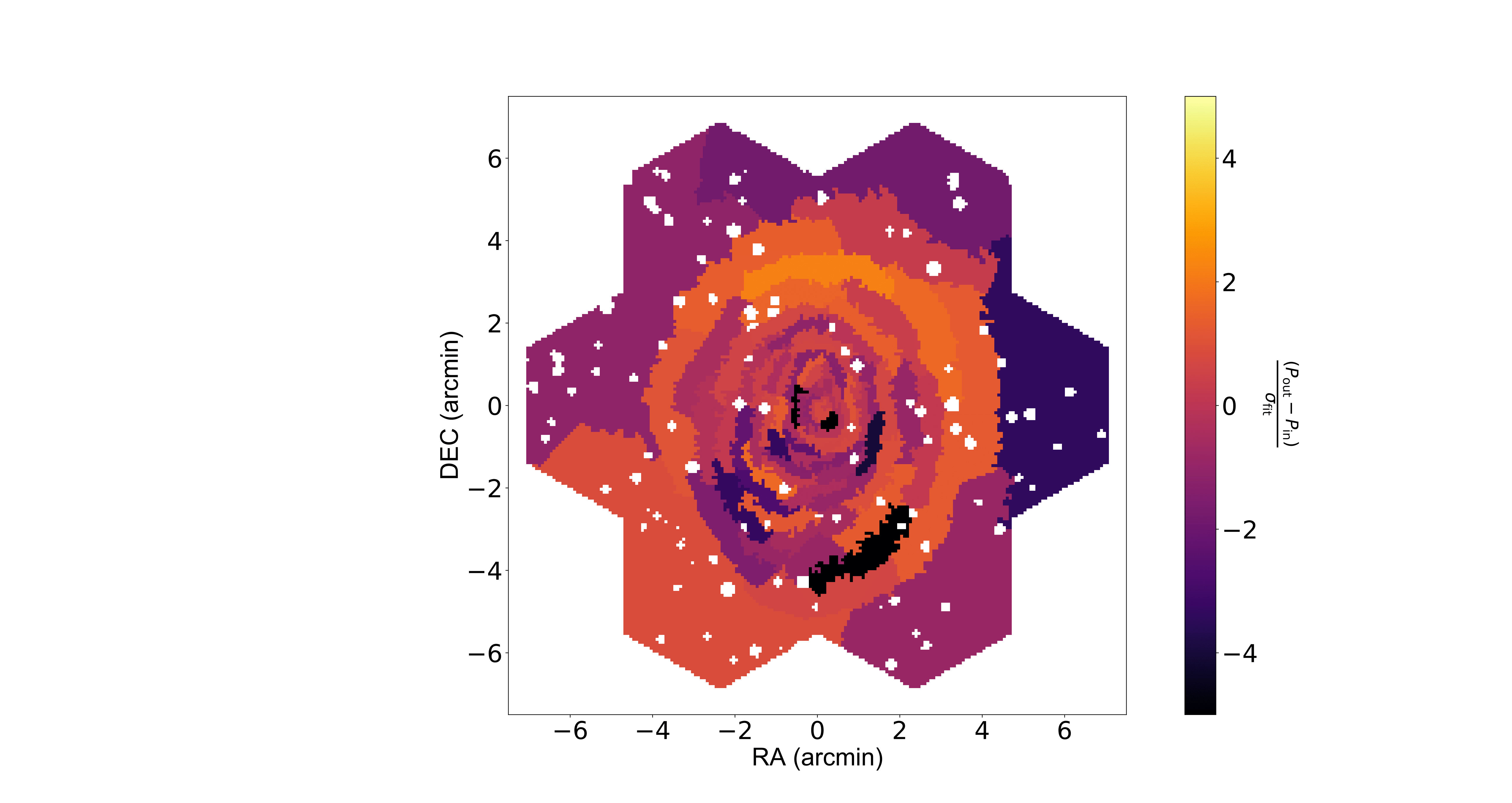}
\includegraphics[width=0.31\textwidth, trim={0 0 0 0.5}, clip]{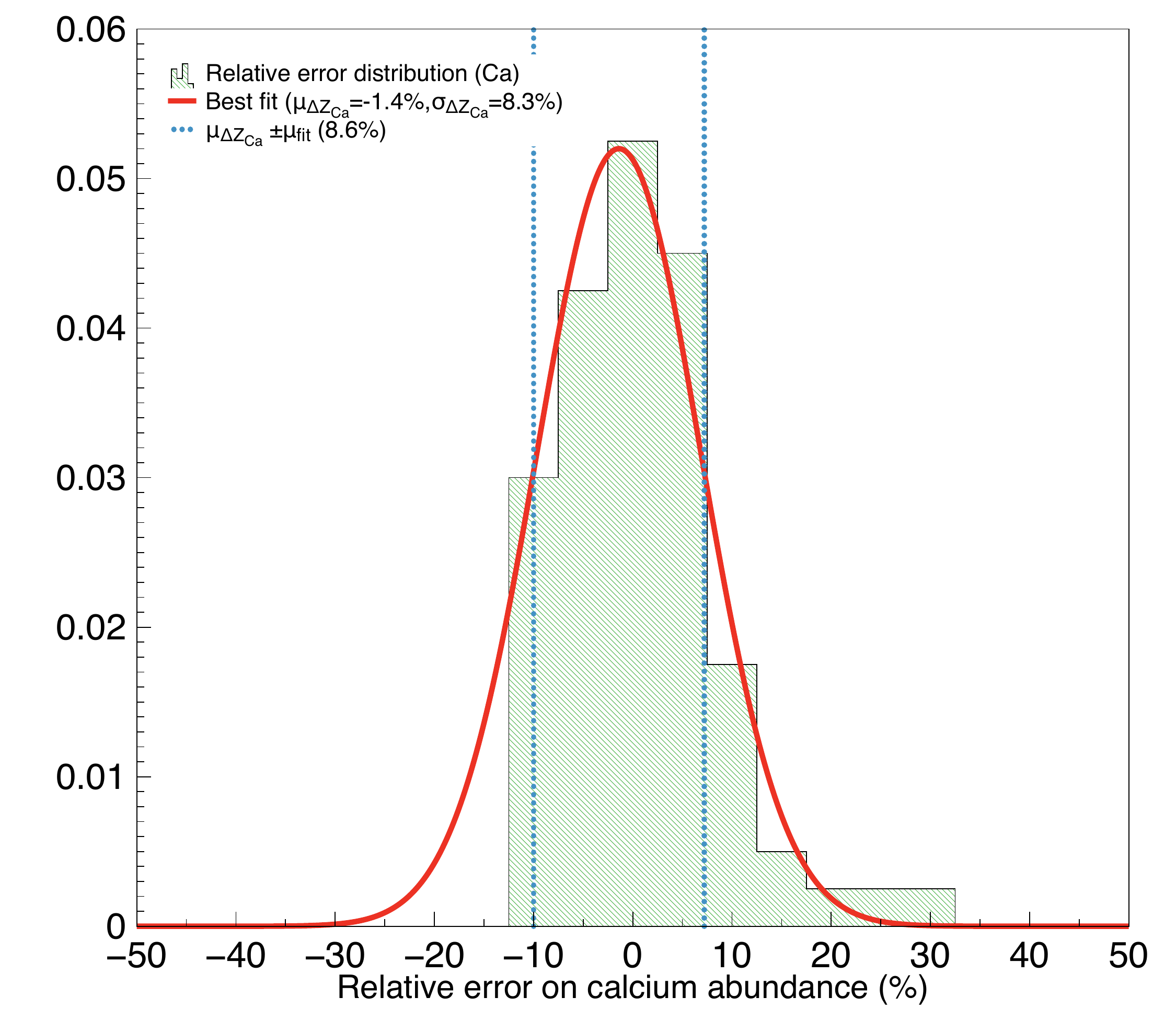}

\includegraphics[width=0.34\textwidth, trim={50cm 2cm 23cm 8cm}, clip]{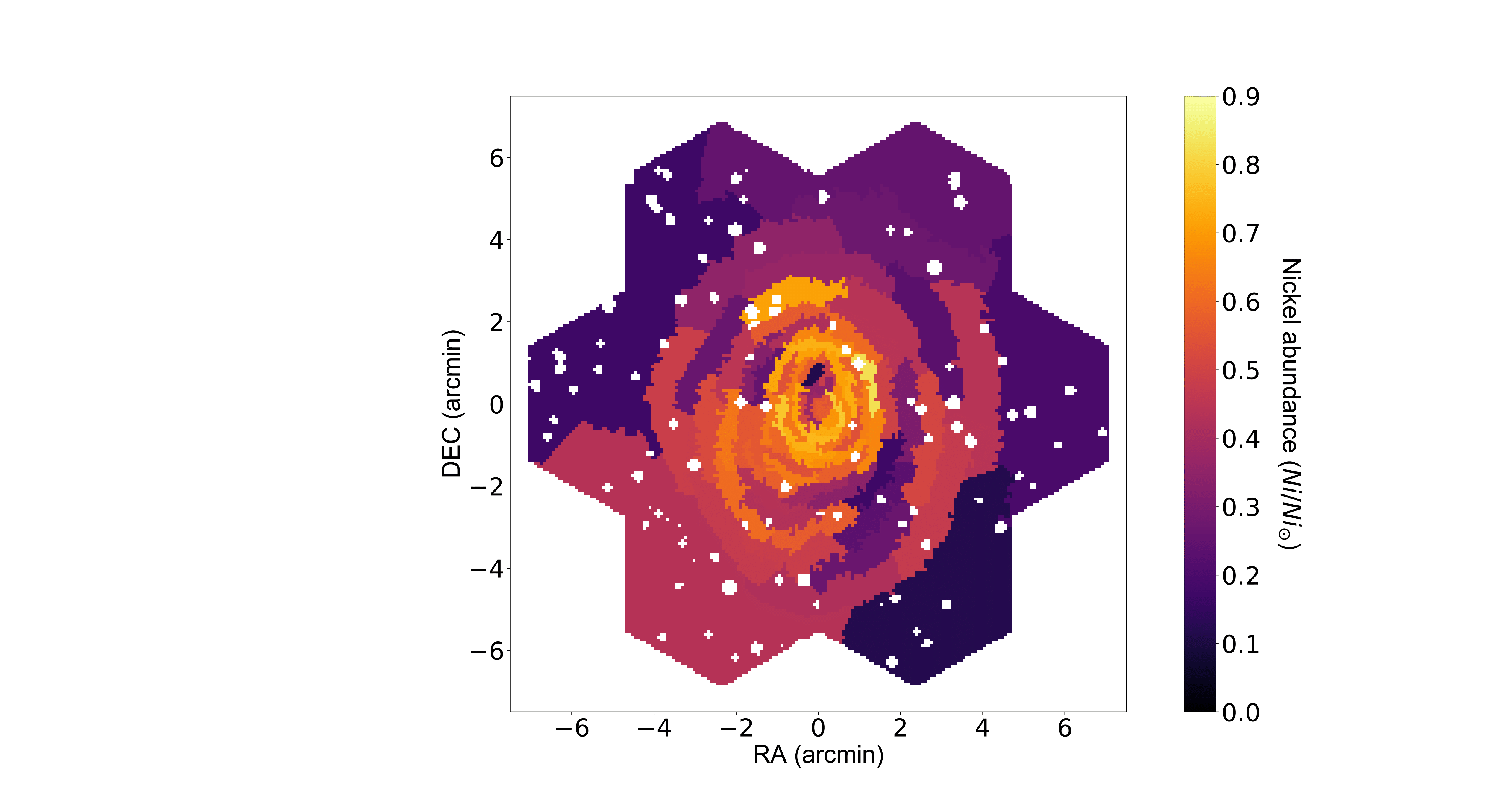}
\vspace{0.6cm}
\includegraphics[width=0.34\textwidth, trim={50cm 2cm 23cm 8cm}, clip]{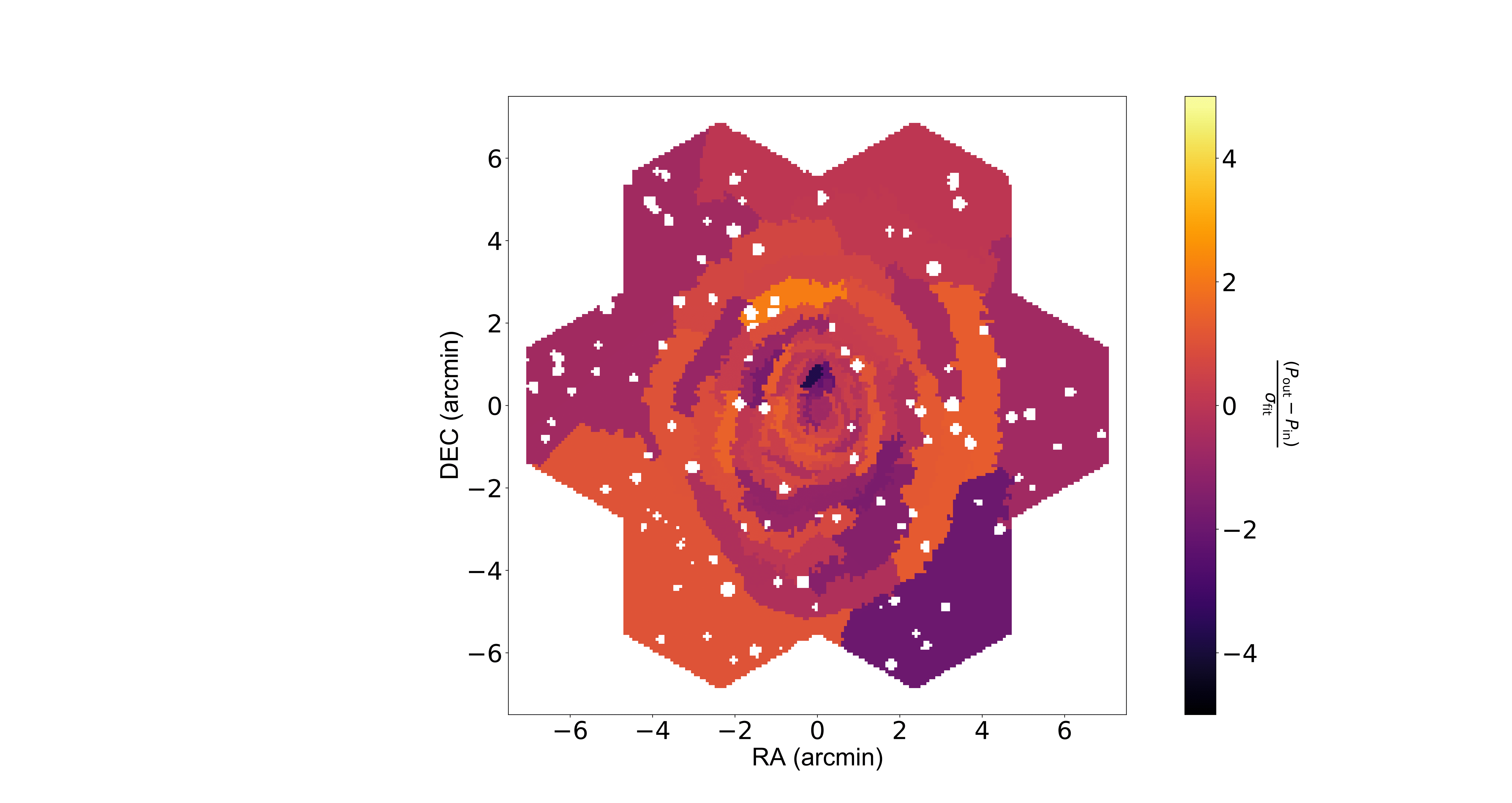}
\includegraphics[width=0.31\textwidth, trim={0 0 0 0.5}, clip]{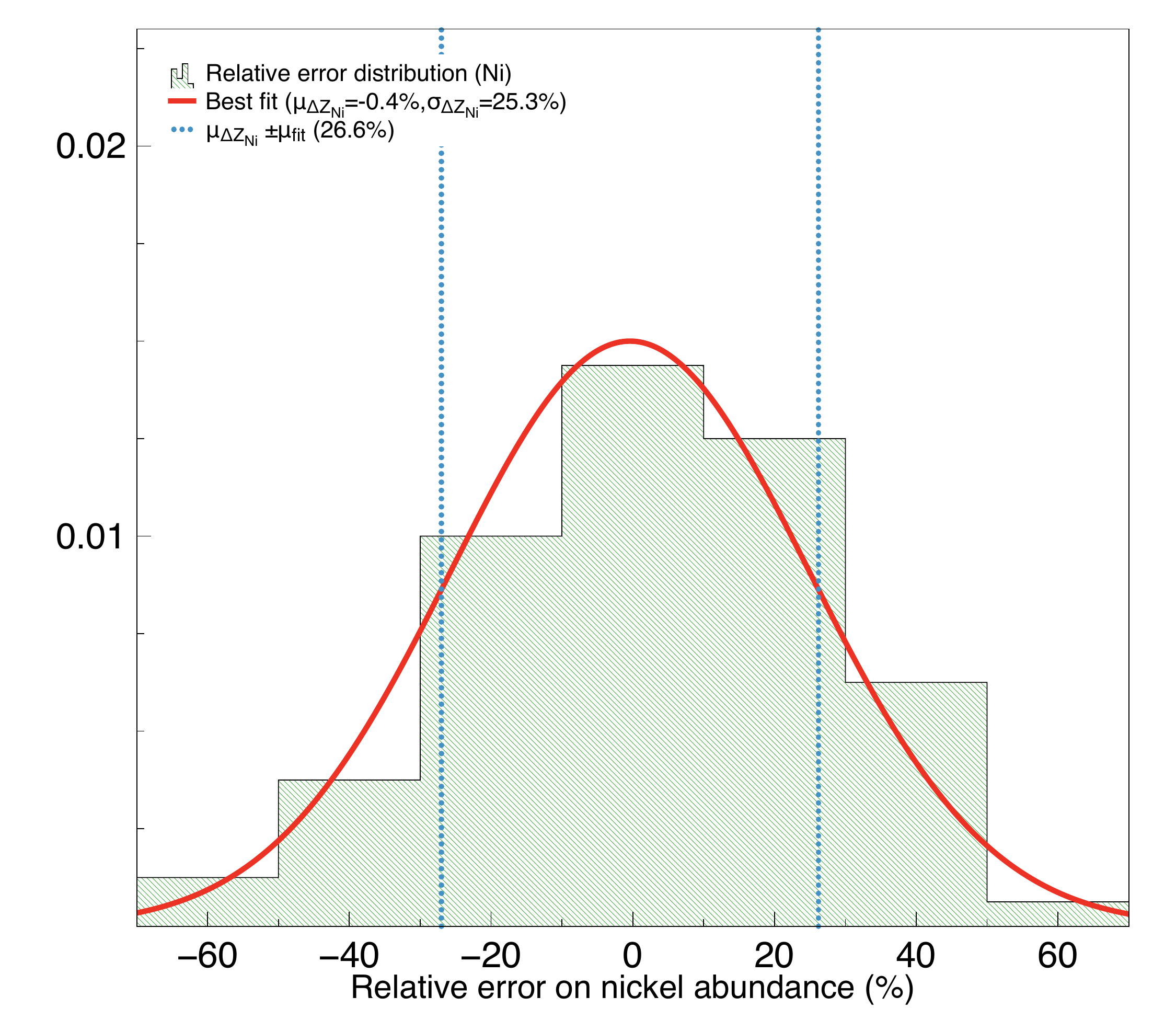}
\caption{Same as Fig.~\ref{fig:maps} \textit{(From top to bottom)} Magnesium, sulphur, calcium and nickel abundances (with respect to solar).}
\label{fig:next}
\end{figure*}

Using the first test, we noticed that the broadband fit initially used introduced biases on the temperature and the abundances recovered for local clusters  ($\sim$\,10\%), when compared to emission-measure-weighted quantities. The use of {\color{black} spectroscopic} temperatures (Eq. \ref{eq:spec}) decreases the biases on temperatures. However, biases on abundances are not accurately corrected by this new approach, underlining a bias in the overall fitting procedure. For our data analysis, we moved therefore to a multi-band fit of the spectrum \citep[following][]{Rasia2008Spectral} including a velocity broadening component to the lines to account for variability and mixing along the line-of-sight (\texttt{bvvapec} model on XSPEC). The fit is performed as follows for each region:
\begin{enumerate}[I)]
\item As a first step, since the cluster sample is relatively hot  ($\geq 4$\,keV in the centre) the temperature is recovered from the high energy band (3.5 -- 12\,keV), then fixed. \vspace{0.3em}
\item Iron metallicity, $Z_{\text{Fe}}$, is recovered by a subsequent broadband fit (i.e., 0.2~--~12\,keV) to estimate the contribution from both the K and L complex, then fixed. \vspace{0.3em}
\item Metallicities of other elements are then computed by fitting specific energy bands (for redshift $z \leq 0.1$): \vspace{0.3em}

\begin{enumerate}[i.)]
\item  Abundances for C, N, O, Ne and Na  are fitted over the 0.2~-~1.2\,keV bandpass.
\item Abundances for  Mg, Si, Ar and Ca are fitted over the 1.2~--~3.5\,keV bandpass.
\item Abundance of Ni is finally fitted over the 0.2~--~12\,keV bandpass. \vspace{0.3em}
\end{enumerate}

\item With all other parameters fixed, redshift, velocity broadening and normalisation are recovered with a broadband fit.
\end{enumerate}

This multi-band post-processing approach is retained in the rest of the paper for local clusters (notably to create Fig.~\ref{fig:maps}~and~\ref{fig:next}), along with the comparison of the fitted output temperature with {\color{black} spectroscopic} input temperature maps. {\color{black} This technique is used under the caveat that despite some fitted parameters are fixed, errors are propagated correctly throughout the fit. Although not perfectly true, this effect remains very limited with respect to the total systematics and to our level of statistics, ensuring a safe application of this method.} For all the fitted parameters, we computed the mean, $\mu_{\rm \Delta P}$, and standard deviation, $\sigma_{\Delta P}$, of the relative errors between the output and input values to the mean of the XSPEC fit errors, $\mu_{\rm fit}$ (see right panels of Fig.~\ref{fig:maps} and \ref{fig:next} for an illustration in the case of cluster C2) and by computing $\chi_{\rm red}^2$ for each parameter. Results of this comparison for the main parameters are given in Table~\ref{table:errors}. In all cases, the mean fit error is consistent with the standard deviation of the error distribution $\sigma_{\rm \Delta P}$, thus excluding large systematic effects. However, in spite of these changes, small biases of the order of a few percent ($\lessapprox 5$\%) are still visible. This is particularly true for the normalisation and the abundances of low mass elements (e.g. O, Si), in which an underestimate of the normalisation directly results in an overestimate of these abundances (see also in Fig.~\ref{fig:maps}). All these systematics are however well inside the statistical deviations. Using the second test, we notice that some of the $\chi_{\rm red}^2$ are not consistently recovered, especially for the normalisation. This can be partially explained by the presence of some outlier regions in the fit, which strongly affect this computation. When the $N_{\rm bad}$ outliers regions for which the parameter is outside a $3 \sigma_{\rm fit}$ level, are removed, the new reduced chi-squared, $\tilde{\chi}_{\rm red}^2$, shows more consistency (Table~\ref{table:errors}), indicating a good agreement between the fitted maps and the input distributions. 

Small errors remaining after post-processing can be attributed to some degeneracy between fitted parameters ($\geq 15$ here) and the spatial binning, which regroups in a relatively large region (a few arcmin$^2$ in the outskirts) many different physical structures (see also the discussion in Appendix \ref{app:2}). The choice of a single temperature model could also introduce some bias accounting for the complex structure over the line-of-sight. Two- or multi-temperature models could bring slight improvements, especially towards the center of the object. Such schemes were investigated over single regions but showed no significant improvement over the entire pointings. The assessment of the improvements introduced by a multi-temperature scheme over the entire temperature distribution as well as the use of line-ratio techniques \citep{Hitomi2017Temperature} would be the object of future improvements of our pipeline, and shall be addressed in future studies. Finally, we also found that small effects related to statistics, to the XSPEC fitting procedure and to the weighting schemes of the input maps partially explain these residual deviations (refer to Appendix\ref{app:2} for a more ample and detailed discussion on the pipeline validation). 

\begin{table} [t]
\centering 
\caption{Values of the mean $\mu_{\Delta P}$ and standard deviation $\sigma_{\Delta P}$ of the best fit of the relative error distributions for the main physical parameters in cluster C2, over 80 spatial regions, with the average XSPEC error of the fits $\mu_{\rm fit}$. The corresponding values of $\chi_{\rm red}^2$ and the corrected $\tilde{\chi}_{\rm red}^2$ are also given, along with the number of outliers regions $N_{\rm bad}$ removed to compute $\tilde{\chi}_{\rm red}^2$.}
\begin{tabular}{c c c c c c c} 
\hline\hline \\[-0.8em] 
 & $\mu_{\Delta P}$ (\%) & $\sigma_{\Delta P}$ (\%)  & $\mu_{\rm fit}$ (\%) & $\chi_{\rm red}^2$ & $\tilde{\chi}_{\rm red}^2$ & $N_{\rm bad}$ \\[0.2em] 
\hline\\[-0.8em] 
$T_{\rm sl}$ & -0.71 & 2.60 & 1.64 & 3.93 & 1.38 & 4\\
$z$ & 0.00 & 0.04 & 0.05 & 1.54 & 1.09 & 3 \\
O & 4.23 & 16.0 & 13.4 & 1.50 & 1.50 & 2 \\
Mg & 5.22 & 20.7 & 18.4 & 2.84 & 1.01 & 1 \\
Si & 3.22 & 11.4 & 10.6 & 1.78 & 1.37 & 3  \\
S & 4.48 & 16.3 & 14.6 & 1.52 & 1.41 & 1 \\
Ca & -1.42 & 8.27 & 8.62 & 2.69 & 1.06 & 3 \\
Fe & 0.94 & 5.73 & 5.22 & 2.71 & 1.49 & 2 \\
Ni & 0.39 & 25.3 & 26.6 & 0.99 & 0.92 & 1 \\
$\mathcal{N}$ & -6.5 & 3.68 & 3.74 & 4.11 & 2.83 & 7 \\
\hline \hline
\end{tabular}
\label{table:errors} 
\end{table}

\section{Properties of the ICM of local clusters}
\label{sec:icm}

We present in this section the results on the ICM properties recovered by the \xifu , starting from the interpretation of the raw output maps to larger studies involving the entire cluster sample.

\subsection{Physical parameters maps}

We show in Fig.~\ref{fig:maps} the reconstructed maps for cluster C2 at $z \sim 0.1$ for the {\color{black} spectroscopic} temperature $T_{\rm sl}$ and the abundances of Fe, Si, and O. Those for Mg, S, Ca and Ni are shown on Fig.~\ref{fig:next}. Similar maps for the other three clusters are provided  in Appendix~\ref{app:3}. Beyond the recovery of abundances, the physical parameter maps and their combination provide a wealth of information on the dynamics of the cluster. For instance, we see from the temperature map of cluster C2 the presence of a hot bubble on the western part of the cluster and a cold arc in the south-eastern region. Interestingly, we also notice the correlation between the presence of low-mass elements (e.g., O, Si) and the temperature of the ICM. Several clumps and small groups are also visible on some of the clusters and cluster C3 exhibits a merging activity with a very bright central object. After post-processing, the redshift of each region is also recovered with excellent accuracy from the fit over a large number of lines and spectral features. The redshift map, once converted into velocities using the mean redshift of the cluster, provides a projected map of the bulk motions within the ICM. Likewise, the velocity broadening of the lines is recovered in the fit. Measuring both bulk motion and turbulent velocities through line shifts and line broadening respectively, is another main scientific objective for \xifu. This goes however beyond the scope of this paper and we redirect the reader to \citet{Roncarelli2018XIFU} for an illustration.
\begin{figure*}[t]
\centering
\includegraphics[width=0.495\textwidth]{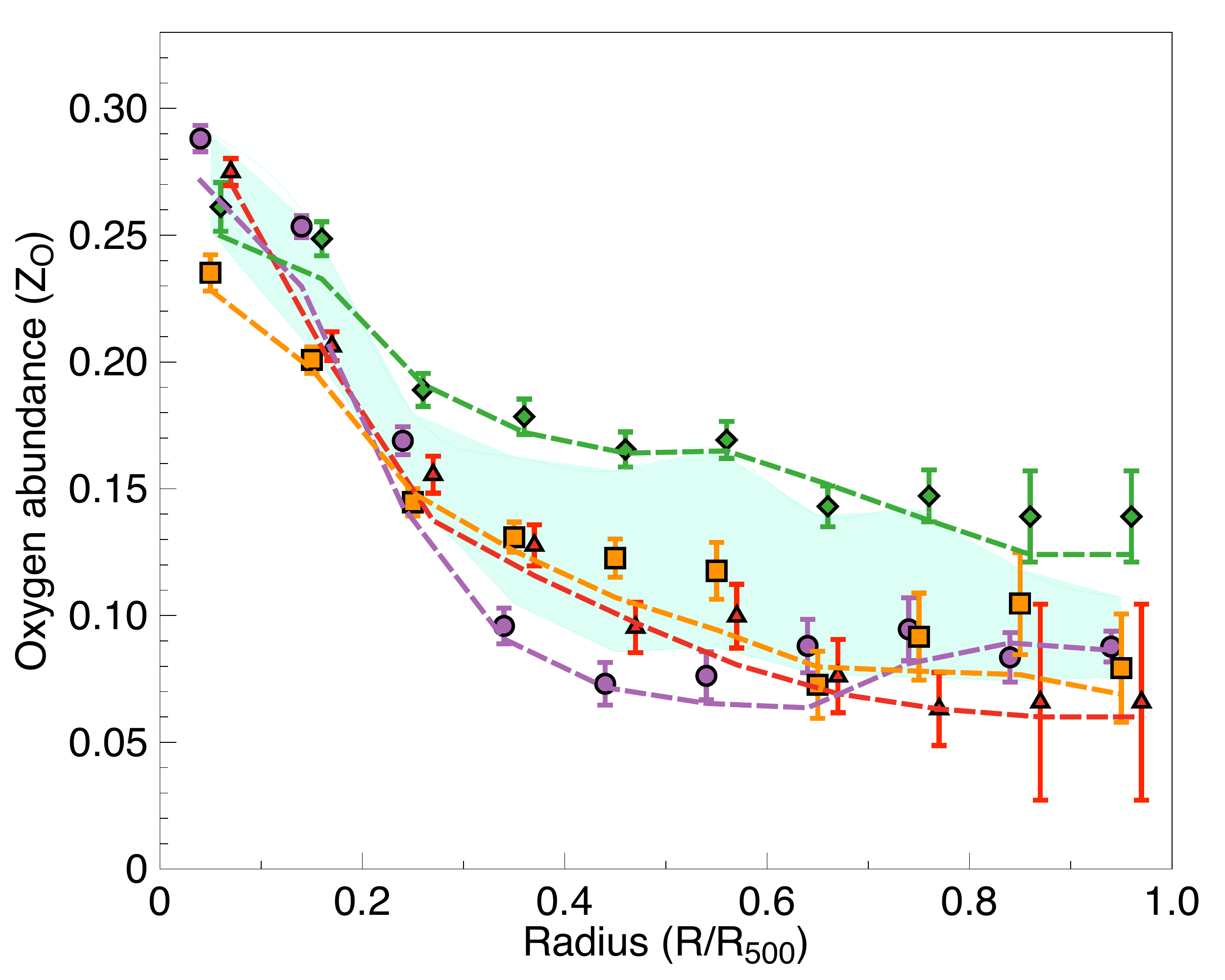}
\includegraphics[width=0.495\textwidth]{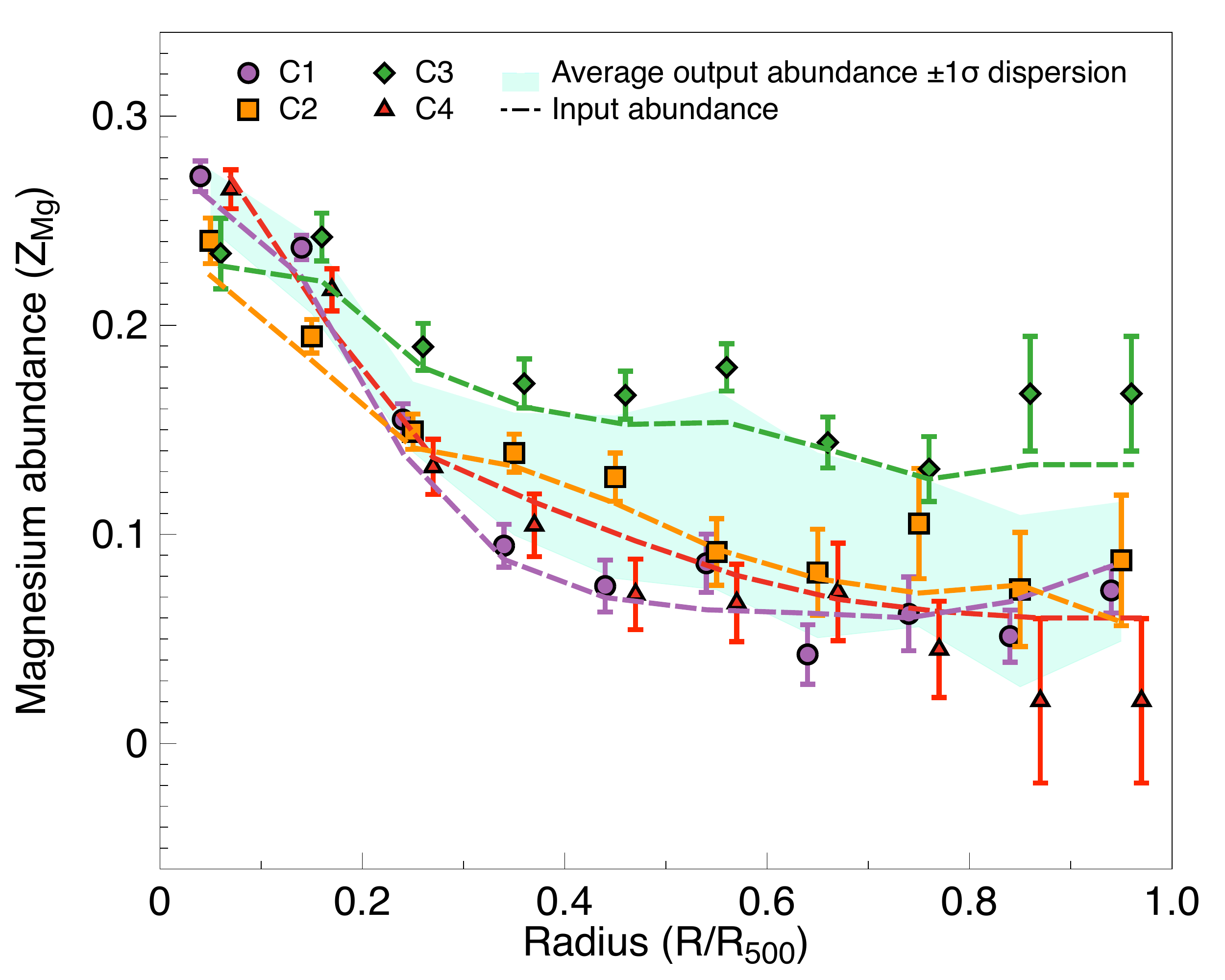}
\includegraphics[width=0.495\textwidth]{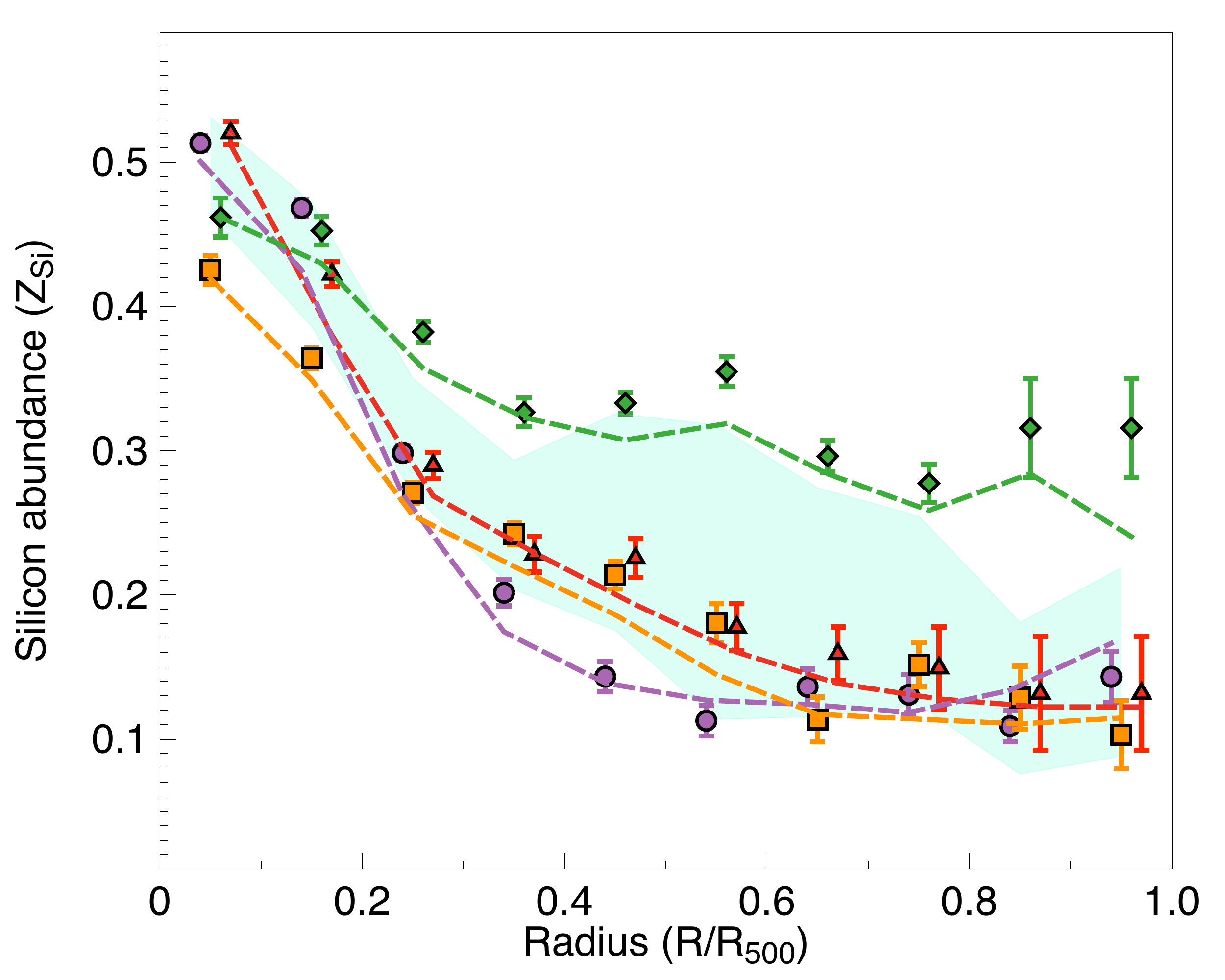}
\includegraphics[width=0.495\textwidth]{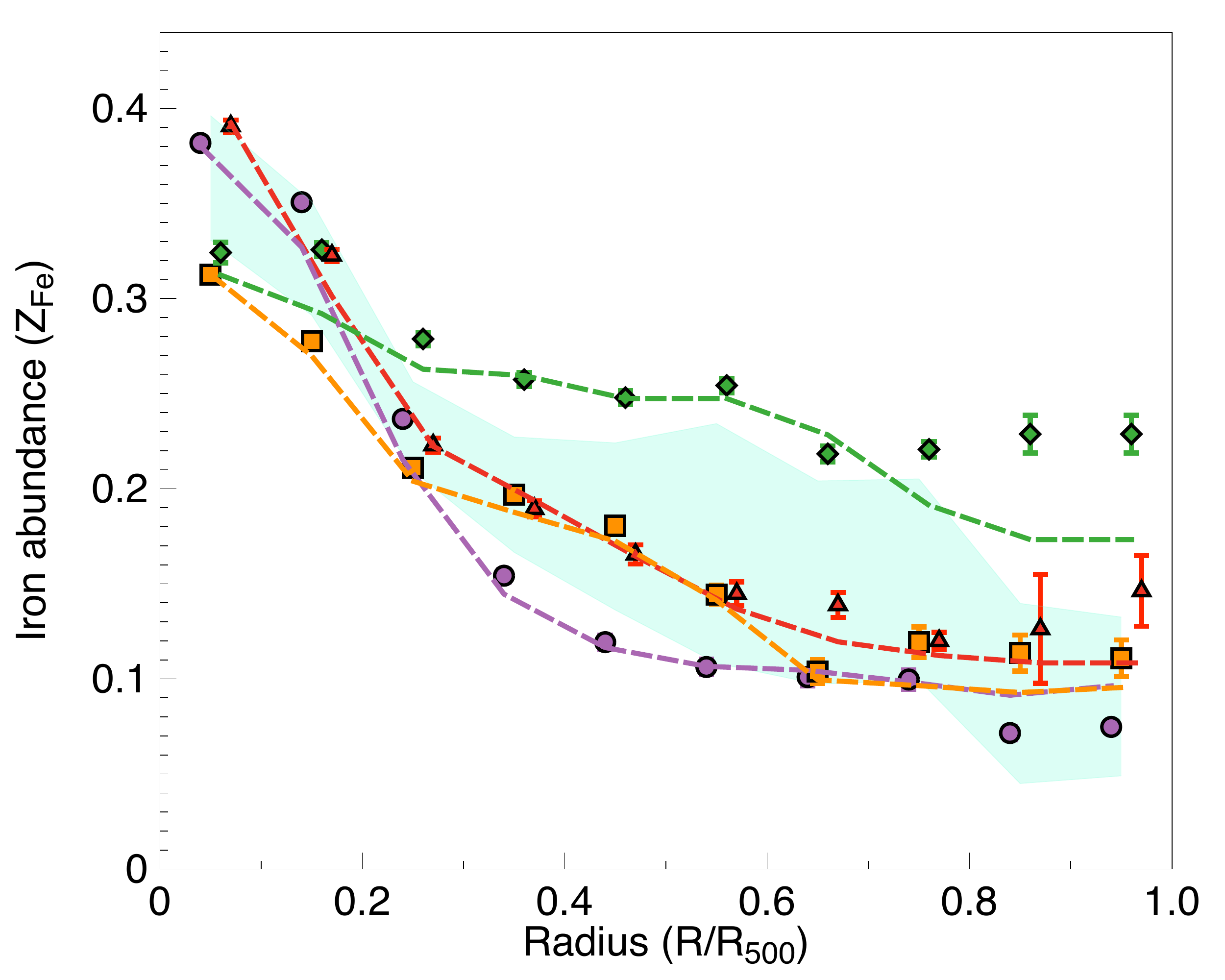}
\caption{Best fit values of the metallicity as a function of radius, up to $R_{500}$ ($0.1~R_{500}$ bins) for the entire sample (C1 - purple dots, C2 - orange squares, C3 - green diamonds, C4 - red triangles). \textit{(From top left to bottom right)} Oxygen, magnesium, silicon and iron abundances with respect to solar. The dashed lines represent instead the profile of the emission-measure-weighted input abundances using the same colors. The cyan shaded envelope represents the $\pm 1 \sigma$ dispersion of the recovered output metallicity for the entire sample. Points are slightly shifted for clarity.}
\label{fig:metals}
\end{figure*}

\subsection{Metallicity profiles of the ICM}
\label{subsec:metals}

The hierarchical formation of galaxy clusters along with processes of production and dispersion of chemical elements within stars and galaxies should lead to self-similar global abundance profiles. The value of the metallicity in the outskirts will depend notably on the enrichment of the intergalactic medium prior to the halo formation \citep[see, e.g.,][]{Biffi2017Clusters}. 

To investigate the capabilities of the \xifu\ in determining abundance profiles over the cluster data set, we computed the radial profiles at $z\sim 0.1$ for some of the main chemical elements, i.e., O, Mg, Si, and Fe (Fig. \ref{fig:metals}). We find that the values of abundance profiles are consistent across the sample, showing a peak near the centre (i.e., up to $0.1$\,$R_{500}$) and a decrease towards a constant value between $\sim 0.1/0.2$\,$Z_{\odot}$ out to $R_{500}$. Cluster C3 shows however values of metallicities systematically higher than the others, which could be caused by the ongoing merging activity visible in Fig.~\ref{fig:clusters} (\textit{Bottom left}) and possibly due to AGN feedback. This dynamic activity also creates a more difficult line-of-sight distribution of the parameters, making the XSPEC fit less accurate. This is particularly the case near the outskirts, where background contributions become relevant and regions are large, thus increasing the deviations from the input maps (especially for iron).  These measured profiles and notably their constant metallicity value in the outskirts suggest, as discussed in \citet{Biffi2017Clusters, Truong2018Simu}, that the enrichment of the cluster in the hydrodynamical simulations pre-dates the large infall towards the central object and is mainly determined by the early enrichment mechanisms of the Universe. Other independent hydrodynamical simulations \citep{Vogelsberger2018Distribution} and recent observational results \citep[e.g.][]{Ezer2017Abund,Mernier2017Radial,Simionescu18Virgo} also argue in favour of this paradigm. 

Overall, the recovered metallicities are consistent with the input metallicity profiles within $3\sigma$ (most well within $1\sigma$), showing the power of the \xifu\ in recovering the properties of the ICM even for typical 100\,ks exposures. This analysis can be compared to a similar study performed by simulating 200\,ks cluster observations with \textit{XMM-Newton}/EPIC MOS1 and MOS2 for the same chemical species (see \citealt{Rasia2008Spectral} for more details, notably Fig.~4) or to current observational data using the CHEERS catalogue \citep{Mernier2017Radial}. {\color{black} For typical exposure times, the \xifu\ will provide accurate measurements of the main metallic content of the ICM, enough to reduce the uncertainties on current observations, even for less abundant elements.} In addition to the individual profiles, the overall dispersion across the sample (Fig.~\ref{fig:metals}) is consistent with the average emission-measure-weighted input distribution. {\color{black} However, the sample considered here is relatively small, and the scatter of our results remains significant.} Namely, we see that the dynamic behaviour of  clusters C3 affects the overall scatter of the sample, otherwise similar for the other three objects (C1, C2 and C4). {\color{black} The values of iron abundance in the outskirts found in this study (between 0.1/0.2\,$Z_{\odot}$) are somewhat lower than measured iron abundances in the outskirts, which range typically around 0.2\,$Z_{\odot}$ \citep{Werner2013Enrich,Mantz2017Enrich,Urban2017Metallicity}. Values remain however consistent with the hydrodynamical inputs, demonstrating that the \xifu\ is able to recover the intrinsic physical parameters used for the clusters. The projection scheme adopted here also provides lower results than e.g., emission-weighted schemes, in which the strongest emission regions (hotter and/or with more metal) will enhance the overall contribution.} 
\begin{table*} [t!]
\centering 
\caption{Mean abundance ratio with respect to iron within $R_{500}$  at $z \sim 0.1$ for each cluster in the sample.} 
\begin{tabular}{c c c c c} 
\hline\hline \\[-0.8em] 
 & Cluster 1 & Cluster 2 & Cluster 3 & Cluster 4 \\[0.2em] 
\hline\\[-0.8em] 
C/Fe & 0.48 $\pm$ 0.34 & 0.35 $\pm$ 0.24 & 0.43 $\pm$ 0.28 & 0.48 $\pm$ 0.31 \\
N/Fe & 0.32 $\pm$ 0.12 & 0.40 $\pm$ 0.15 & 0.52 $\pm$ 0.30 & 0.57 $\pm$ 0.20 \\
O/Fe & 0.73 $\pm$ 0.04 & 0.68 $\pm$ 0.04 & 0.67 $\pm$ 0.06 & 0.69 $\pm$ 0.06 \\
Ne/Fe & 0.10 $\pm$ 0.03 & 0.12 $\pm$ 0.04 & 0.14 $\pm$ 0.06 & 0.20 $\pm$ 0.06\\
Na/Fe & 0.17 $\pm$ 0.12 & 0.34 $\pm$ 0.24 & 0.20 $\pm$ 0.13 & 0.21 $\pm$ 0.17 \\
Mg/Fe & 0.63 $\pm$ 0.05 & 0.58 $\pm$ 0.06 & 0.59 $\pm$ 0.08 & 0.53 $\pm$ 0.08 \\
Al/Fe & 0.16 $\pm$ 0.08 & 0.18 $\pm$ 0.10 & 0.21 $\pm$ 0.18 & 0.20 $\pm$ 0.13 \\
Si/Fe & 1.24 $\pm$ 0.06 & 1.18 $\pm$ 0.05 & 1.27 $\pm$ 0.08 & 1.27 $\pm$ 0.09 \\
S/Fe & 1.33 $\pm$ 0.08 & 1.30 $\pm$ 0.08 & 1.26 $\pm$ 0.12 & 1.18 $\pm$ 0.12 \\
Ar/Fe & 0.19 $\pm$ 0.09 & 0.22 $\pm$ 0.11 & 0.20 $\pm$ 0.15 & 0.30 $\pm$ 0.17 \\
Ca/Fe & 0.94 $\pm$ 0.20 & 0.71 $\pm$ 0.16 & 0.71 $\pm$ 0.24 & 0.99 $\pm$ 0.25 \\
Ni/Fe & 2.12 $\pm$ 0.20 & 1.91 $\pm$ 0.21 & 2.11 $\pm$ 0.29 & 2.12 $\pm$ 0.32 \\
\hline \hline
\end{tabular}
\label{table:ratios} 
\end{table*}

\begin{figure*}[t]
\centering
\includegraphics[width=0.495\textwidth, trim={0 0 0 0}, clip]{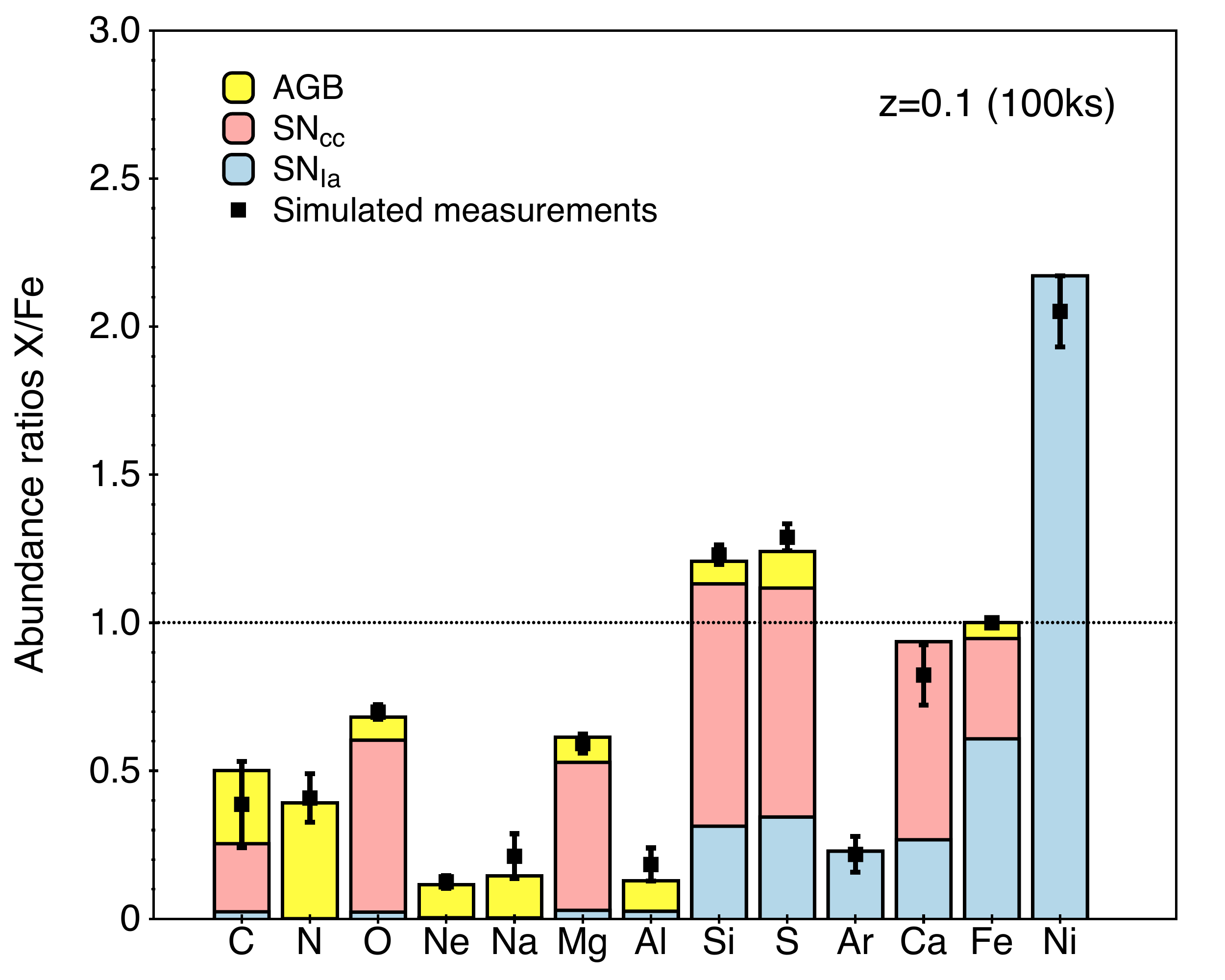}
\includegraphics[width=0.495\textwidth, trim={0 0 0 0}, clip]{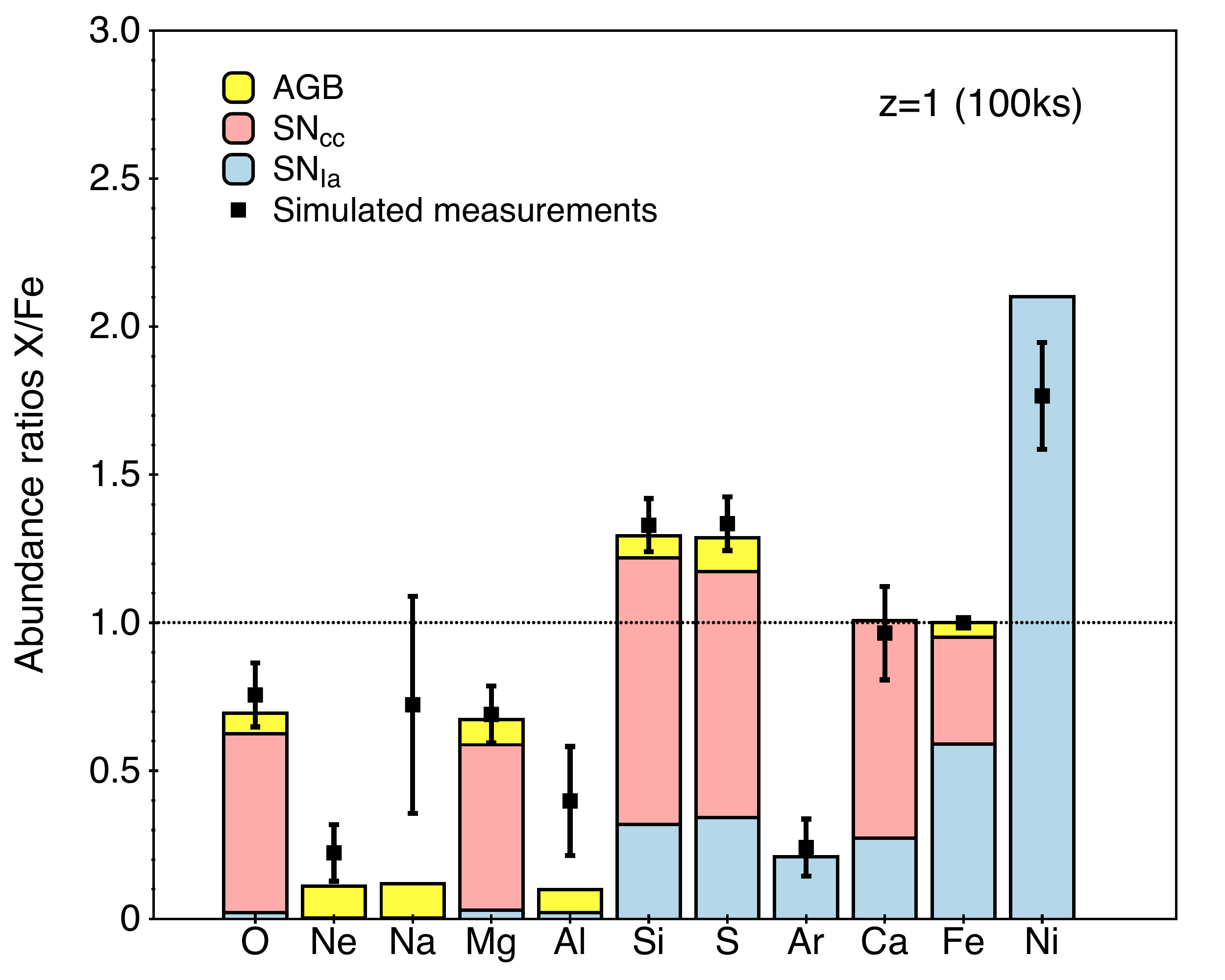}
\caption{Average abundance ratio with respect to iron within  $R_{500}$  at $z~\sim~0.1$ over the cluster sample for $z = 0.1$ (\textit{Left}) and $z = 1$ (\textit{Right}) recovered using 100\,ks observations. The input abundance ratio are shown as histogram bars filled with the respective contributions of SN$_{\text{Ia}}$ (blue), SN$_{\text{cc}}$ (magenta) and AGB stars (yellow), computed from the outputs of the hydrodynamical simulation presented in Sec.~\ref{sec:sim}. For $z = 1$, carbon (C) and nitrogen (N) are not shown as the lines are outside the energy bandpass of the instrument (not fitted).} 
\label{fig:ratio}
\end{figure*}

Once in orbit, the \xifu\ will probe a much larger number of galaxy clusters ($\geq 10$ per mass and redshift bins), therefore reducing the sample variance of these profiles even further, especially near the outskirts of the clusters. A more accurately constrained scatter will provide important information on the metallicity distribution of the ICM and firm observational confirmation of the nature of the enrichment scenario during the early phases of the Universe. These results highlight the sensitivity of the \xifu\ to constrain with high accuracy the chemical enrichment pattern in cluster outskirts, and, therefore, to fully exploit its potential as a fossil record of the star formation history and feedback in the proto-cluster ecosystem.

\subsection{Constraints on the chemical enrichment model}
\label{subsec:const}

As chemical elements are trapped within the ICM, they represent a fossil record of the integrated history of chemical enrichment of the cluster. Strong constraints on the relative contribution of the various enrichment mechanisms (notably SN$_{\text{cc}}$ and SN$_{\text{Ia}}$) could be given by accurate measurements of abundance ratios of elements within clusters. Using the small sample at our disposal, we estimated the capabilities of the \xifu\ in recovering this information in the input hydrodynamical simulations within a radius of $R_{500}$. A first noticeable result is that for the entire cluster sample, recovered abundance ratios are consistent for all the elements (see Table \ref{table:ratios}). Using the average abundance ratio profile and the corresponding production yields of each element in the input hydrodynamical simulations, we computed the input fraction of each of the enrichment mechanisms (SN$_{\text{cc}}$, SN$_{\text{Ia}}$ or AGB) and compared it to the overall output abundance ratio given by our E2E simulations (Fig.~\ref{fig:ratio} -- \textit{Left}). For this study a small fraction of outlier regions ($\lesssim$ 5\%) was excluded, as results clearly showed incompatible values with respect to emission-measure-weighted inputs for rare elements (N, Na and Al). All of the elements present in the simulations are comparable to the inputs within their statistical error bars. Most of all, the ratios of the main elements of the ICM (e.g., O/Fe, Mg/Fe and Si/Fe) are very accurately recovered with a significance of the detection $ \geq 10$ (i.e., ratio between the value and the error). Rarer elements (typically Ne, Ar, S, Ca) are also very consistent with the hydrodynamical simulations. Less abundant elements (Na and Al) have looser constraints in the fitted regions due to the considered exposure time (low S/N of the lines) and seem slightly overestimated in their reconstruction. 

\begin{figure*}[t!]
\centering
\includegraphics[width=0.495\textwidth,  trim={50cm 0 22cm 8cm}, clip]{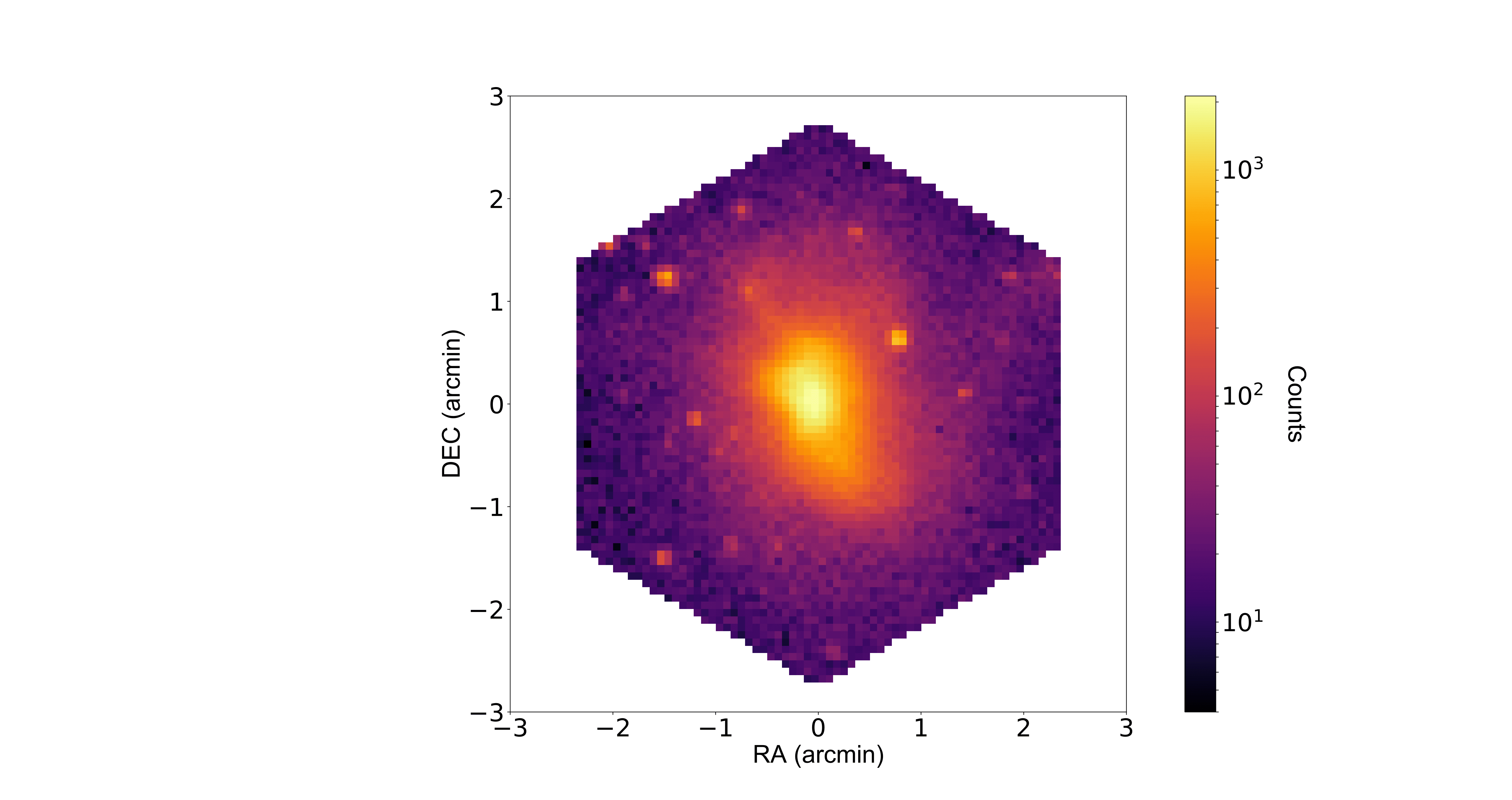}
\includegraphics[width=0.495\textwidth,  trim={50cm 0 22cm 8cm}, clip]{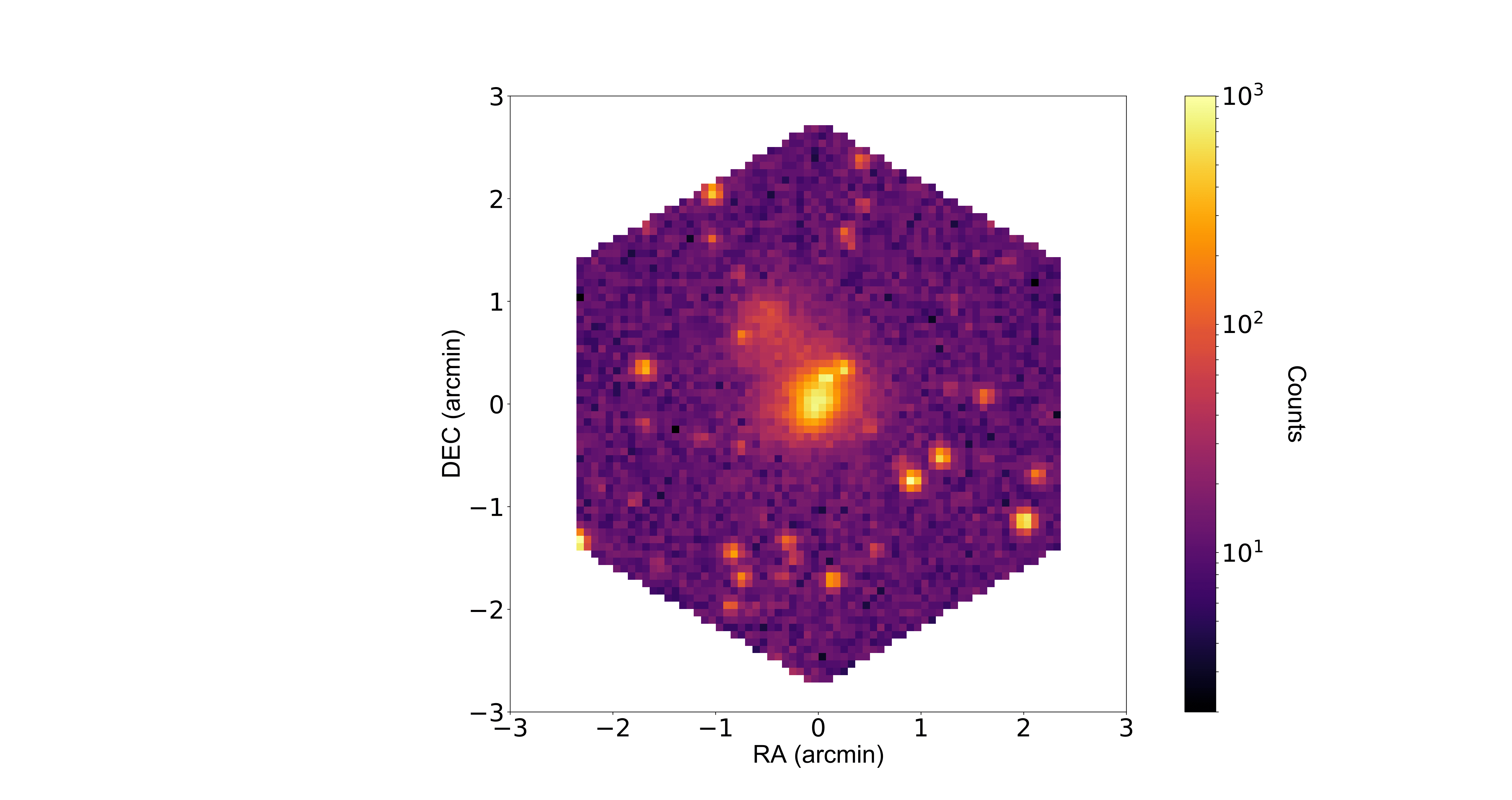}
\includegraphics[width=0.495\textwidth,  trim={50cm 0 22cm 8cm}, clip]{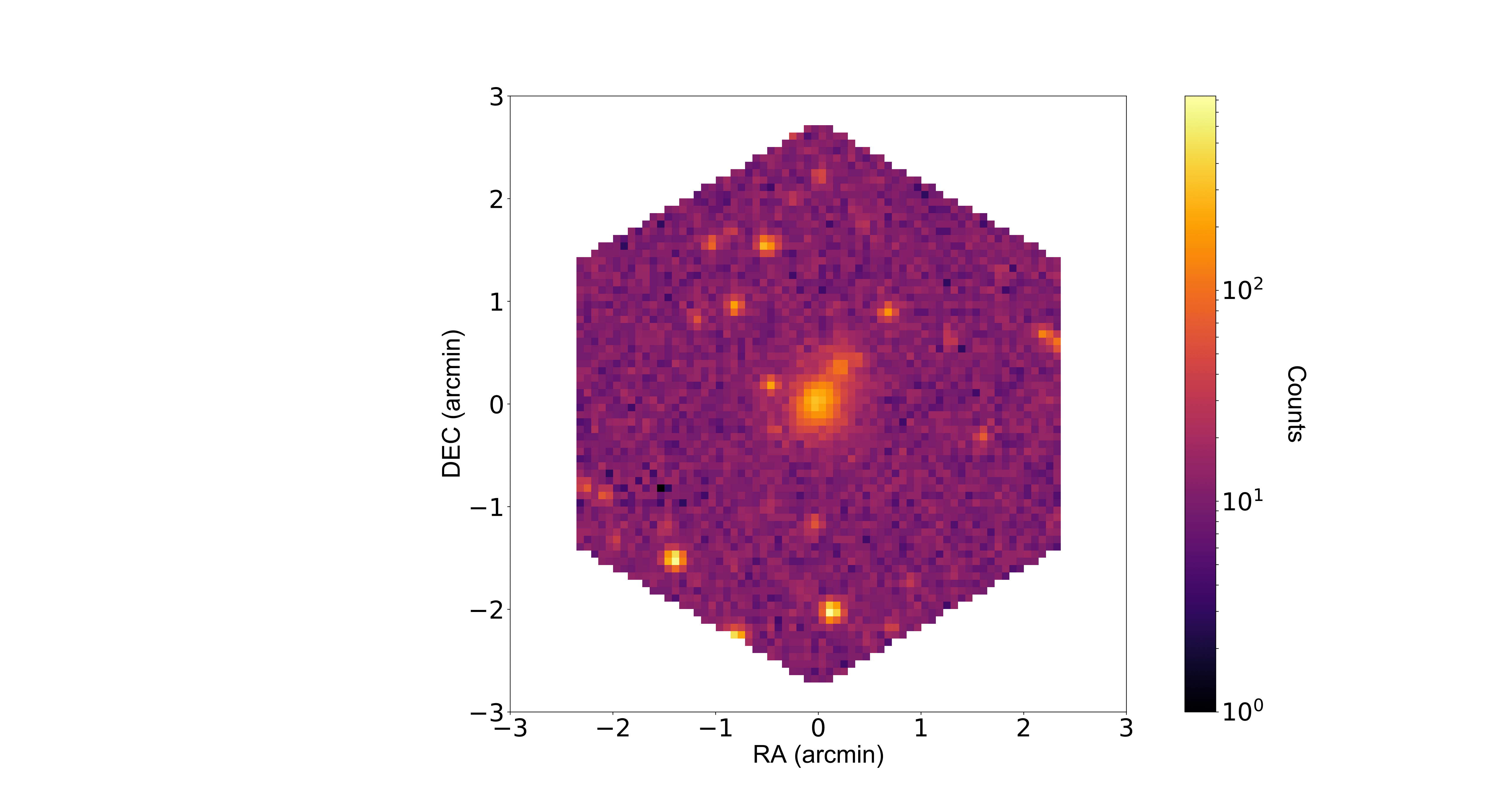}
\includegraphics[width=0.495\textwidth,  trim={50cm 0 22cm 8cm}, clip]{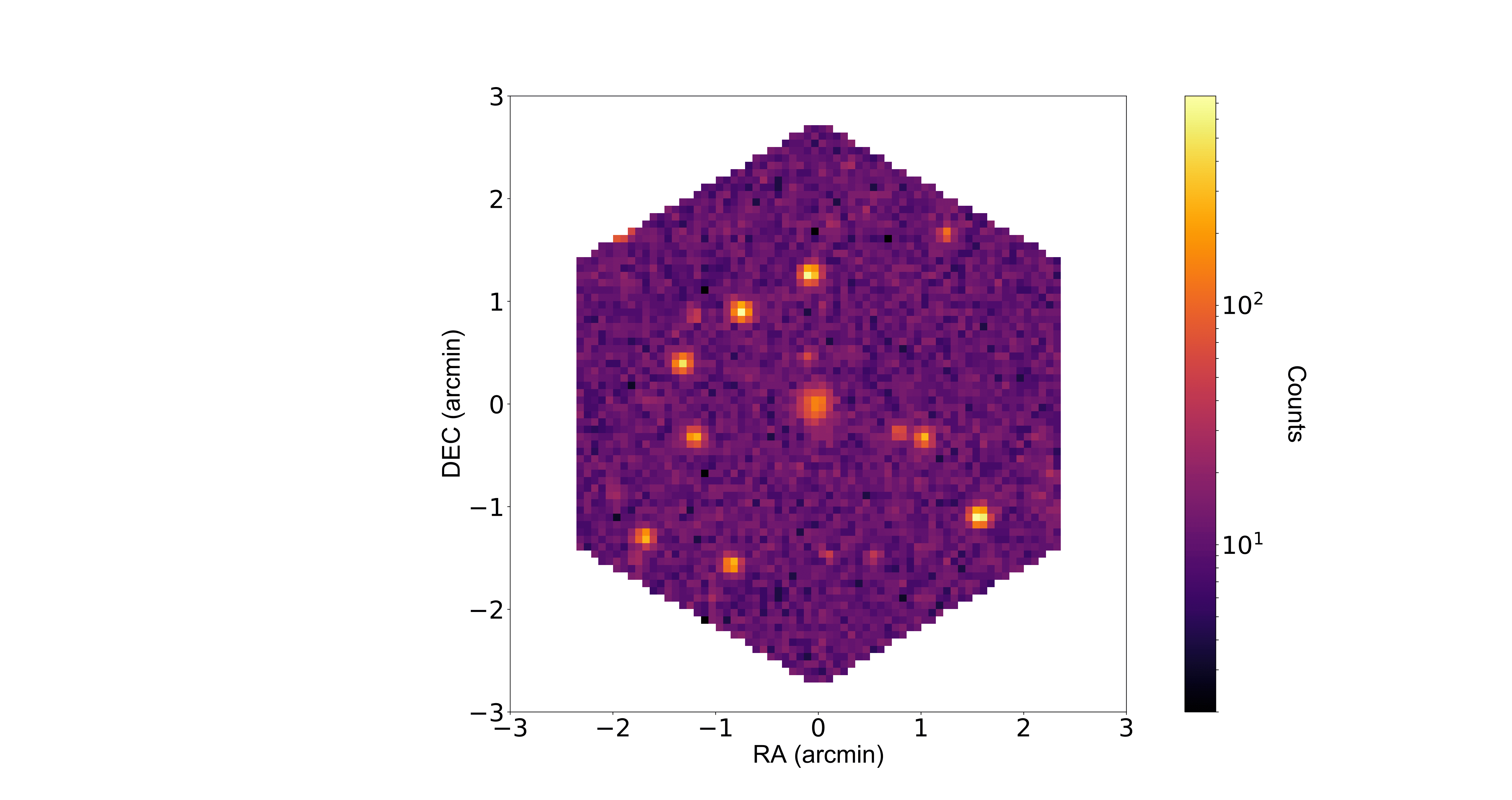}
\caption{(\textit{From top left to bottom right}) Maps in number of counts per \xifu\ pixel (249\,\textmu m pitch) for cluster C2 (see Table \ref{table:1}) simulated with the end-to-end simulator SIXTE for redshift $z \sim 0.5$ {\color{black}(\textit{Top left})}, 1  {\color{black}(\textit{Top right})}, 1.5 {\color{black}(\textit{Bottom left})} and 2 {\color{black}(\textit{Bottom right})}, for an exposure time of 100\,ks.}
\label{fig:clusters_evo}
\end{figure*}

With the small sample used here, we demonstrated that the \xifu\ is able to provide robust estimations of the abundance ratios. Given the low errors on the measurements, these results can be used to distinguish the contributions of the various mechanisms at play in the cosmological simulations, by comparing them to metal production theoretical models. Notably, using elements produced by single mechanisms (e.g., Ne, Na for AGB, or Ar, Ni for SN$_{\text{Ia}}$), accurate measurements of abundance ratios will provide strong constraints to the IMF and the contribution of each mechanism at local redshift. The ability to recover the corresponding supernov\ae\ yield and to distinguish between multiple other models shall be addressed in a forthcoming study. The current observational strategy of the \xifu\ plans to use at least 40 clusters of galaxies to investigate the chemical enrichment of the Universe. With such a large sample and by giving unprecedented information on other rare elements (e.g., Mn, Co, that could not be tested here since these metals are not separately traced in the hydrodynamical simulations), the \xifu\ will undoubtedly provide new constraints to the global chemical enrichment models.

\section{Chemical enrichment through cosmic time}
\label{sec:evo}

Element abundances in local clusters embed the integrated chemical enrichment of the Universe up to this day. However, to understand how and when the ICM was enriched, the evolution of production sources with time and how the overall enrichment processes relate (e.g.,  the star formation history and the initial mass function) need to be assessed. To do so, we analysed synthetic observations of the same four clusters taken at different stages of their evolution, hence considering five redshift values up to $z = 2$. We chose to keep a realistic exposure time fixed to 100\,ks, regardless of the redshift. This exercise tested the capabilities of the \xifu\ in a regime of low statistics, thus preventing the full spatial analysis presented above. For the highest redshift clusters we complied strictly to the definition of the \textit{Athena} science case on chemical abundances and performed measurement in two different annuli from the cluster center between 0~--~0.3\,$R_{500}$ and 0.3 -- 1\,$R_{500}$. 
\begin{figure*}[p]
\centering
\includegraphics[width=0.49\textwidth, trim={0 0 0 0}, clip]{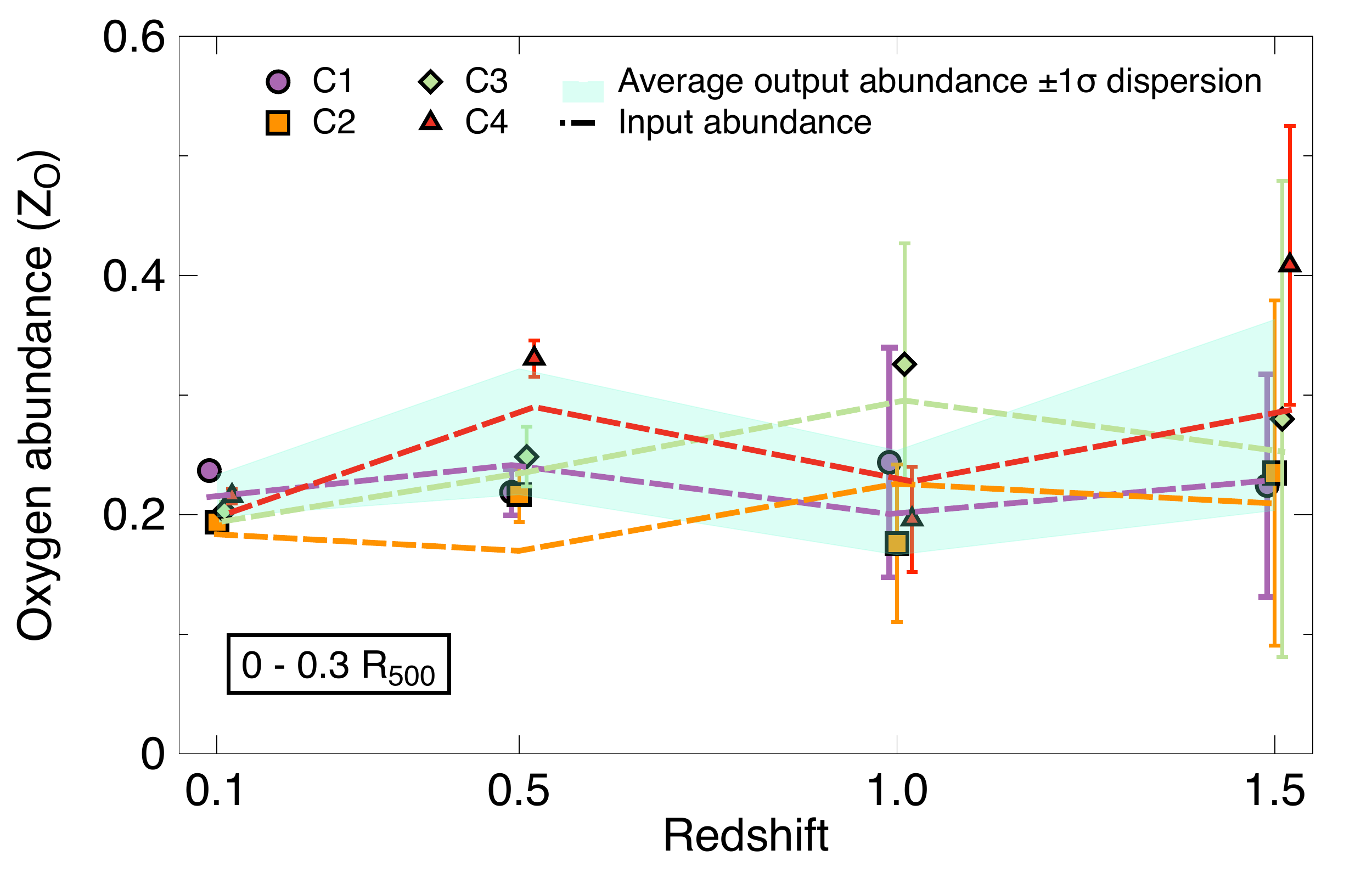}
\includegraphics[width=0.49\textwidth, trim={0 0 0 0}, clip]{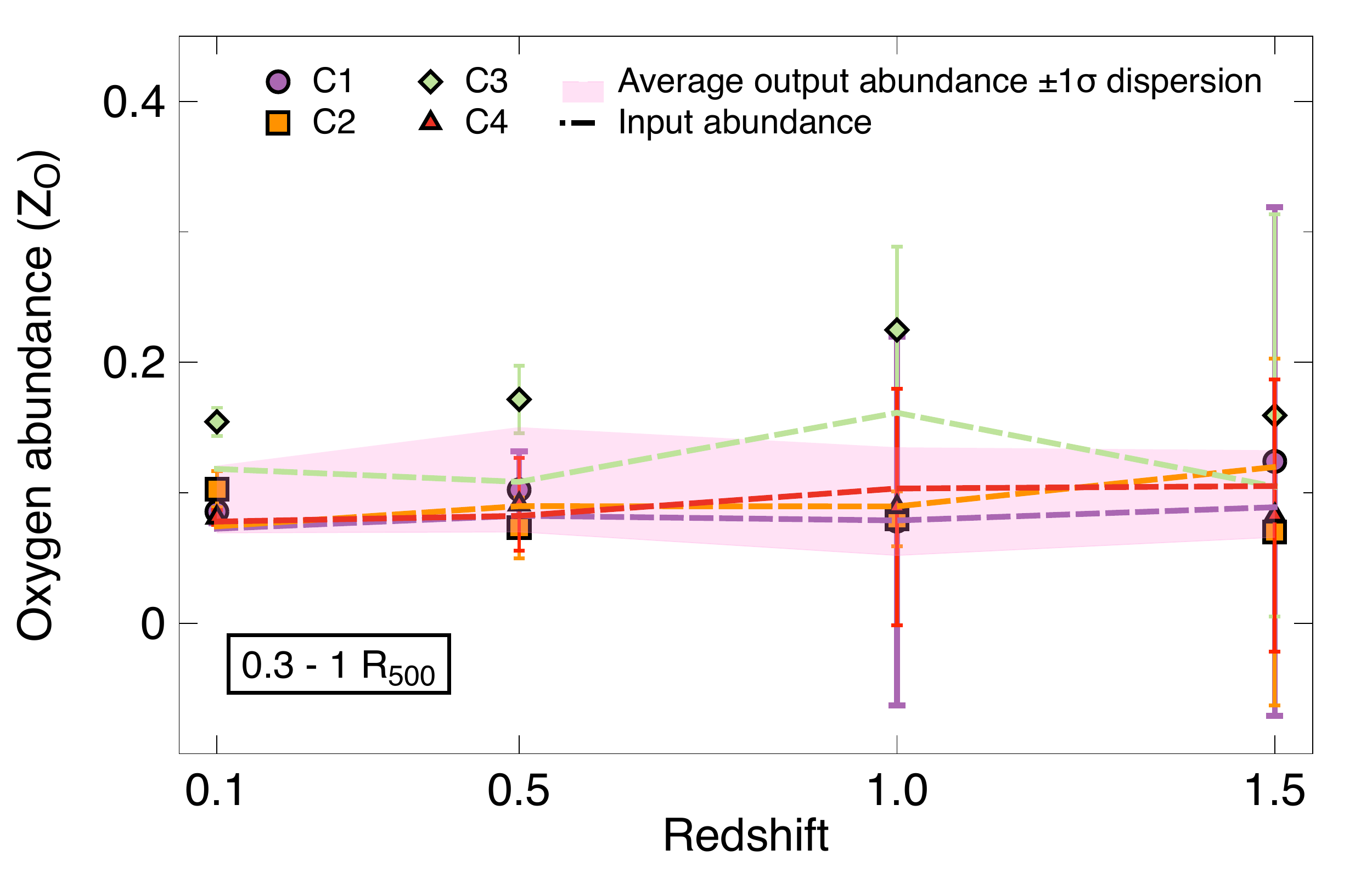}
\includegraphics[width=0.49\textwidth, trim={0 0 0 0}, clip]{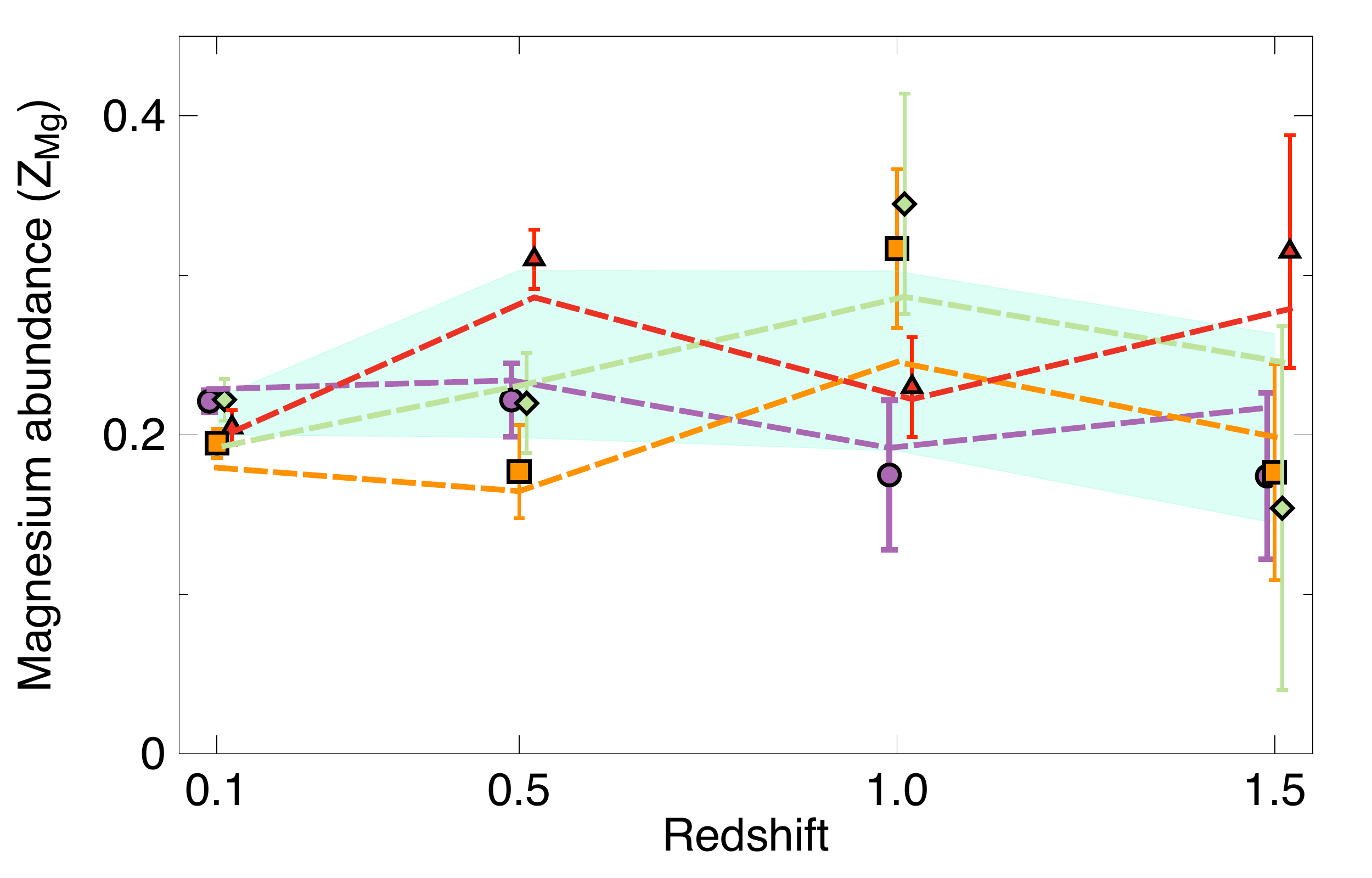}
\includegraphics[width=0.49\textwidth, trim={0 0 0 0}, clip]{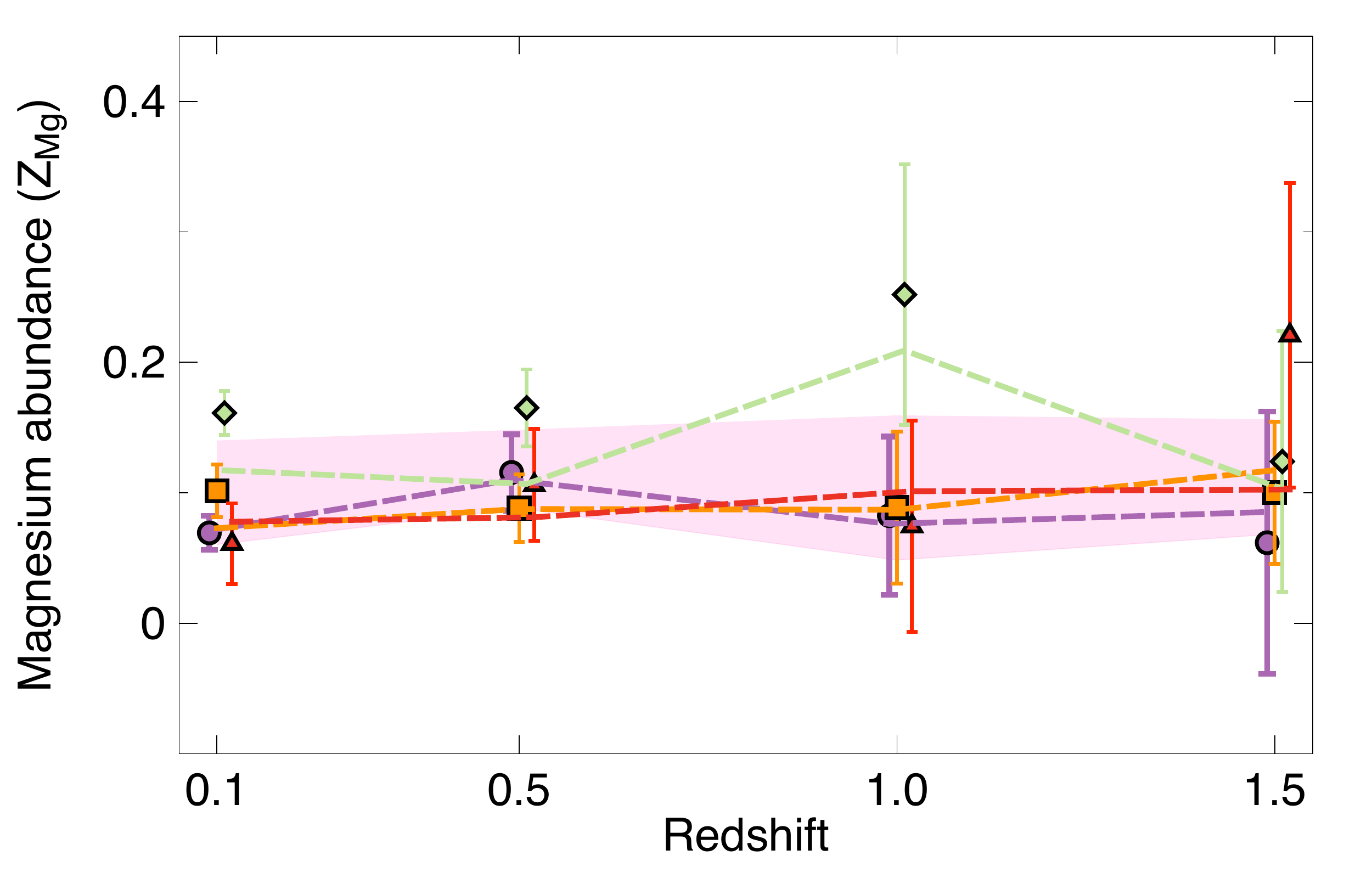}
\includegraphics[width=0.49\textwidth, trim={0 0 0 0}, clip]{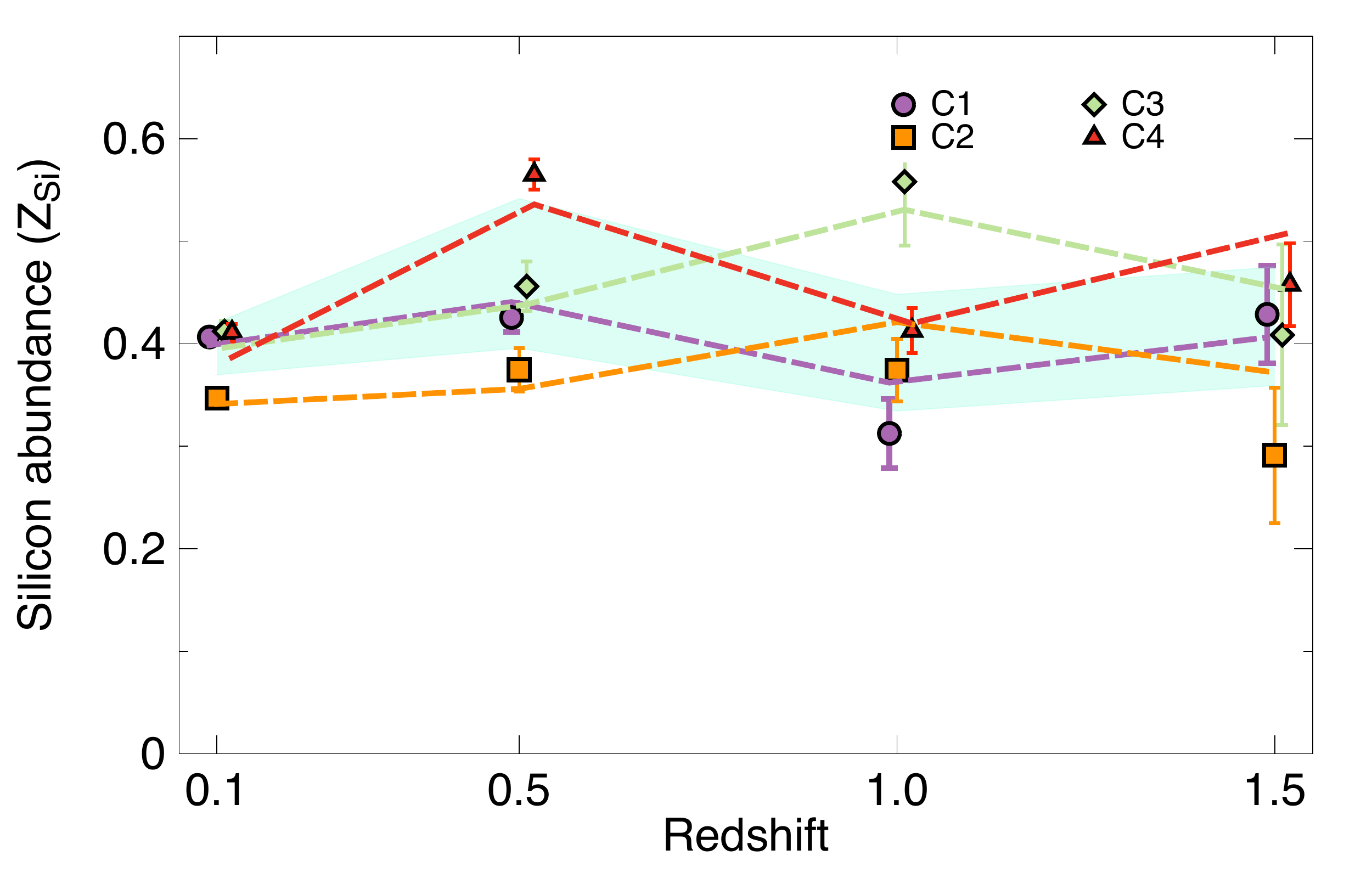}
\includegraphics[width=0.49\textwidth, trim={0 0 0 0}, clip]{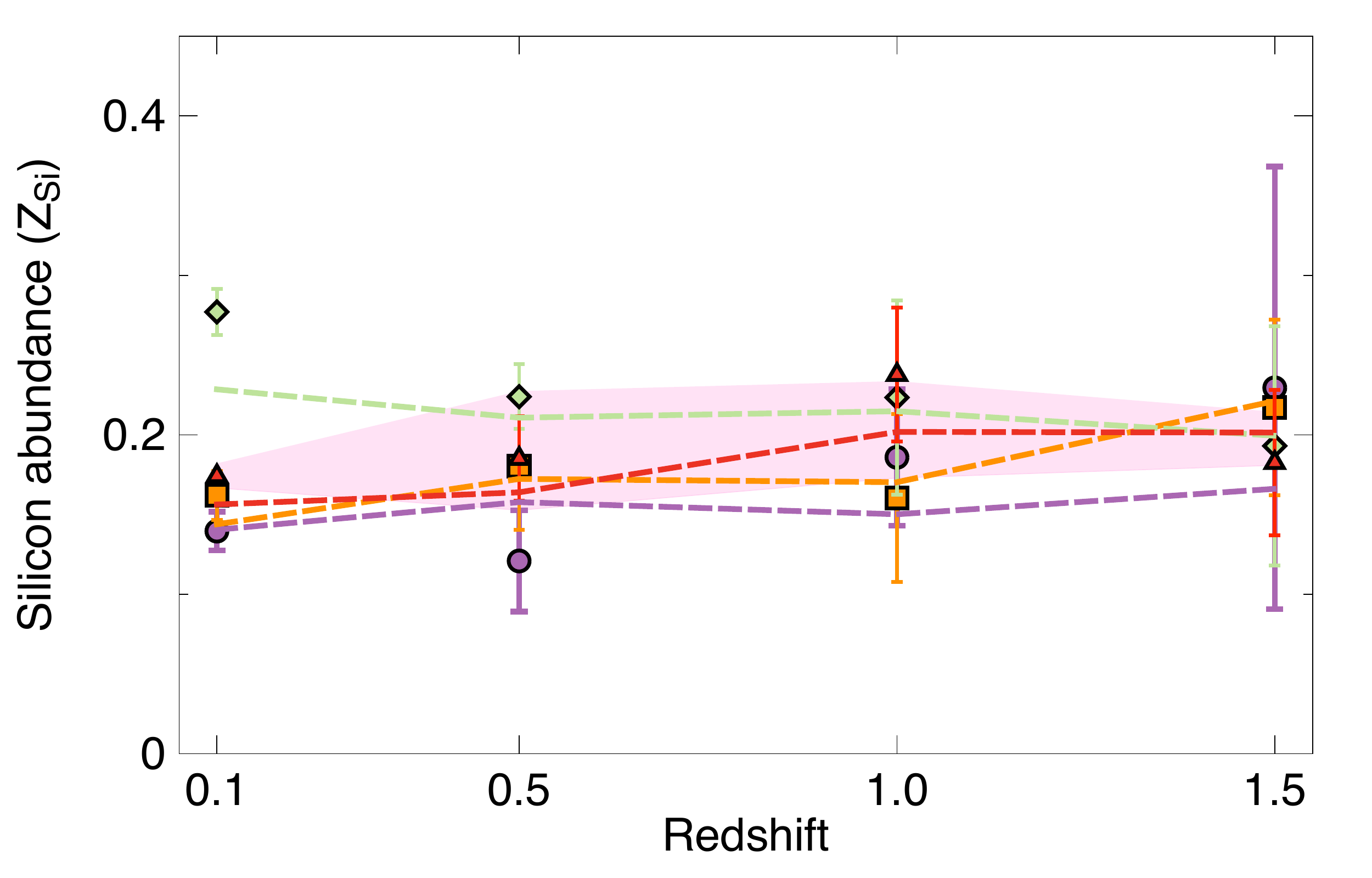}
\includegraphics[width=0.49\textwidth, trim={0 0 0 0}, clip]{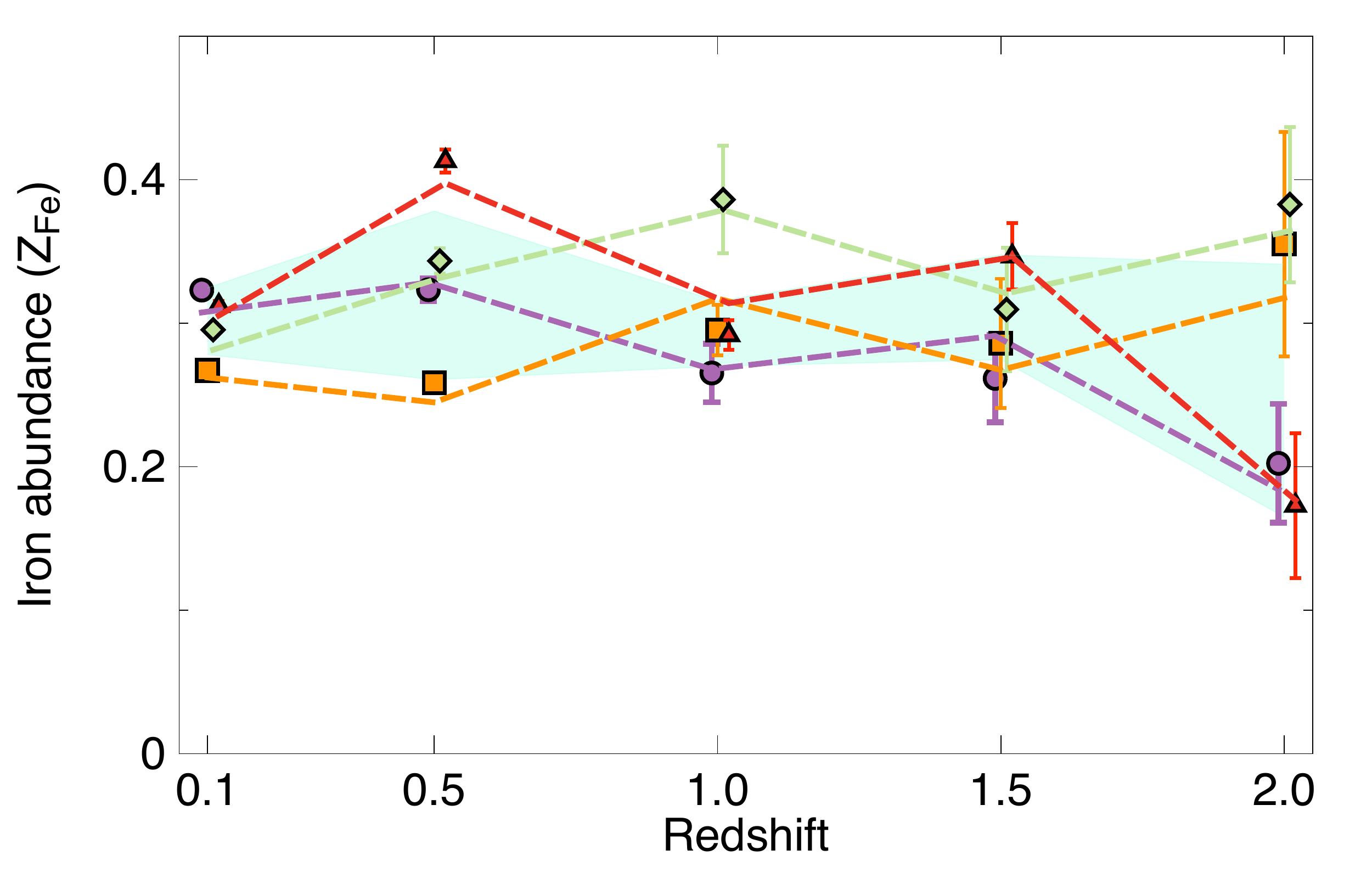}
\includegraphics[width=0.49\textwidth, trim={0 0 0 0}, clip]{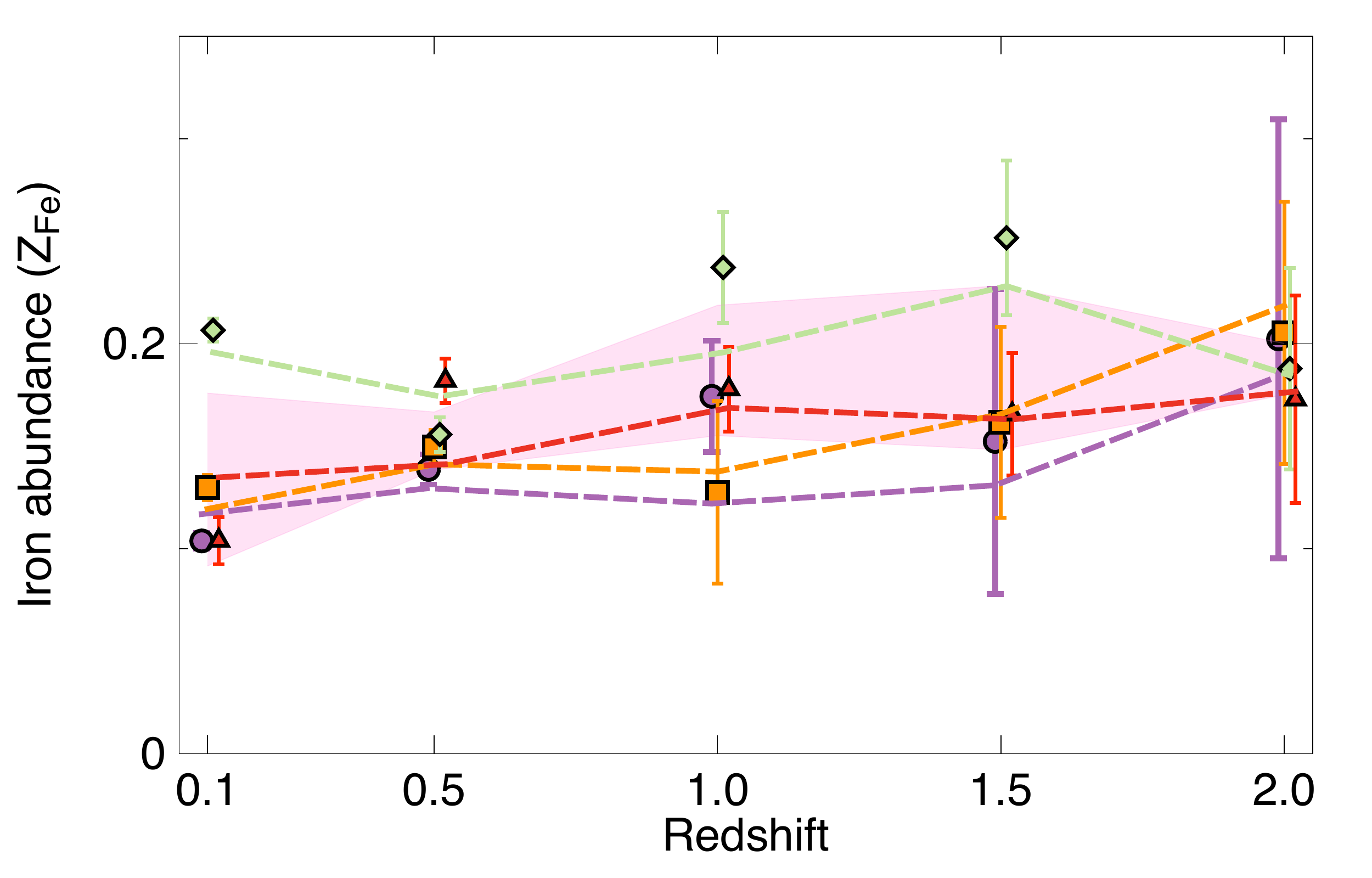}
\caption{Evolution of the average abundance of the cluster sample (C1 - purple dots, C2 - orange squares, C3 - green diamonds, C4 - red triangles) recovered via XSPEC as a function of the redshift between 0 -- 0.3 $R_{500}$ (\textit{Left}) and 0.3 -- 1 $R_{500}$ (\textit{Right}). (\textit{From top to bottom}) Oxygen, magnesium, silicon and iron abundances with respect to solar. The dashed lines represent the profile of the emission-measure-weighted input abundances using the same colors. The cyan (resp. magenta) shaded envelope represents the $\pm 1 \sigma$ dispersion of the output metallicity over the sample. Points are slightly shifted for clarity.} 
\label{fig:evo}
\end{figure*}

Fig.~\ref{fig:clusters_evo} shows the mock surface brightness maps for C2 at the various redshift snapshots and illustrates the assembly history of halos through, e.g., merging events. Fig.~\ref{fig:evo} shows the evolution of the mean cluster abundance over our whole sample  as a function of the redshift in the two aforementioned annular regions for O, Mg, Si and Fe. Despite the lower source-to-background level of some objects, input metallicity values are recovered accurately within the statistical uncertainties of the measurements even for high-redshift clusters, although for $z \geq 1$, error bars start to be significant for elements such as O and Mg. In the case of low-mass elements, abundances are not measurable up to a redshift of $z = 2$, as lines are redshifted outside the instrument energy band (e.g. O and Mg for $z \geq 1.5$) or are too weak to be disentangled from the foreground/background (e.g. Si at $ z = 2$). As expected, measurements in the central parts of the cluster are more accurate, due to the higher level of background in the outskirts with respect to the cluster emission, especially for $z \geq 1.5$.

Through these measurements, we find that in the central parts of the cluster, metallicity hardly changes across time, even at a redshift of $z = 2$, once again consistently with the analysis by \cite{Biffi2017Clusters}. This indicates that {\color{black} most of the enrichment occurs in the early days of the cluster.} Interestingly, we  notice that the iron abundance in the center of the cluster slightly decreases with redshift, which could be explained for instance by an increase in time of iron production mechanisms (e.g., SN$_{\text{Ia}}$) or by the time delay with which long-lived SN$_{\text{Ia}}$ release Fe. Abundances in the outskirts show a similar trend, with near-constant values up to local redshift values. Similar observational evidences, as e.g., reported in \cite{Ettori2015evol} for iron abundance are consistent with these conclusions. The dynamic history of the clusters (mergers, shocks) is visible over time (data points are taken from the same cluster at different time steps), displaying local peaks of abundance, e.g., C3 at $z=1$ or C4 at $z=0.5$. Given the sparse number of redshift points, turbulence or mixing within the cluster (whose eddy turn over time scale is of the order of a few Gyr over scales of$\sim$1\,Mpc for typical $\sim$500\,km/s velocities) create a more homogeneous distribution of metals in the structure, returning abundances to typical values after mergers. 

Using the accuracy of the \xifu\ abundance measurements for high-redshift objects, we can also analyse changes in the metal enrichment mechanisms by performing a similar study as the one presented in Sec.~\ref{subsec:const}. In the case of redshift $z = 1$ (see Fig.~\ref{fig:ratio} -- \textit{Right}) for 100\,ks observations, we find that the \xifu\ will still be capable of accurately recovering abundance ratios within $R_{500}$ with excellent accuracy. Most of the main elements have in fact no significant changes between both redshift values, consistently with our previous conclusions. In the case of high-redshift objects however, low-mass elements such as carbon and nitrogen can no longer be detected (lines outside the energy bandpass) and rarer elements (e.g., Ne, Na, Al) have large uncertainties due to the low S/N of the observations. Ni also tends to be underestimated (mostly in the outskirts) likely due to the low $\textrm{S/N}$ of the line with respect to the high-energy background. This calls for more adapted exposure strategies to optimise the results for distant objects and further investigate the chemical enrichment across cosmic time. 

\section{Summary and discussion}
\label{sec:dis}

In this paper, we addressed the feasibility of constraining the chemical enrichment of the Universe, which will be one of the key science objectives {\color{black} and a main driver of the performances} of the future mission \textit{Athena}. Notably, we investigated and quantified the capabilities of the \xifu\ in accurately recovering metal abundances of the ICM across cosmic time.  To this end, we developed a full end-to-end pipeline, which creates synthetic \xifu\ observations using the instrument simulator SIXTE.  We used as input of this pipeline a sample of four clusters generated using state-of-the-art cosmological simulations presented in \cite{Rasia2015Simu} and \cite{Biffi2017Clusters} to create realistic event lists. All the relevant instrumental effects such as the convolution of spectra with the instrument spatial and spectral responses, realistic sources of foreground/background, and detector geometry were also included to obtain observations as realistic as possible.
 
The sample of four clusters was simulated at five different redshift values, for a fixed exposure time of 100\,ks in order to achieve abundance measurements out to  $R_{500}$. The accuracy of the pipeline was quantified by comparing our synthetic observations to weighted inputs quantities (e.g., {\color{black} spectroscopic} temperature, emission-measure-weighted abundances). We found that a straightforward approach of a broadband fit created systematic biases above 10\% in a number of physical parameters. Rather a multi-band energy fitting procedure ensured more accurate recovery by optimising the extraction of the several chemical abundances and other physical parameters of interest (notably temperature). After post-processing, distributions are accurately recovered (almost always within the $3\sigma$ of the measurement error) with little to no systematic biases (of the order of a 5\%, see Sec.~\ref{sub:sys}) found mainly between the low-mass element abundances (e.g., O, Si) and the normalisation. The comparison of the relative distribution between outputs and inputs with respect to the XSPEC statistical error also showed reduced chi-squared values $\tilde{\chi}_{\rm red}^2$ close to 1 when a small fraction of outlier regions ($\leq 5\%$) is removed indicating a good accuracy in the fits. Remaining errors and biases can be linked to correlation between parameters (notably the normalisation), to the choice of the input weighting scheme, and to mixing effects along the line-of-sight in view of the single plasma temperature model used here. We also found that most of these errors decrease when statistics are strongly increased (biases below 2\% at $1$\,Ms for the same spatial regions), suggesting that these effects may simply be related to statistics (see Appendix \ref{app:2}). {\color{black} Studies conducted by decreasing the statistics of the runs (typically by decreasing the $\textrm{S/N}$ of the regions to 50 or 100) provided equally encouraging results. Despite larger statistical errors (up to 10\% higher), the main parameters (temperature, redshift, O, Si or Fe abundance) were accurately recovered. Some fainter lines (e.g., Na, N or Al) become however very difficult to constrain in this case.}

For local clusters ($z \sim 0.1$), we demonstrated the power of the \xifu\ in accurately recovering spatially resolved parameter maps, along with abundance profiles (Sec.~\ref{subsec:metals}) and abundance ratios (Sec.~\ref{subsec:const}). The study was then extended at different redshift values, up to $z \sim 2$. By probing the chemical enrichment for very distant clusters and despite the lack of an optimised observation strategy (i.e.,  non optimised  exposure time), we also show the power of the \xifu\ in investigating the ICM properties and the chemical enrichment of the distant Universe. 

The binning and fitting procedures used here comprise ``classical'' approaches to X-ray data analysis, using $\textrm{S/N}$ binning and fits through instrumental response matrices in XSPEC. Despite our efforts, the fitting procedures remain slightly biased ($\leq 5\%$) and small changes in the fitting approach can impact the overall results of the simulation (of the order of a few \%). More accurate results may be achieved using e.g. Monte-Carlo (MC) fitting approaches, but unfortunately remain computationally cumbersome to be used on our large set of spatial regions. More optimised binning techniques could also be investigated for future applications \citep{Kaastra2016Bin}. The access to high-resolution spectra will provide new proxies to estimate quantities such as the temperature by using e.g. line-ratio techniques. Eventually, hyper-spectral methods (e.g. Blind Source Separation algorithms) or machine-learning-based fitting techniques \citep[see, e.g.,][]{Ichinohe2018Neural} could open new perspectives for the post-processing of high-resolution X-ray spectra. We would like to underline that, even though not applicable in our simulation case, the expected level of spectral resolution of the \xifu\ will challenge our current knowledge accuracy of the spectral lines (centroid energies and intrinsic widths), which is critical to allow a meaningful interpretation of the results \citep[as demonstrated in][for line ratios]{Hitomi2017Spectral} and to disentangle fine spectroscopic effects (such as resonant scattering, \citealt{Hitomi2017Resonant}). This emphasises the need for dedicated tools able to process and analyse future \xifu\ high-resolution spectroscopy data-cube. On this regard, the \textit{Athena} mission will certainly benefit from the advances expected in processing tools, fitting methods and atomic databases, from the future \textit{{\color{black} XRISM}} mission \citep{Ishisaki2018XARM}.

Not only do these E2E simulations allow to explore the capabilities of the future \xifu\ instrument, but they also give crucial information on the effect of instrumental parameters in science observations. In this study for instance, the spectral shape of all the foreground/background components were assumed to be perfectly known. For the more local and massive clusters however, the field-of-view of the \xifu\ will easily be encompassed within the angular extension of $R_{500}$. Cluster emission-free regions might thus be unavailable for local background calibration. The spectral resolution of the \xifu\ will help mitigate this effect, by allowing to disentangle various components through the characteristics of their spectral energy distribution. The instrument background may also contaminate the observation of faint sources, as the level of precision to which \xifu\ is expected to perform requires its accurate and reproducible knowledge in flight. This may be achieved, e.g., through in-flight cross-correlation with the WFI or the \xifu\ cryogenic anti-coincidence detector \citep{Cucchetti2018NXB}. Future developments could take advantage of this simulation pipeline to test other realistic instrumental effects (e.g. stray-light for galaxy cluster outskirts observations). More detailed studies of the abundance ratios recovered here will also be at the center of a forthcoming study to highlight the capabilities of the \xifu\ in constraining the ICM chemical enrichment, and notably to disentangle between the contributions of the various mechanisms of chemical enrichment (e.g. SN, AGB) throughout cosmic time.

Our study underlines the revolutionary capabilities brought by the \xifu\ in future X-ray spectroscopy. With typical routine observations, the \xifu\ will drastically change our understanding of ICM mechanisms and provide a quantum leap forward in X-ray astronomy.

\small{\paragraph{\textit{Acknowledgements}}
V. Biffi, S. Borgani and E. Rasia acknowledge financial contribution from the contract ASI-INAF n.2017-14-H.0. E.~Rasia acknowledges the ExaNeSt and Euro Exa projects, funded by the European Union's Horizon 2020 research and innovation programme, under grant agreement No 754337. S~.Borgani and L.~Tornatore acknowledge support from the EU H2020 Research and Innovation Programme under the ExaNeSt project (Grant Agreement No. 671553).  S.~Borgani also acknowledge support from the INFN INDARK grant. S.~Ettori acknowledges financial  contribution from the contracts NARO15 ASI-INAF I/037/12/0, ASI 2015-046-R.0 and ASI-INAF n.2017-14-H.0. M.~Gaspari is supported by NASA through Einstein Postdoctoral Fellowship Award Number PF5-160137 issued by the Chandra X-ray Observatory Center, which is operated by the SAO for and on behalf of NASA under contract NAS8-03060. Support for this work was also provided by Chandra grant GO7-18121X. The authors would like to extend the thanks to the anonymous referee for the suggestions and helpful comments.}

\bibliographystyle{aa}
\bibliography{paper}

\appendix

\section{Spatial binning algorithm comparison}
\label{app:1}

Spatial binning of the data is used to increase the signal-to-noise (S/N) ratio of the considered regions, to have higher significance spectra. This is of particular interest whenever fine structures (e.g. line ratios, line doublets, absorption features) need to be observed in a spectrum. Multiple methods can be used to bin spatially, in this study two of them were considered:
\begin{itemize}
\item \texttt{contbin} tool developed by J. Sanders \citep{Sanders2006Contour}. The contbin scheme was run for a constrain fill value of two, which represents the maximum ratio length/width of a region. 
\item Voronoï tessellation, as defined by \citep{Cappellari2003Voronoi}
\end{itemize} 
\begin{figure} [hbp]
\centering
\includegraphics[height=3.5cm, trim={50cm 0 22cm 8cm}, clip]{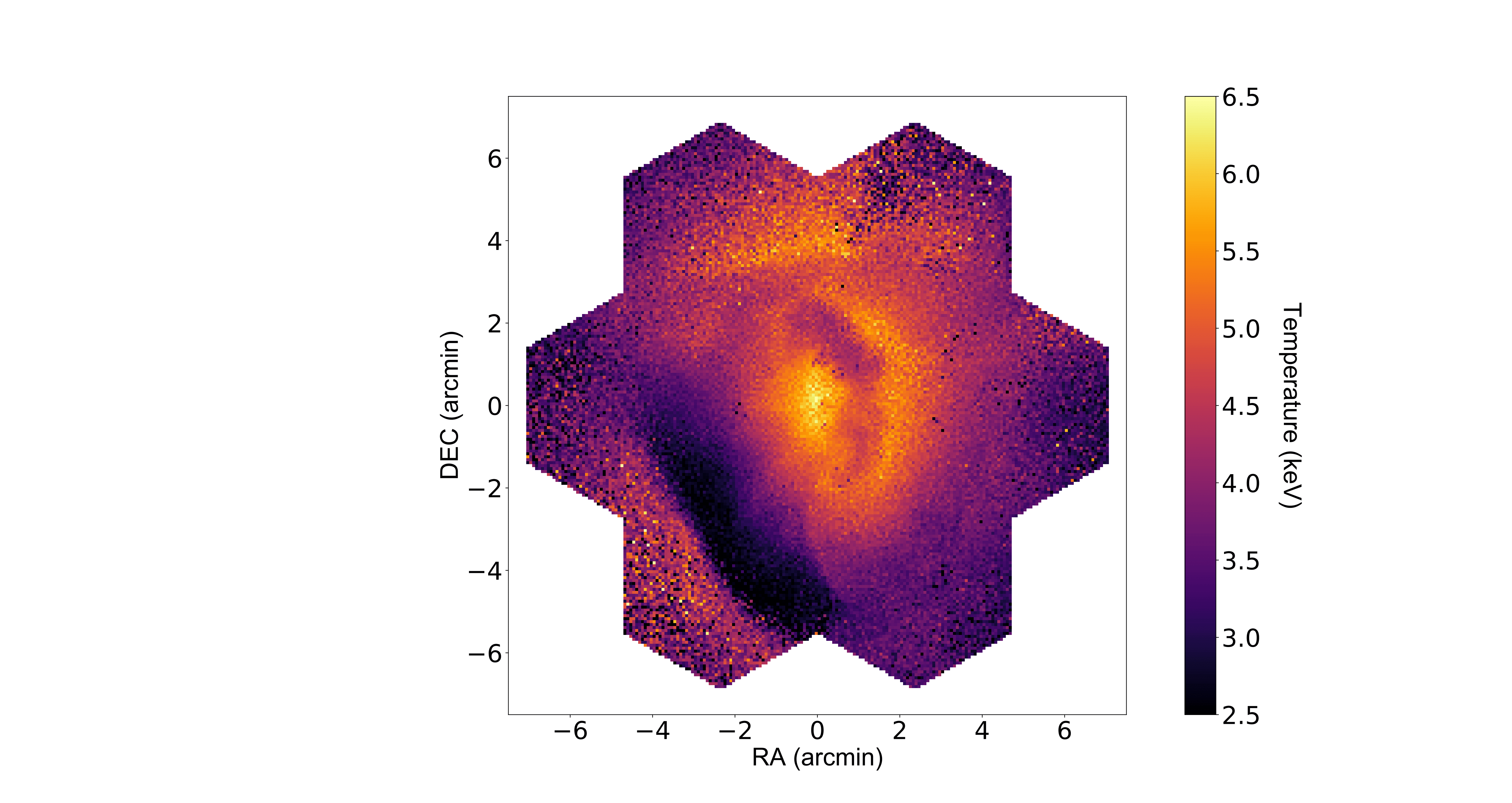}
\includegraphics[height=3.5cm, trim={50cm 0 22cm 8cm}, clip]{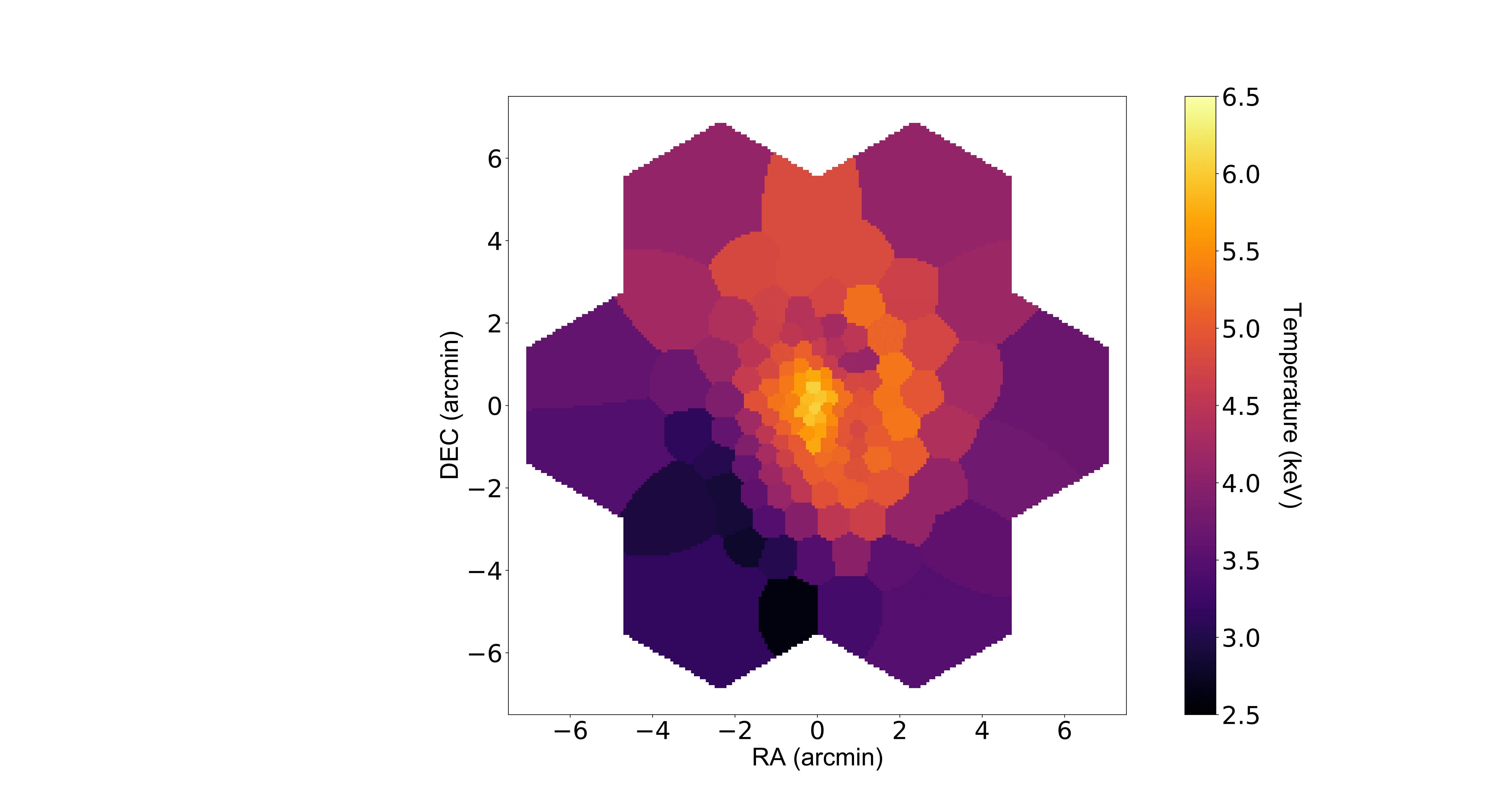}
\includegraphics[height=3.5cm, trim={50cm 0 22cm 8cm}, clip]{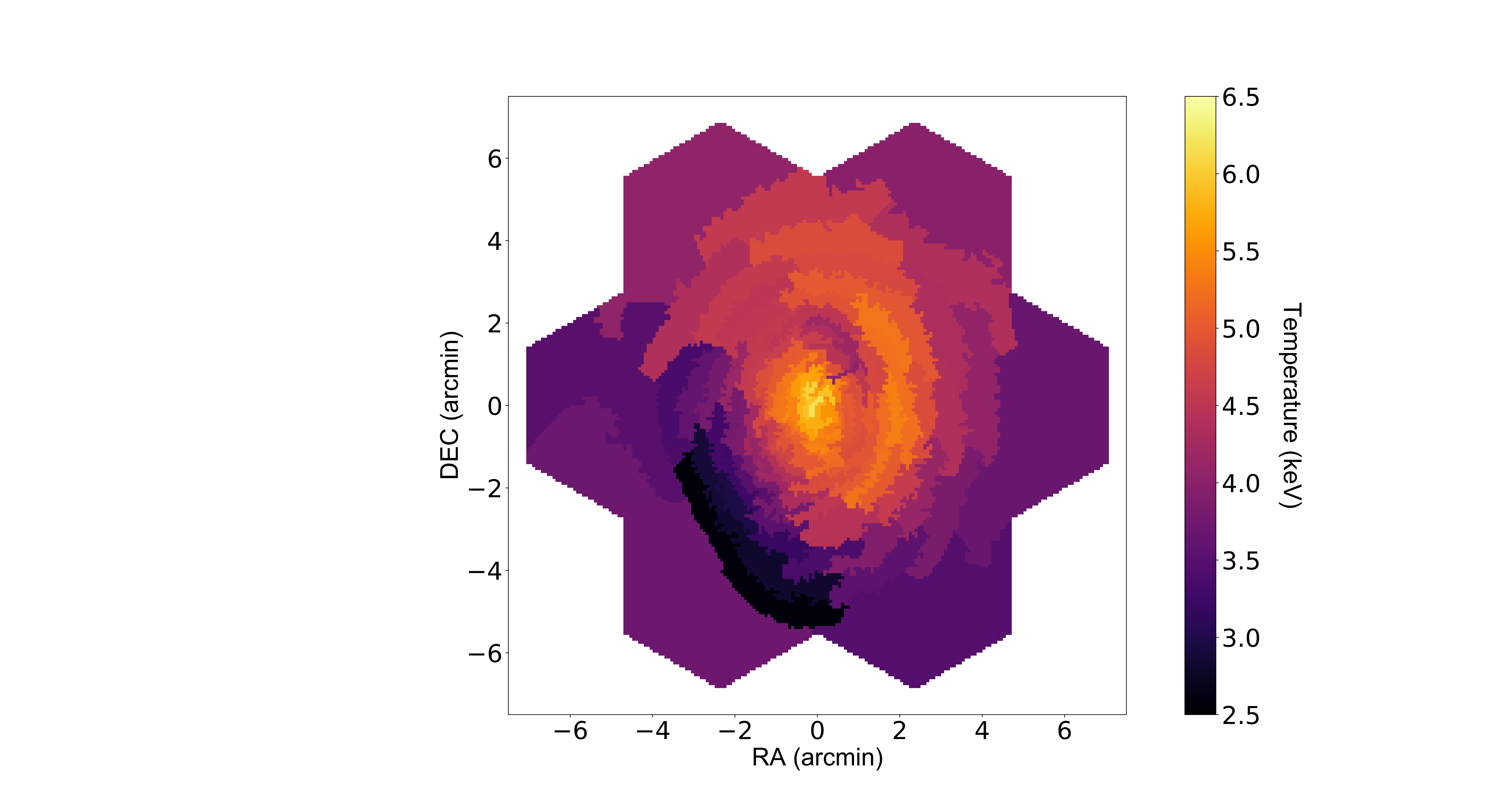}
\includegraphics[height=3.5cm, trim={50cm 0 22cm 8cm}, clip]{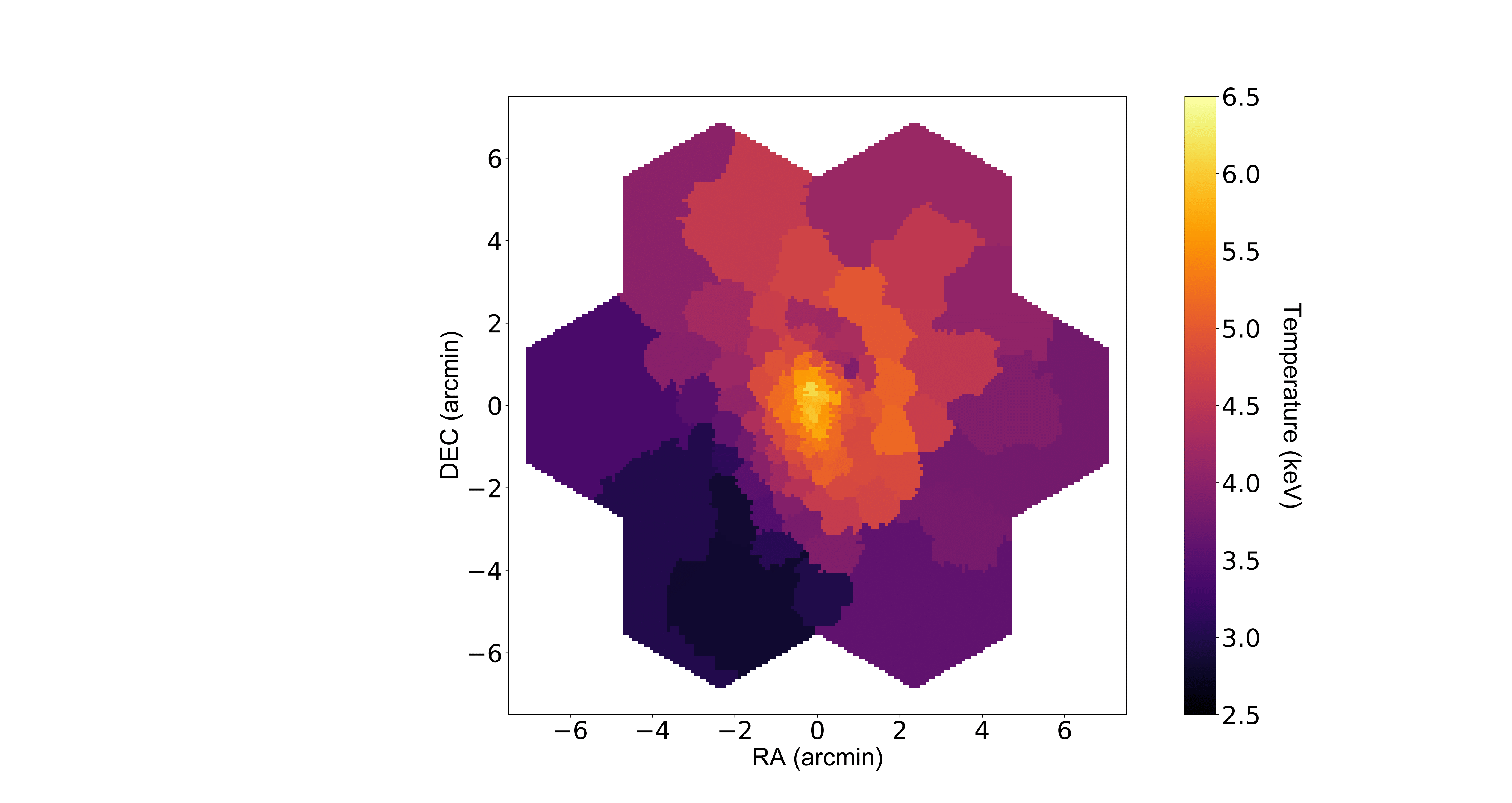}
\caption{Spatial binning scheme comparison for cluster 4 {\color{black} spectroscopic} temperature map (keV) without background, at redshift $z = 0.105$. (\textit{Top left}) Unbinned raw input map from the hydrodynamical simulation. (\textit{Top right}) Voronoi tessellation map using the algorithms described in \citep{Cappellari2003Voronoi}. (\textit{Bottom left}) Contour map using \texttt{contbin} tool \citep{Sanders2006Contour} with an aspect ratio constraint, $C_{\text{fill}} = 2$. (\textit{Bottom right}) Same as bottom left, with $C_{\text{fill}} = 1$.}
\label{fig:spatbin}
\end{figure}
Both these methods were tested on the same surface brightness map to estimate their performances. Various criteria were compared, such as their ability to reproduce spatial features of the cluster or the mean S/N ratio of the regions created. Figure \ref{fig:spatbin} shows a comparison of the cluster 2 regions at $z = 0.105$ without any foreground/background component (to investigate purely the binning effects) computed using either of these methods. Visually, we notice that \texttt{contbin} provides very similar results to Voronoï whenever the aspect ratio of the region is constrained to $C_{\text{fill}} = 1$ (Fig.~\ref{fig:spatbin} -- \textit{Right}). When a slightly higher value of the aspect ratio is allowed (Fig.~\ref{fig:spatbin} -- \textit{Bottom left}), we notice that \texttt{contbin} is able to reproduce more accurately the radial contours of the cluster, notably the cold arc visible in the south-east corner of input data or the hot bubble rising west of the cluster (Fig.~\ref{fig:spatbin} -- \textit{Top left}). Further, no difference is found on the average S/N ratio of the regions, which is always above the required level. Finally, the methods give very close results in terms of number of regions (85 for Voronoï vs 87 for \texttt{contbin} for a S/N of 300), thus being equivalent computationally. For our purposes, \texttt{contbin} tool provides a more suitable binning algorithm than Voronoï. More ample tests also showed no significant difference in the recovery of the physical parameters between both techniques.

\section{Validation of the simulation pipeline}
\label{app:2}

\subsection*{Test hypothesis} 

The accuracy of the simulation pipeline needs to be verified using the cluster inputs provided. These inputs are 3D cubes of data, which needs to be projected along the line-of-sight of the instrument to be compared to the outputs of the end-to-end simulations. Ideally, this projection should be deterministic and give an unequivocal results. However, since the parameter distribution along the line-of-sight cannot be perfectly integrated (we only measure discrete number of counts, affected by statistics and background sources), multiple schemes exist to compare inputs and outputs depending on the physical quantity we wish to compare. Among those, the most widely used include:
\begin{itemize}
\item Emission-weighted projection, using the product $\rho^2 \sqrt{T}$ of each element along the line-of-sight. 
\item Mass-weighted projection, using the mass of each element
\item Emission-measure-weighted projection, using the emission-measure of a given line derived from Eq.~\ref{eq:norm}
\item {\color{black} Spectroscopic} schemes, as defined in \citep{Mazzotta2004Spectral}
\end{itemize} 

\begin{figure} [!b]
\centering
\includegraphics[width=0.45\textwidth, trim={0cm 0 0cm 0.5cm}, clip]{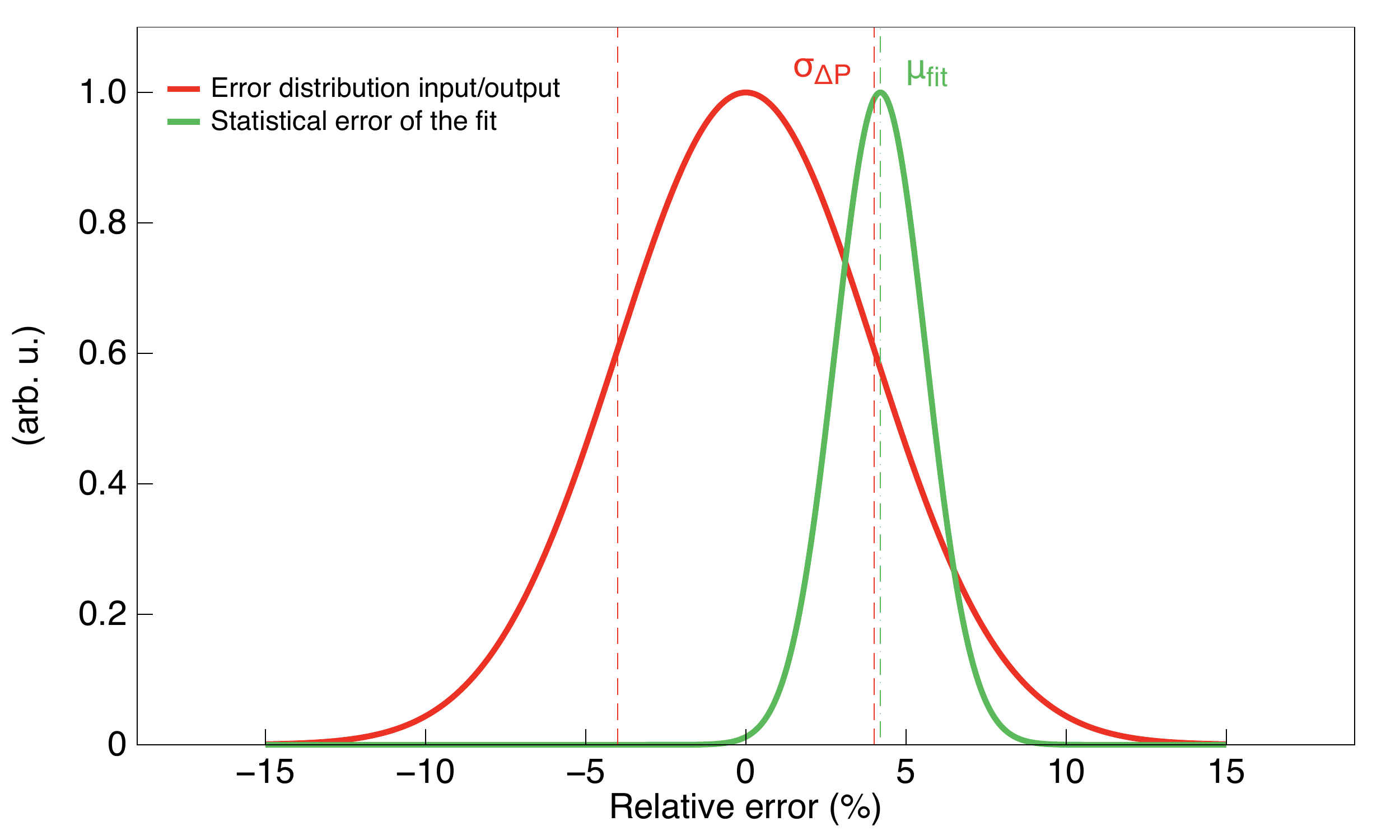}
\caption{Comparison between the relative error distribution of the parameters (red) and the fitting error distribution (green - symmetric but not shown here for clarity). Ideally, if no biases are present relative error distribution should be centred and its standard deviation $\sigma_{\Delta P}$ should be very close to the mean value of the fitting error distribution $\mu_{\rm fit}$.}
\label{fig:distrib}
\end{figure}

The accuracy of the simulated measurements was first tested by taking as estimator the relative error distribution of the output map, using emission-weighted-input maps as proxy for the input parameters. Being each region much larger than the telescope PSF they are considered independent on a strict statistical term. The relative error is assumed to follow a Gaussian distribution given the sufficiently high number of regions considered ($\geq 80$), with a mean $\mu_{\Delta P}=0$ (if no biases are present) and a standard deviation of $\sigma_{\Delta P}$, which indicates the total error on the parameter. The fitting error returned by XSPEC should also be Gaussian, and centred on given a value $\mu_{\rm fit}$ which depends on the exposure time and the emission model parameters. For an accurate measurement, the value of  $\mu_{\rm fit}$ should be comparable to $\sigma_{\Delta P}$ do all parameter (Figure~\ref{fig:distrib}).  A second test can also be performed using as estimator the ratio between the relative difference and the XSPEC error $\sigma_{\rm fit}$ for each region, i.e., $\chi_j=(P_{\text{fit}, j}-P_{\text{in}, j})/\sigma_{\text{fit}, j}$ and the corresponding reduced chi-square. Using the emission-measure-weighted input and the output distributions, let us take as null hypothesis (H$_0$): \textit{"The measurements obtained with the pipeline are consistent with the statistical errors for a given exposure time"} and (H$_1$): \textit{"The measurements are biased"} with a threshold $p_{\alpha} = 5$\% (i.e., 97.5\% of the Gaussian distribution, or $\sim$\,2.5\,$\sigma$). 
\vspace{-0.5cm}

\paragraph{\textbf{H$_0$:}} (H$_0$) can be rejected if the value of $\sigma_{\Delta P}$ is outside $\mu_{\rm fit} \pm 2.5 \sigma_{\rm fit}$ for all the parameters. This can also be seen on the $\chi_j$ maps, if the number of regions with  $|\chi_j| \geq 2.5$ is high. For all runs, the value of the error dispersion is always within this threshold and the number of outliers regions is small (see also Appendix~\ref{app:3} and Table~\ref{table:errors}), indicating that the (H$_0$) is valid and errors are consistent with the corresponding statistics.

\begin{figure} [t]
\centering
\includegraphics[width=0.45\textwidth, trim={0cm 0 0cm 0.5cm}, clip]{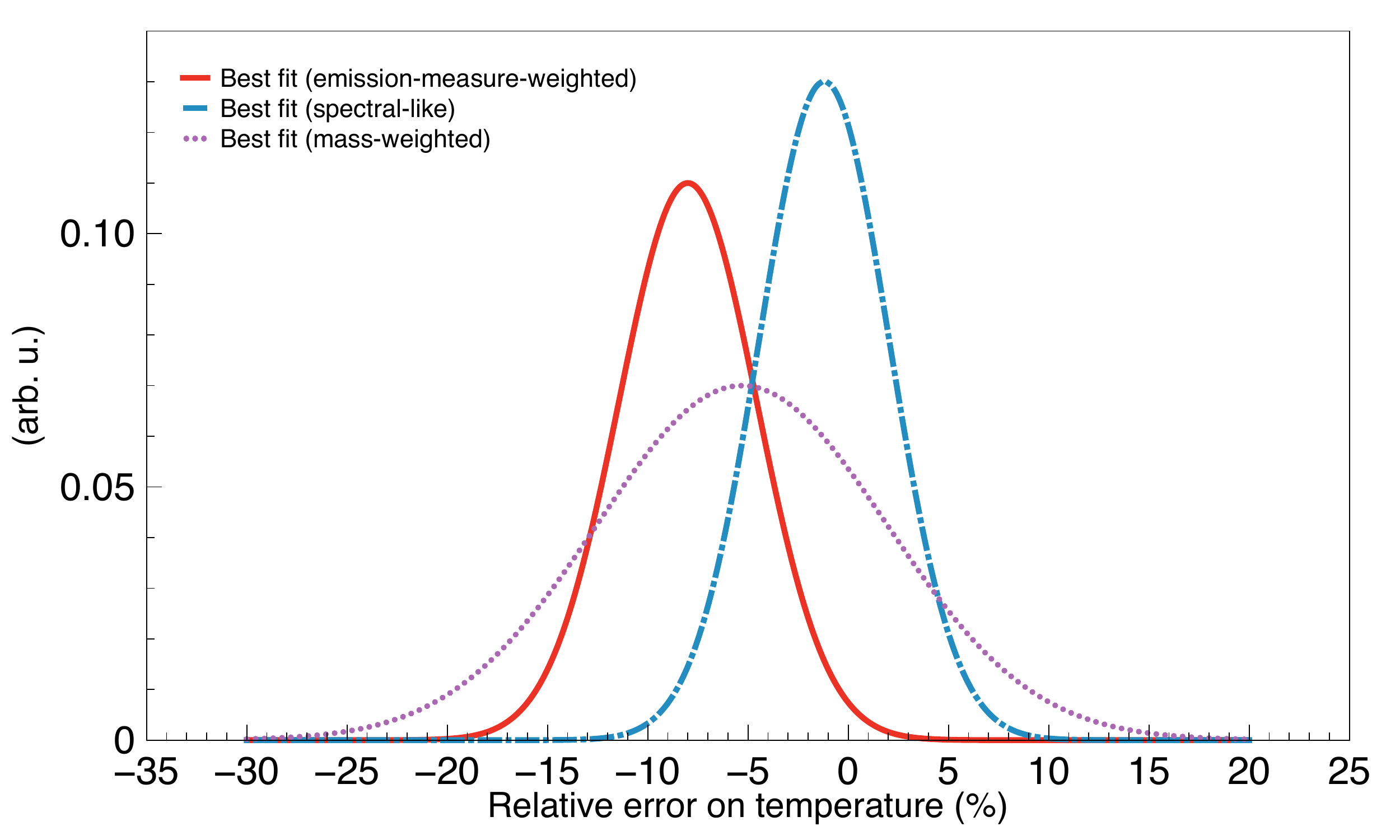}
\caption{Gaussian best fits of the normalised relative error distribution on the measured temperature for different input map weighting scheme for cluster C4. The red solid curve indicates the emission-weighted best fit of the distribution ($\mu_{\rm \Delta T}=-8.0\%$, $\sigma_{\rm \Delta T}=3.4\%$). The blue dash-dotted line indicates the {\color{black} spectroscopic} temperature best fit ($\mu_{\rm \Delta T}=-1.2\%$, $\sigma_{\rm \Delta T}=2.2\%$), while the dotted violet line indicates the best fit for a mass-weighted input ($\mu_{\rm \Delta T}=-5.2\%$, $\sigma_{\rm \Delta T}=7.2\%$).}
\label{fig:bias}
\end{figure}

\begin{figure} [!t]
\centering
\includegraphics[width=0.45\textwidth, trim={0cm 0 0cm 0cm}, clip]{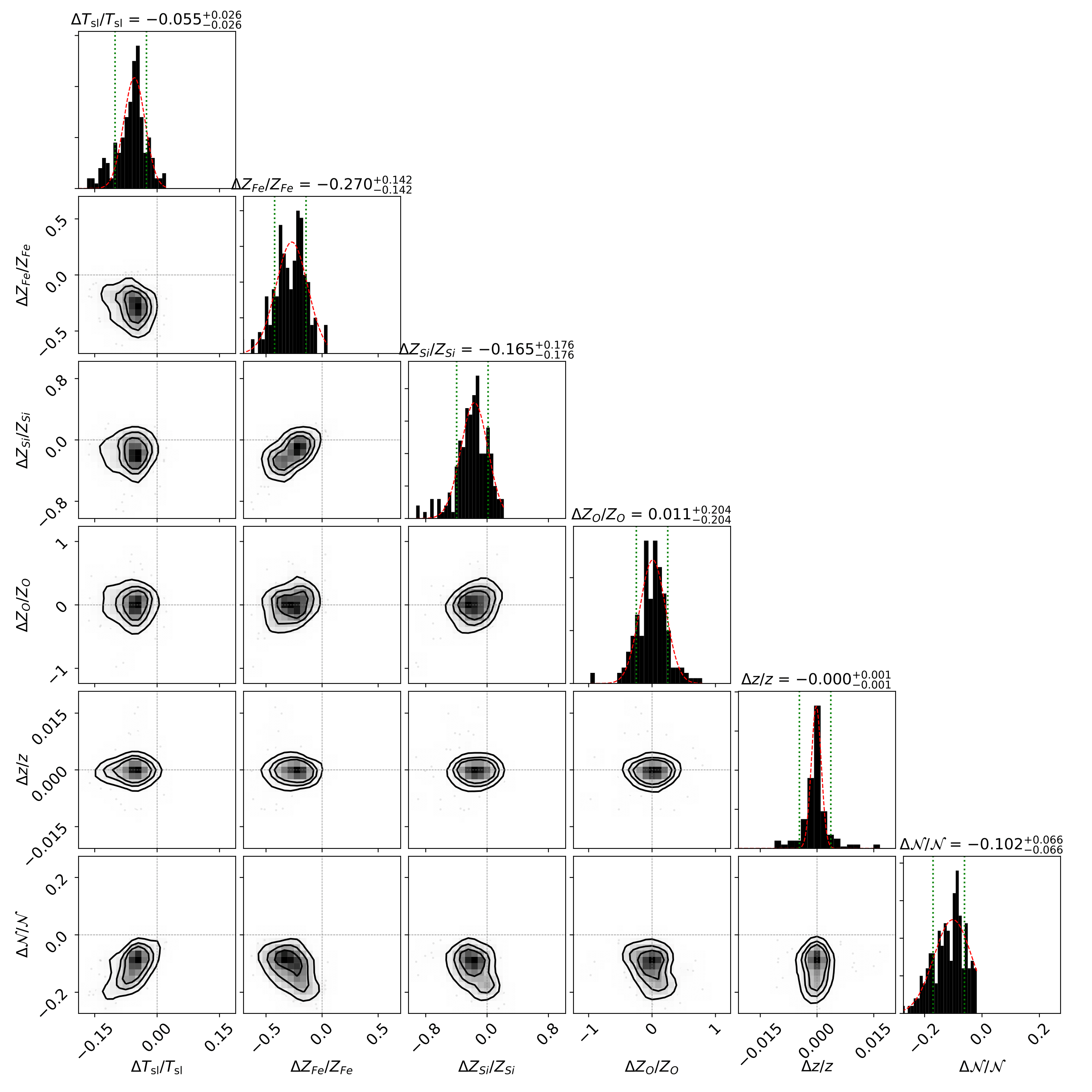}
\includegraphics[width=0.45\textwidth, trim={0cm 0 0cm 0cm}, clip]{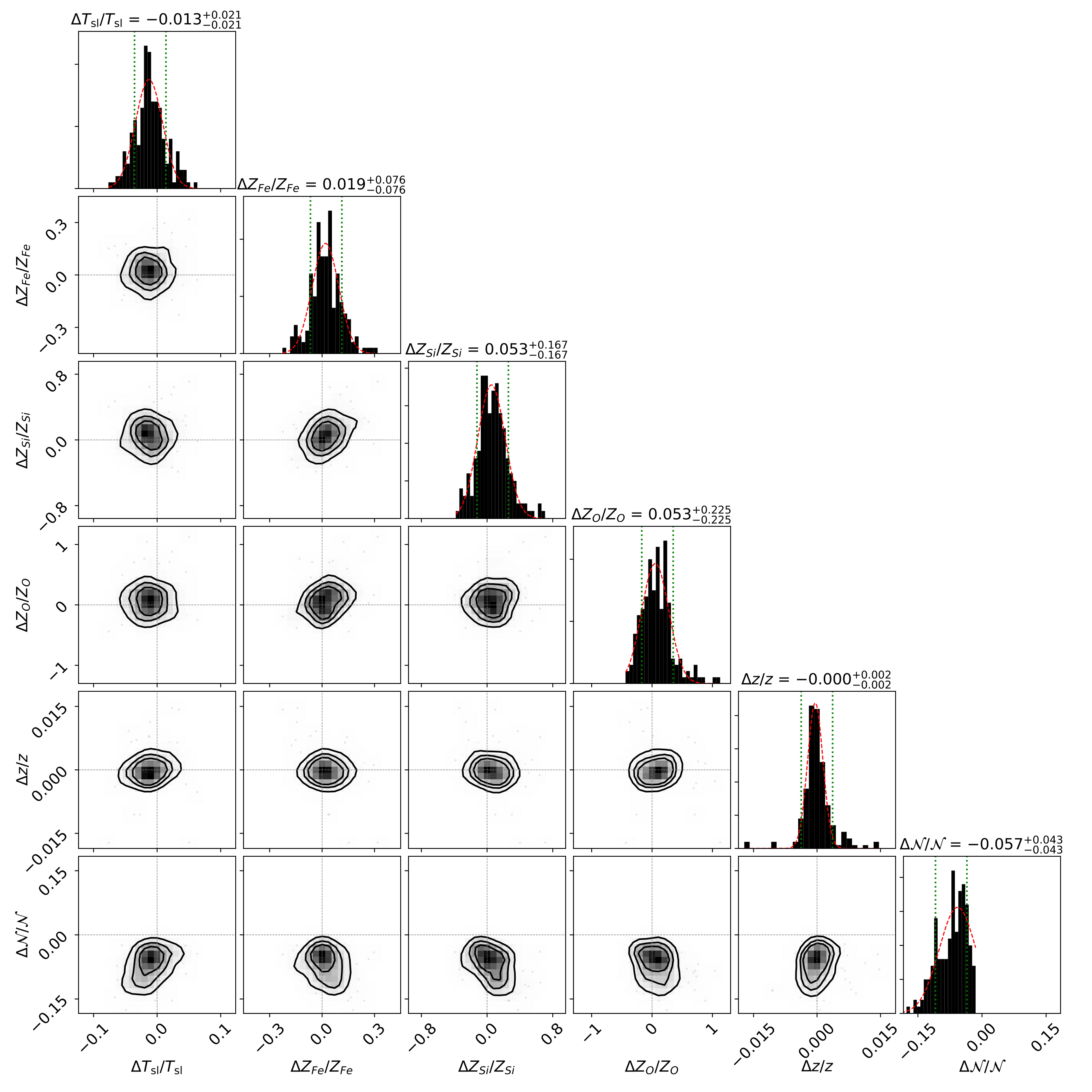}
\caption{Corner plots of the relative error on parameter $\nicefrac{(X_{fit} - X_{inp})}{X_{inp}} = \nicefrac{\Delta X}{X_{inp}}$ as function of parameter $\nicefrac{(Y_{fit} - Y_{inp})}{Y_{inp}} = \nicefrac{\Delta Y}{Y_{inp}}$, for the {\color{black} spectroscopic} temperature, $T_{\rm sl}$, oxygen, silicon and iron abundance, redshift, $z$, and the normalisation $\mathcal{N}$, for cluster C4. The diagonal panels are the corresponding relative error distribution, where the red solid line indicates the Gaussian best fit of the distribution (parameters $\mu_{\Delta P}$, $\sigma_{\Delta P}$ given above) and the dotted line is the value of $\mu_{\Delta P} \pm \mu_{\rm fit}$. (\textit{Top}) Broadband fit. (\textit{Bottom}) Multi-band fit (Sec.~\ref{sub:sys}) considering a {\color{black} spectroscopic} temperature.}
\label{fig:corner}
\end{figure}

\paragraph{\textbf{H$_1$:}} (H$_1$) can be rejected if $\mu_{\Delta P} \sim 0$, within $\sim 2.5$ times the mean standard error of the distribution (due to the finite size of the sample). The samples are generally composed of $N_{\text{reg}} \sim$ 80 regions and the standard error on $\mu_{\Delta P}$ is given by $\sigma_{\Delta P}/\sqrt N_{\text{reg}}$. Under these assumptions, clear biases are visible in the reconstructed emission-weighted and mass-weighted temperature maps, which presented a systematic underestimation of $\sim$ 5-10\% (Figure \ref{fig:bias}). This bias is explained by mixing effects along the line-of-sight and complexity to disentangle multi-temperature plasma with a single plasma model \citep{Mazzotta2004Spectral}. It can be reduced using {\color{black} spectroscopic} temperature maps (Fig.~\ref{fig:bias}). The use of the broadband fit induced in fact many other visible biases, notably between abundances (O, Si, Fe) and temperature (Fig.~\ref{fig:corner} -- \textit{Upper panel}). The use of multi-band fitting (detailed in  Sec.~\ref{sub:sys}) significantly reduces these biases (Fig.~\ref{fig:corner} -- \textit{Lower panel}) within the statistical variations of the parameters. Despite this improvement, small correlations are visible between abundances and temperature ($\sim$ 1\%) and between the normalisation and all other parameters ($\sim 5$\%). Efforts to reduce this bias were conducted by fixing or releasing various fitting parameters, without significant success. A clear rejection of the null hypothesis (H$_1$) cannot be performed, although results suggest that small residual biases and correlations remain in the current fitting procedure, mainly on normalisation. 
\vspace{-0.5cm}

\subsection*{Influence of the projection scheme}
Depending on the weighting schemes (mass-weighted or emission-measure-weighted) different error distributions of the same quantities can be obtained. These discrepancies add a further complexity to evaluate any potential bias in the pipeline (Figure \ref{fig:bias}).

\subsection*{Accuracy in terms of probability distributions}
The previous test is only valid whenever the error distributions are assumed to be Gaussian, which is unfortunately not always the case (slight deviations from a Gaussian behaviour are observed). If so, the accuracy of our method needs to be tested in the sense of the statistical distributions performing for instance a Kolmogorov-Smirnov (KS) test over the output/input distributions. The KS test compares the probability $p_{KS}$  for two random variables to be drawn from the same data set (i.e., same probability density). Using both the input and output distribution, this test showed high $p_{KS}$ values (between 0.6 and 1 for the different parameters), which gives strong hints that the output distribution indeed matches the input. Statistically speaking, no real conclusion can be achieved with a single realization of the observation. To fully validate the pipeline, a large number of observations of the same cluster (either along the same line-of-sight or by taking multiple lines-of-sight) would be needed to perform a meaningful comparison using a KS method. Unfortunately, one full simulation of a cluster takes the order of a day, making it computationally cumbersome to carry out this test.  For simplicity, a very high exposure time simulation of these extended sources were carried out instead. Although beyond the scope of this paper, such observations ($\geq 1$\,Ms) with the same binning regions decrease most of the biases below 2\% and create distributions which much more alike ($p_{KS} \approx 0.8/1$), suggesting that the residual errors are in part related to statistics and to the fitting scheme. 

\section{Further results of the simulations}
\label{app:3}
We show here the reconstructed maps of the main physical parameters ({\color{black} spectroscopic} temperature, oxygen, silicon and iron) for clusters C1, C3 and C4.

\begin{figure*}[h]
\centering
\includegraphics[width=0.34\textwidth, trim={50cm 2cm 23cm 8cm}, clip]{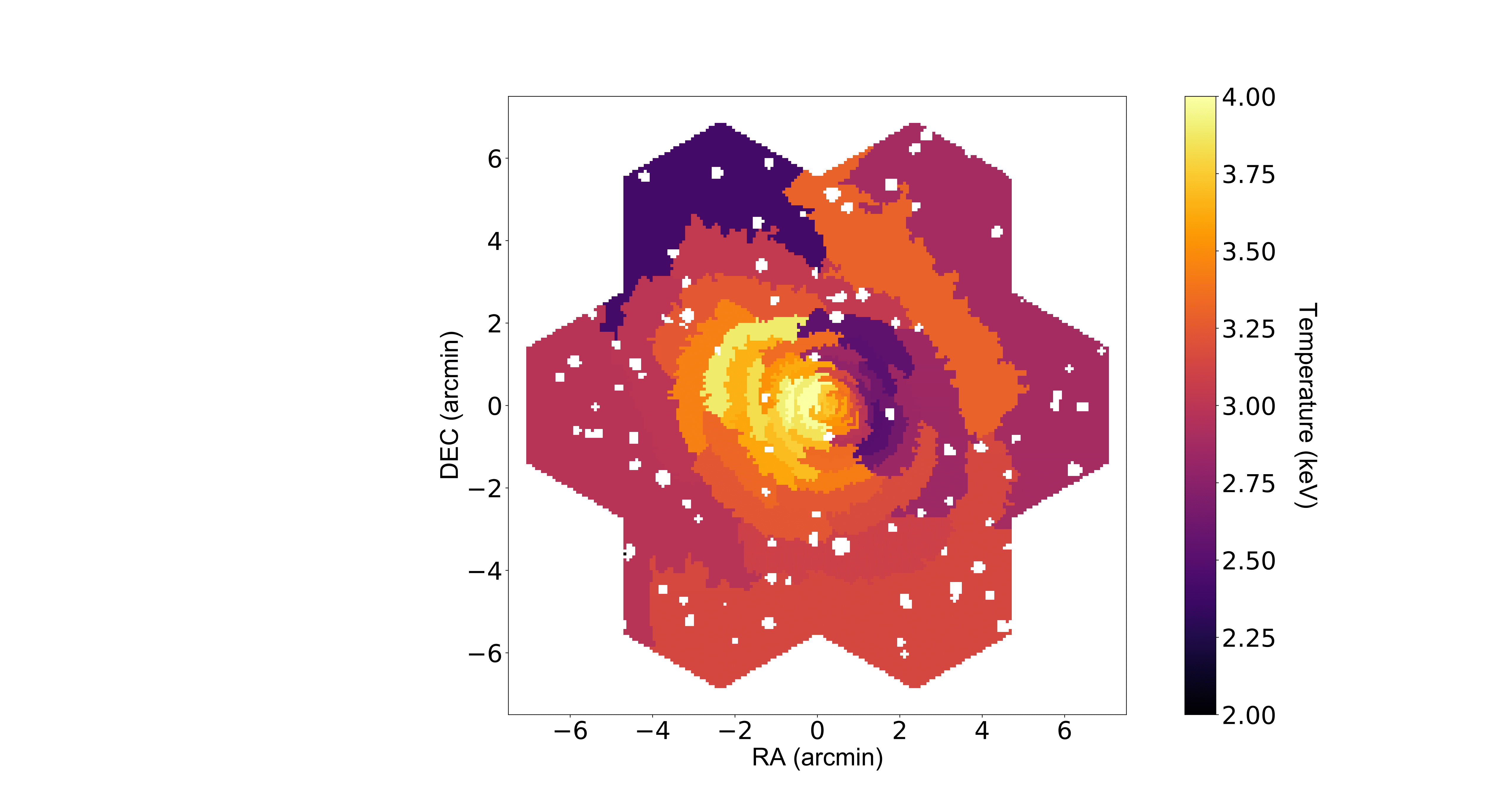}
\vspace{0.6cm}
\includegraphics[width=0.34\textwidth, trim={50cm 2cm 23cm 8cm}, clip]{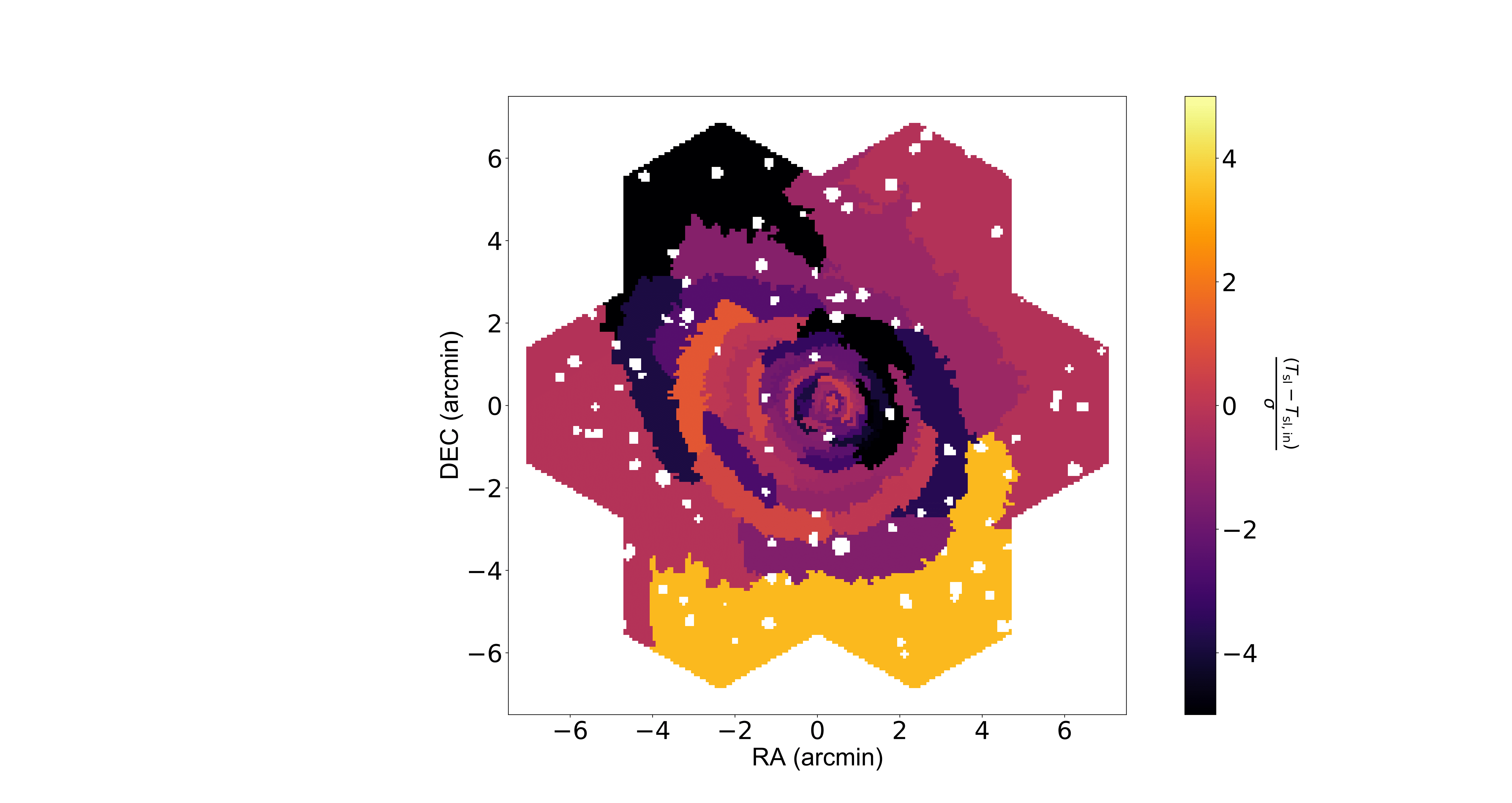}
\includegraphics[width=0.31\textwidth, trim={0 0 0 0.5}, clip]{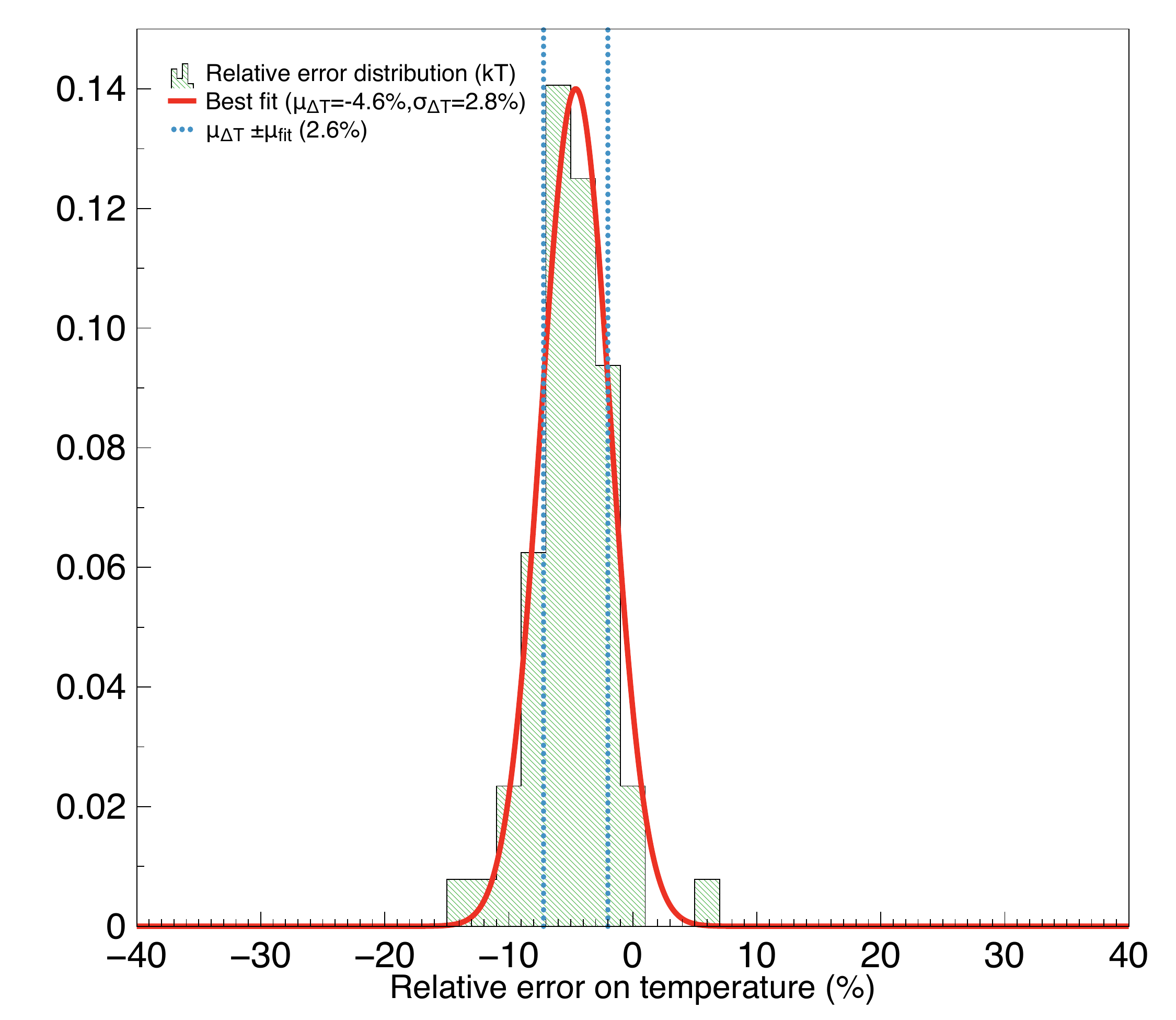}

\includegraphics[width=0.34\textwidth, trim={50cm 2cm 23cm 8cm}, clip]{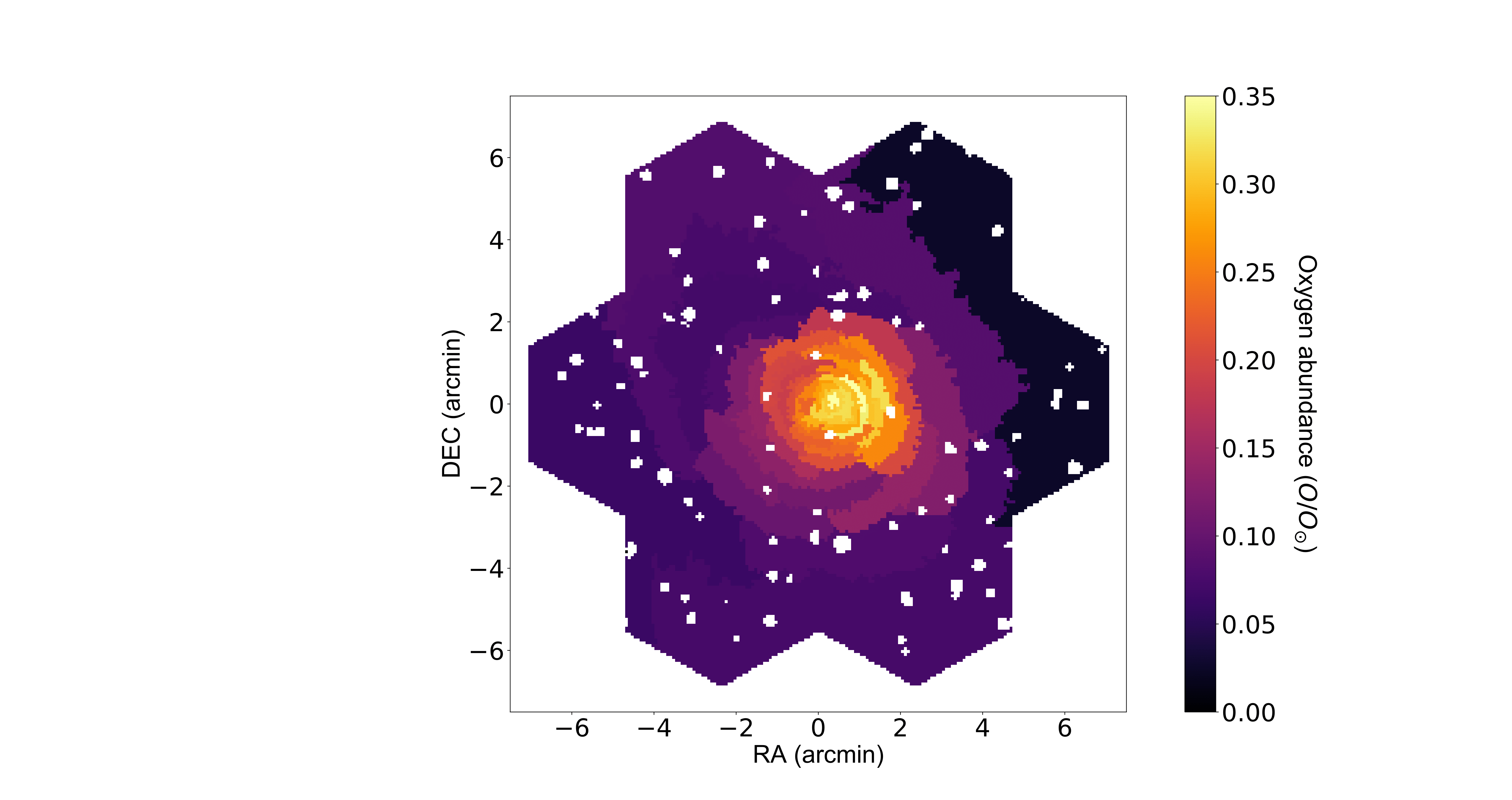}
\vspace{0.6cm}
\includegraphics[width=0.34\textwidth, trim={50cm 2cm 23cm 8cm}, clip]{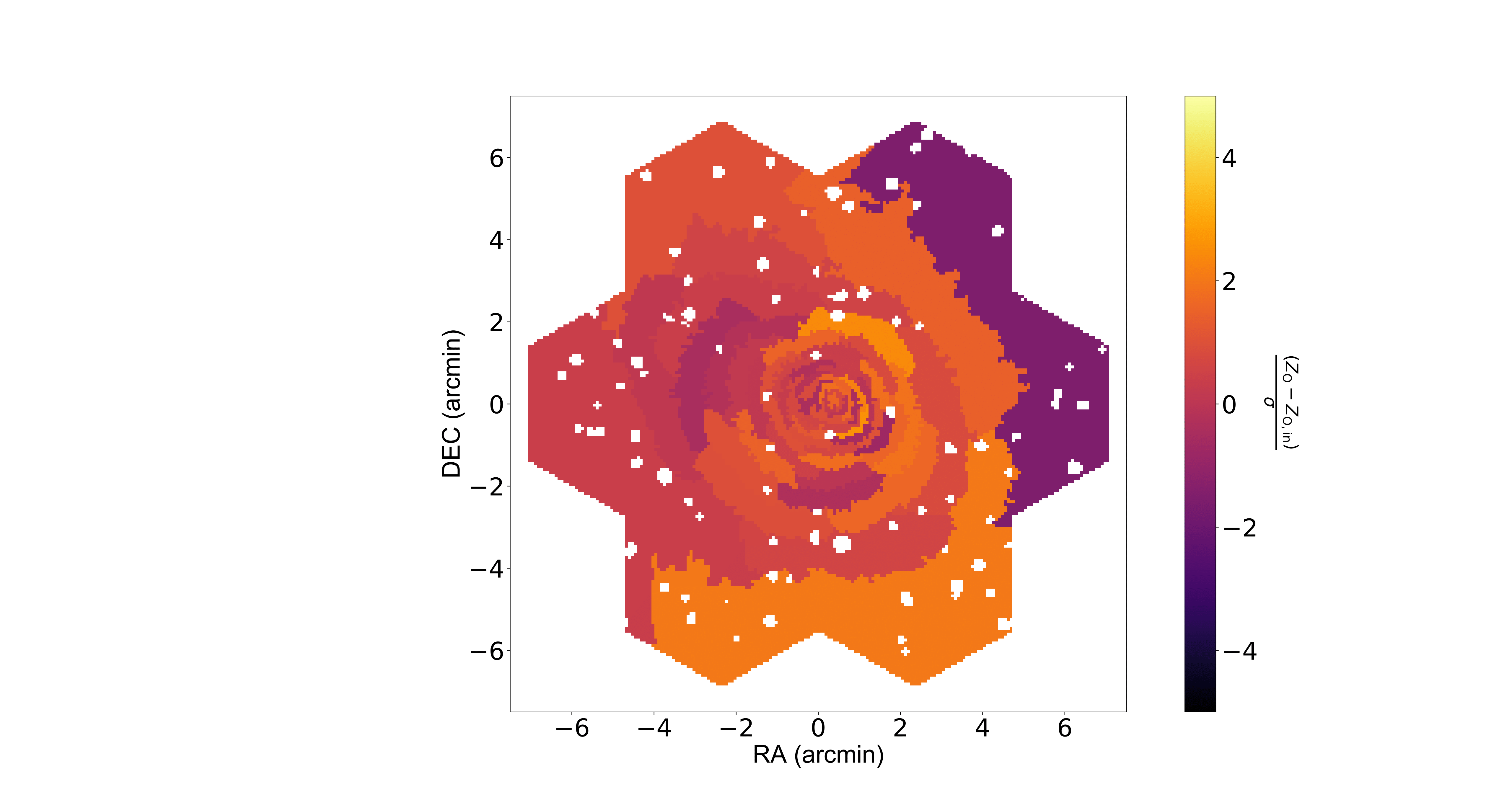}
\includegraphics[width=0.31\textwidth, trim={0 0 0 0.5}, clip]{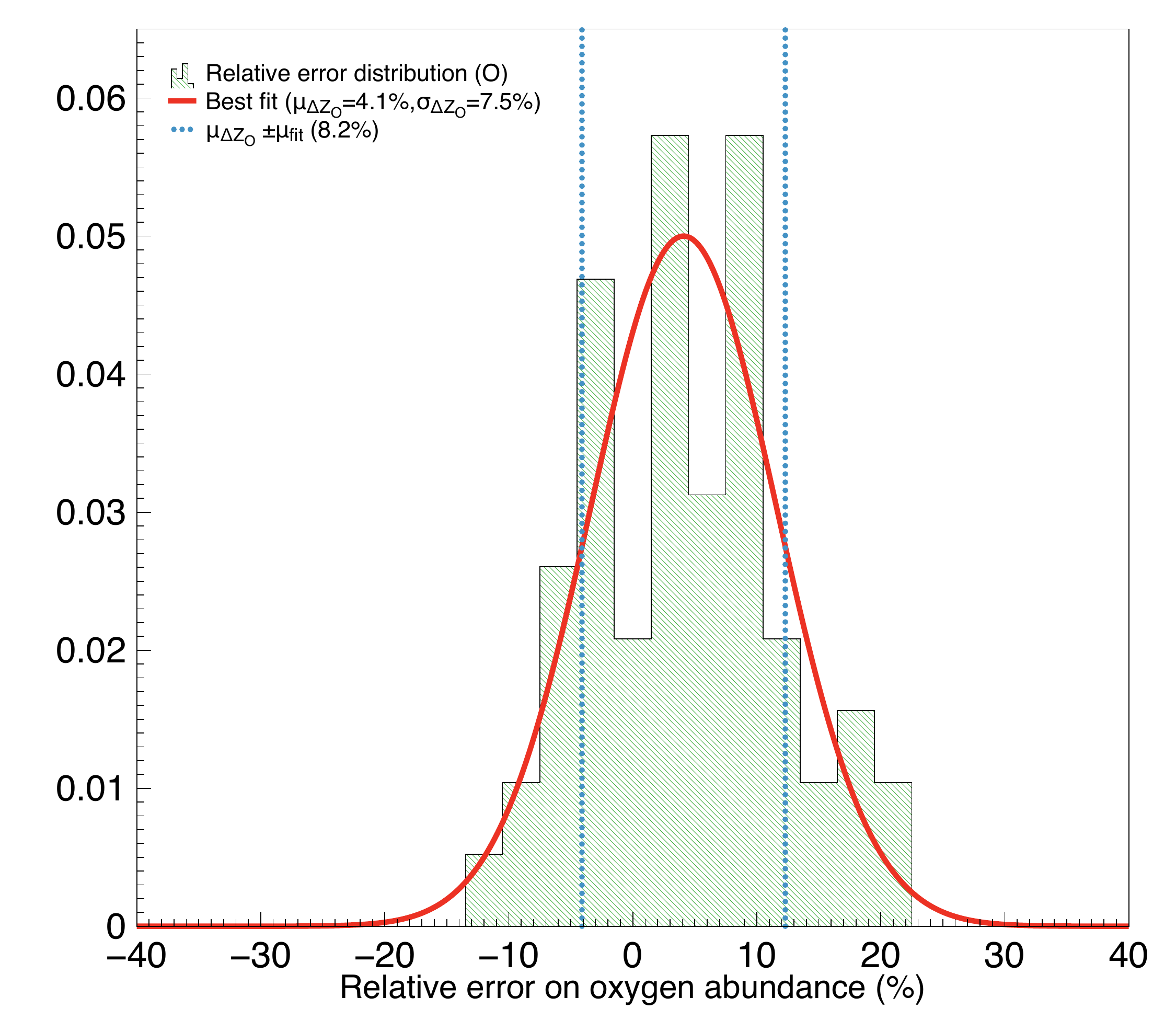}

\includegraphics[width=0.34\textwidth, trim={50cm 2cm 23cm 8cm}, clip]{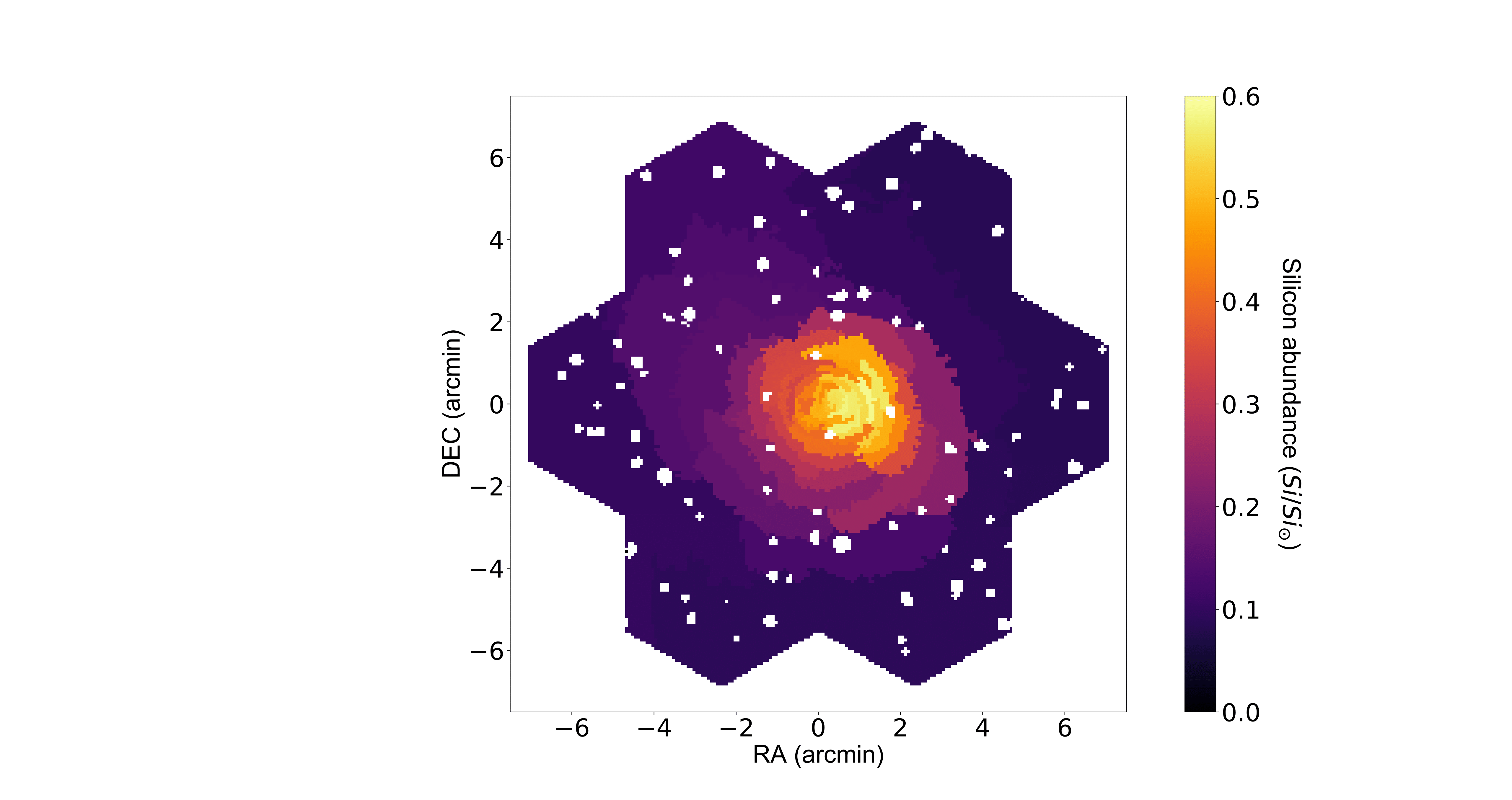}
\vspace{0.6cm}
\includegraphics[width=0.34\textwidth, trim={50cm 2cm 23cm 8cm}, clip]{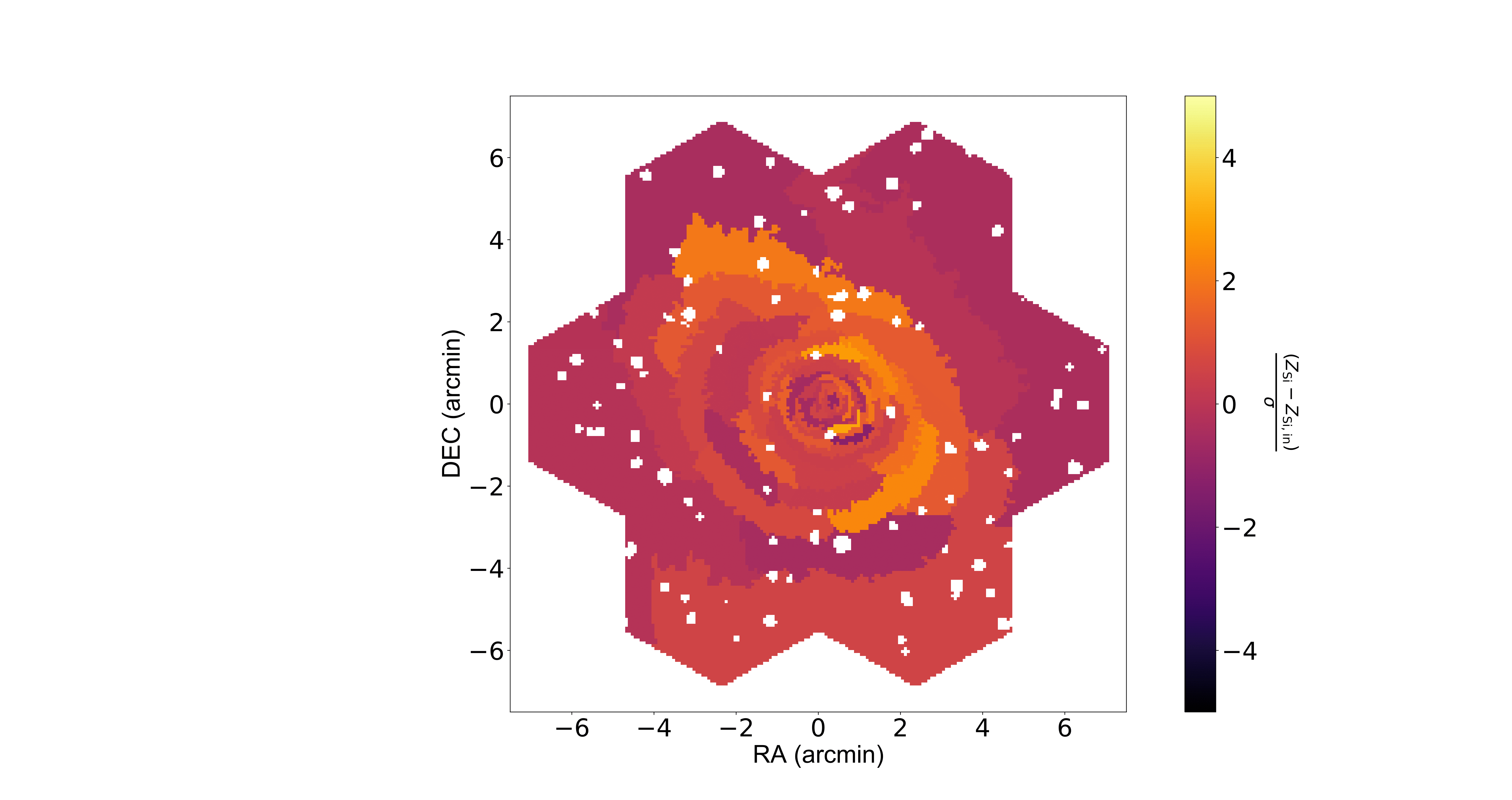}
\includegraphics[width=0.31\textwidth, trim={0 0 0 0.5}, clip]{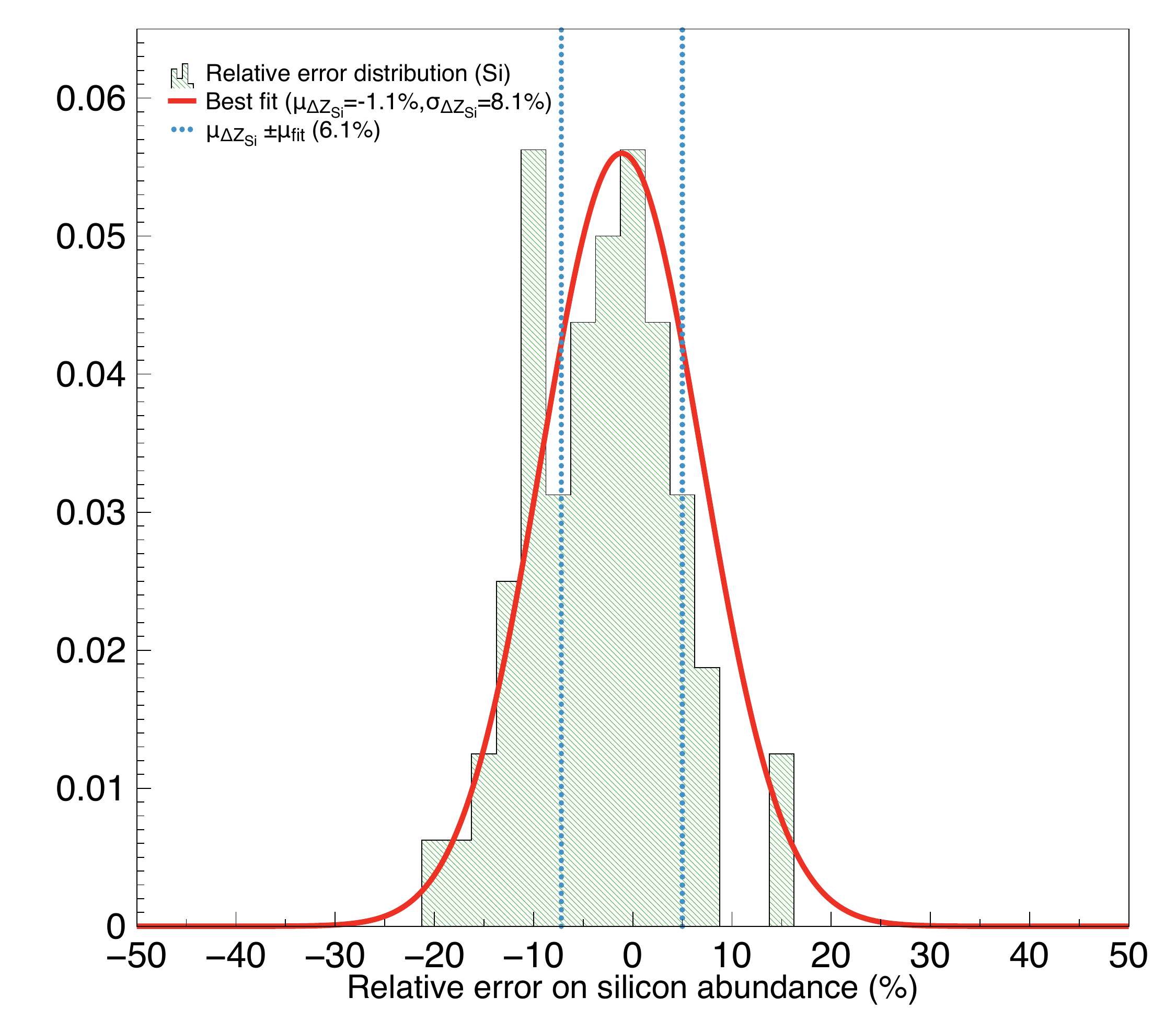}

\includegraphics[width=0.34\textwidth, trim={50cm 2cm 23cm 8cm}, clip]{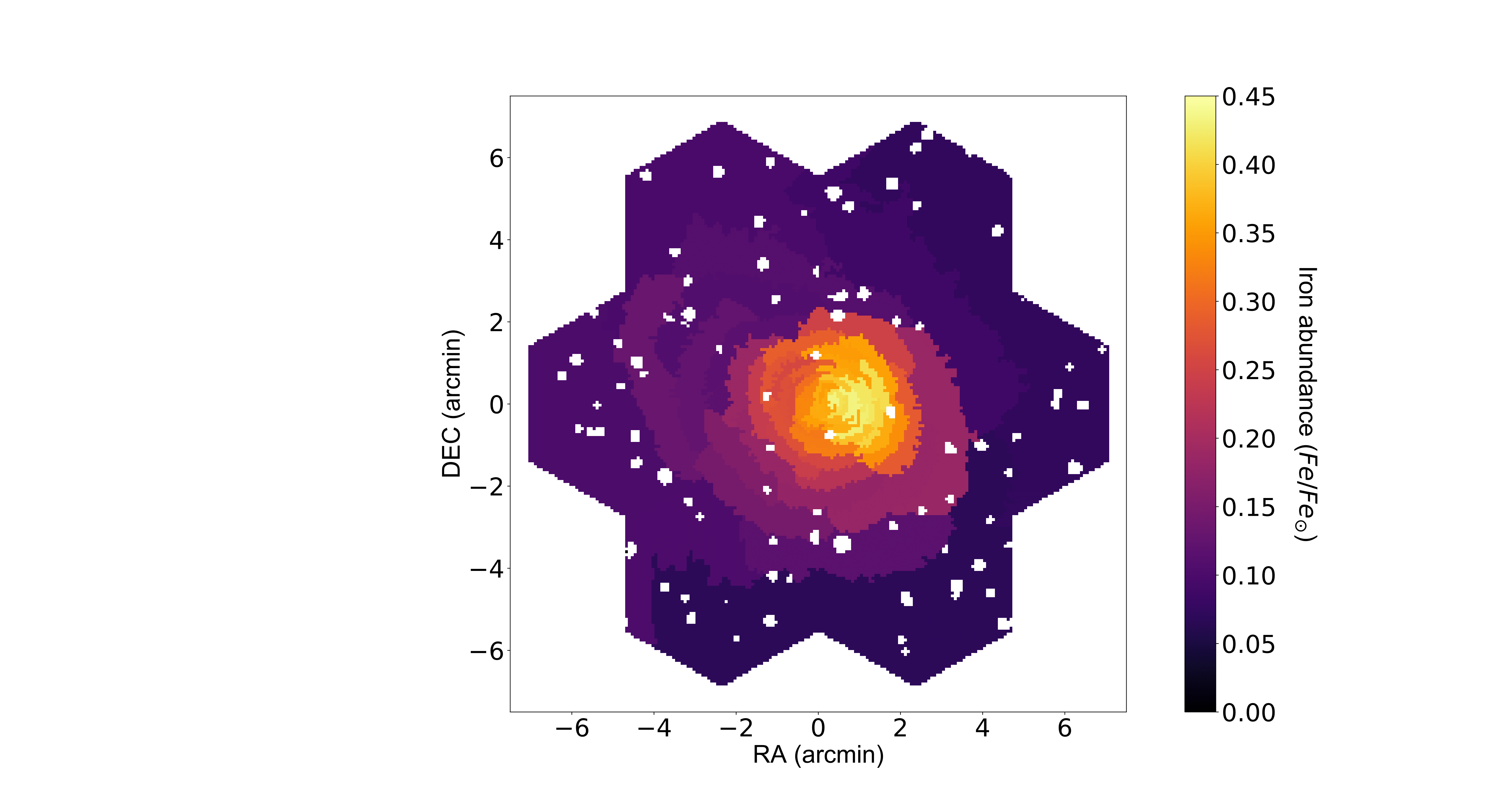}
\vspace{0.6cm}
\includegraphics[width=0.34\textwidth, trim={50cm 2cm 23cm 8cm}, clip]{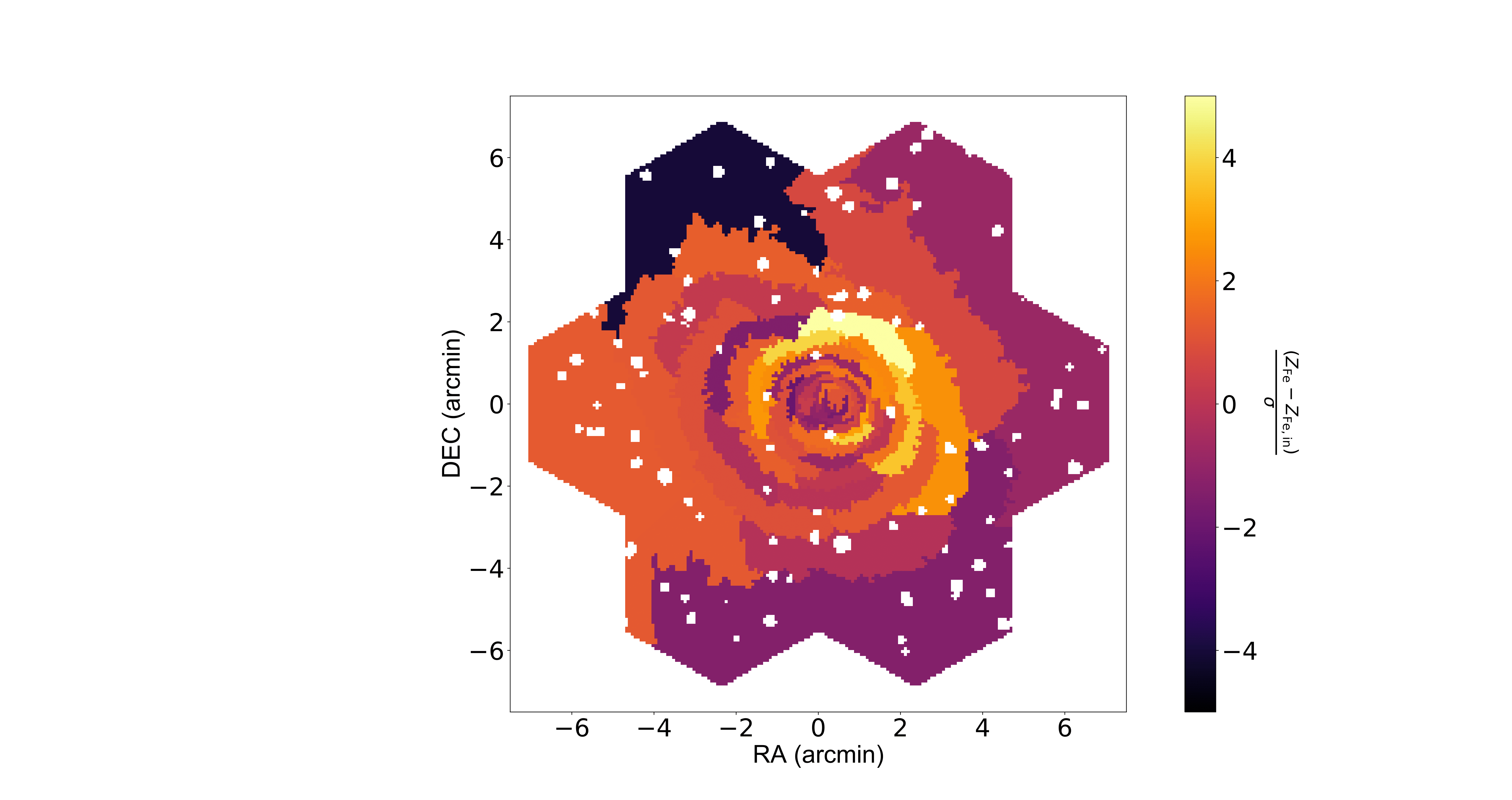}
\includegraphics[width=0.31\textwidth, trim={0 0 0 0.5}, clip]{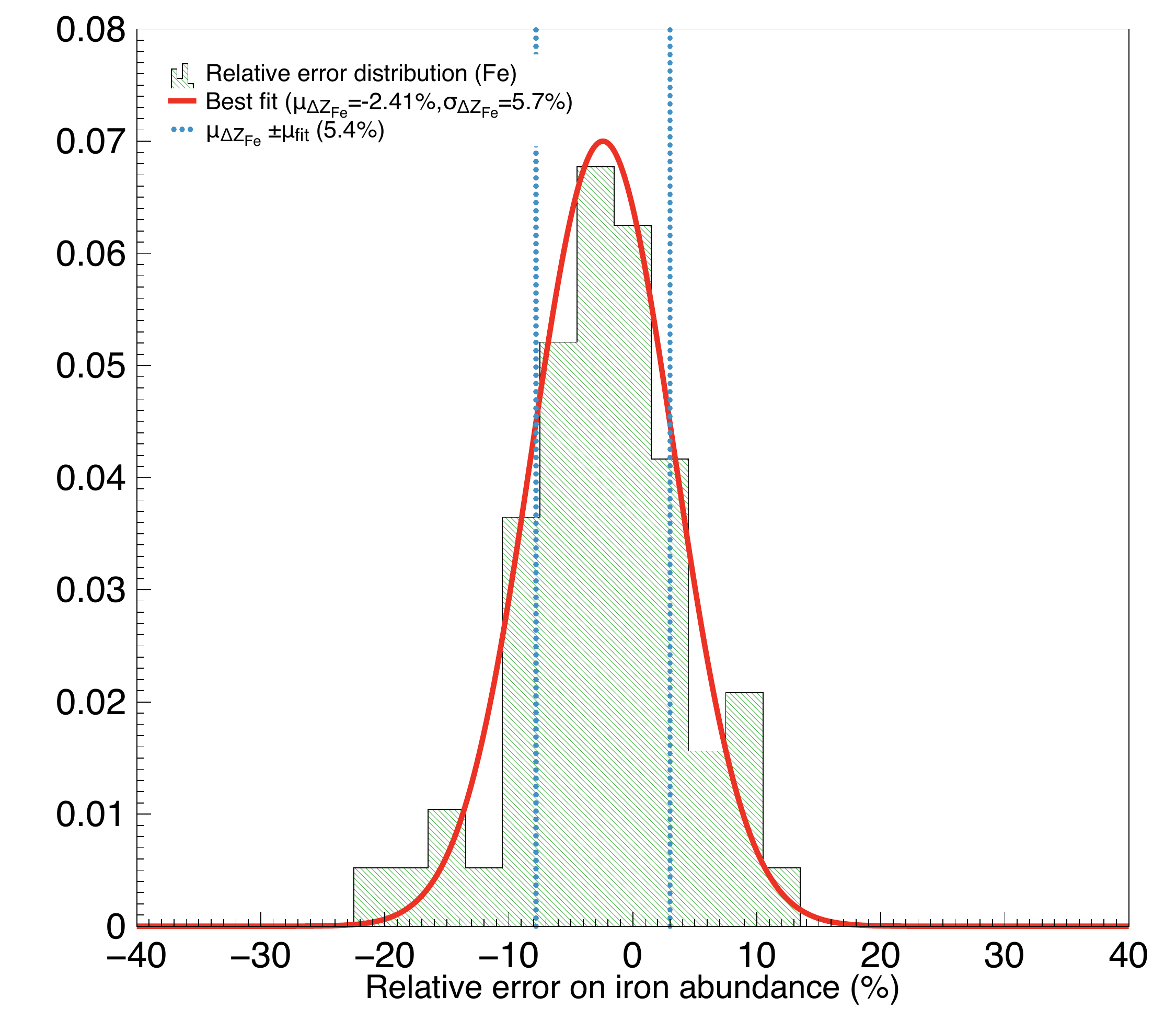}
\caption{Same as Fig.~\ref{fig:maps} for cluster C1.}
\label{fig:mapsC1}
\end{figure*}

\FloatBarrier

\begin{figure*}[p]
\centering
\includegraphics[width=0.34\textwidth, trim={50cm 2cm 23cm 8cm}, clip]{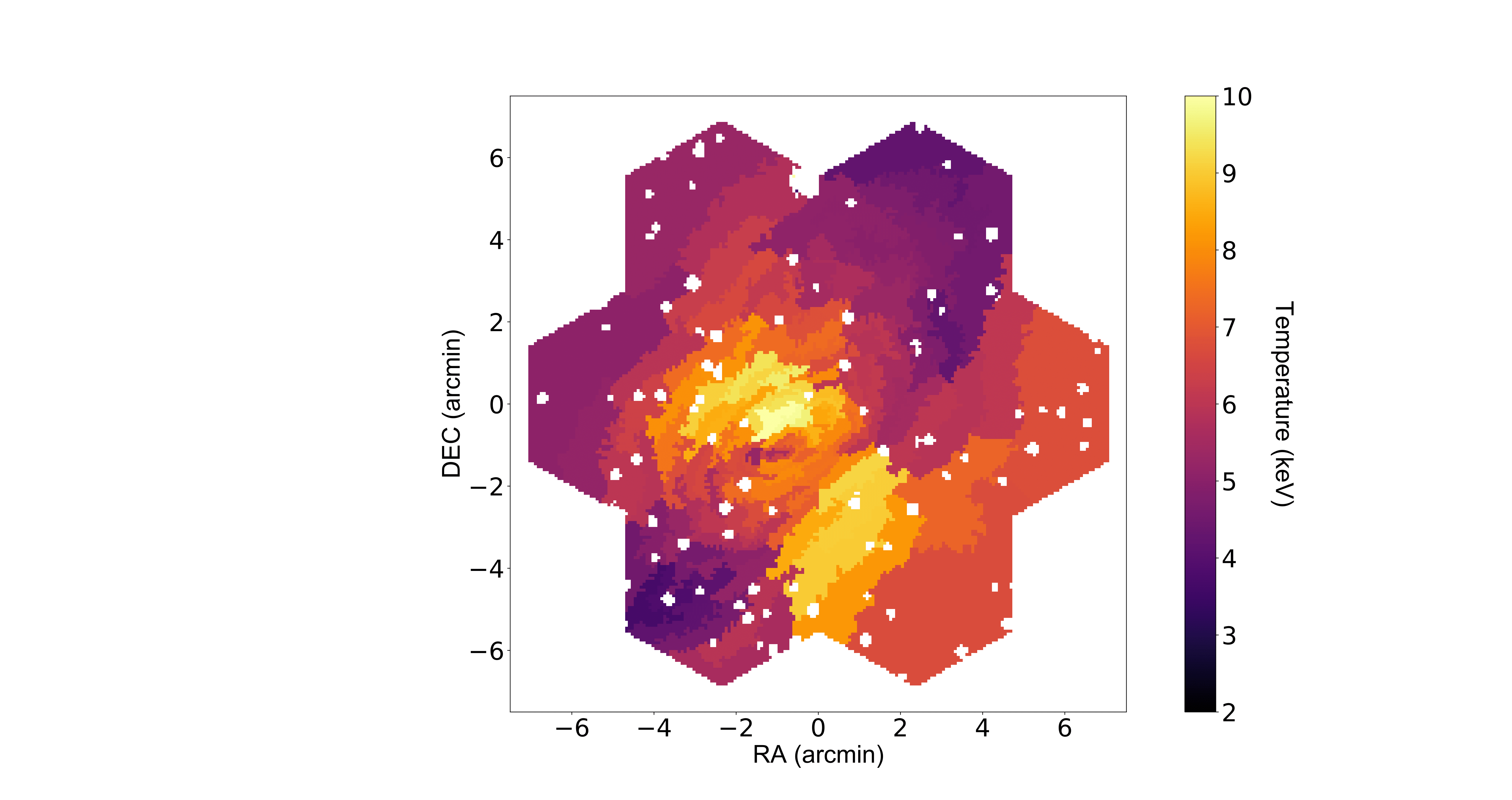}
\vspace{0.6cm}
\includegraphics[width=0.34\textwidth, trim={50cm 2cm 23cm 8cm}, clip]{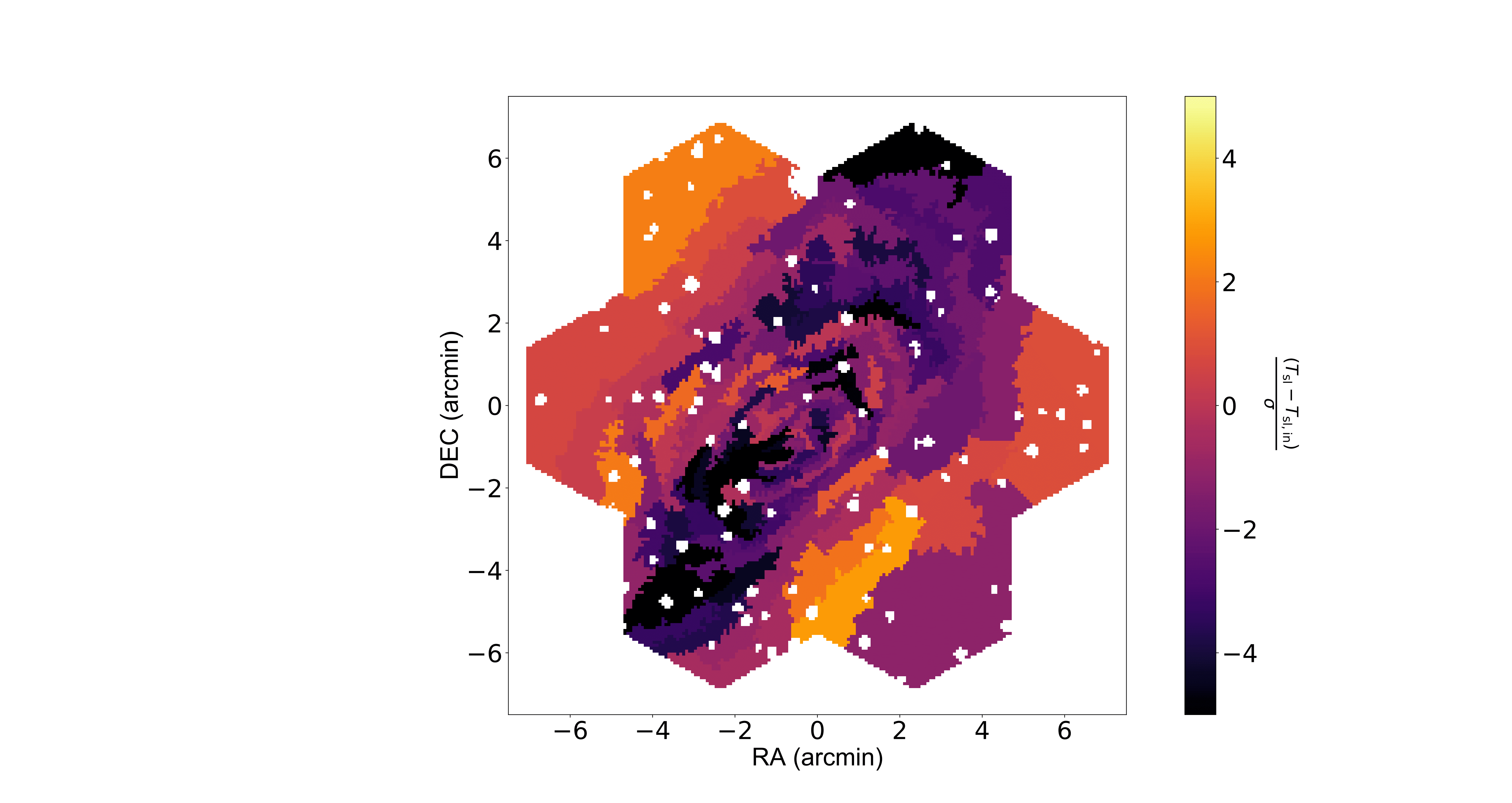}
\includegraphics[width=0.31\textwidth, trim={0 0 0 0.5}, clip]{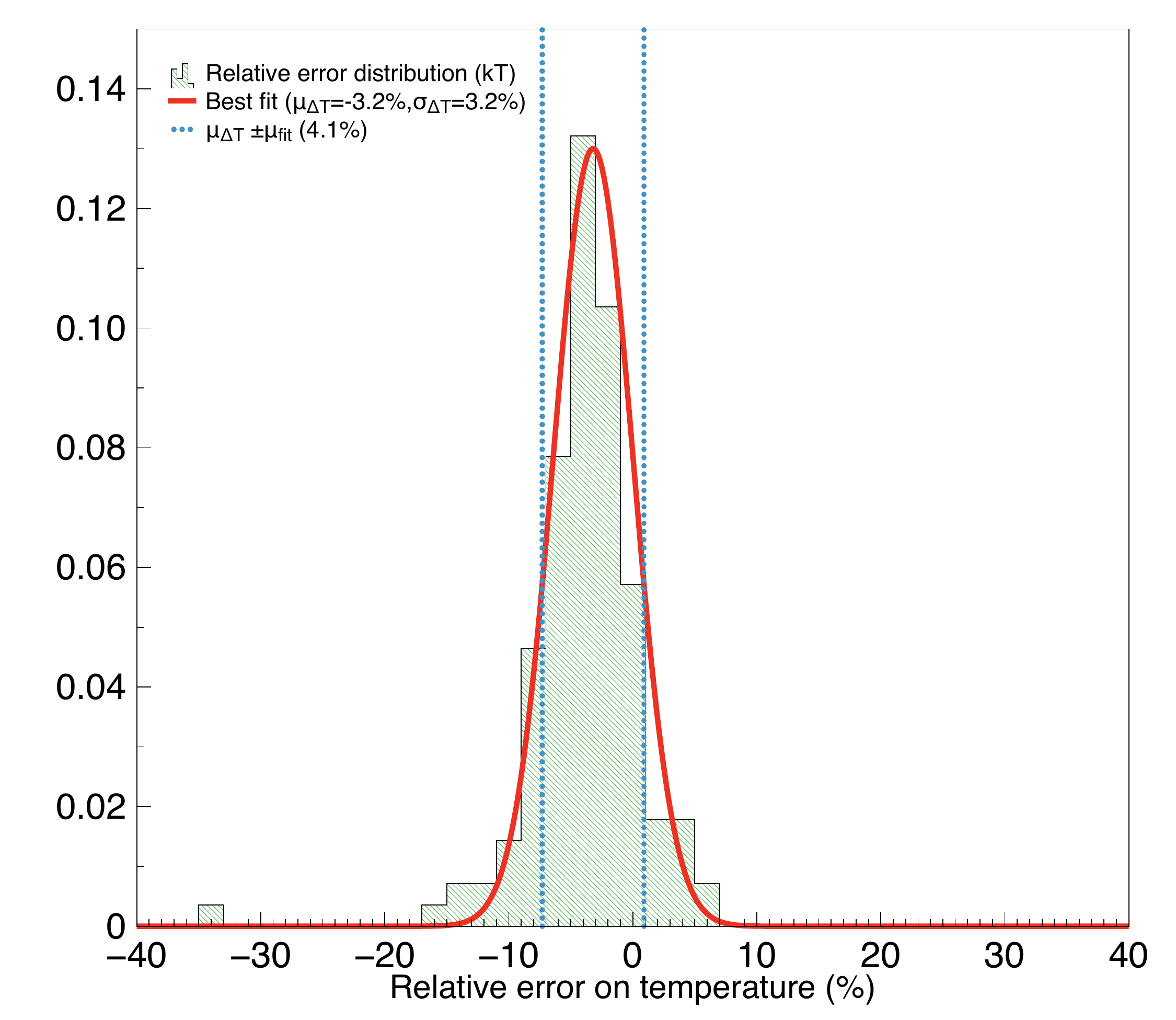}

\includegraphics[width=0.34\textwidth, trim={50cm 2cm 23cm 8cm}, clip]{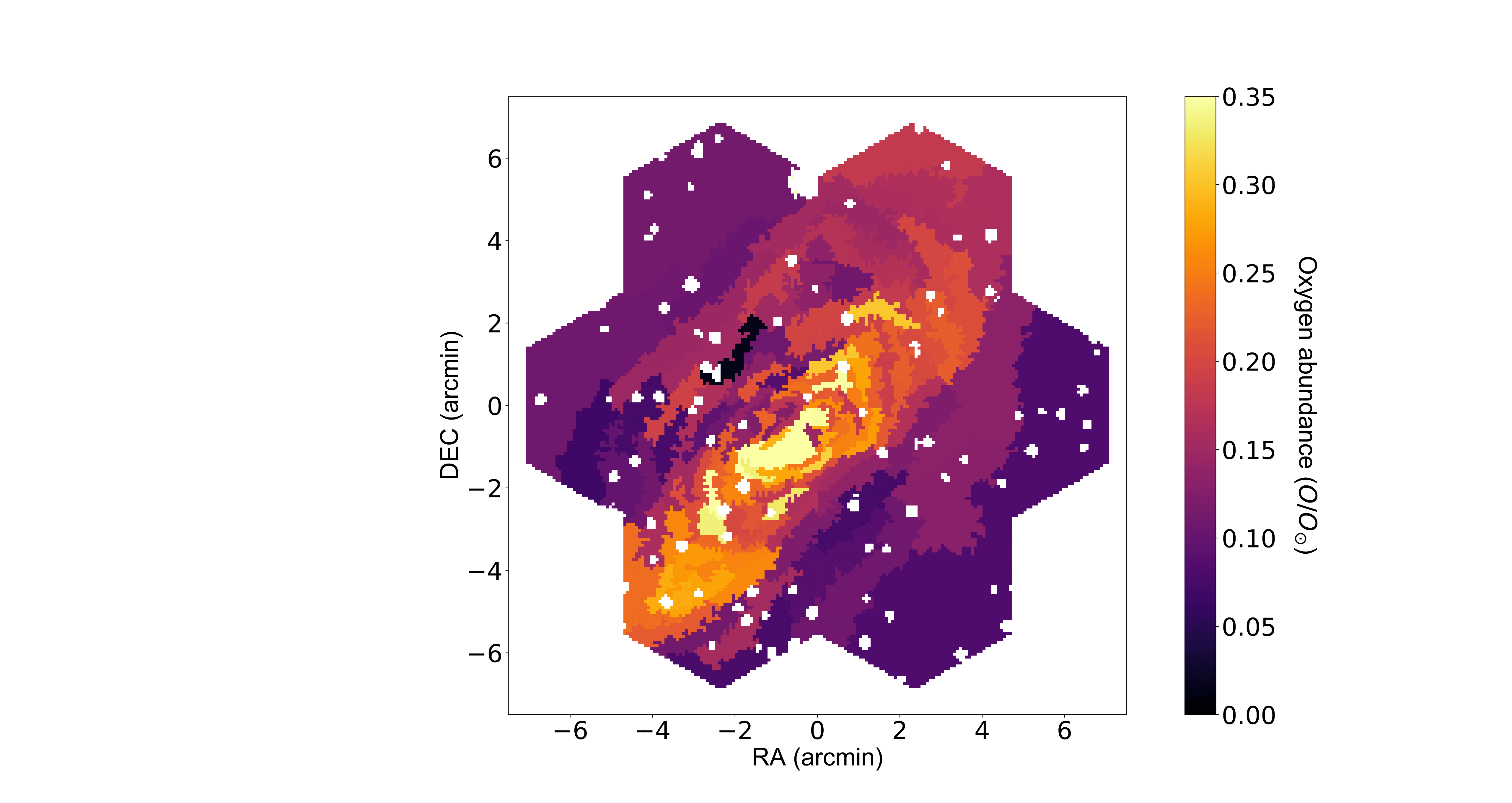}
\vspace{0.6cm}
\includegraphics[width=0.34\textwidth, trim={50cm 2cm 23cm 8cm}, clip]{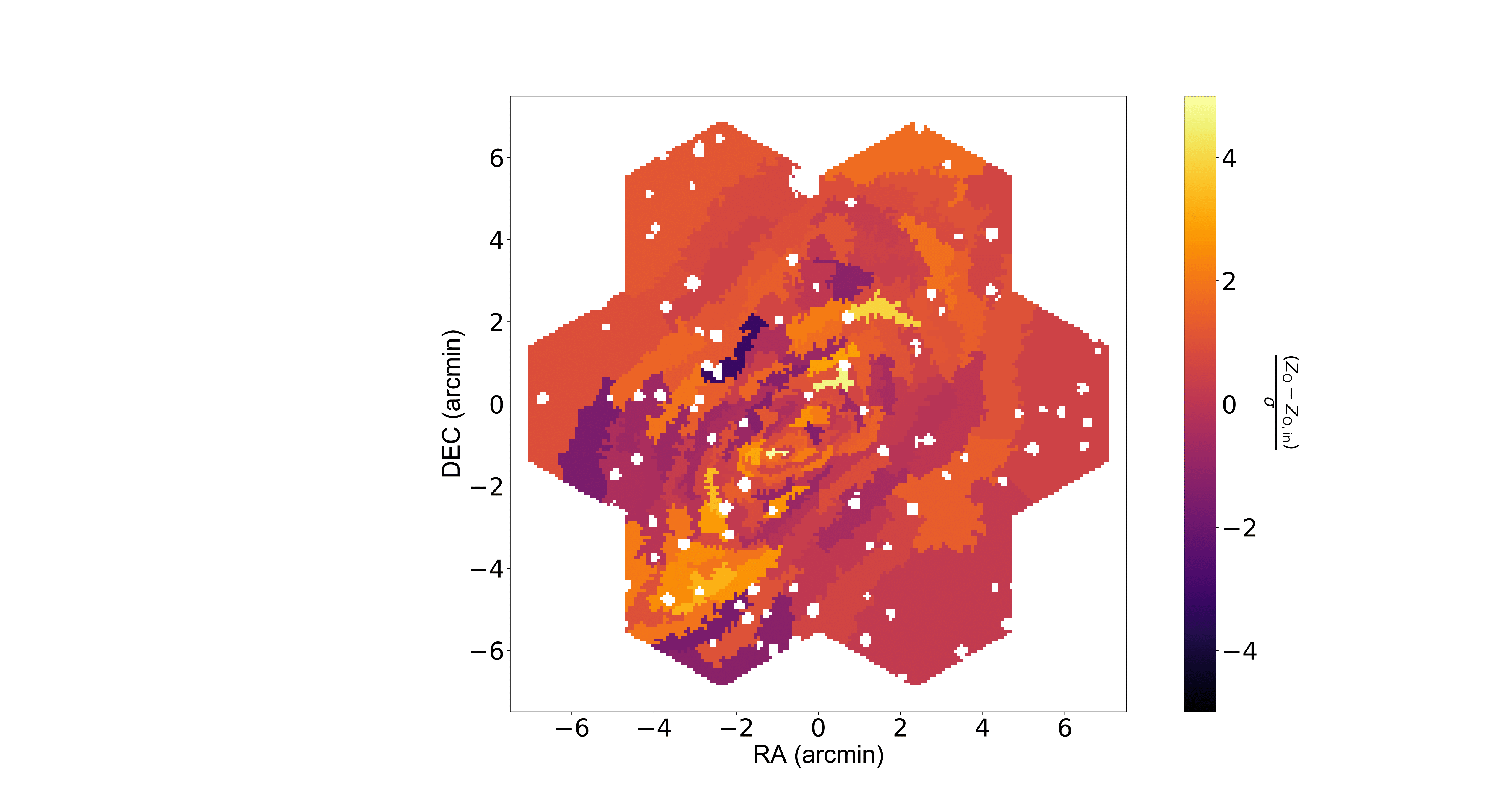}
\includegraphics[width=0.31\textwidth, trim={0 0 0 0.5}, clip]{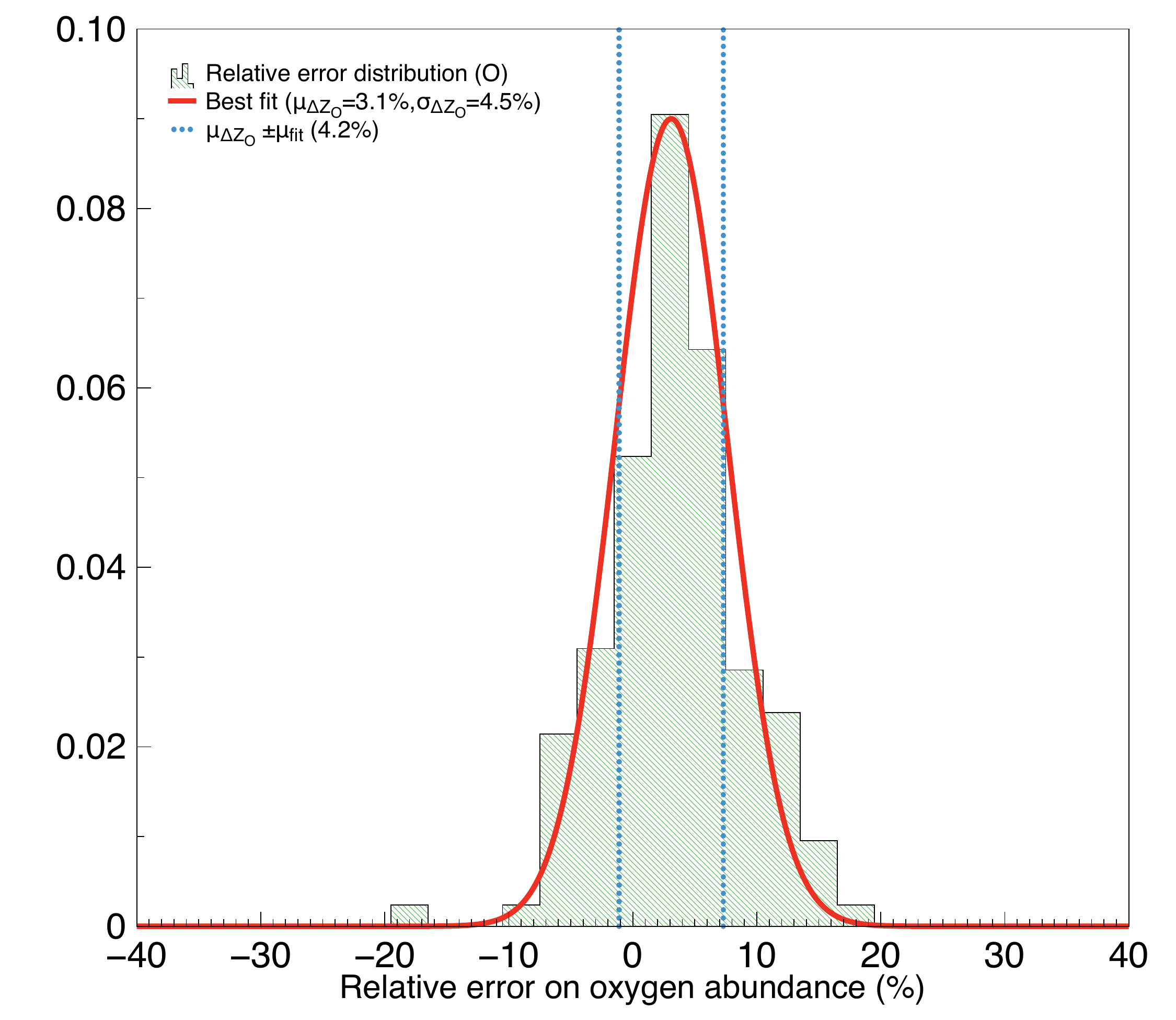}

\includegraphics[width=0.34\textwidth, trim={50cm 2cm 23cm 8cm}, clip]{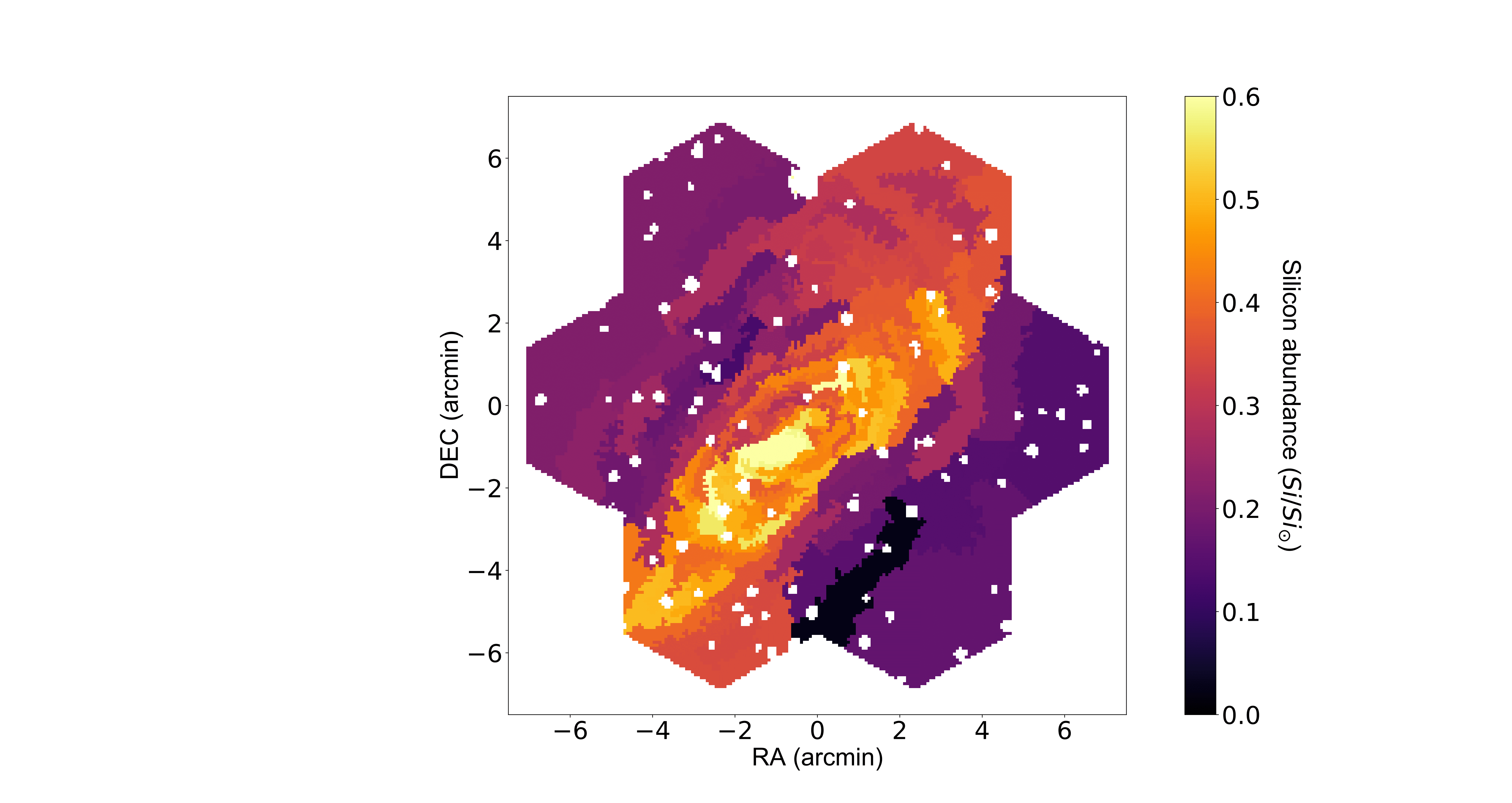}
\vspace{0.6cm}
\includegraphics[width=0.34\textwidth, trim={50cm 2cm 23cm 8cm}, clip]{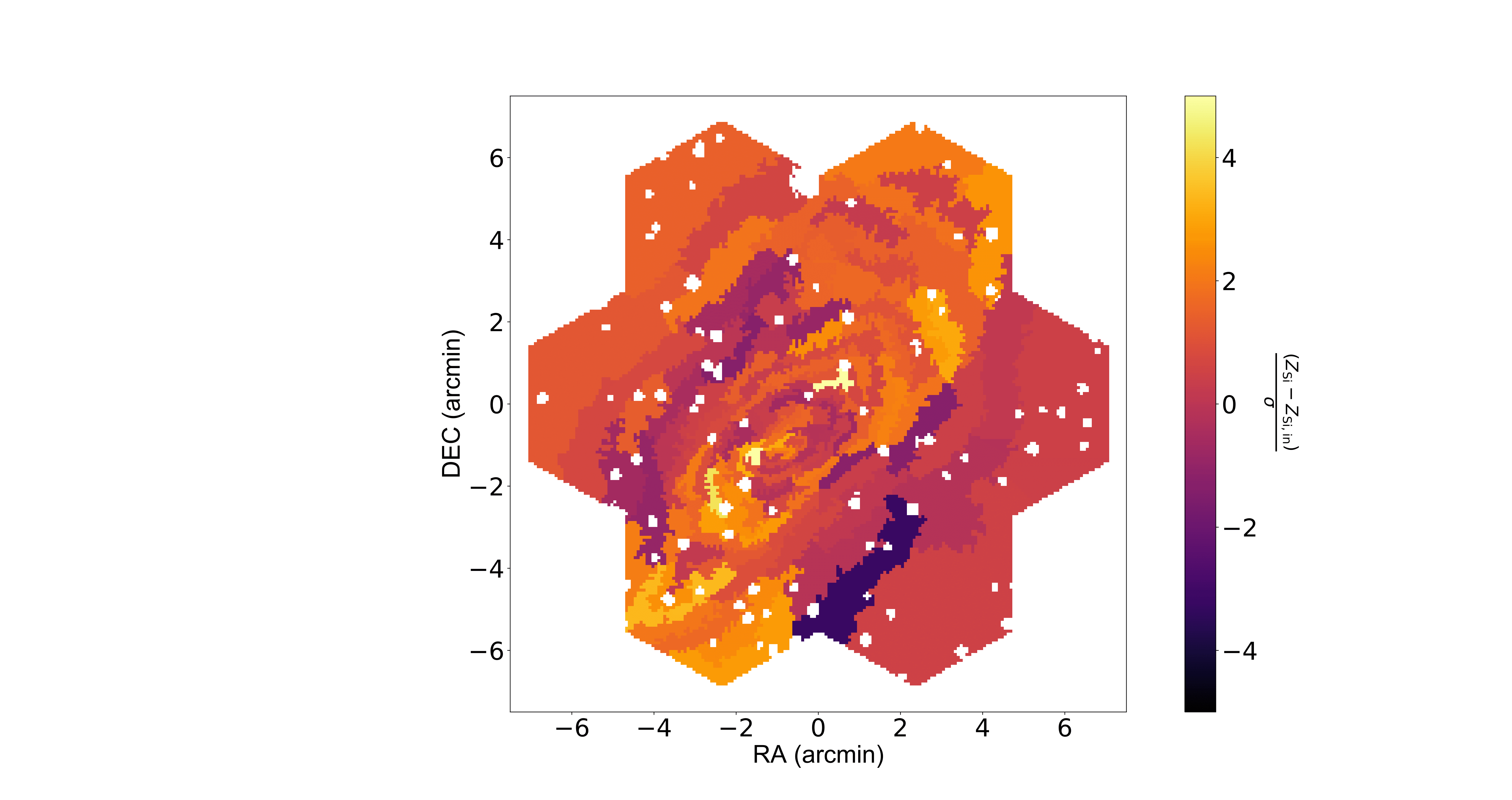}
\includegraphics[width=0.31\textwidth, trim={0 0 0 0.5}, clip]{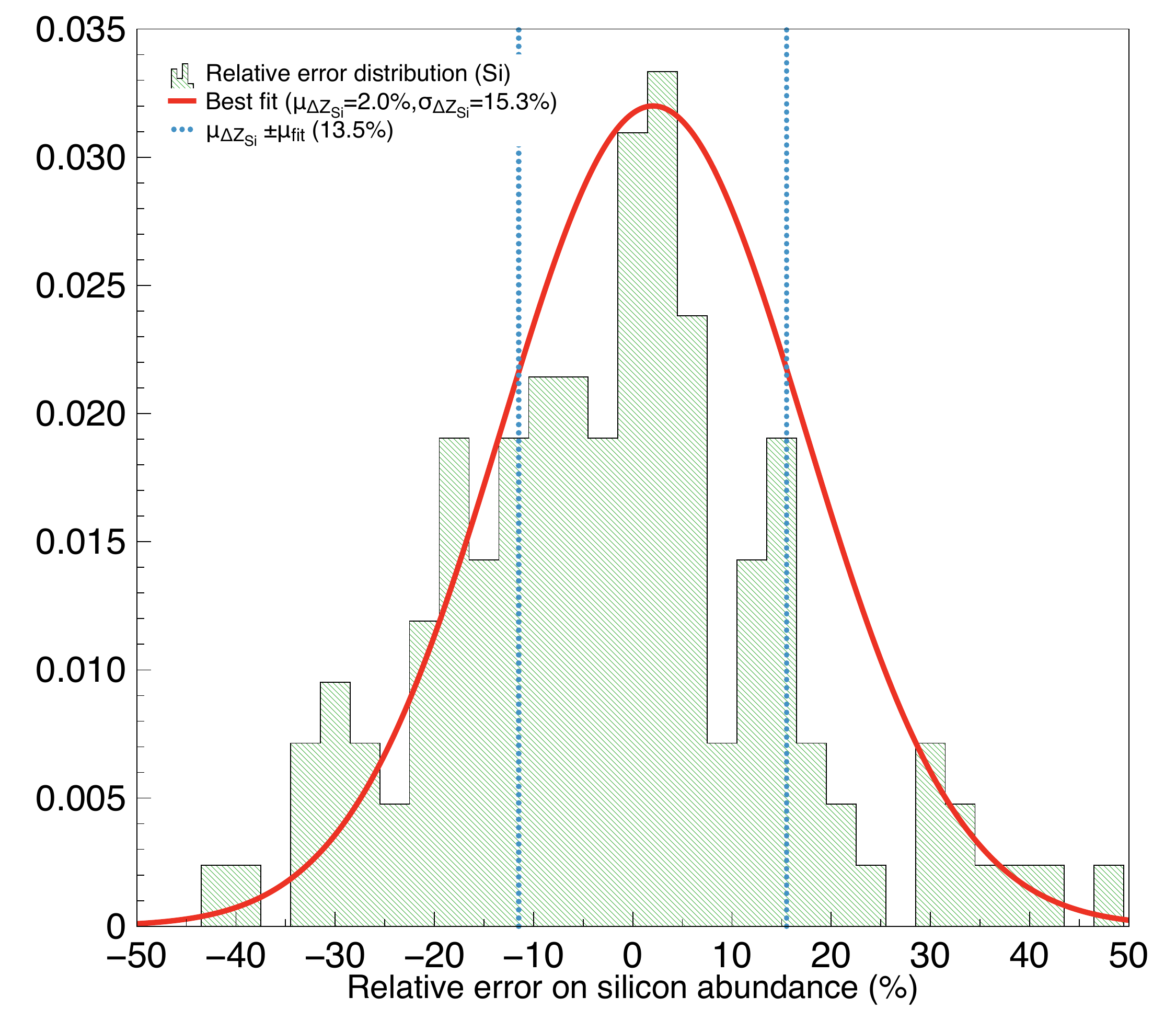}

\includegraphics[width=0.34\textwidth, trim={50cm 2cm 23cm 8cm}, clip]{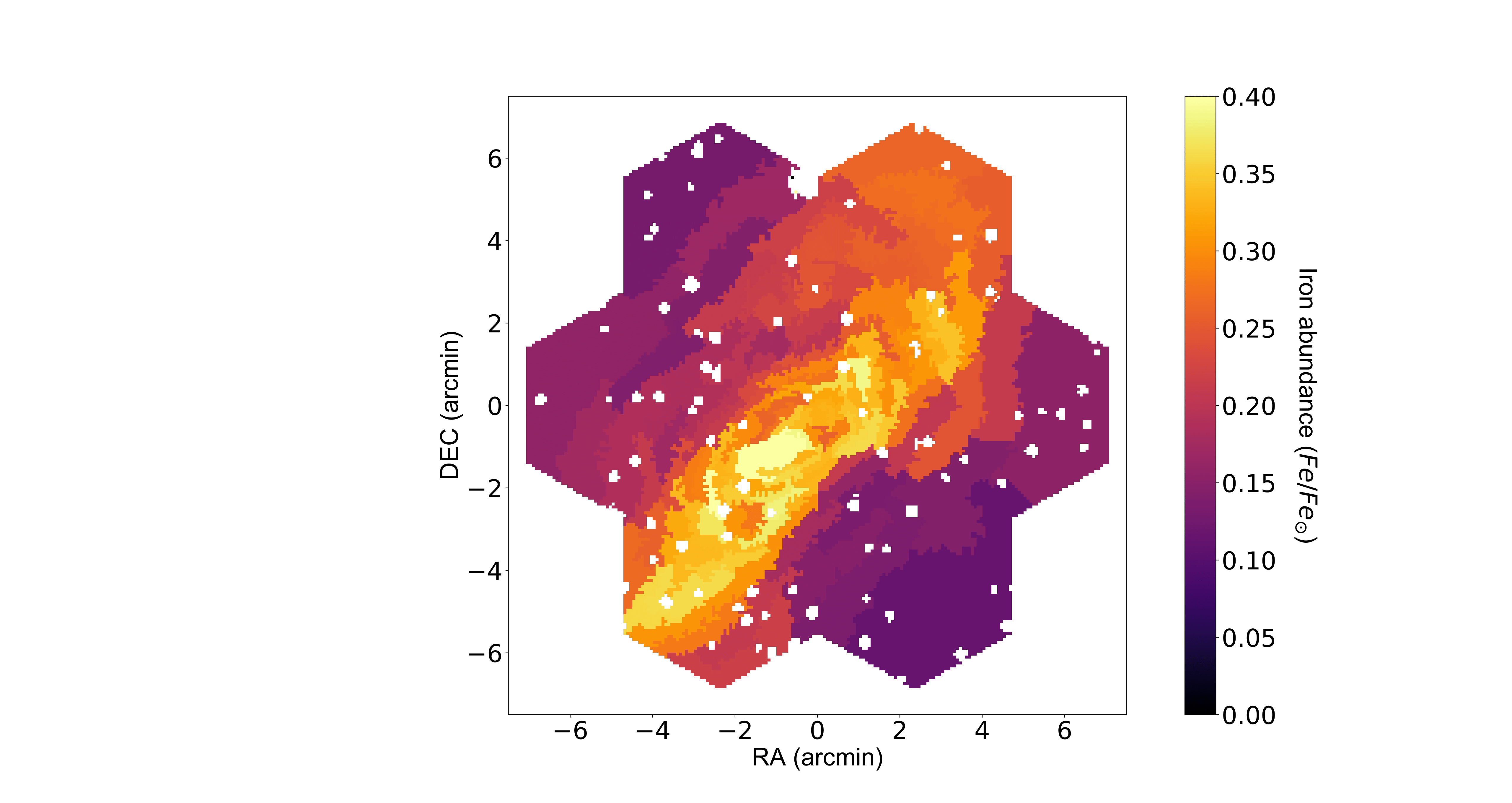}
\vspace{0.6cm}
\includegraphics[width=0.34\textwidth, trim={50cm 2cm 23cm 8cm}, clip]{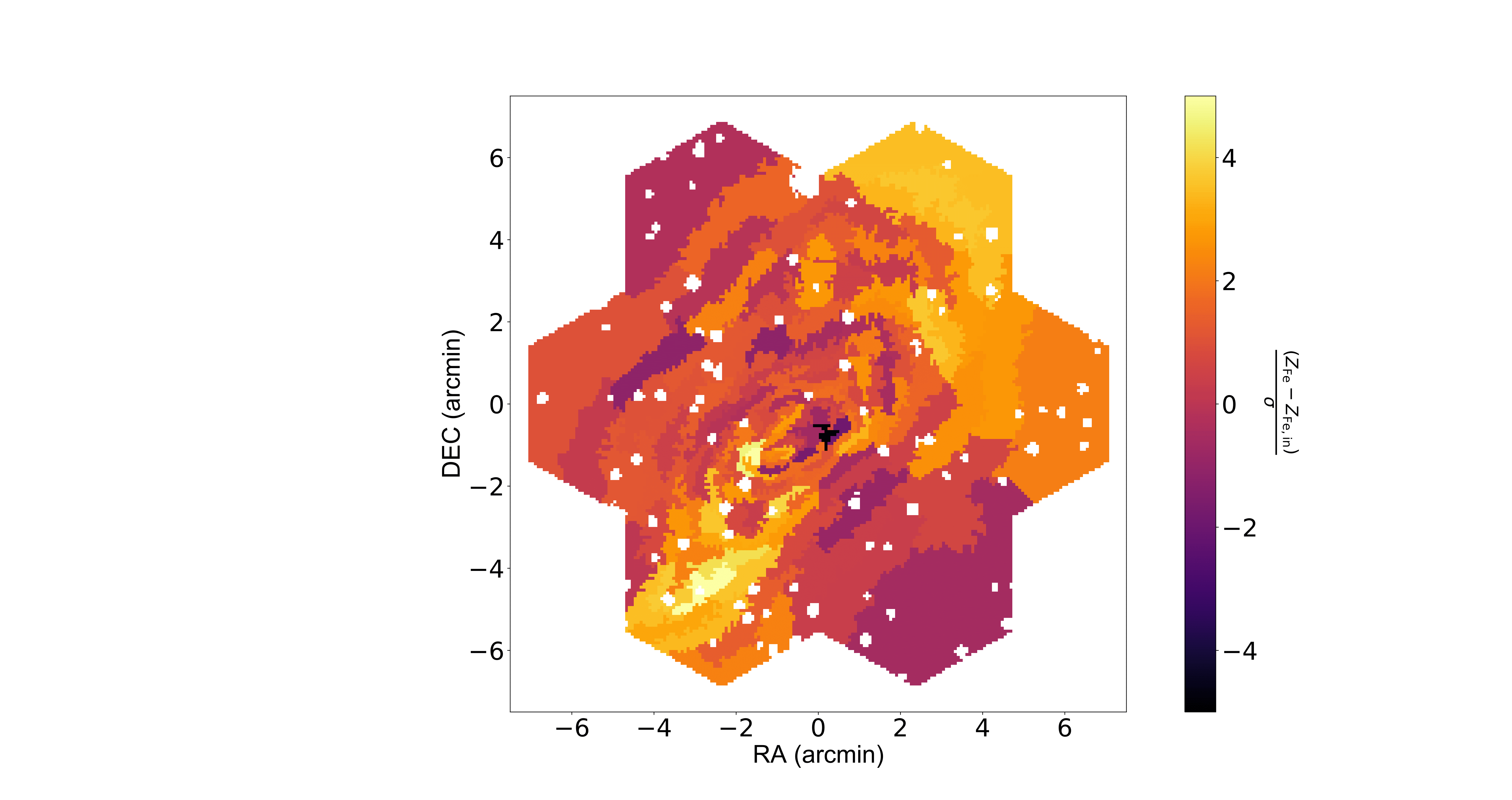}
\includegraphics[width=0.31\textwidth, trim={0 0 0 0.5}, clip]{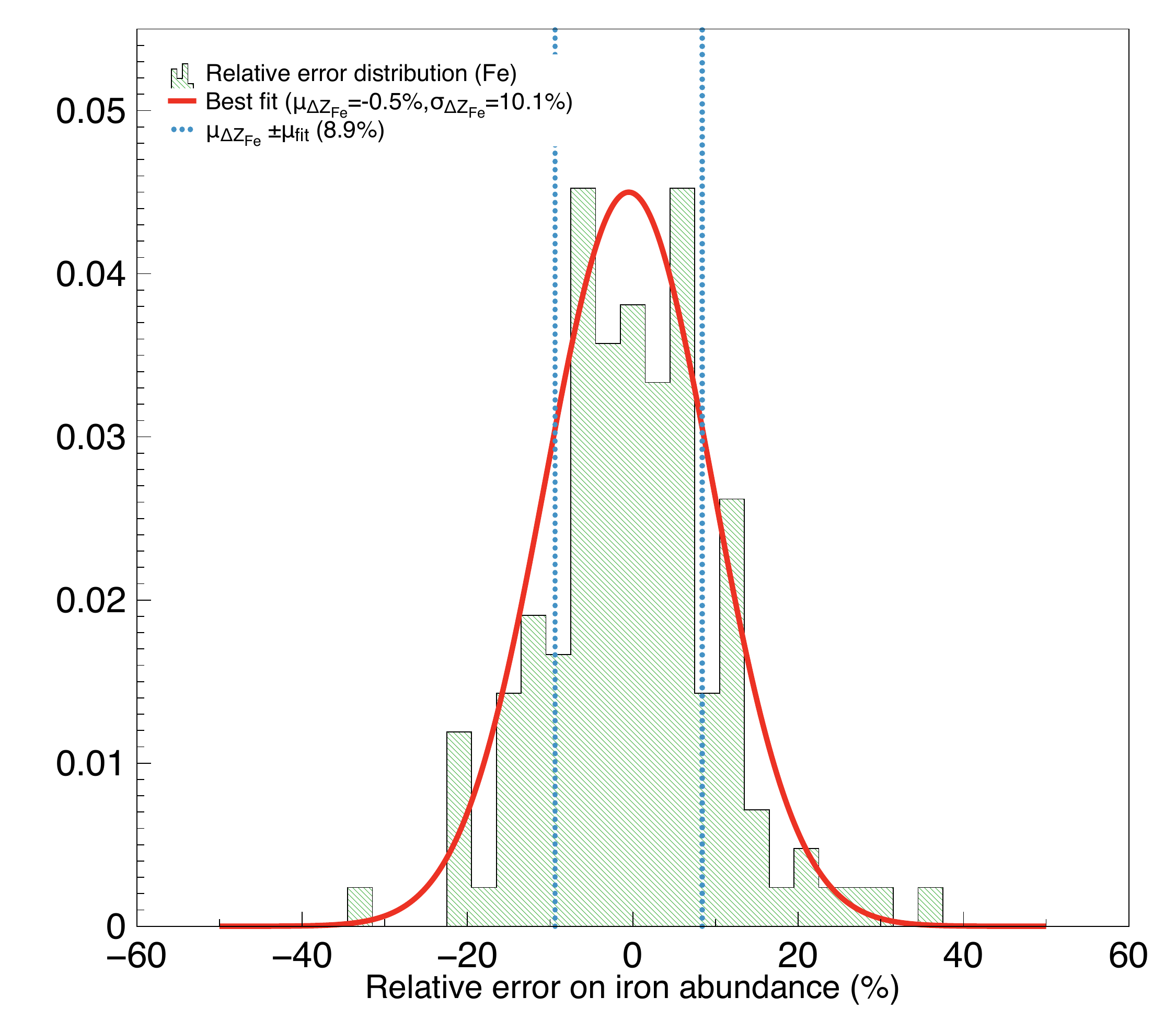}
\caption{Same as Fig.~\ref{fig:maps} for cluster C3.}
\label{fig:mapsC3}
\end{figure*}

\begin{figure*}[p]
\centering
\includegraphics[width=0.34\textwidth, trim={50cm 2cm 23cm 8cm}, clip]{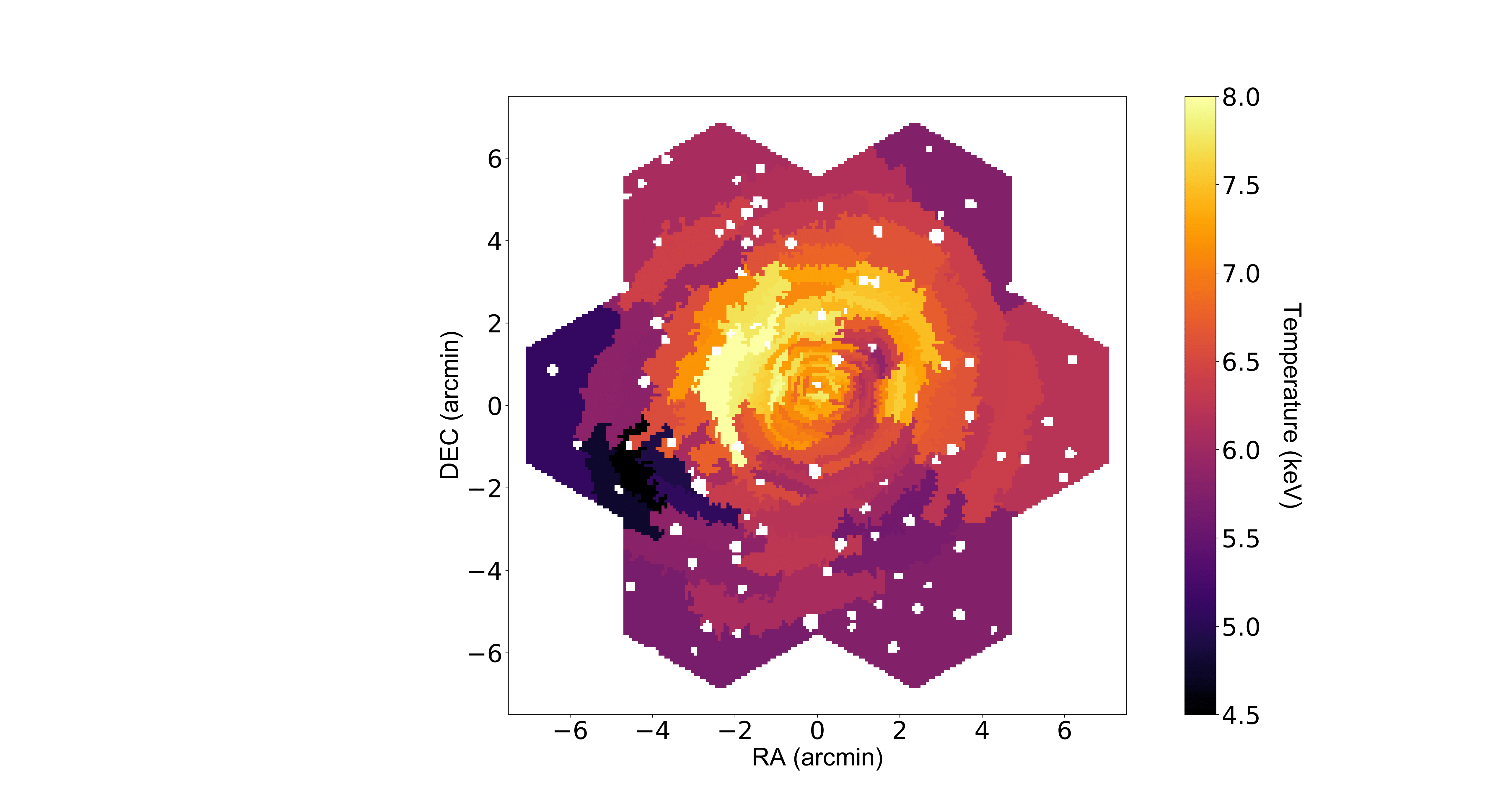}
\vspace{0.6cm}
\includegraphics[width=0.34\textwidth, trim={50cm 2cm 23cm 8cm}, clip]{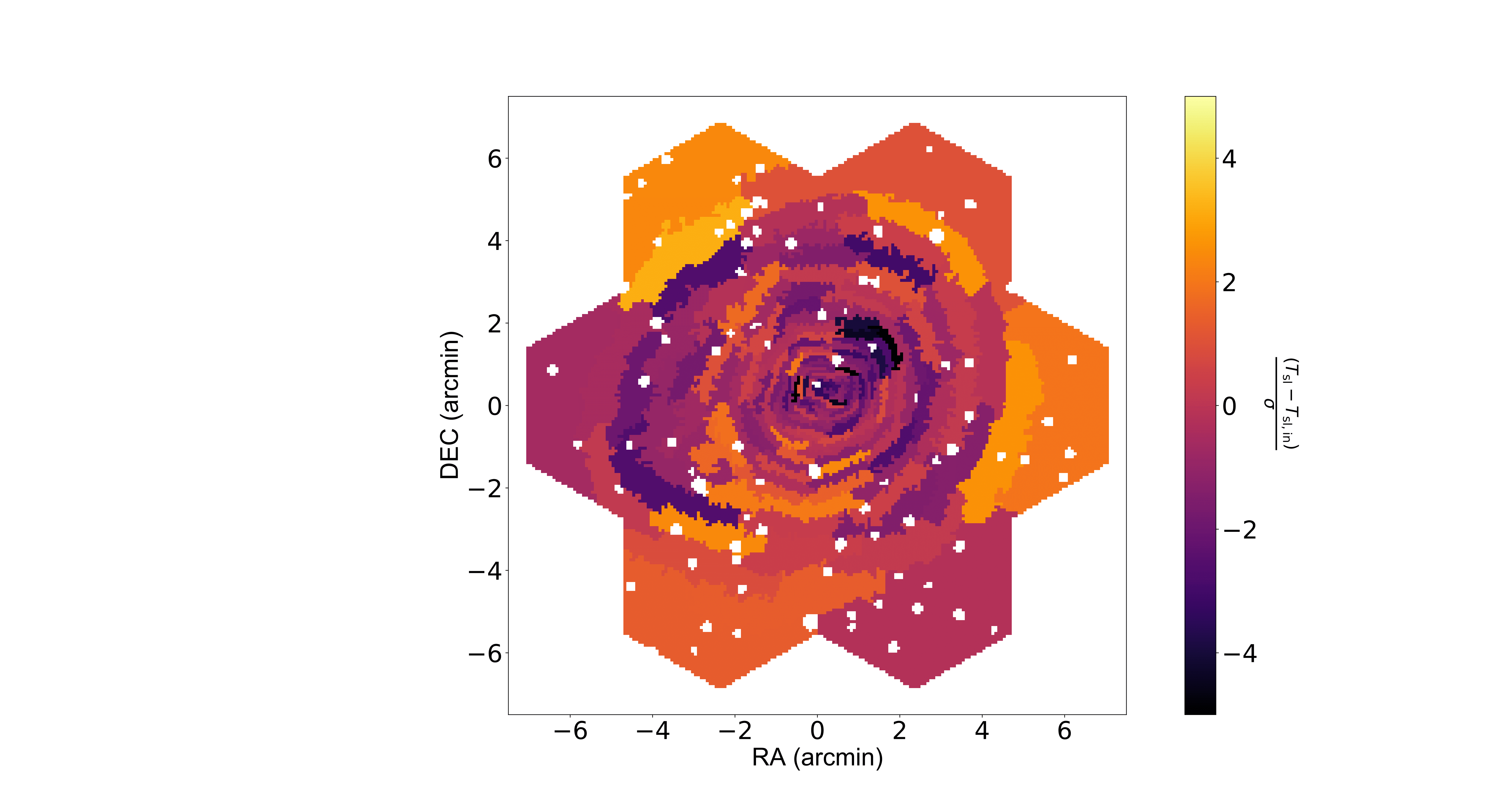}
\includegraphics[width=0.31\textwidth, trim={0 0 0 0.5}, clip]{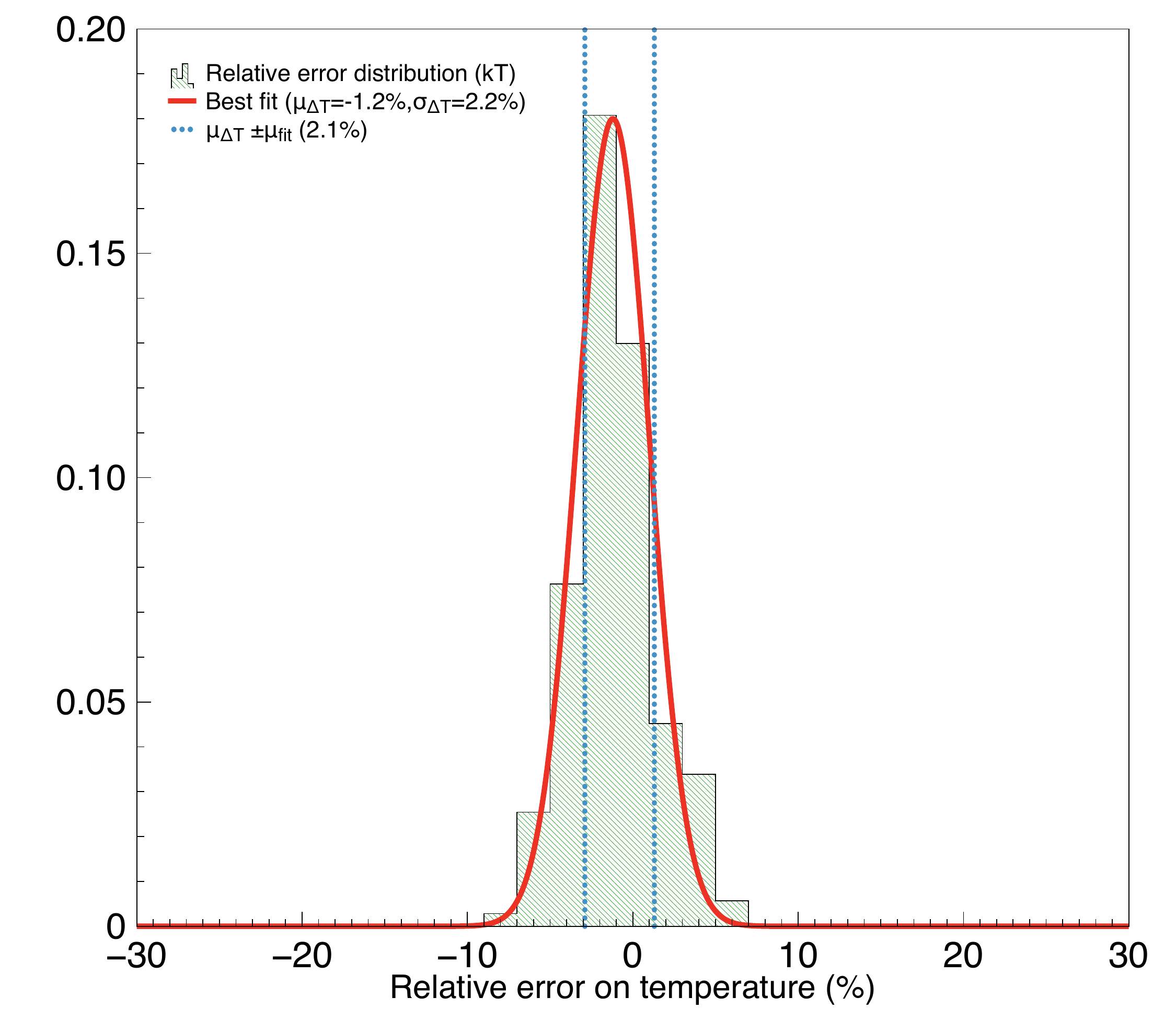}

\includegraphics[width=0.34\textwidth, trim={50cm 2cm 23cm 8cm}, clip]{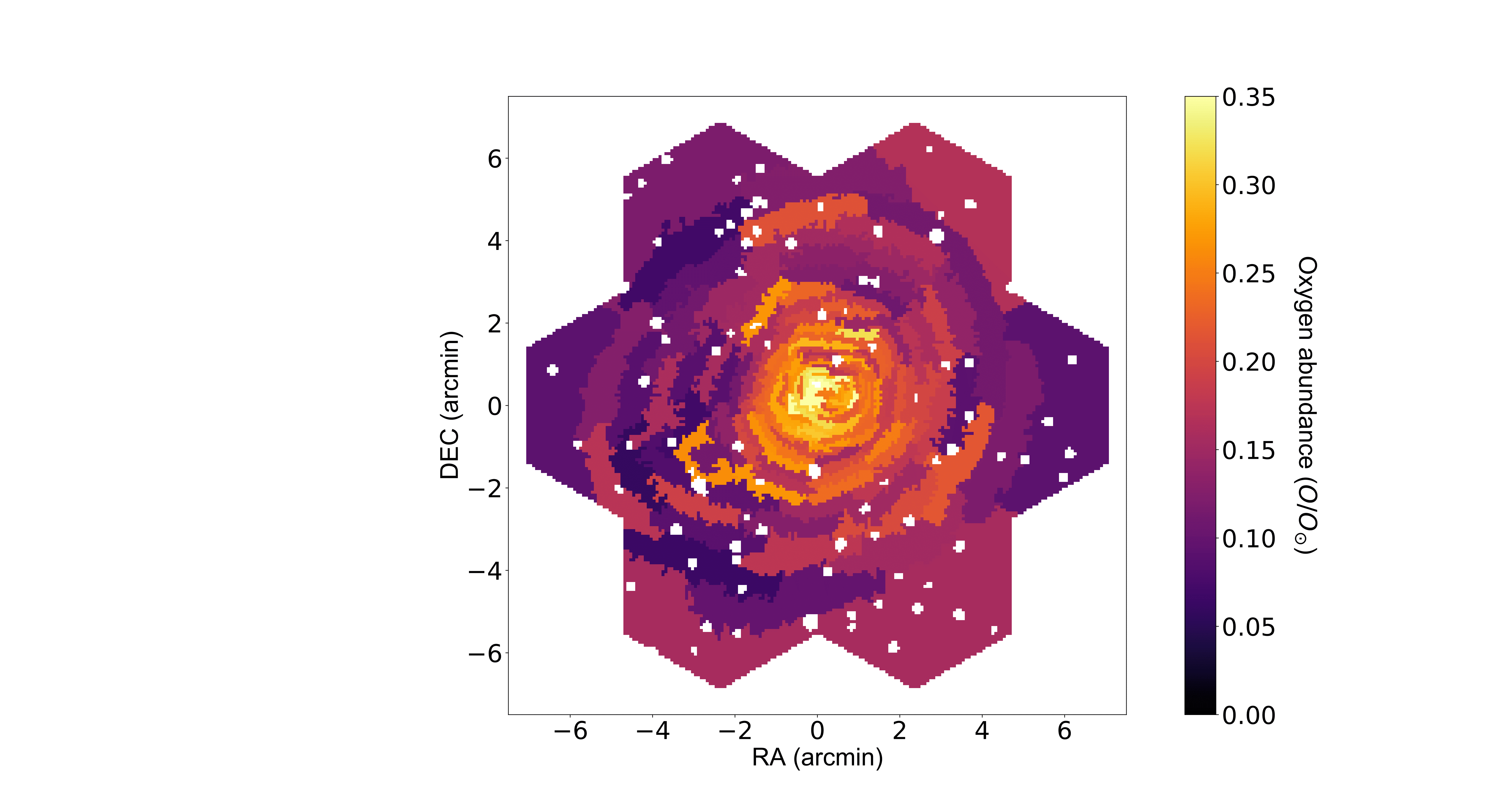}
\vspace{0.6cm}
\includegraphics[width=0.34\textwidth, trim={50cm 2cm 23cm 8cm}, clip]{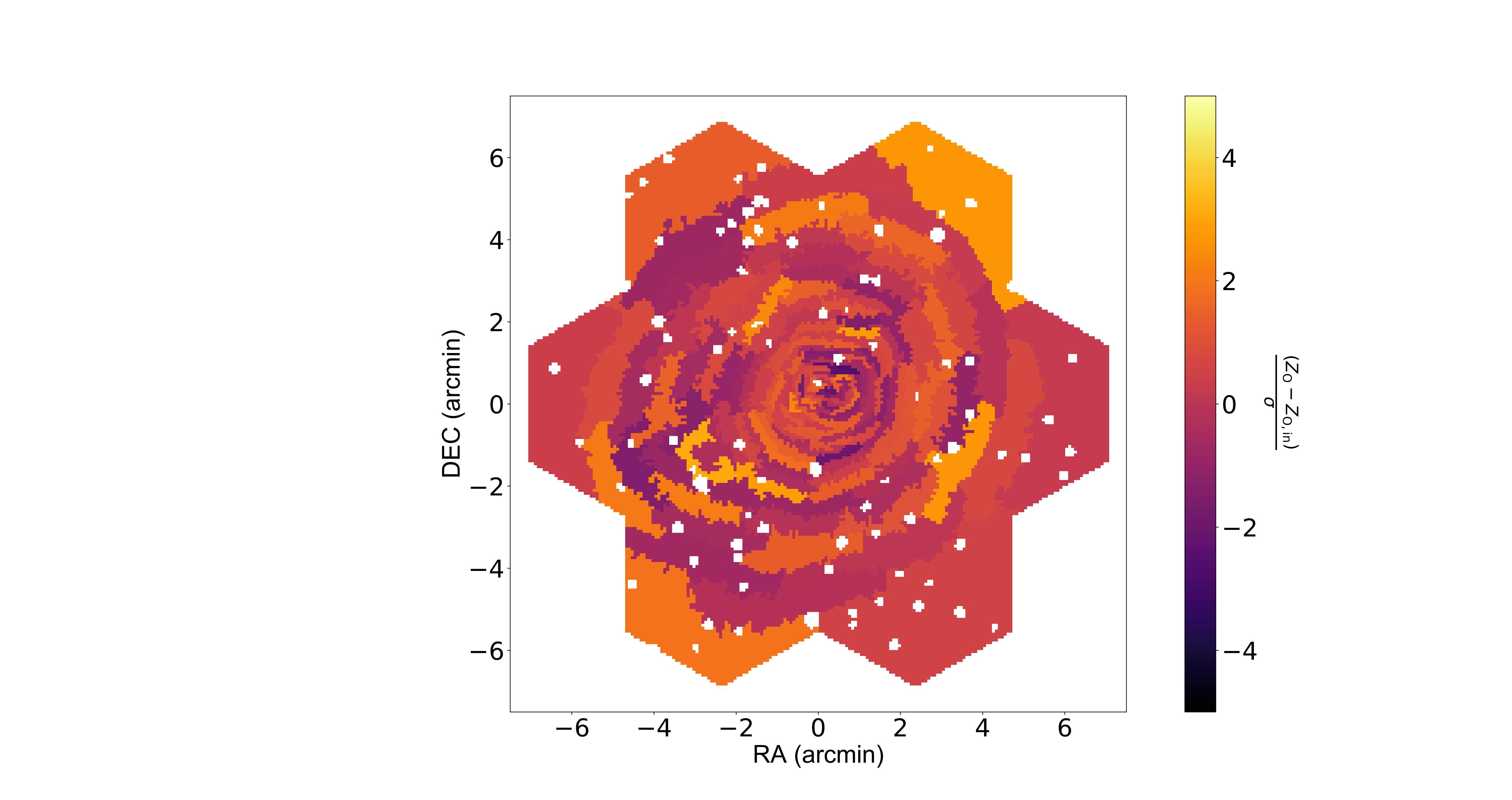}
\includegraphics[width=0.31\textwidth, trim={0 0 0 0.5}, clip]{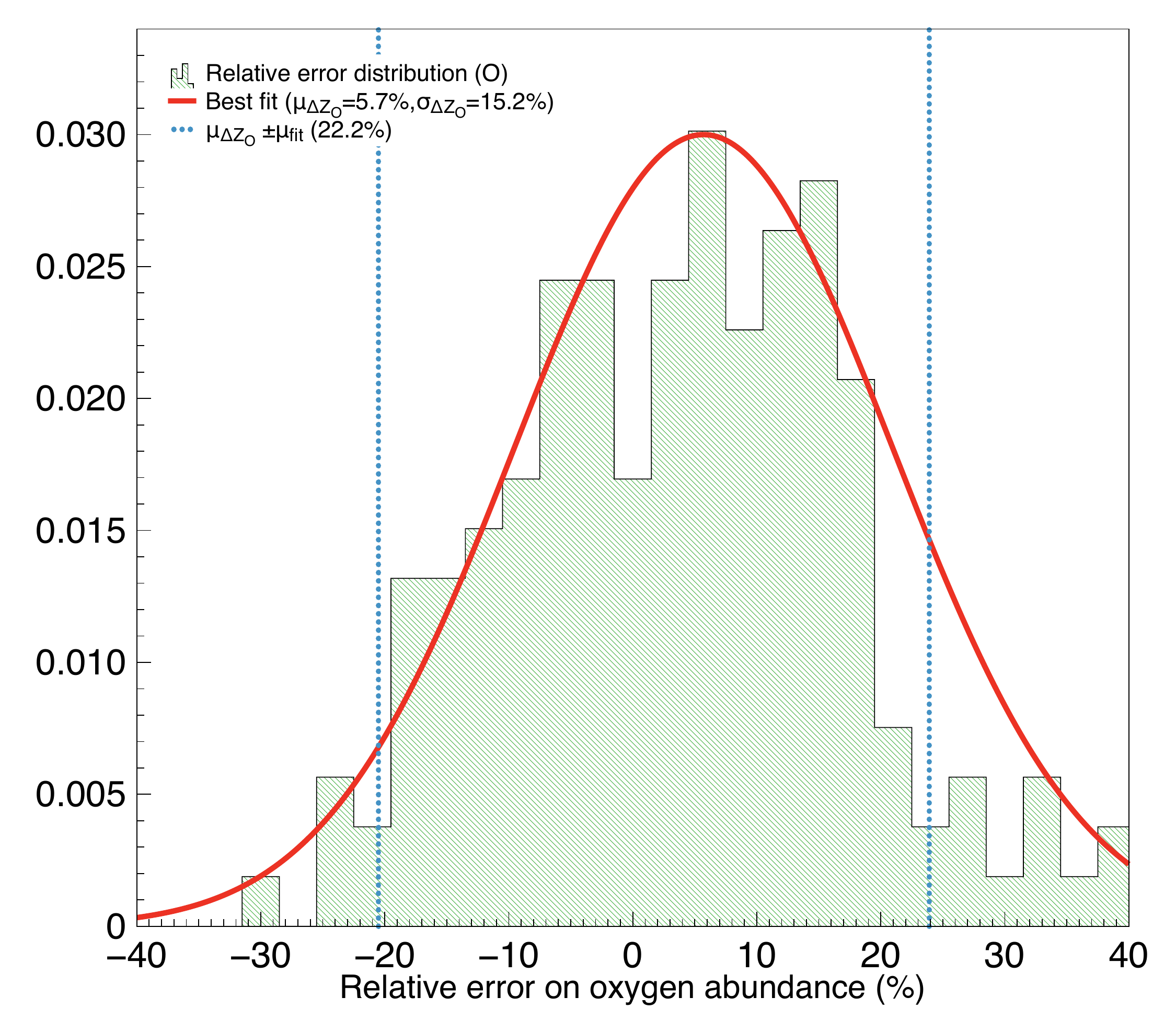}

\includegraphics[width=0.34\textwidth, trim={50cm 2cm 23cm 8cm}, clip]{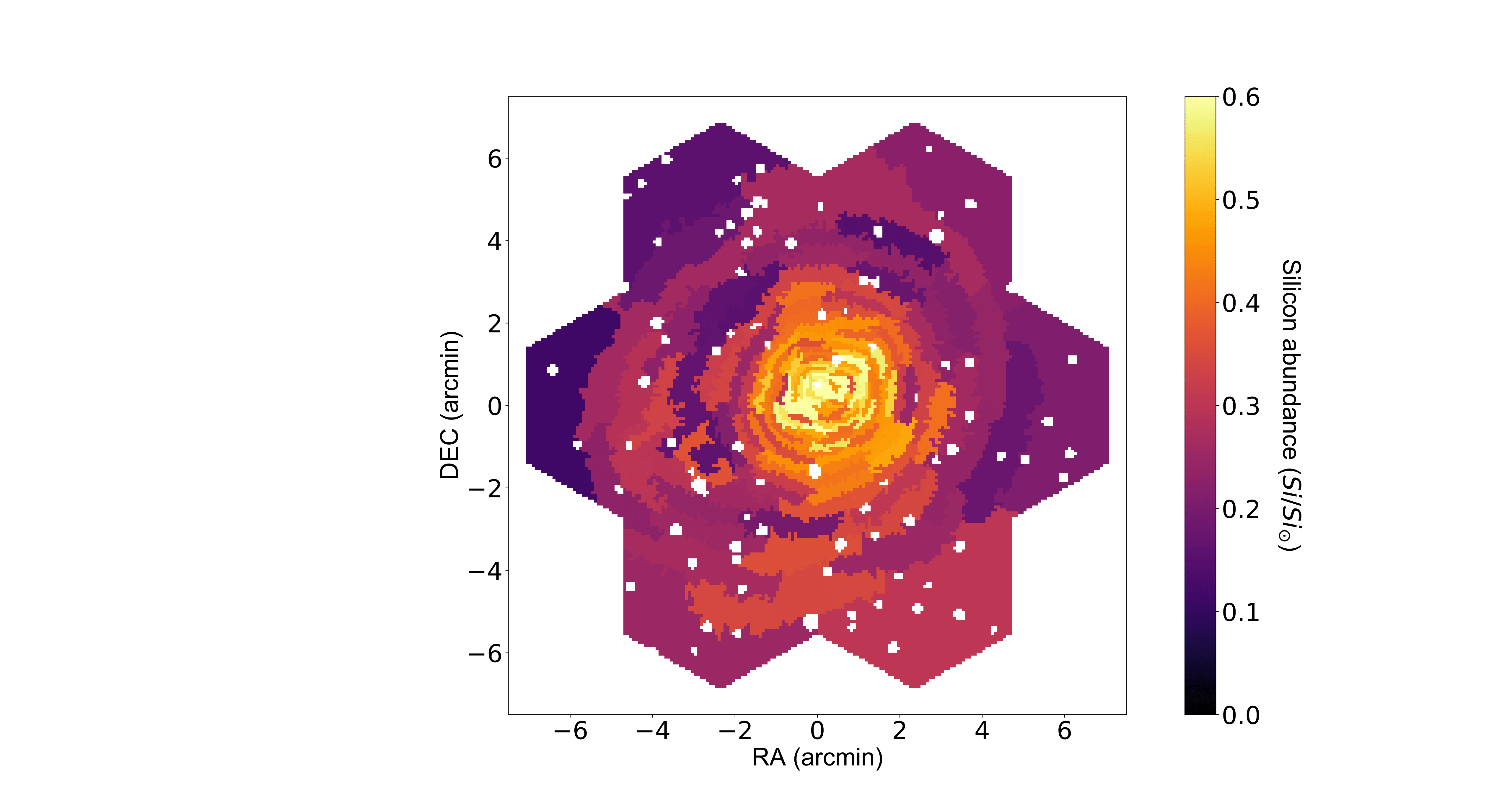}
\vspace{0.6cm}
\includegraphics[width=0.34\textwidth, trim={50cm 2cm 23cm 8cm}, clip]{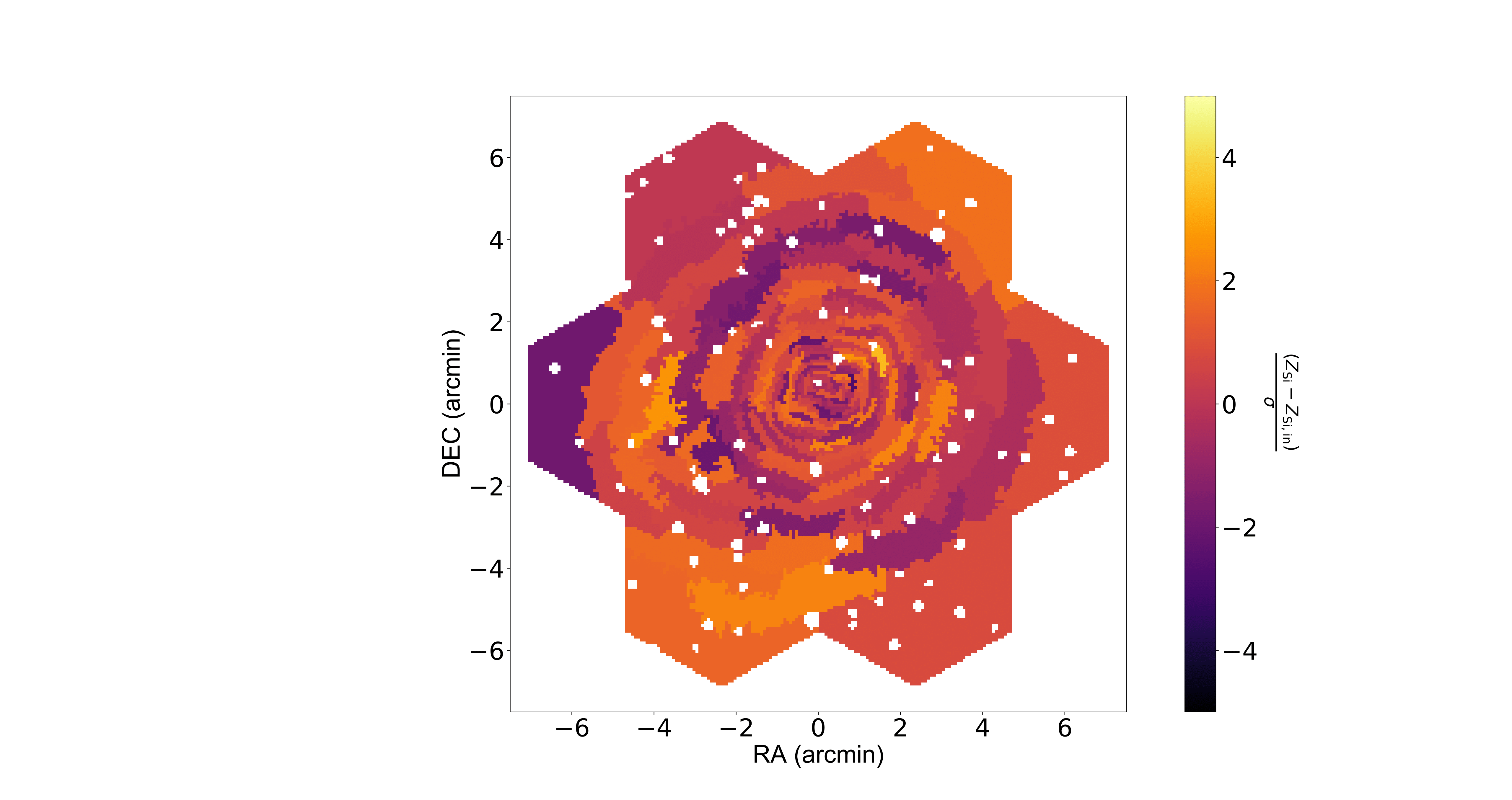}
\includegraphics[width=0.31\textwidth, trim={0 0 0 0.5}, clip]{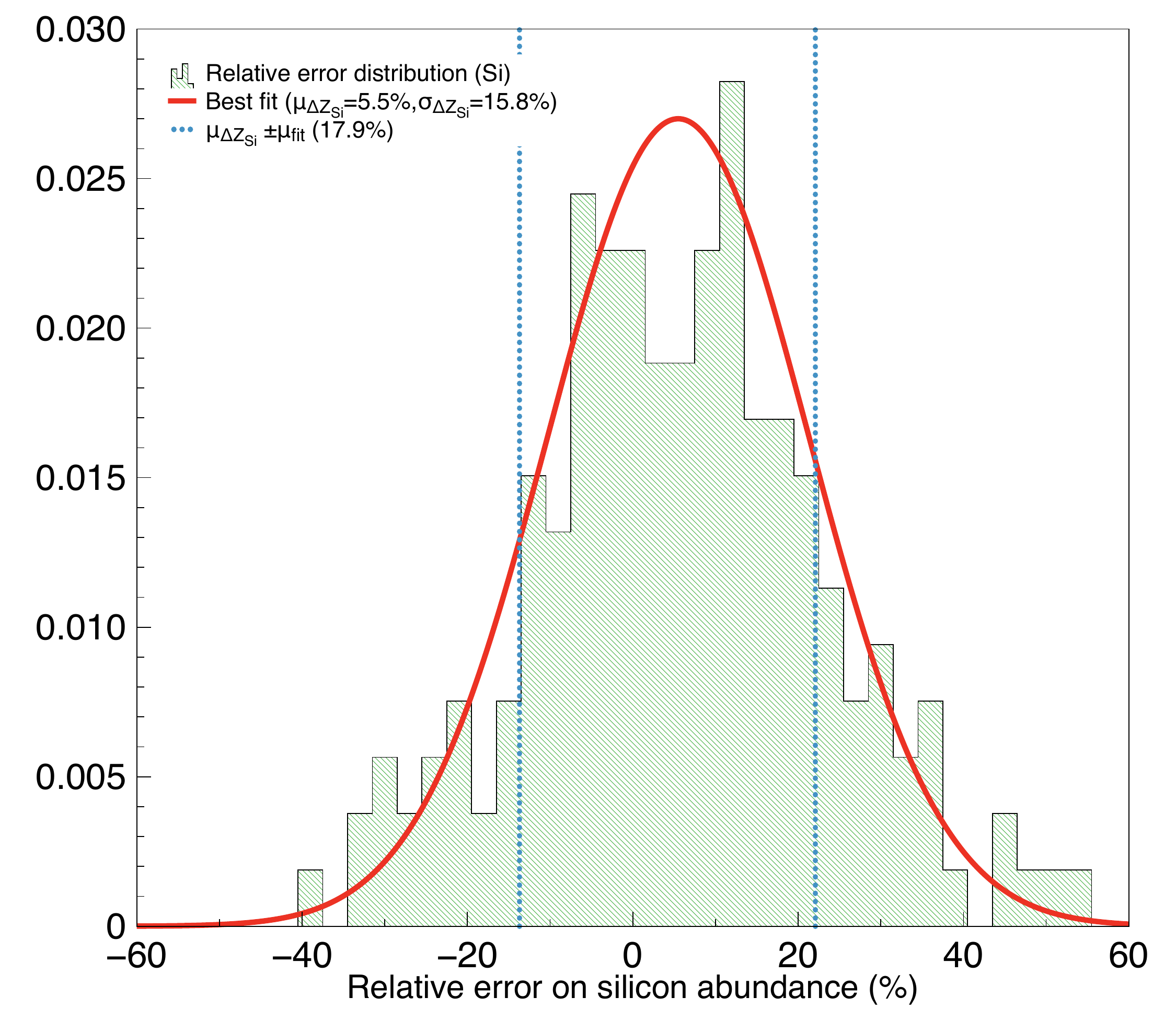}

\includegraphics[width=0.34\textwidth, trim={50cm 2cm 23cm 8cm}, clip]{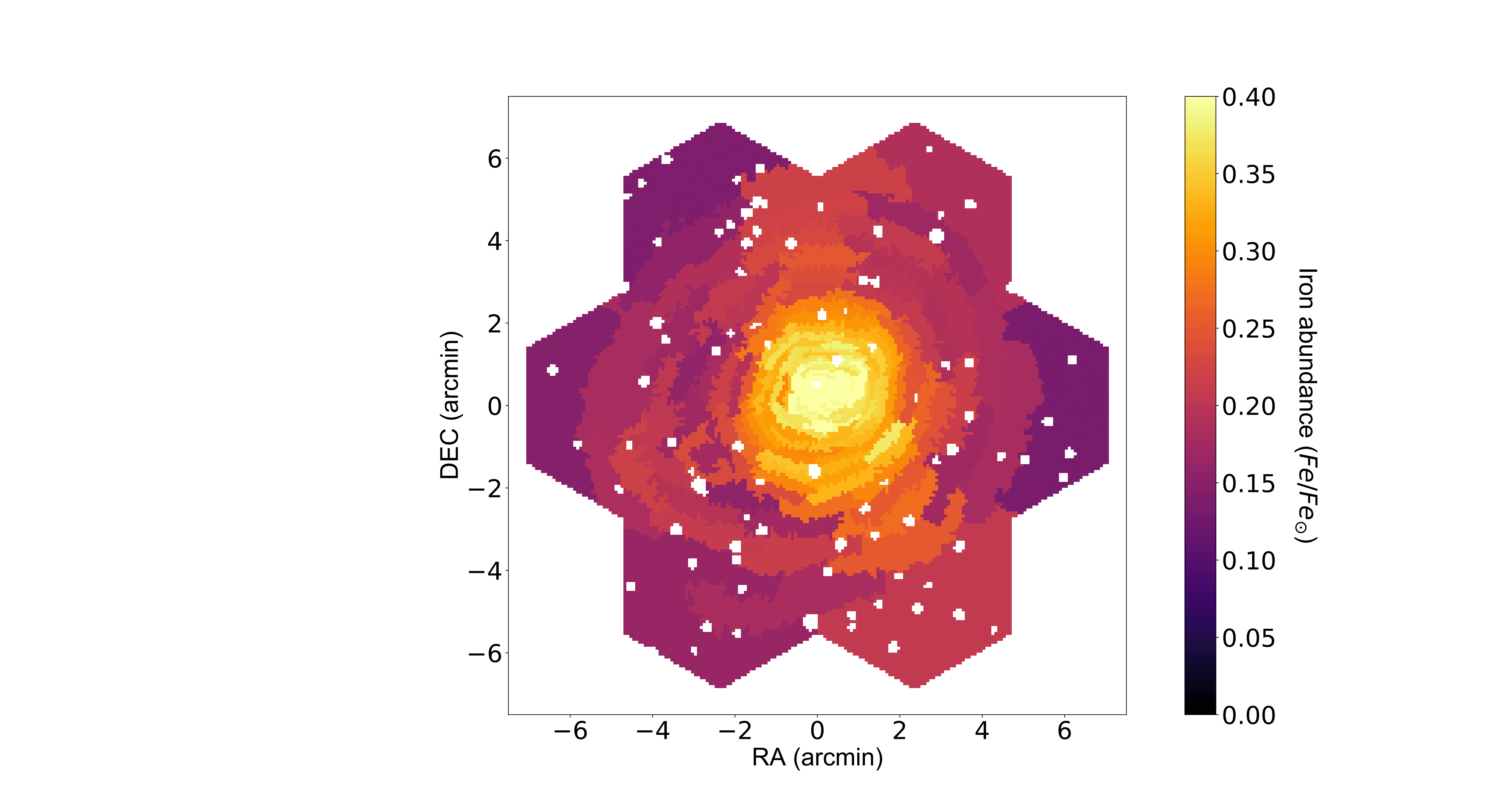}
\vspace{0.6cm}
\includegraphics[width=0.34\textwidth, trim={50cm 2cm 23cm 8cm}, clip]{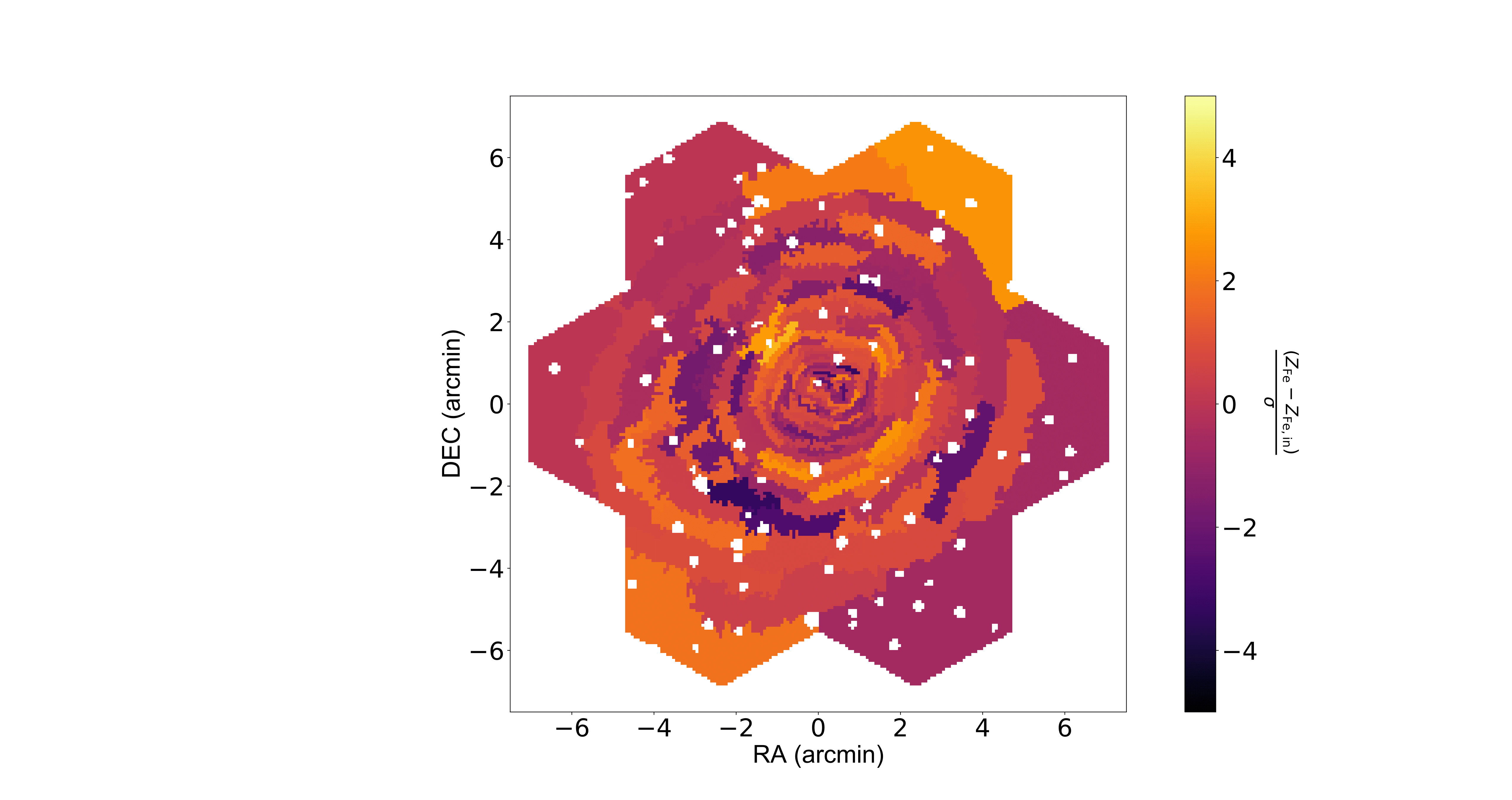}
\includegraphics[width=0.31\textwidth, trim={0 0 0 0.5}, clip]{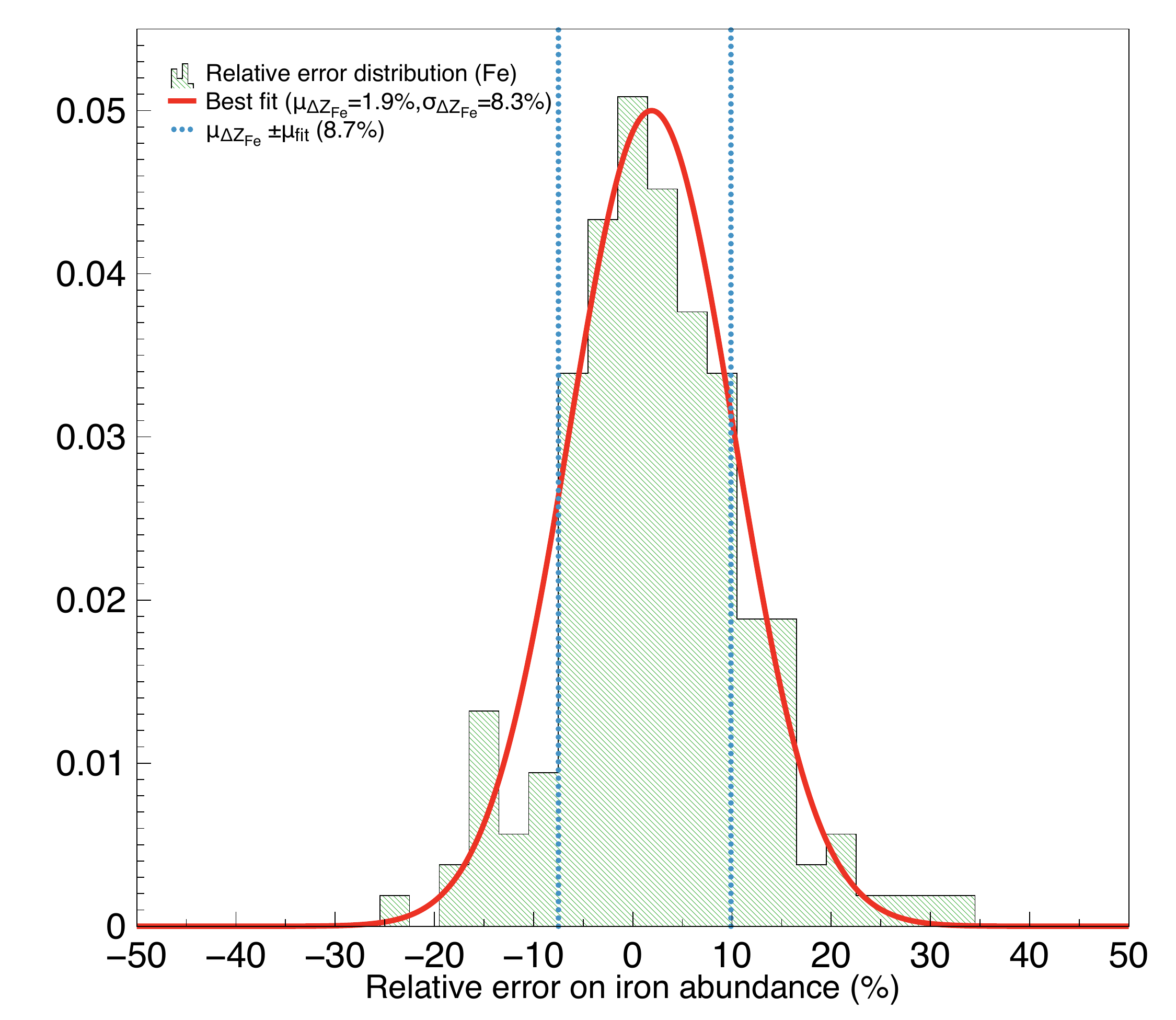}
\caption{Same as Fig.~\ref{fig:maps} for cluster C4.}
\label{fig:mapsC4}
\end{figure*}

\end{document}